\documentclass[aps,prd,superscriptaddress,10pt,showpacs,notitlepage,  
nofootinbib
]{revtex4-1}
\usepackage{bm}
\usepackage{amsmath,amssymb,amsthm}
\usepackage{latexsym,graphicx,color,subfigure}
\usepackage{enumerate}
\usepackage{amssymb}
\usepackage{hyperref}
\usepackage{mathtools}

\usepackage{graphicx}
\graphicspath{ {images/} }

\usepackage{color}

\newcommand{\drm}{\mathrm{d}}

\newcommand{\grad}{\mathrm{grad}}
\newcommand{\bulk}{\mathrm{bulk}}

\newcommand{\be}{\begin{equation}}
\newcommand{\ee}{\end{equation}} 
\newcommand{\beq}{\begin{eqnarray}}
\newcommand{\eeq}{\end{eqnarray}}

\newcommand{\bea}{\begin{eqnarray}}
\newcommand{\eea}{\end{eqnarray}}

\def\tc{\textcolor{red}}
\def\tc2{\textcolor{blue}}
 
\renewcommand{\vec}[1]{\boldsymbol{#1}}

\allowdisplaybreaks[3]
\def\simge{\mathrel{
   \rlap{\raise 0.511ex \hbox{$>$}}{\lower 0.511ex \hbox{$\sim$}}}}
\def\simle{\mathrel{
   \rlap{\raise 0.511ex \hbox{$<$}}{\lower 0.511ex \hbox{$\sim$}}}}
\def\bigs{\mathrel{
   \rlap{\raise 0.531ex \hbox{$>$}}{\lower 0.531ex \hbox{$<$}}}}

\usepackage[normalem]{ulem}

\renewcommand\sout{\bgroup \color{blue} \ULdepth=-.5ex \ULset}

\newcommand{\fa}{f_{1}^{\mathrm{bulk}}}
\newcommand{\fb}{f_{2}^{\mathrm{bulk}}}

\begin{document}
\title{Topological defects at the boundary of neutron $^{3}P_{2}$ superfluids in neutron stars}
\author{Shigehiro Yasui}
\email{yasuis@keio.jp}
\affiliation{Department of Physics $\&$ Research and Education Center for Natural Sciences,\\ Keio University,Hiyoshi 4-1-1, Yokohama, Kanagawa 223-8521, Japan}
\author{Chandrasekhar Chatterjee}
\email{chandra@phys-h.keio.ac.jp}
\affiliation{Department of Physics $\&$ Research and Education Center for Natural Sciences,\\ Keio University,Hiyoshi 4-1-1, Yokohama, Kanagawa 223-8521, Japan}
\author{Muneto Nitta}
\email{nitta(at)phys-h.keio.ac.jp}
\affiliation{Department of Physics $\&$ Research and Education Center for Natural Sciences,\\ Keio University,Hiyoshi 4-1-1, Yokohama, Kanagawa 223-8521, Japan}
\begin{abstract}
We study surface effects of neutron $^{3}P_{2}$ superfluids in neutron stars.
$^{3}P_{2}$ superfluids are in uniaxial nematic (UN), D$_{2}$ biaxial nematic (BN), or 
D$_{4}$ BN phase, 
depending on the strength of magnetic fields from small to large.
We suppose a neutron $^{3}P_{2}$ superfluid in a ball with a spherical boundary.
Adopting a suitable boundary condition for  $^{3}P_{2}$ condensates,
we solve the Ginzburg-Landau equation 
to find several surface properties for the neutron $^{3}P_{2}$ superfluid.
First,  the phase on the surface can be different from that of the bulk, 
and symmetry restoration or breaking occurs in general on the surface.
Second, the distribution of the surface energy density has an anisotropy depending on the polar angle in the sphere, which may lead to the deformation of the geometrical shape of the surface.
Third, 
the order parameter manifold induced on the surface, which is described by 
two-dimensional vector fields induced on the surface from the condensates,
allows topological defects (vortices) on the surface, 
and there must exist such defects even in the ground state 
 thanks to the Poincar\'{e}-Hopf theorem:
although the numbers of the vortices and antivortices
depend on the bulk phases, 
the difference between them is topologically invariant (the Euler number $\chi=2$) irrespective of the bulk phases.
These vortices, which are not extended to the bulk,
are called boojums in the context of liquid crystals 
and helium-3 superfluids.
The surface properties of the neutron $^{3}P_{2}$ superfluid found in this paper may provide us useful information to study neutron stars.
\end{abstract}
\maketitle


\section{Introduction}

Neutron stars provide us extreme environments for nuclear physics, such as high density state, rapid rotation, strong magnetic field, strong gravitational field, and so on, leading to the unveiling of new phases of nuclear systems (see Refs.~\cite{Graber:2016imq,Baym:2017whm} for recent reviews). 
A variety of phases are considered to be inside neutron stars: there can be neutron rich gas and crusts at the surface, and there can be neutron superfluidity, hyperon matter, $\pi$ and/or $K$ condensates, and quark matter in the inner core.
The most recent observational developments include
the observations of massive neutron stars whose masses are almost twice as large as the solar mass~\cite{Demorest:2010bx,Antoniadis1233232} and
the observation of gravitational waves from a binary neutron star merger~\cite{TheLIGOScientific:2017qsa}. 
One of the bulk phases inside neutron stars is neutron superfluidity, in which 
neutron pairs make a condensate in the ground state. 
The neutron superfluidity is directly related to astrophysical phenomena (see Refs.~\cite{Chamel2017,Haskell:2017lkl,Sedrakian:2018ydt} for recent reviews).
For example, it has been discussed that the neutron superfluidity affects relaxation time after pulsar glitches ({\it i.e.}, sudden speed-up events of neutron-star rotation)~\cite{Reichely1969},\footnote{We notice that pulsar glitch phenomena can be explained by unpinning of a large amount of superfluid vortices~\cite{Anderson:1975zze}.} and the neutron superfluidity enhances rapid cooling by neutrino emissions from the inside of neutron stars (the modified Urca process)~\cite{Yakovlev:1999sk}.

From the viewpoint of nuclear physics,
it is an interesting property that nuclear forces have different attractive channels depending on the baryon number density, from low to high density~(see Ref.~\cite{Dean:2002zx} for a recent review). 
In the early stage, Migdal considered the $^{1}S_{0}$ channel to be the most attractive one at low density~\cite{Migdal:1960}, though this channel turns out to be repulsive due to the strong core repulsion at higher densities~\cite{1966ApJ...145..834W}.
At high density, instead,
 the attraction is supplied by the $^{3}P_{2}$ channel stemming from the $LS$ potential, leading to neutron $^{3}P_{2}$ superfluidity~\cite{Tabakin:1968zz,Hoffberg:1970vqj,Tamagaki1970,Takatsuka1971,Takatsuka1972,Takatsuka:1992ga,Amundsen:1984qc,Baldo:1992kzz,Elgaroy:1996hp,Khodel:1998hn,Baldo:1998ca,Khodel:2000qw,Zverev:2003ak,Maurizio:2014qsa,Bogner:2009bt,Srinivas:2016kir}.\footnote{We notice that the $^{3}P_{0}$ and $^{3}P_{1}$ channels are repulsive from the $LS$ potential, and hence that those channels do not contribute to the neutron pairing. More precisely, the neutron $^{3}P_{2}$ superfluidity exists with a small fraction of superconducting protons and normal electrons.
However, such a mixture effect can be neglected in most cases.
}
Neutron $^{3}P_{2}$ superfluidity has important impacts on the observations of neutron stars.
One example is the rapid neutrino cooling in neutron stars, as studied for Cassiopeia A~\cite{Heinke2010,Shternin2011,Page:2010aw}, although the existence of $^{3}P_{2}$ superfluidity is still elusive. 
Another example is the tolerance of the spin-triplet pairing in $^{3}P_{2}$ superfluidity against strong magnetic fields,
while the spin-singlet pairing in the $^{1}S_{0}$ superfluidity is easily broken due to the Zeeman effect.
The tolerance may give us a chance to observe $^{3}P_{2}$ superfluidity in magnetars, in which the strength of the magnetic field reaches about $10^{15}$ G at the surface and possibly about $10^{18}$ G in the inside.\footnote{In the literature, the origin of the strong magnetic fields has been studied in terms of several mechanisms such as spin-dependent interactions between two neutrons~\cite{Brownell1969,RICE1969637,Silverstein:1969zz,Haensel:1996ss}, pion domain walls~\cite{Eto:2012qd,Hashimoto:2014sha}, spin polarizations in the quark-matter core~\cite{Tatsumi:1999ab,Nakano:2003rd,Ohnishi:2006hs} and so on. However, this problem remains still hard to deal with. It may be significant that the recent many-body calculation indicates a negative result for the generation of strong magnetic fields~\cite{Bordbar:2008zz}.
}

In the pairing in neutron $^{3}P_{2}$ superfluidity, there are a variety of pairing combinations of relative angular momentum and  total spin, leading to the existence of different phases which have symmetry breaking from the $\mathrm{U}(1)_{B} \times \mathrm{SO}(3)_{S} \times \mathrm{SO}(3)_{L} \times T \times P$ symmetry ($B$ for the baryon number, $S$ for spin rotation, $L$ for spatial rotation, $T$ for the time-reversal symmetry, and $P$ for the parity symmetry)~\cite{Fujita1972,Richardson:1972xn,Sauls:1978lna,Muzikar:1980as,Sauls:1982ie,Vulovic:1984kc,Masuda:2015jka,Masuda:2016vak}.
Examples of the symmetry breaking pattern will be presented shortly.
As a consequence of the symmetry breaking, there appear low-energy excitations which affect the cooling process by neutrino emission~\cite{Bedaque:2003wj,Bedaque:2012bs,Bedaque:2013fja,Bedaque:2013rya,Bedaque:2014zta,Leinson:2009nu,Leinson:2010yf,Leinson:2010pk,Leinson:2010ru,Leinson:2011wf,Leinson:2011jr,Leinson:2012pn,Leinson:2013si,Leinson:2014cja}.\footnote{It should be noted that the cooling process is related not only to low-energy excitations but also to quantum vortices~\cite{Shahabasyan:2011zz}.}
Recently, studies of neutron $^{3}P_{2}$ superfluidity have been also devoted to understanding its topological properties: topological superfluidity and 
gapless Majorana fermions on the boundary of $^{3}P_{2}$ superfluids~\cite{Mizushima:2016fbn}, 
a quantized vortex~\cite{Muzikar:1980as,Sauls:1982ie,Masuda:2015jka,Masaki:2019rsz}, 
a soliton on it~\cite{Chatterjee:2016gpm},  
and a half-quantized non-Abelian vortex~\cite{Masuda:2016vak}.
Those states have relevance to analogous states in condensed matter physics, 
such as $D$-wave superconductivity~\cite{Mermin:1974zz}, 
$P$-wave superfluidity in $^{3}$He liquid~\cite{vollhardt2013superfluid}, 
chiral $P$-wave superconductivity, e.g., in Sr$_2$RuO$_4$~\cite{RevModPhys.75.657}, 
spin-2 Bose-Einstein condensates~\cite{2010arXiv1001.2072K}, and so on.

Neutron $^{3}P_{2}$ superfluidity can be described by the Bogoliubov--de-Gennes (BdG) equation as the fundamental equation~\cite{Tabakin:1968zz,Hoffberg:1970vqj,Tamagaki1970,Takatsuka1971,Takatsuka:1992ga,Amundsen:1984qc,Baldo:1992kzz,Elgaroy:1996hp,Khodel:1998hn,Baldo:1998ca,Khodel:2000qw,Zverev:2003ak,Maurizio:2014qsa,Bogner:2009bt,Srinivas:2016kir,Mizushima:2019spl}.
In fact, the BdG equation was successfully applied to study phase structures and topological properties~\cite{Mizushima:2016fbn}.
As a special case, the BdG equation can be reduced to the Ginzburg-Landau (GL) equation as the low-energy bosonic effective theory near the critical temperature~\cite{Fujita1972,Richardson:1972xn,Sauls:1978lna,Muzikar:1980as,Sauls:1982ie,Vulovic:1984kc,Masuda:2015jka,Masuda:2016vak,Yasui:2018tcr,Yasui:2019unp,Mizushima:2019spl}.
At the weak-coupling limit in the GL equation,
the ground state is in the nematic phase, {\it i.e.}, uniaxial nematic (UN) phase or biaxial nematic (BN) phase~\cite{Sauls:1978lna}.
Each phase has different patterns of the symmetry breaking.
The UN phase has an unbroken O(2) symmetry.
The BN phase is separated furthermore into D$_{2}$-BN and D$_{4}$-BN phases, which have unbroken dihedral symmetries, D$_{2}$ and D$_{4}$, respectively. 
The phase realized in the ground state depends on the temperature and the magnetic field.
The UN phase is favored at zero magnetic field, and the BN phases are favored at finite magnetic fields~\cite{Masuda:2015jka,Mizushima:2016fbn},
More precisely, the D$_{2}$-BN phase is favored at weak magnetic field, while the D$_{4}$-BN phase is favored at strong magnetic field. 
Thus, the D$_{2}$-BN and D$_{4}$-BN phases would be relevant for magnetars.
The GL equation has been also applied to study vortices in neutron $^{3}P_{2}$ superfluids:
vortex structures of neutron $^{3}P_{2}$ superfluidity~\cite{Richardson:1972xn,Muzikar:1980as,Sauls:1982ie} and spontaneous magnetization in the core of the vortices~\cite{Sauls:1982ie,Masuda:2015jka,Masuda:2016vak,Chatterjee:2016gpm}. 

It should be noted that the GL equation is an expansion series for small magnitude of the order parameter near the critical temperature.
For terms up to the fourth order for the order parameter,
the UN, D$_{2}$-BN, and D$_{4}$-BN phases are degenerate, and hence the ground state cannot be determined uniquely.\footnote{At the fourth order, an $\mathrm{SO}(5)$ symmetry happens to exist as an extended symmetry in the potential term, which is absent in the original Hamiltonian. In this case, the spontaneous breaking eventually leads to a quasi-Nambu-Goldstone mode~\cite{Uchino:2010pf}.
}
Such degeneracy is resolved in the sixth order, where one of the BN, D$_{2}$-BN, and D$_{4}$-BN phases is realized in the ground state according to the temperature and the magnetic field.
However, the sixth-order term cannot fully support the stability of the ground state.
The next-to-leading order terms of the magnetic field were investigated for strong magnetic fields~\cite{Yasui:2018tcr}.
Recently, it was shown that the stability of the ground state is guaranteed at the eighth order of the condensate~\cite{Yasui:2019unp}.
Interestingly, there exists a first-order phase transition at low temperature and weak magnetic field.
The existence of first-order phase transitions is consistent with the result from the BdG equation~\cite{Mizushima:2016fbn}.

So far many studies of neutron $^{3}P_{2}$ superfluidity have been devoted to the bulk properties 
(except for Majorana fermions on the surface
\cite{Mizushima:2016fbn}). 
In neutron stars, however, the neutron $^{3}P_{2}$ superfluid interfaces with the other phases, 
such as a neutron $^{1}S_{0}$ phase and nuclear crusts near the surface.
Thus, it is an important question to ask
how properties of the neutron $^{3}P_{2}$ superfluid change at the boundary.
As an analogous situation, there have been several studies of surface effects of 
liquid crystals~\cite{Urbanski_2017} and 
$^{3}$He superfluids~\cite{PhysRevA.9.2676,Buchholtz1979,PhysRevB.23.5788,PhysRevB.33.1579,Nagai1988} (see also Refs.~\cite{vollhardt2013superfluid,Volovik:2003fe} and the references therein), and applications of the geometrical structure of $^{3}$He liquid were developed: droplets~\cite{Barton1975}, slabs~\cite{PhysRevB.14.2801}, pore geometry~\cite{10.1143/PTP.78.562}, and so on.
In neutron stars, the neutron $^{3}P_{2}$ superfluid will necessarily interface with neutron $^{1}S_{0}$ superfluidity at low density.
As the most simple situation, however, we suppose that 
neutron $^{3}P_{2}$ superfluid interfaces with other phases at a sharp boundary at the surface of the neutron star.
We notice that the phase with which the neutron $^{3}P_{2}$ superfluid interfaces is not necessarily the vacuum, but is possibly another phase such as the normal phase of  neutron gas, the neutron $^{1}S_{0}$ superfluid phase, and so on.
In spite of the simple situation, we will elucidate that neutron $^{3}P_{2}$ superfluidity exhibits nontrivial properties of symmetry breaking and topology at the surface.
We first find that  the phase structure on the surface can be different from that in the bulk, and therefore symmetry restoration or breaking can occur in the vicinity of or on the surface.
We also find that 
the distribution of the surface energy density has an anisotropy depending on the polar angle in the sphere. 
Thus, this may lead to the deformation of the geometrical shape of the surface.
Also, the order parameter manifold (OPM) induced on the surface, which is described by 
two-dimensional vector fields induced on the surface from the condensates,
allows topological defects (vortices) on the surface.
We show that there must exist such defects even in the ground state 
 due to the Poincar\'{e}-Hopf theorem:
The numbers of the vortices and antivortices
depend on the bulk phases, but 
the difference between them is topologically invariant (the Euler number $\chi=2$) irrespective to the bulk phases.
We point out that
these vortices are not extended to the bulk 
since the first homotopy group is trivial for the OPM in the bulk while it is
nontrivial for the OPM reduced on the boundary. 
Such defects are called boojums, which were named after Lewis Carroll's poem about imaginary monsters, in the context of liquid crystals~\cite{Urbanski_2017}
and helium-3 superfluids~\cite{vollhardt2013superfluid,Volovik:2003fe}.
The surface properties of the neutron $^{3}P_{2}$ superfluids that we find in this paper will hopefully provide us useful information about neutron stars.

This paper is organized as follows.
In Sec.~\ref{sec:formalism}, we summarize the GL equation up to the eighth order. 
On the surface, we introduce the boundary condition for condensates of neutron $^{3}P_{2}$ superfluids.
In Sec.~\ref{sec:numerical}, we perform numerical analyses by solving the GL equations with a spherical boundary condition.
After summarizing the symmetry breaking in the bulk space, we show results of new patterns of the symmetry breaking near the surface.
Furthermore, we show anisotropic distributions of the surface energy density on the neutron star, and present the emergence of  topological  defects (vortices or boojums) on the surface.
The final section is devoted to our conclusion and perspectives.
In Appendix~\ref{sec:EL_appendix}, we show explicit forms of the Euler-Lagrange equation from the GL equation.
In Appendix~\ref{sec:symmetries}, we briefly summarize the symmetries of neutron $^{3}P_{2}$ superfluidity.
In Appendix~\ref{sec:symmetry_surface_results}, we show the numerical results of the profile functions in the condensate.

\section{Formalism}
\label{sec:formalism}

\subsection{Ginzburg-Landau equation}

The condensate of the neutron $^{3}P_{2}$ superfluidity 
can be expressed by a symmetric and traceless three-by-three tensor $A$ as an order parameter of the symmetry breaking.
The components of $A$ are denoted by $A^{ab}$ with the indices $a=1,2,3$ for the spin indices and $b=1,2,3$ for the space indices.
The Ginzburg-Landau (GL) equation can be obtained by integrating out the neutron degrees of freedom and by adopting the loop expansion for the small coupling strength in the $^{3}P_{2}$ interaction for two neutrons~\cite{Fujita1972,Richardson:1972xn,Sauls:1978lna,Muzikar:1980as,Sauls:1982ie,Vulovic:1984kc,Masuda:2015jka,Masuda:2016vak,Yasui:2018tcr,Yasui:2019unp}.
The GL equation is valid in the region in which the temperature $T$ is close to the critical temperature $ T_{c0}$; $|1-T/T_{c0}| \ll 1$,
where $T_{c0}$ is determined at zero magnetic field.
The concrete form of the GL free energy reads 
\begin{eqnarray}
  f[{A}] = f_{0} + f_{\grad}[A] + f_{8}^{(0)}[{A}] + f_{2}^{(\le4)}[{A}] + f_{4}^{(\le2)}[{A}] + {\cal O}(B^{m}{A}^{n})_{m+n\ge7},
\label{eq:eff_pot_coefficient02_f}
\end{eqnarray}
as an expansion in terms of the condensate $A$ and the magnetic field $\vec{B}$.
Each term is explained as follows.
The first term $f_{0}$ is the sum of the free part and the spin-magnetic coupling term
\begin{eqnarray}
f_{0}
=
  - T
    \int \frac{\mathrm{d}^{3}\vec{p}}{(2\pi)^{3}}
    \ln \Bigl(
             \bigl( 1+e^{-\xi_{\vec{p}}^{-}/T} \bigr)
             \bigl( 1+e^{-\xi_{\vec{p}}^{+}/T} \bigr)
         \Bigr),
\label{eq:eff_pot_free_magneticfield}
\end{eqnarray}
with $\displaystyle \xi_{\vec{p}}^{\pm} = \xi_{\vec{p}} \pm |\vec{\mu}_{n}||\vec{B}|$ and $\displaystyle \xi_{\vec{p}}=\vec{p}^{2}/(2m)-\mu$ for the neutron three-dimensional momentum $\vec{p}$, the neutron mass $m$, and the neutron chemical potential $\mu$.
The bare magnetic moment of a neutron is $\vec{\mu}_{n}=-(\gamma_{n}/2)\vec{\sigma}$ 
with the gyromagnetic ratio $\gamma_{n}=1.2 \times 10^{-13}$ MeV/T (in natural units, $\hbar = c=1$) and the Pauli matrices for the neutron spin $\vec{\sigma}$.
The following terms include the condensate $A$:
$f_{8}^{(0)}[{A}]$ consists of the terms including the field $A$ up to the eighth order with no magnetic field,
$f_{2}^{(\le4)}[{A}]$ consists of the terms including the field $A$ up to the second order with the magnetic field up to $|\vec{B}|^{4}$, 
and $f_{4}^{(\le2)}[{A}]$ consists of the terms including the field $A$ up to the fourth order with the magnetic field up to $|\vec{B}|^{2}$.
We show their explicit forms,
\begin{eqnarray}
 f_{\rm grad}[{A}]
=
 K^{(0)}
  \Bigl(
        \nabla_{i} {A}^{ba\ast}
        \nabla_{i} {A}^{ab}
     + \nabla_{i} {A}^{ia\ast}
        \nabla_{j} {A}^{aj}
     + \nabla_{i} {A}^{ja\ast}
        \nabla_{j} {A}^{ai}
  \Bigr),
\label{eq:eff_pot_kin}
\end{eqnarray}
for the gradient term (summed over the repeated indices) and 
\begin{widetext}
\begin{eqnarray}
 f_{8}^{(0)}[{A}]
&=&
 \alpha^{(0)}
   \bigl(\mathrm{tr} {A}^{\ast} {A} \bigr)
\nonumber \\ &&
+ \beta^{(0)}
   \Bigl(
        \bigl(\mathrm{tr} \, {A}^{\ast} {A} \bigr)^{2}
      - \bigl(\mathrm{tr} \, {A}^{\ast 2} {A}^{2} \bigr)
   \Bigr)
\nonumber \\ &&
+ \gamma^{(0)}
   \Bigl(
         - 3  \bigl(\mathrm{tr} \, {A}^{\ast} {A} \bigr) \bigl(\mathrm{tr} \, {A}^{2} \bigr) \bigl(\mathrm{tr} \, {A}^{\ast 2} \bigr)
        + 4 \bigl(\mathrm{tr} \, {A}^{\ast} {A} \bigr)^{3}
        + 6 \bigl(\mathrm{tr} \, {A}^{\ast} {A} \bigr) \bigl(\mathrm{tr} \, {A}^{\ast 2} {A}^{2} \bigr)
      + 12 \bigl(\mathrm{tr} \, {A}^{\ast} {A} \bigr) \bigl(\mathrm{tr} \, {A}^{\ast} {A} {A}^{\ast} {A} \bigr)
              \nonumber \\ && \hspace{3em} 
         - 6 \bigl(\mathrm{tr} \, {A}^{\ast 2} \bigr) \bigl(\mathrm{tr} \, {A}^{\ast} {A}^{3} \bigr)
         - 6 \bigl(\mathrm{tr} \, {A}^{2} \bigr) \bigl(\mathrm{tr} \, {A}^{\ast 3} {A} \bigr)
       - 12 \bigl(\mathrm{tr} \, {A}^{\ast 3} {A}^{3} \bigr)
      + 12 \bigl(\mathrm{tr} \, {A}^{\ast 2} {A}^{2} {A}^{\ast} {A} \bigr)
        + 8 \bigl(\mathrm{tr} \, {A}^{\ast} {A} {A}^{\ast} {A} {A}^{\ast} {A} \bigr)
   \Bigr)
\nonumber \\ &&
 + \delta^{(0)}
\Bigl(
       \bigl( \mathrm{tr}\,A^{\ast 2} \bigr)^{2} \bigl( \mathrm{tr}\, A^{2} \bigr)^{2}
 + 2 \bigl( \mathrm{tr}\,A^{\ast 2} \bigr)^{2} \bigl( \mathrm{tr}\, A^{4} \bigr)
  - 8 \bigl( \mathrm{tr}\,A^{\ast 2} \bigr) \bigl( \mathrm{tr}\,A^{\ast}AA^{\ast}A \bigr) \bigl( \mathrm{tr}\,A^{2} \bigr)
  - 8 \bigl( \mathrm{tr}\,A^{\ast 2} \bigr) \bigl( \mathrm{tr}\,A^{\ast}A \bigr)^{2} \bigl( \mathrm{tr}\,A^{2} \bigr)
       \nonumber \\ && \hspace{3em}
 - 32 \bigl( \mathrm{tr}\,A^{\ast 2} \bigr) \bigl( \mathrm{tr}\,A^{\ast}A \bigr) \bigl( \mathrm{tr}\,A^{\ast}A^{3} \bigr)
 - 32 \bigl( \mathrm{tr}\,A^{\ast 2} \bigr) \bigl( \mathrm{tr}\,A^{\ast}AA^{\ast}A^{3} \bigr)
 - 16 \bigl( \mathrm{tr}\,A^{\ast 2} \bigr) \bigl( \mathrm{tr}\,A^{\ast}A^{2}A^{\ast}A^{2} \bigr)
       \nonumber \\ && \hspace{3em}
  + 2 \bigl( \mathrm{tr}\,A^{\ast 4} \bigr) \bigl( \mathrm{tr}\,A^{2} \bigr)^{2}
  + 4 \bigl( \mathrm{tr}\,A^{\ast 4} \bigr) \bigl( \mathrm{tr}\,A^{4} \bigr)
  - 32 \bigl( \mathrm{tr}\,A^{\ast 3}A \bigr) \bigl( \mathrm{tr}\,A^{\ast}A \bigr) \bigl( \mathrm{tr}\,A^{2} \bigr)
       \nonumber \\ && \hspace{3em}
  - 64 \bigl( \mathrm{tr}\,A^{\ast 3}A \bigr) \bigl( \mathrm{tr}\,A^{\ast}A^{3} \bigr)
  - 32 \bigl( \mathrm{tr}\,A^{\ast 3}AA^{\ast}A \bigr) \bigl( \mathrm{tr}\,A^{2} \bigr)
  - 64 \bigl( \mathrm{tr}\,A^{\ast 3}A^{2}A^{\ast}A^{2} \bigr)
  - 64 \bigl( \mathrm{tr}\,A^{\ast 3}A^{3} \bigr) \bigl( \mathrm{tr}\,A^{\ast}A \bigr)
       \nonumber \\ && \hspace{3em}
  - 64 \bigl( \mathrm{tr}\,A^{\ast 2}AA^{\ast 2}A^{3} \bigr)
  - 64 \bigl( \mathrm{tr}\,A^{\ast 2}AA^{\ast}A^{2} \bigr) \bigl( \mathrm{tr}\,A^{\ast}A \bigr)
 + 16 \bigl( \mathrm{tr}\,A^{\ast 2}A^{2} \bigr)^{2}
 + 32 \bigl( \mathrm{tr}\,A^{\ast 2}A^{2} \bigr) \bigl( \mathrm{tr}\,A^{\ast}A \bigr)^{2}
       \nonumber \\ && \hspace{3em}
 + 32 \bigl( \mathrm{tr}\,A^{\ast 2}A^{2} \bigr) \bigl( \mathrm{tr}\,A^{\ast}AA^{\ast}A \bigr)
 + 64 \bigl( \mathrm{tr}\,A^{\ast 2}A^{2}A^{\ast 2}A^{2} \bigr)
  -16 \bigl( \mathrm{tr}\,A^{\ast 2}AA^{\ast 2}A \bigr) \bigl( \mathrm{tr}\,A^{2} \bigr)
   + 8 \bigl( \mathrm{tr}\,A^{\ast}A \bigr)^{4}
       \nonumber \\ && \hspace{3em}
 + 48 \bigl( \mathrm{tr}\,A^{\ast}A \bigr)^{2} \bigl( \mathrm{tr}\,A^{\ast}AA^{\ast}A \bigr)
 +192 \bigl( \mathrm{tr}\,A^{\ast}A \bigr) \bigl( \mathrm{tr}\,A^{\ast}AA^{\ast 2}A^{2} \bigr)
 + 64 \bigl( \mathrm{tr}\,A^{\ast}A \bigr) \bigl( \mathrm{tr}\,A^{\ast}AA^{\ast}AA^{\ast}A \bigr)
       \nonumber \\ && \hspace{3em}
  -128 \bigl( \mathrm{tr}\,A^{\ast}AA^{\ast 3}A^{3} \bigr)
 + 64 \bigl( \mathrm{tr}\,A^{\ast}AA^{\ast 2}AA^{\ast}A^{2} \bigr)
 + 24 \bigl( \mathrm{tr}\,A^{\ast}AA^{\ast}A \bigr)^{2}
 +128 \bigl( \mathrm{tr}\,A^{\ast}AA^{\ast}AA^{\ast 2}A^{2} \bigr)
       \nonumber \\ && \hspace{3em}
 + 48 \bigl( \mathrm{tr}\,A^{\ast}AA^{\ast}AA^{\ast}AA^{\ast}A \bigr)
\Bigr),
\label{eq:eff_pot_w0_coefficient02_f}
\\
   f_{2}^{(\le4)}[{A}]
&=&
      \beta^{(2)}
      \vec{B}^{t} {A}^{\ast} {A} \vec{B}
+ \beta^{(4)}
   |\vec{B}|^{2}
   \vec{B}^{t} {A}^{\ast} {A} \vec{B},
\label{eq:eff_pot_B4w2_coefficient02_f}
\\
   f_{4}^{(\le2)}[{A}]
&=&
  \gamma^{(2)}
  \Bigl(
       - 2 \, |\vec{B}|^{2} \bigl(\mathrm{tr} \, {A}^{2} \bigr) \bigl(\mathrm{tr} \, {A}^{\ast 2} \bigr)
       - 4 \, |\vec{B}|^{2} \bigl(\mathrm{tr} \, {A}^{\ast} {A} \bigr)^{2}
      + 4 \, |\vec{B}|^{2} \bigl(\mathrm{tr} \, {A}^{\ast} {A} {A}^{\ast} {A} \bigr)
      + 8 \, |\vec{B}|^{2} \bigl(\mathrm{tr} \, {A}^{\ast 2} {A}^{2} \bigr)
            \nonumber \\ && \hspace{2em}
        + \vec{B}^{t} {A}^{2} \vec{B} \bigl(\mathrm{tr} \, {A}^{\ast 2} \bigr)
       - 8 \, \vec{B}^{t} {A}^{\ast} {A} \vec{B} \bigl(\mathrm{tr} \, {A}^{\ast} {A} \bigr)
         + \vec{B}^{t} {A}^{\ast 2} \vec{B} \bigl(\mathrm{tr} \, {A}^{2} \bigr)
      + 2 \, \vec{B}^{t} {A} {A}^{\ast 2} {A} \vec{B}
            \nonumber \\ && \hspace{2em}
      + 2 \, \vec{B}^{t} {A}^{\ast} {A}^{2} {A}^{\ast} \vec{B}
       - 8 \, \vec{B}^{t} {A}^{\ast} {A} {A}^{\ast} {A} \vec{B}
       - 8 \, \vec{B}^{t} {A}^{\ast 2} {A}^{2} \vec{B}
  \Bigr),
\label{eq:eff_pot_B2w4_coefficient02_f}
\end{eqnarray}
\end{widetext}
for the potential and interaction terms.\footnote{In the present paper, we separate the derivative term and the nonderivative term for later convenience. This convention is different from the previous works by the present authors.}
The trace ($\mathrm{tr}$) is taken over the spin and space indices of $A$.
The coefficients are given by
\begin{eqnarray}
  K^{(0)}
&=&
   \frac{7 \, \zeta(3)N(0) p_{F}^{4}}{240m^{2}(\pi T_{c0})^{2}},
\nonumber \\ 
   \alpha^{(0)}
&=&
   \frac{N(0)p_{F}^{2}}{3} 
   \frac{T-T_{c0}}{T_{c0}},
\nonumber \\ 
  \beta^{(0)}
&=&
   \frac{7\,\zeta(3)N(0)p_{F}^{4}}{60\,(\pi T_{c0})^{2}},
\nonumber \\ 
  \gamma^{(0)}
&=&
   - \frac{31\,\zeta(5)N(0)p_{F}^{6}}{13440\,(\pi T_{c0})^{4}},
\nonumber \\ 
  \delta^{(0)}
&=&
  \frac{127\,\zeta(7)N(0)p_{F}^{8}}{387072\,(\pi T_{c0})^{6}},
\nonumber \\ 
   \beta^{(2)}
&=&
    \frac{7\,\zeta(3)N(0)p_{F}^{2}\gamma_{n}^{2}}{48(1+F_{0}^{a})^{2}(\pi T_{c0})^{2}},
\nonumber \\ 
   \beta^{(4)}
&=&
    - \frac{31\,\zeta(5)N(0)p_{F}^{2}\gamma_{n}^{4}}{768(1+F_{0}^{a})^{4}(\pi T_{c0})^{4}},
\nonumber \\ 
  \gamma^{(2)}
&=&
  \frac{31\,\zeta(5)N(0)p_{F}^{4}\gamma_{n}^{2}}{3840(1+F_{0}^{a})^{2}(\pi T_{c0})^{4}}.
\label{eq:eff_pot_coefficient0_parameters_FL_f}
\end{eqnarray}
We denote $N(0)=m\,p_{F}/(2\pi^{2})$ for the state-number density at the Fermi surface and $|\vec{\mu}_{n}^{\ast}|=(\gamma_{n}/2)/(1+F_{0}^{a})$ for the magnitude of the magnetic momentum of a neutron modified by the Landau parameter $F_{0}^{a}$.
Notice that $\vec{\mu}_{n}^{\ast}$ is different from the bare magnetic moment of the neutron $\vec{\mu}_{n}$.
It may be worthwhile to remember that the interaction between the neutron and the magnetic field ($\vec{B}$) supplies the energy splitting by the interaction Hamiltonian $-\vec{\mu}_{n}^{\ast}\cdot\vec{B}$.
We notice that the Landau parameter stems from the Hartree-Fock approximation, which is not taken into account explicitly in the present mean-field approximation.
In the expressions in Eq.~(\ref{eq:eff_pot_coefficient0_parameters_FL_f}),
$\zeta(n)$ is the zeta function.
In the present study, we will focus on the energy difference between the state with the superfluid state ($A\neq0$) and the normal state ($A=0$).
Thus, we do not consider explicitly the contribution from $f_{0}$ in Eq.~\eqref{eq:eff_pot_free_magneticfield}, and hence we will neglect $f_{0}$ in the following discussion.

\subsection{Boundary condition on the surface}

\begin{figure}[tb]
\begin{center}
\includegraphics[scale=0.25]{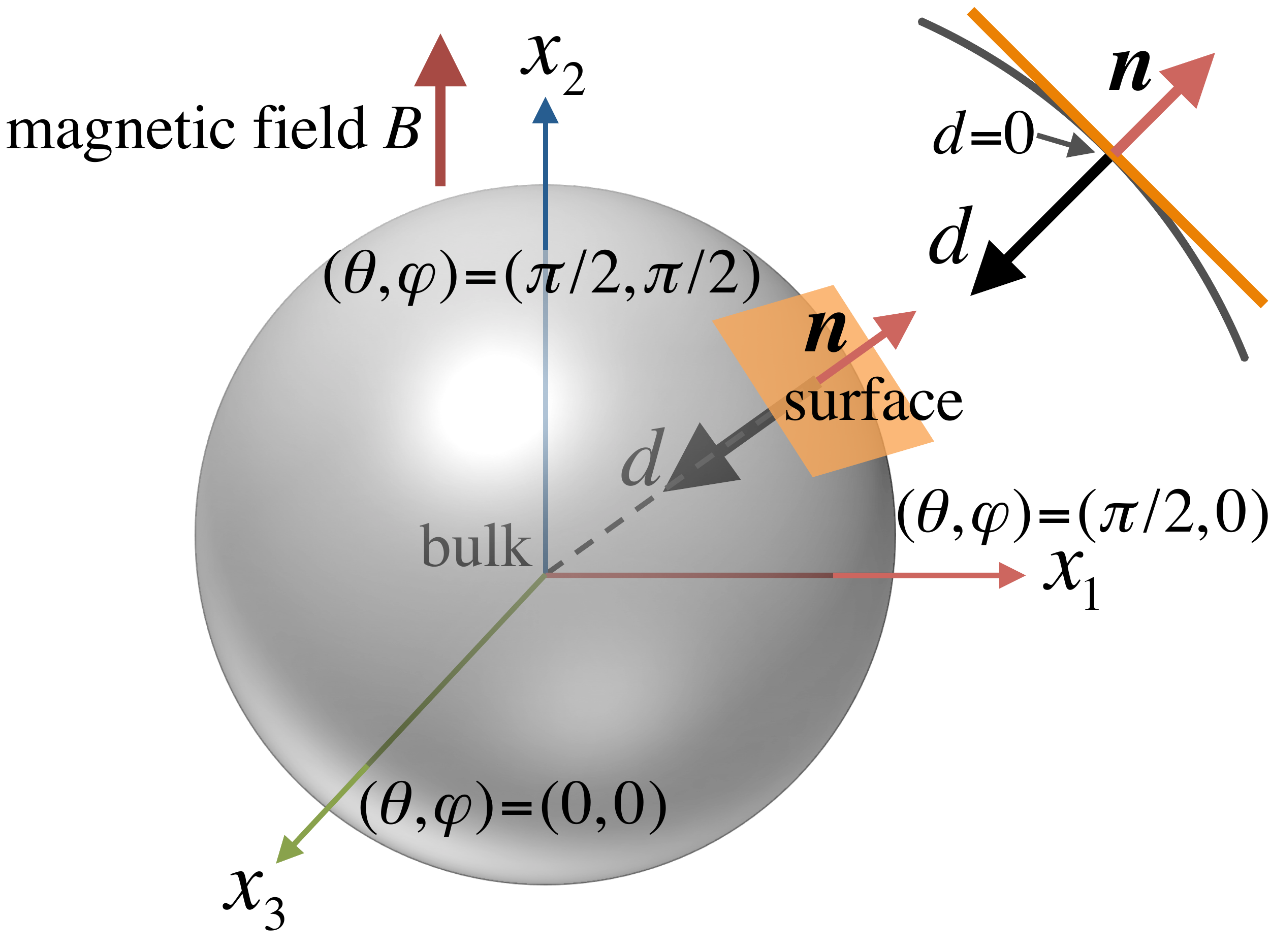}
\caption{The plane on the surface for the normal vector $\vec{n}=(n_{1},n_{2},n_{3})$ with $n_{1}=\sin\theta\cos\varphi$, $n_{1}=\sin\theta\sin\varphi$, and $n_{3}=\cos\theta$ is presented for neutron star. $d$ is the distance from the surface position toward the center of the neutron star. The direction of the magnetic field is along the $x_{2}$ axis.}
\label{fig:Fig_181229_d}
\end{center}
\end{figure}

We consider the boundary of a neutron $^{3}P_{2}$ superfluid 
and introduce the normal vector perpendicular to the surface: $\vec{n}=(n_{1},n_{2},n_{3})$ with $n_{1}=\sin\theta\cos\varphi$, $n_{2}=\sin\theta\sin\varphi$, and $n_{3}=\cos\theta$ as shown in Fig.~\ref{fig:Fig_181229_d}.
We assume that the geometrical shape of the surface can be locally approximated by a tangent plane, 
where the curvature can be neglected. 
This simplification should be justified when the curvature radius is sufficiently larger than the coherence length of the neutron $^{3}P_{2}$ superfluidity.
We introduce the $d$ axis from the surface toward the center of the neutron star (Fig.~\ref{fig:Fig_181229_d}).
Then, the condensate $A$ is a $3\times3$ matrix whose components are functions of $d(\ge0)$:
\begin{eqnarray}
   A(d;\vec{n})
=
\left(
\begin{array}{ccc}
 -F_{1}(d;\vec{n}) & G_{3}(d;\vec{n}) & G_{2}(d;\vec{n}) \\
 G_{3}(d;\vec{n}) & -F_{2}(d;\vec{n}) & G_{1}(d;\vec{n}) \\
 G_{2}(d;\vec{n}) & G_{1}(d;\vec{n}) & F_{1}(d;\vec{n})+F_{2}(d;\vec{n})  
\end{array}
\right).
\end{eqnarray}
For precision, we should include $\vec{n}$ for the variables, because $A$ should depend on the choice of the surface with the normal vector $\vec{n}$.
In the following equations, however, we will sometimes omit $\vec{n}$ for shorter notation: $A(d)=A(d;\vec{n})$, $F_{\alpha}(d)=F_{\alpha}(d;\vec{n})$, and $G_{\beta}(d)=G_{\beta}(d;\vec{n})$ with $\alpha=1,2$ and $\beta=1,2,3$.
With the above setup,
 we rewrite the gradient terms in \eqref{eq:eff_pot_kin} as
\begin{eqnarray}
   {f}_{\grad}[A]
&=&
   \frac{{K}^{(0)}}{4}
   \biggl(
           \bigl(
                    2
                 - \sin^{2}\theta \sin^{2}\varphi
           \bigr)
           \bigl(\nabla_{\!d}F_{1}\bigr)^{2}
        + \bigl(
                    2
                  - \sin^{2}\theta \cos^{2}\varphi
           \bigr)
           \bigl(\nabla_{\!d}F_{2}\bigr)^{2}
        + \bigl(
                    1
                 + 2\cos^{2}\theta
           \bigr)
           \bigl(\nabla_{\!d} F_{1}\bigr) \bigl(\nabla_{\!d}F_{2}\bigr)
   \nonumber \\ && \hspace{1em} 
+ 
           2\cos\theta\sin\theta\sin\varphi \bigl(\nabla_{\!d}F_{1}\bigr) \bigl(\nabla_{\!d}G_{1}\bigr)
        + 2\cos\theta\sin\theta\cos\varphi \bigl(\nabla_{\!d}F_{2}\bigr) \bigl(\nabla_{\!d}G_{2}\bigr)
           \nonumber \\ && \hspace{1em} 
         - 2\sin^{2}\theta\cos\varphi\sin\varphi \bigl(\nabla_{\!d}F_{1}+\nabla_{\!d}F_{2}\bigr) \bigl(\nabla_{\!d}G_{3}\bigr)
   \nonumber \\ && \hspace{1em} 
+ 
           \bigl(
                   2
                 - \sin^{2}\theta \cos^{2}\varphi
           \bigr)
           \bigl(\nabla_{\!d}G_{1}\bigr)^{2}
        + \bigl(
                    2
                  - \sin^{2}\theta \sin^{2}\varphi
           \bigr)
           \bigl(\nabla_{\!d}G_{2}\bigr)^{2}
        + \bigl(
                    1
                 + \sin^{2}\theta
           \bigr)
           \bigl(\nabla_{\!d}G_{3}\bigr)^{2}
           \nonumber \\ && \hspace{1em} 
        + 2\sin^{2}\theta\cos\varphi\sin\varphi \bigl(\nabla_{\!d}G_{1}\bigr) \bigl(\nabla_{\!d}G_{2}\bigr)
        + 2\cos\theta\sin\theta \bigl(\cos\varphi \nabla_{\!d}G_{1}+\sin\varphi \nabla_{\!d}G_{2}\bigr) \nabla_{\!d}G_{3}
   \biggr),
\label{eq:kin_eff_pot}
\end{eqnarray}
with $\nabla_{\!d}=\partial/\partial d$.
We emphasize that $\vec{n}$ is assumed to be a constant vector and hence the derivatives with respect to $\theta$ and $\varphi$ are not included in Eq.~\eqref{eq:kin_eff_pot}.
With the above-mentioned coordinate setting, we obtain the Euler-Lagrange (EL) equations for $A$,
\begin{eqnarray}
   -\nabla_{\!d} \frac{\delta {f}[A]}{\delta (\nabla_{\!d}F_{\alpha})}
   + \frac{\delta {f}[A]}{\delta F_{\alpha}} &=& 0,
   \label{eq:EL_f} \\ 
   -\nabla_{\!d} \frac{\delta {f}[A]}{\delta (\nabla_{\!d}G_{\beta})}
   + \frac{\delta {f}[A]}{\delta G_{\beta}} &=& 0.
   \label{eq:EL_g}
\end{eqnarray}
The concrete expressions of the left-hand sides are presented in detail in Appendix~\ref{sec:EL_appendix}.
We impose the boundary conditions at $d=0$ and $d\rightarrow\infty$ for the EL equations.
On the surface (at $d=0$), we adopt the condition that $A(0)$ satisfies
\begin{eqnarray}
   \vec{n}^{t}A(0)\vec{n} \equiv \sum_{i,j=1,2,3} n_{i} A_{ij}(0) n_{j} = 0
\label{eq:boundary_condition_1}
\end{eqnarray}
as the boundary condition.
This is consistent with the condition presented by Ambegaokar, de~Gennes, and Rainer~\cite{PhysRevA.9.2676}.\footnote{In  Ref.~\cite{PhysRevA.9.2676}, the authors considered the surface with $\vec{n}=(0,0,1)$ and concluded that, among the components in the matrix $A_{ij}(d)$ ($i,j=1,2,3$), $A_{11}(d)$ and $A_{22}(d)$ are symmetric functions for transforming $d$ to $-d$, hence they can continue to have finite values at the surface, while $A_{33}(d)$ is an antisymmetric function for transforming $d$ to $-d$, and hence it should vanish at the surface.
This property is reflected in Eq.~\eqref{eq:boundary_condition_1} for arbitrary $\vec{n}$.
}
At the center of the $^3P_2$ condensate (at $d\rightarrow\infty$), on the other hand, we require that the state approaches the bulk state, and hence that the condensate values should satisfy $F_{1}(d)\rightarrow F_{1}^{\bulk}$, $F_{2}(d)\rightarrow F_{2}^{\bulk}$, $G_{1}(d)\rightarrow G_{1}^{\bulk}=0$, $G_{2}(d)\rightarrow G_{2}^{\bulk}=0$, and $G_{3}(d)\rightarrow G_{3}^{\bulk}=0$, where $F_{\alpha}^{\bulk}$ ($\alpha=1,2$) and $G_{\beta}^{\bulk}$ ($\beta=1,2,3$) are the values in the ground state in the bulk.

\subsection{Dimensionless form}

For the convenience of the analysis, we introduce the dimensionless quantities $\tilde{A}$, $f_{\alpha}$ ($\alpha=1,2$), $g_{\beta}$ ($\beta=1,2,3$), $\vec{b}$, and $x$ defined by
\begin{eqnarray}
   \tilde{A} \equiv \frac{p_{F}}{T_{c0}} A, \hspace{1em}
   f_{\alpha} \equiv \frac{p_{F}}{T_{c0}} F_{\alpha}, \hspace{1em}
   g_{\beta} \equiv \frac{p_{F}}{T_{c0}} G_{\beta}, \hspace{1em}
   \vec{b} \equiv \frac{\gamma_{n}}{(1+F_{0}^{a})T_{c0}} \vec{B}, \hspace{1em}
   x \equiv \frac{m T_{c0}}{p_{F}}d,
\label{eq:dimensionless}
\end{eqnarray}
instead of $A$, $F_{\alpha}$, $G_{\beta}$, $\vec{B}$, and $d$.
Notice that $\tilde{A}=\tilde{A}(x)$, $f_{\alpha}=f_{\alpha}(x)$, and $g_{\beta}=g_{\beta}(x)$ are regarded as functions of the dimensionless distance $x$.
With these new variables, we obtain the dimensionless form of the GL free energy, $\tilde{f}[\tilde{A}]$, which is given from $f[A]$ by replacing $A$, $F_{\alpha}$, $G_{\beta}$, $\vec{B}$, and $d$ with $\tilde{A}$, $f_{\alpha}$, $g_{\beta}$, $\vec{b}$, and $x$ as well as by replacing the dimensionful coefficients $K^{0}$, $\alpha^{(0)}$, $\beta^{(0)}$, $\gamma^{(0)}$, $\delta^{(0)}$, $\beta^{(2)}$, $\beta^{(4)}$, and $\gamma^{(2)}$ with the dimensionless ones
\begin{eqnarray}
  \tilde{K}^{(0)}
&=&
   \frac{7 \, \zeta(3)}{240\,\pi^{2}},
\nonumber \\ 
   \tilde{\alpha}^{(0)}
&=&
   \frac{1}{3} \bigl( t-1 \bigr),
\nonumber \\ 
  \tilde{\beta}^{(0)}
&=&
   \frac{7\,\zeta(3)}{60\,\pi^{2}},
\nonumber \\ 
  \tilde{\gamma}^{(0)}
&=&
   - \frac{31\,\zeta(5)}{13440\,\pi^{4}},
\nonumber \\ 
  \tilde{\delta}^{(0)}
&=&
  \frac{127\,\zeta(7)}{387072\,\pi^{6}},
\nonumber \\ 
   \tilde{\beta}^{(2)}
&=&
    \frac{7\,\zeta(3)}{48\,\pi^{2}},
\nonumber \\ 
   \tilde{\beta}^{(4)}
&=&
    - \frac{31\,\zeta(5)}{768\,\pi^{4}},
\nonumber \\ 
  \tilde{\gamma}^{(2)}
&=&
  \frac{31\,\zeta(5)}{3840\,\pi^{4}}.
\label{eq:eff_pot_coefficient0_parameters_FL_f_dimensionless}
\end{eqnarray}
with the normalized temperature $t=T/T_{c0}$.
$f[A]$ and $\tilde{f}[\tilde{A}]$ are related by
\begin{eqnarray}
   f[A] = N(0)T_{c0}^{2} \tilde{f}[\tilde{A}].
\end{eqnarray}
The boundary condition at $x=0$ ({\it i.e.}, $d=0$) is expressed by $\vec{n}^{t}\tilde{A}(0)\vec{n}=0$.

\section{Numerical results}
\label{sec:numerical}

In the numerical calculation, we use the following parameter settings:
the critical temperature $T_{c0}=0.2$ MeV,
the neutron number density $n=0.17$ fm$^{-3}$ (the Fermi momentum $p_{F}=338$ MeV), and
the Landau parameter $F_{0}^{a}=-0.75$.
The value of $F_{0}^{a}$ is that of $^{3}$He liquid at low temperature.
We consider without loss of generality that the magnetic field is applied along the $x_{2}$ axis: $\vec{B}=(0,B,0)$ or $\vec{b}=(0,b,0)$.
This direction is chosen to minimize the total energy when the matrix $A$ ($\tilde{A}$) is expressed in the conventional form, as will be shown in Eq.~\eqref{eq:A_gs}.

\subsection{Phase diagram in bulk space}

\begin{table*}[t!]
\begingroup
\renewcommand{\arraystretch}{1.2}
 \begin{tabular}{|c|c|c|c|}
 \hline 
    $r$ & Phase & $H$ & $M \simeq G/H$ \\
    \hline \hline 
    $-1/2$ & UN & $\mathrm{O}(2)$ &$ \mathrm{U}(1) \times [\mathrm{SO}(3) / \mathrm{O}(2)]$ \\
    $-1 < r < -1/2$ & ${\rm D}_2$-BN  & ${\rm D}_2$ & $ \mathrm{U}(1) \times [\mathrm{SO}(3) / {\rm D}_2]$  \\
    $-1$ & ${\rm D}_4$-BN & ${\rm D}_4$ & $ [\mathrm{U}(1) \times \mathrm{SO}(3)] / {\rm D}_{4}$  \\
    \hline
 \end{tabular}
\endgroup
\caption{The classification of nematic phases (a table taken from Ref.~\cite{Masuda:2015jka}).
We show the range of $r$, phases, unbroken symmetries $H$, 
and order parameter manifolds $M \simeq G/H$.}
\label{table-sym}
\end{table*}

Before investigating the surface effect, in this subsection, 
we analyze the phase diagram in the ground (uniform) state in the bulk space ($d\rightarrow\infty$), by neglecting the gradient term in the GL free energy \eqref{eq:eff_pot_coefficient02_f}.  
In the bulk space, we can express conventionally the order parameter $\tilde{A}$ as a constant matrix in diagonal form:
\begin{eqnarray}
   \tilde{A}^{\bulk}
=
\left(
\begin{array}{ccc}
 -\fa & 0 & 0 \\
 0 & -\fb & 0 \\
 0 & 0 & \fa+\fb
\end{array}
\right),
\end{eqnarray}
where $f_{\alpha}^{\bulk}$ ($\alpha=1,2$) are values in the bulk space and 
the off-diagonal components are zero: $g_{\beta}^{\bulk}=0$ ($\beta=1,2,3$).
Without loss of generality, as a convention, we may restrict the range of the  values of $\fa$ and $\fb$ to satisfy
$|\fa| \le |\fb| \le |\fa+\fb|$,
{\it i.e.}, $\fb \le \fa$
and $\fb\ge0$.
In the literature, $\tilde{A}^{\bulk}$ is often expressed as
\begin{eqnarray}
   \tilde{A}^{\bulk}
=
\tilde{A}_{0}
\left(
\begin{array}{ccc}
 r & 0 & 0 \\
 0 & -1-r & 0 \\
 0 & 0 & 1
\end{array}
\right),
\label{eq:A_gs}
\end{eqnarray}
with the magnitude $\tilde{A}_{0}\ge0$ and the internal parameter $-1\le r \le -1/2$.
Both expressions are related through $\fa=\tilde{A}_{0}r$ and $\fa+\fb=\tilde{A}_{0}$.

According to the values of $\fa$ and $\fb$,
the order parameter $\tilde{A}^{\bulk}$ expresses different phases with various symmetries, as summarized in Table~\ref{table-sym} and in Appendix~\ref{sec:symmetries}.
When $\fa=\fb$ ($r=-1/2$),
there is an O(2) symmetry around the $x_{3}$ axis, and this phase is called the uniaxial-nematic (UN) phase.
When $\fa<\fb$ ($-1\le r<-1/2$),
the continuous symmetry is lost, and there remains only the discrete (dihedral) symmetry, whose phase is called the biaxial-nematic (BN) phase.
The BN phase is furthermore classified into the case with $-1<r<-1/2$ and the case with $r=-1$.
The former ($-1<r<-1/2$) has D$_{2}$ symmetry and is called the D$_{2}$-BN phase, 
while 
the latter ($r=-1$) has D$_{4}$ symmetry and is called the D$_{4}$-BN phase.
The ground state is determined to minimize the GL free energy density \eqref{eq:eff_pot_coefficient02_f} with respect to the variations of $f_{1}^{\bulk}$ and $f_{2}^{\bulk}$ (or $\tilde{A}_{0}$ and $r$)\footnote{It may be significant that, for the terms up to and including ${\cal O}(A^{4})$ at zero magnetic field in the GL free energy density \eqref{eq:eff_pot_coefficient02_f}, there happens to appear degeneracy for the UN phase, the D$_{2}$-BN phase, and the D$_{4}$-BN phase. In this situation, there is the SO(5) symmetry which is absent in the original Lagrangian, and the spontaneous breaking of the symmetry eventually leads to the existence of a quasi-Nambu-Goldstone mode \cite{Uchino:2010pf}.}.

\begin{figure}[tb]
\begin{center}
\includegraphics[scale=0.3]{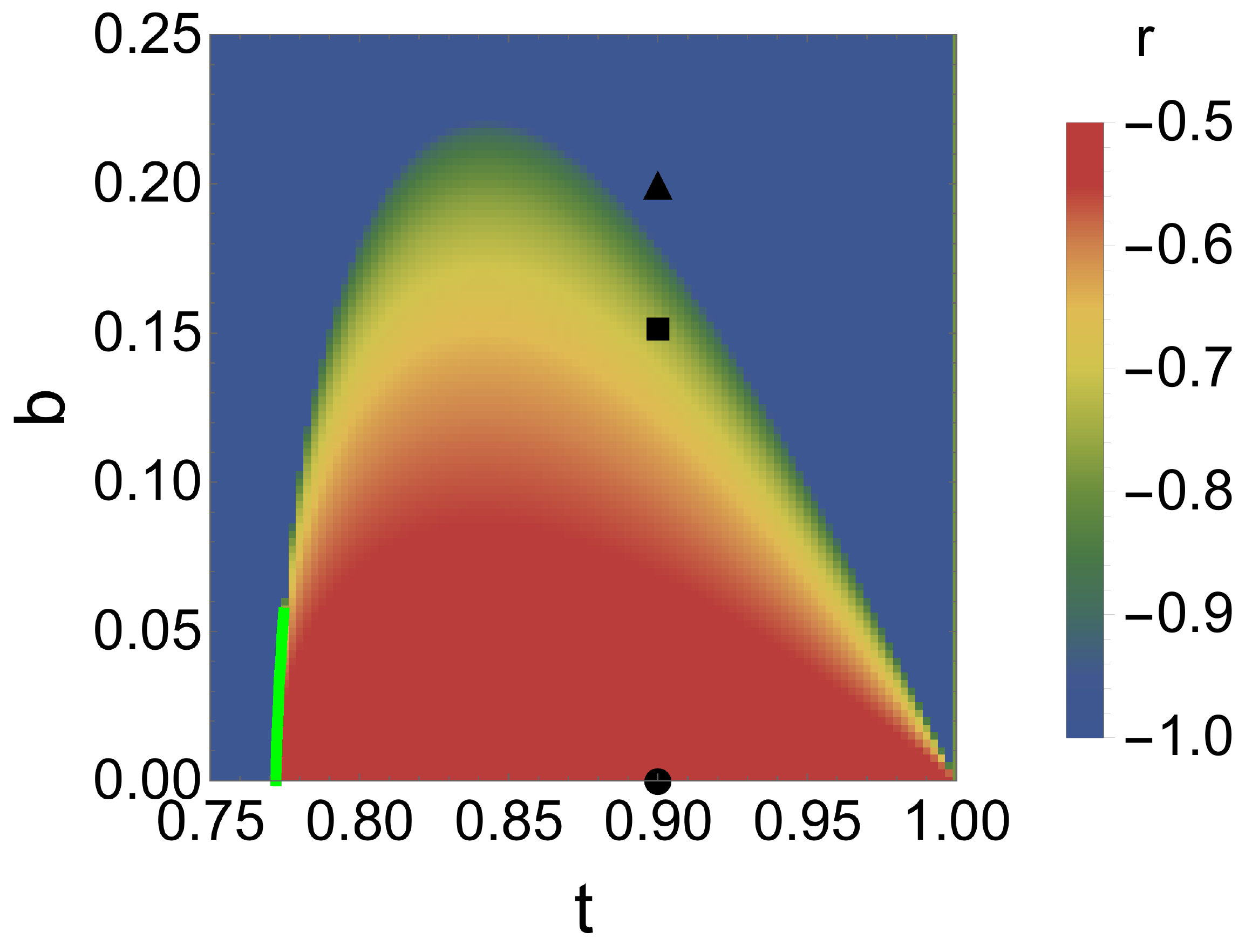}
\caption{The phase diagram of the neutron $^{3}P_{2}$ superfluidity in the bulk space. The value of $r$ is plotted as a function of the normalized temperature ($t$) and the normalized magnetic field ($b$). The magnetic field is applied in the direction of the $x_{2}$ axis. The typical locations of each phase displayed at $t=0.9$: (i) the UN phase at $b=0$ (the circle), (ii) the D$_{2}$-BN phase at $b=0.15$ (the square), and (iii) the D$_{4}$-BN phase at $b=0.2$ (the triangle). The green line around $t \approx 0.772$ indicates the first-order phase transition.}
\label{fig:phase_diagram_dimensionless_1}
\end{center}
\end{figure}

We show the obtained phase diagram in Fig.~\ref{fig:phase_diagram_dimensionless_1}.
We observe that there exist the UN phase at zero magnetic field ($b=0$), the D$_{2}$-BN phase at weak magnetic field, and the D$_{4}$-BN phase at strong magnetic field.
Their phase boundaries are the second-order phase transition, except for the first-order phase transition at  low temperature ($t \approx 0.772$) indicated by the green line in Fig.~\ref{fig:phase_diagram_dimensionless_1}.
Such a first-order phase transition was found recently when the eighth-order term ($\delta_{0}$ term) was included in the GL equation~\cite{Yasui:2019unp}.\footnote{The eighth-order term is important also to give stability in the ground state.}
It is worthwhile to mention that the existence of the first-order phase transition in the GL equation qualitatively agrees with the result from the analysis using the BdG equation \cite{Mizushima:2016fbn}.

In the following subsections, we will consider the surface effects on the neutron $^{3}P_{2}$ superfluidity.
We suppose three cases as the typical phases for the bulk space at $x \rightarrow \infty$ ({\it i.e.}, $d \rightarrow \infty$). We choose the temperature $t=0.9$ and change the values of the magnetic field for each phase.
We show the numerical parameter used for the magnetic field strength ($b$) and the values of $f_{1}^{\bulk}$, $f_{2}^{\bulk}$, and $r$ for each bulk phase:
\begin{description}
\item[(i) UN phase] $b=0$; $(f_{1}^{\bulk},f_{2}^{\bulk})=(0.64,0.64)$ and $r=-1/2$.
\item[(ii) D$_{2}$-BN phase] $b=0.15$; $(f_{1}^{\bulk},f_{2}^{\bulk})=(0.92,0.32)$ and $r=-0.74$.
\item[(iii) D$_{4}$-BN phase] $b=0.2$; $(f_{1}^{\bulk},f_{2}^{\bulk})=(1.12,0)$ and $r=-1$.
\end{description}
We have considered that the strength of the magnetic field can reach maximally $B \approx 10^{15} \,\mathrm{G}$ ($10^{11} \, \mathrm{T}$) at the surface of magnetars.\footnote{Notice the unit conversion $1\,\mathrm{T}=10^{4}\,\mathrm{G}$ for the strength of a magnetic field.}
From Eq.~\eqref{eq:dimensionless}, we obtain $b=0.24$ for $B=10^{15}$ G.
We denote these three cases (i), (ii), and (iii) by circle, square, and triangle, respectively, in the phase diagram in Fig.~\ref{fig:phase_diagram_dimensionless_1}.

\subsection{Symmetry near the surface}
\label{sec:symmetry_surface}

For the situations (i), (iii), and (iii) in the previous subsection,
we solve the EL equations \eqref{eq:EL_f} and \eqref{eq:EL_g} with the boundary conditions, {\it i.e.}, $\vec{n}^{t}\tilde{A}(0)\vec{n}=0$ at $x=0$ from Eq.~\eqref{eq:boundary_condition_1} and 
$\tilde{A}(x)\rightarrow\tilde{A}^{\bulk}$
 in the bulk space ($x\rightarrow\infty$).
We plot the obtained profile functions $f_{1}(x)$, $f_{2}(x)$, $g_{1}(x)$, $g_{2}(x)$, and $g_{3}(x)$ for several choices of the surface direction $\vec{n}=(n_{1},n_{2},n_{3})$ ($\theta$ and $\varphi$) in Figs.~\ref{fig:profile_functions_3P2_bulk1}, \ref{fig:profile_functions_3P2_bulk2}, and \ref{fig:profile_functions_3P2_bulk3} in Appendix~\ref{sec:symmetry_surface_results}.
We observe that the profile functions approach constant values in the bulk space as the boundary condition at $x \rightarrow \infty$, while they change drastically in the region $x \simle 1$ near the surface.
We notice that the typical value of the healing distance from the surface is expressed by $r \approx \xi$ ({\it i.e.}, $x \approx 1$) with $\xi=p_{F}/(m T_{c0})\simeq360$ fm for $p_{F}=338$ MeV and $T_{c0}=0.2$ MeV.
The changes of the profile functions induce the symmetry breaking near the surface.
This can be seen directly by transforming $\tilde{A}(x)$ into diagonal form with some appropriate unitary matrix $U(x)$ at each position $x$:
\begin{eqnarray}
   \tilde{A}(x)
&\rightarrow&
   U(x) \tilde{A}(x) U(x)^{-1}
=
\tilde{A}_{0}(x)
\left(
\begin{array}{ccc}
 r(x) & 0 & 0 \\
 0 & -1-r(x) & 0 \\
 0 & 0 & 1  
\end{array}
\right),
\end{eqnarray}
with $\tilde{A}_{0}(x) \ge 0$ and $-1\le r(x) \le -1/2$.
Notice that $\tilde{A}_{0}$ and $r$ are functions of the position $x$, as they have been constant values in the bulk space [cf.~Eq.~\eqref{eq:A_gs}].
We plot $r(x)$ in Figs.~\ref{fig:r_3P2_bulk1}, \ref{fig:r_3P2_bulk2}, and \ref{fig:r_3P2_bulk3} in Appendix~\ref{sec:symmetry_surface_results} for the bulk conditions (i), (ii), and (iii).
We confirm that $r(x)$ is dependent on $x$, and that there exist various phases, {\it i.e.}, the UN phase [$r(x)=-1/2$], the D$_{2}$-BN phase [$-1<r(x)<-1/2$], and the D$_{4}$-BN phase [$r(x)=-1$], that are different from those in the bulk.

The symmetry restoration or breaking near the surface can be understood in a reasonable way.
First of all, we notice that the boundary condition at the surface, $\vec{n}^{t}\tilde{A}(0)\vec{n}=0$ in Eq.~\eqref{eq:boundary_condition_1}, does not allow the general transformation of spin and space for $\tilde{A}(0)$, because $\vec{n}$ is a fixed vector in space.
Thus, the symmetry cannot be generally maintained from the bulk to the surface, and it should exhibit variation near the surface.
However, there are some exceptional cases in which the symmetry is maintained from the bulk to the surface: $\vec{n}=(0,0,1)$ ($\theta=0$ and $\varphi$=0) with the bulk UN phase and $\vec{n}=(0,\pm1,0)$ ($\theta=\pi/2$ and $\varphi=\pi/2$, $3\pi/2$) with the bulk D$_{4}$-BN phase.
In both cases, the profile functions are constant at any $x$, and hence the symmetry is kept invariant, as shown in Figs.~\ref{fig:profile_functions_3P2_bulk1} and \ref{fig:profile_functions_3P2_bulk3} as well as in Figs.~\ref{fig:r_3P2_bulk1} and \ref{fig:r_3P2_bulk3}.
In order to understand this invariance,
we remember that the UN phase possesses the O(2) symmetry around the $x_{3}$ axis, and the D$_{4}$-BN phase has the D$_{4}$ symmetry around the $x_{2}$ axis.
In the former case, we find that the boundary condition $\vec{n}^{t}\tilde{A}(0)\vec{n}=0$ with $\vec{n}=(0,0,1)$ holds for any transformation of $\tilde{A}(0)$ under the U(1) symmetry.
In the latter case, we also find that the boundary condition $\vec{n}^{t}\tilde{A}(0)\vec{n}=0$ with $\vec{n}=(0,1,0)$ holds for any  transformation of $\tilde{A}(0)$ under the D$_{4}$ symmetry.

\begin{figure}[tb]
\begin{center}
\includegraphics[scale=0.22]{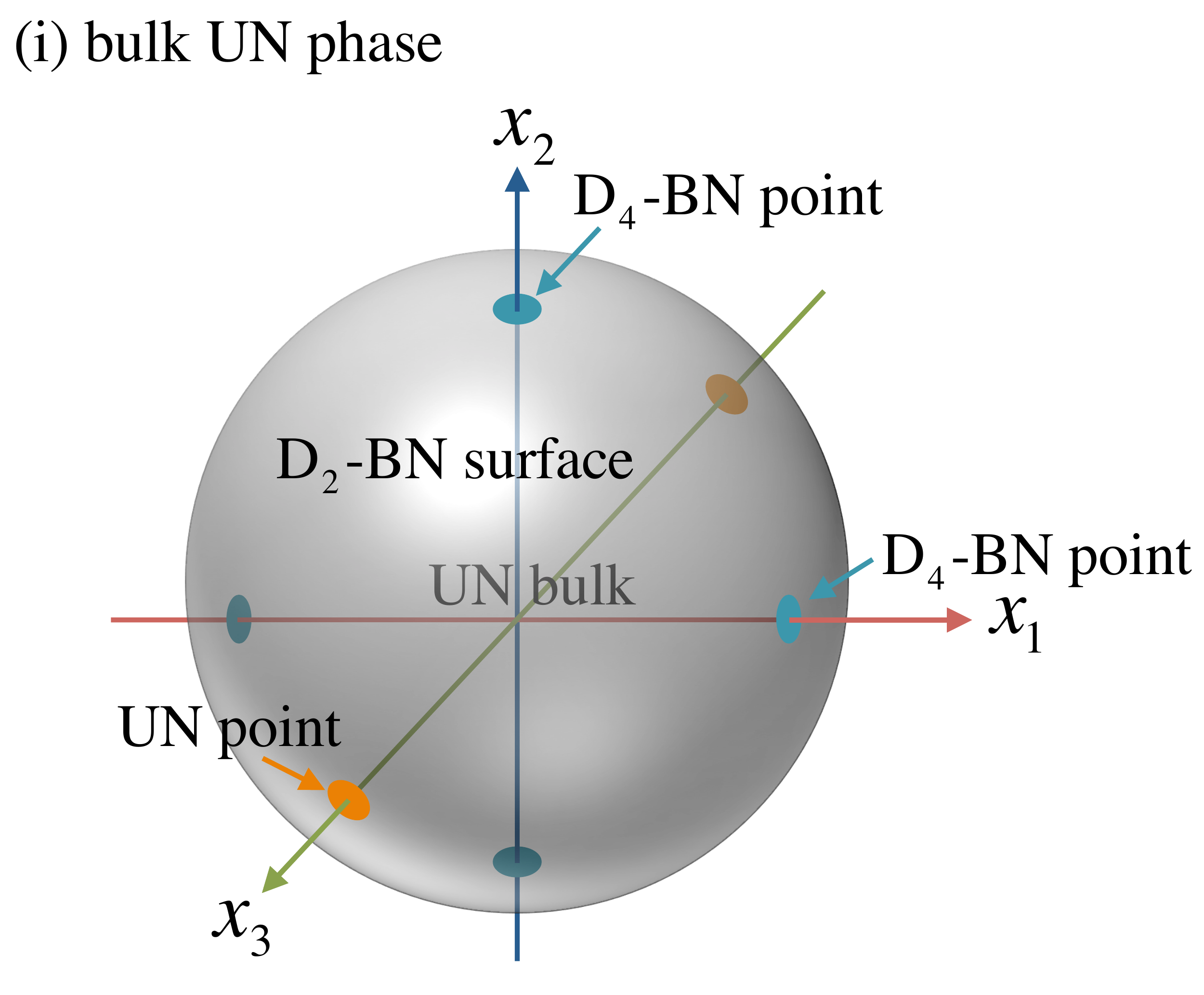}
\hspace{0.5em}
\includegraphics[scale=0.22]{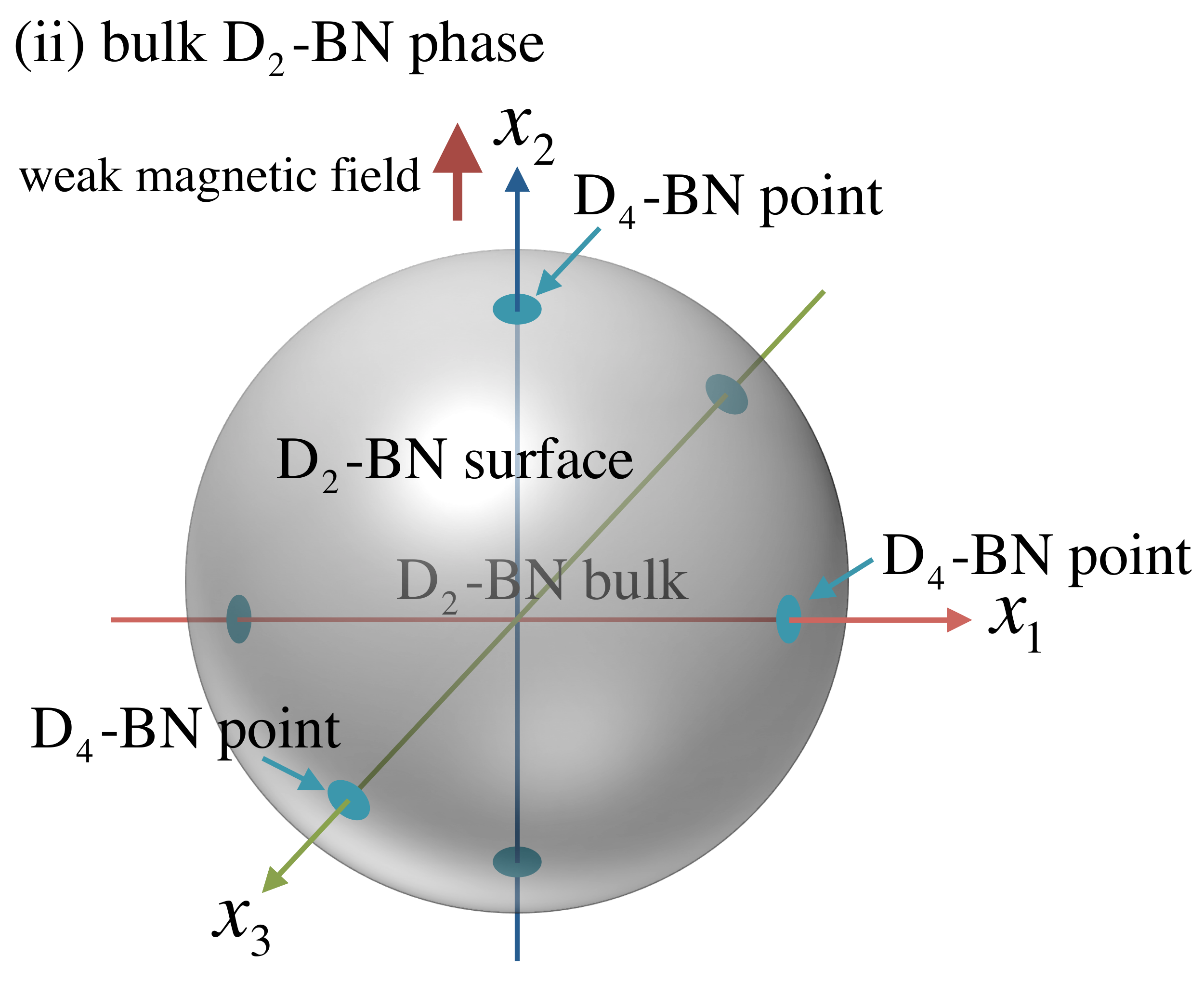}
\hspace{0.5em}
\includegraphics[scale=0.22]{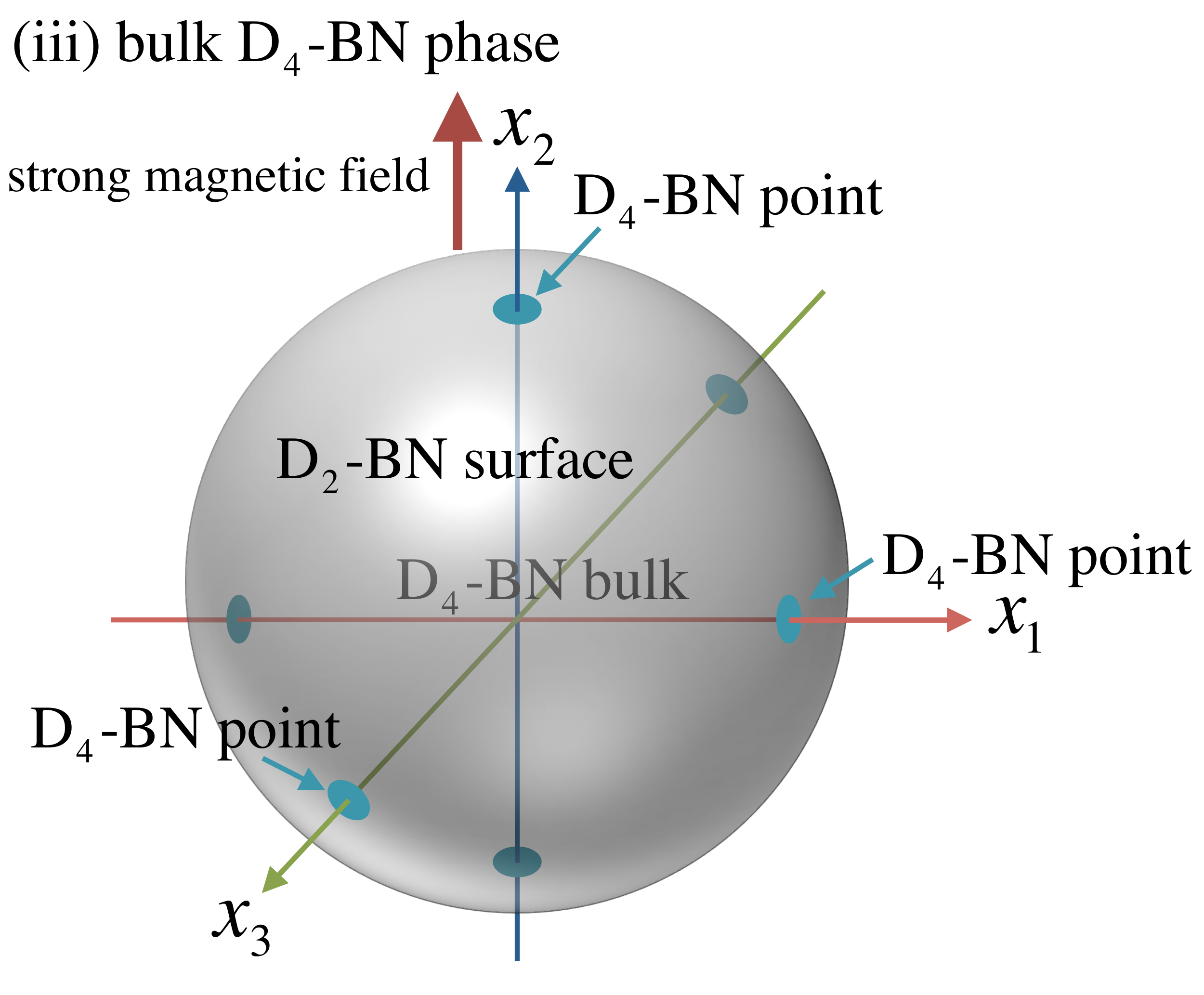}
\caption{The special points on the surface, {\it i.e.}, the UN points and the D$_{4}$-BN points, are shown for several bulk phases: (i) the bulk UN phase, (ii) the bulk D$_{2}$-BN phase, and (iii) the bulk D$_{4}$-BN phase. See the text for more explanation.}
\label{fig:Fig_190520}
\end{center}
\end{figure}

From Figs.~\ref{fig:profile_functions_3P2_bulk1}, \ref{fig:profile_functions_3P2_bulk2}, and \ref{fig:profile_functions_3P2_bulk3} and Figs.~\ref{fig:r_3P2_bulk1}, \ref{fig:r_3P2_bulk2}, and \ref{fig:r_3P2_bulk3},
we notice that the D$_{2}$-BN phase [$-1<r(0)<-1/2$] is realized at most points on the surface.
However, there are special points, $(\theta,\varphi)=(0,0)$, $(\pi/2,0)$, $(\pi/2,\pi/2)$, $(\pi/2,\pi)$, $(\pi/2,3\pi/2)$, and $(\pi,0)$, where either of the UN phase [$r(0)=-1/2$] or the D$_{4}$-BN phase [$r(0)=-1$] is realized.
We call those points the UN points or the D$_{4}$-BN points, and they appear differently for each bulk phase:
\begin{description}
\item[(i) UN phase]
the UN points at $(\theta,\varphi)=(0,0)$, $(\pi,0)$ and the D$_{4}$-BN points at $(\theta,\varphi)=(\pi/2,0)$, $(\pi/2,\pi/2)$, $(\pi/2,\pi)$, $(\pi/2,3\pi/2)$,
\item[(ii) D$_{2}$-BN phase]
the D$_{4}$-BN points at $(\theta,\varphi)=(0,0)$, $(\pi/2,0)$, $(\pi/2,\pi/2)$, $(\pi/2,\pi)$, $(\pi/2,3\pi/2)$, $(\pi,0)$,
\item[(iii) D$_{4}$-BN phase]
the D$_{4}$-BN points at $(\theta,\varphi)=(0,0)$, $(\pi/2,0)$, $(\pi/2,\pi/2)$, $(\pi/2,\pi)$, $(\pi/2,3\pi/2)$, $(\pi,0)$,
\end{description}
as displayed graphically in Fig.~\ref{fig:Fig_190520}.
We notice that the UN points appear only for the bulk UN phase, and not  for the bulk D$_{2}$-BN phase or the bulk D$_{4}$-BN phase.

\subsection{Energy density and topological defects on the surface}

\begin{figure}[!tb]
\begin{center}
\includegraphics[scale=0.185]{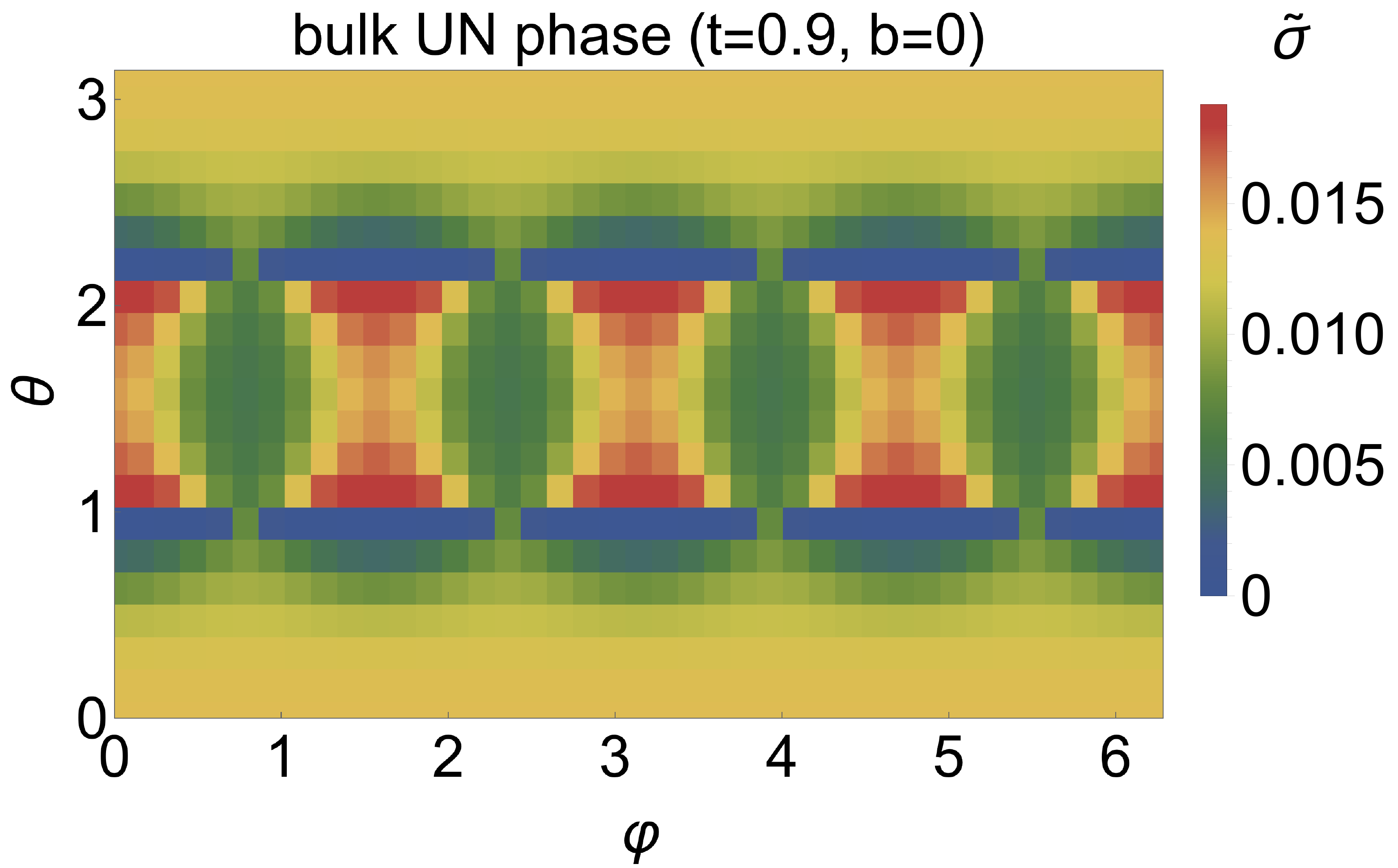}
\includegraphics[scale=0.185]{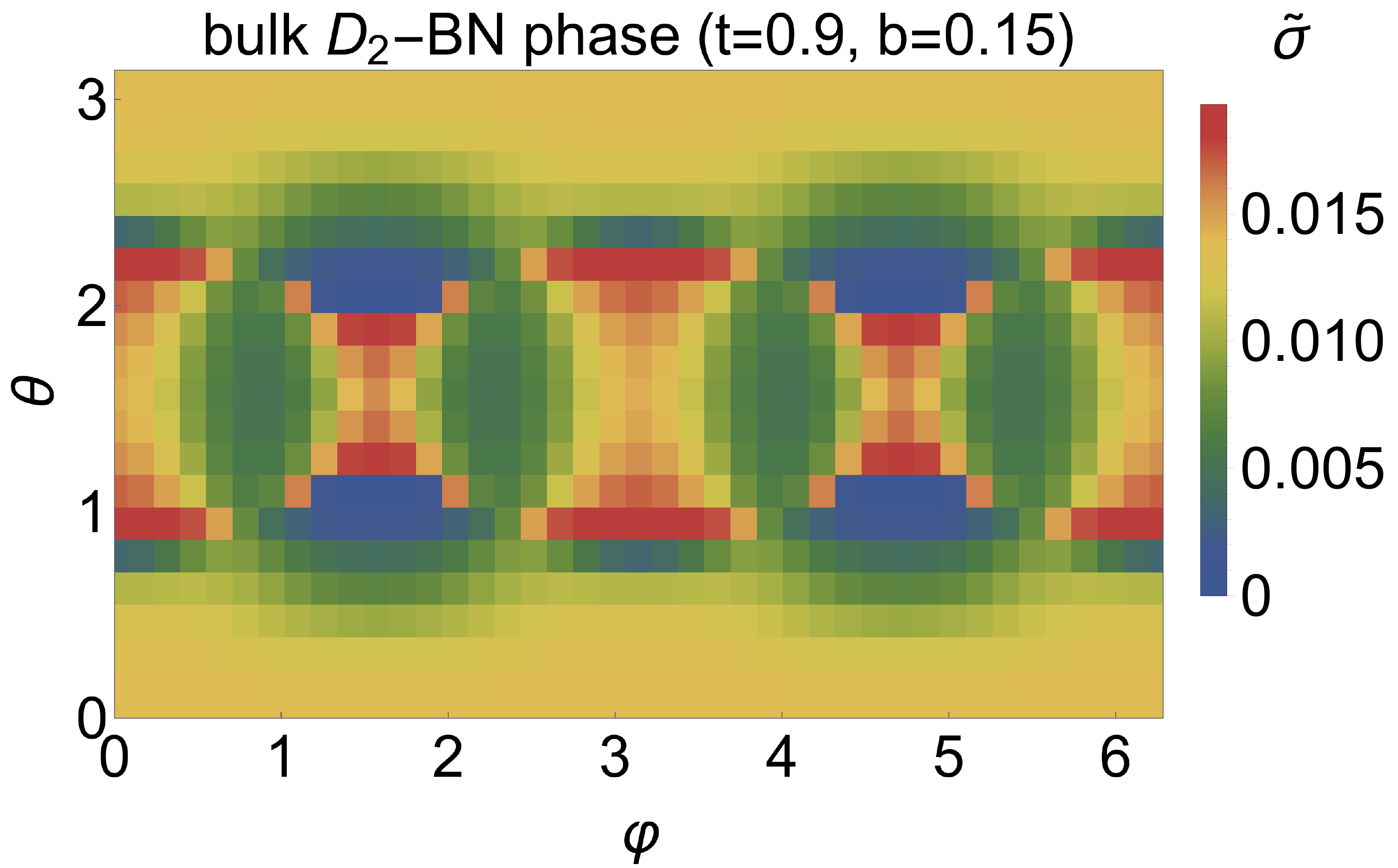}
\includegraphics[scale=0.185]{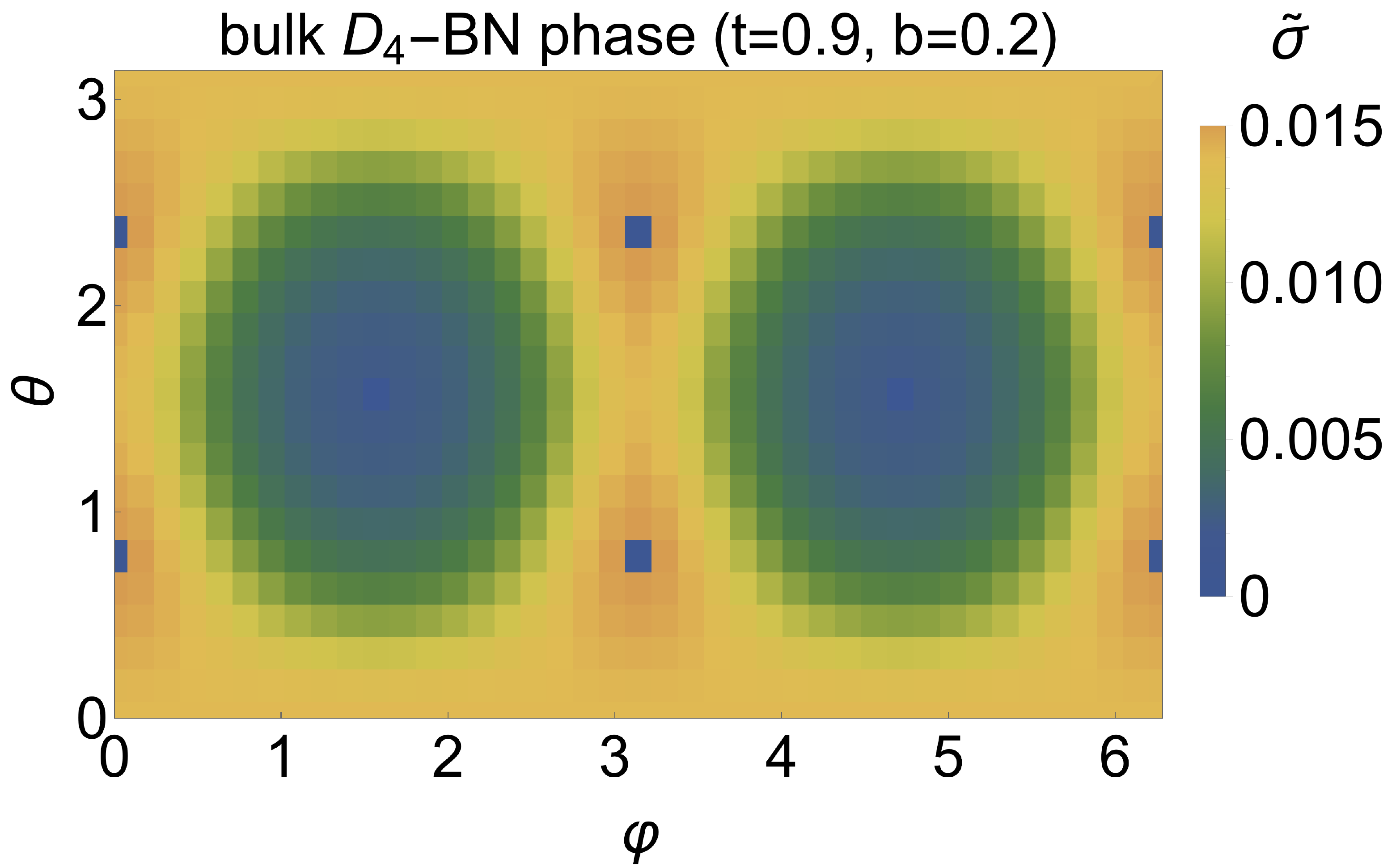}
\\
\includegraphics[scale=0.16]{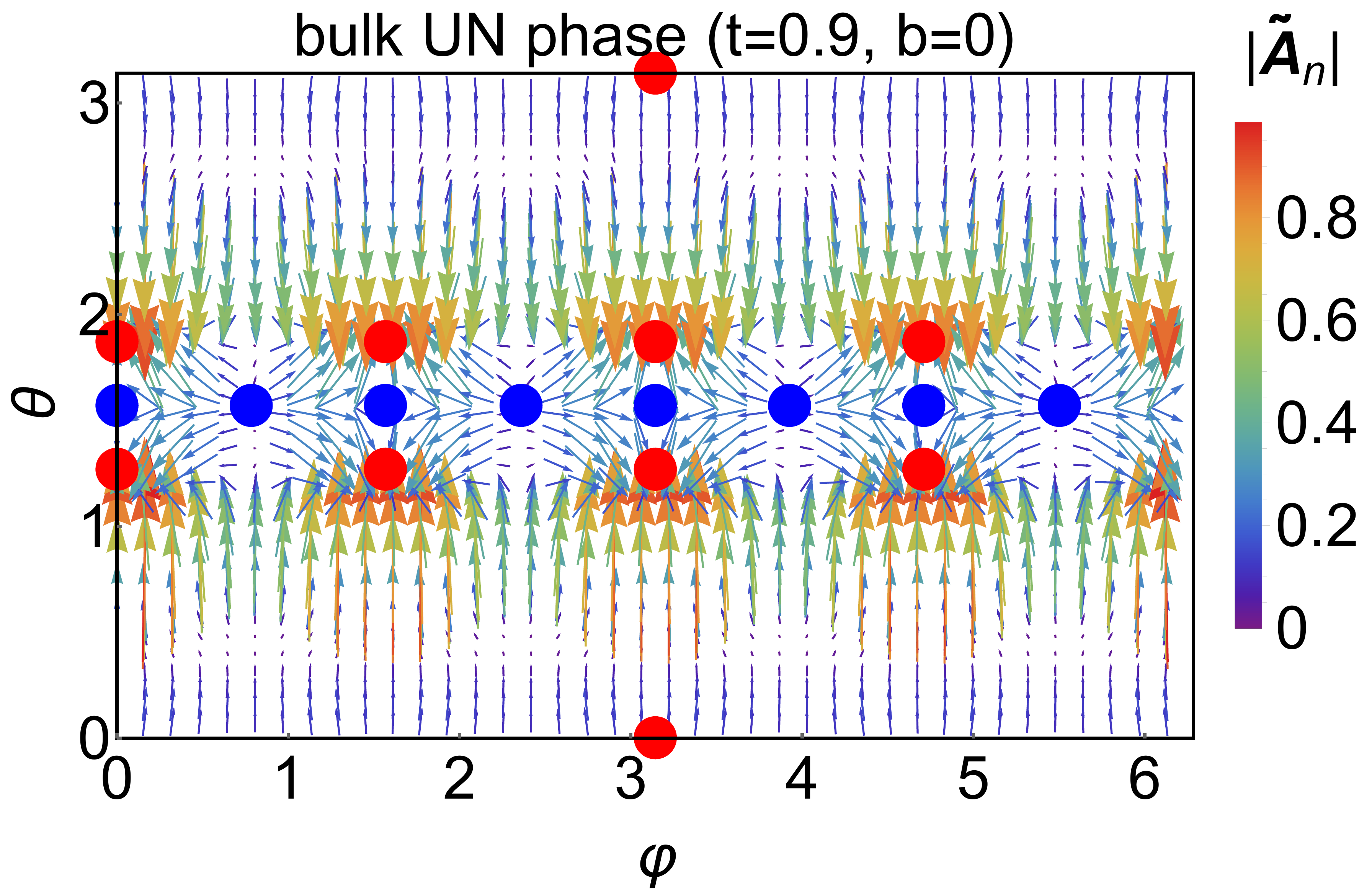}
\includegraphics[scale=0.16]{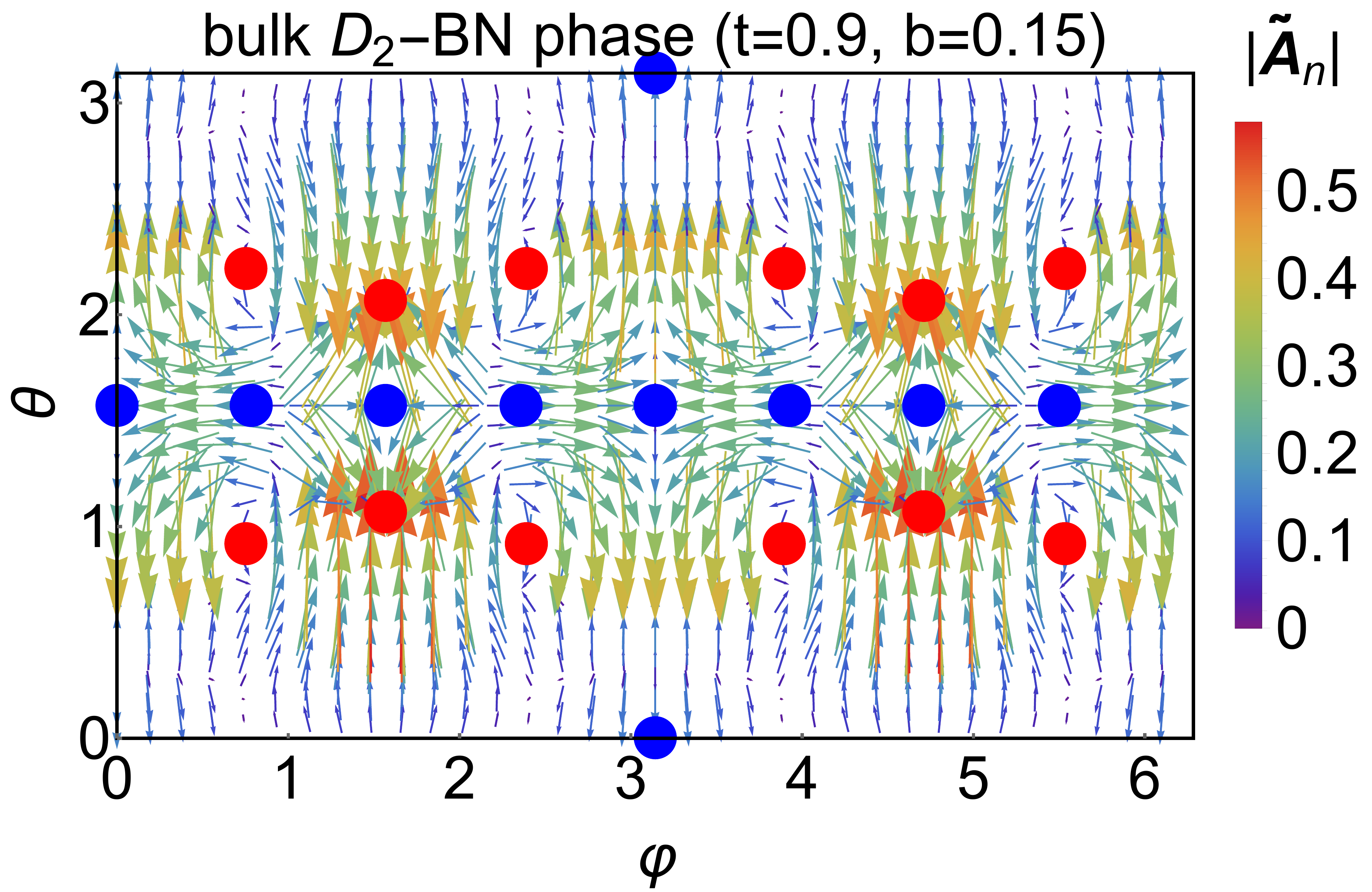}
\includegraphics[scale=0.16]{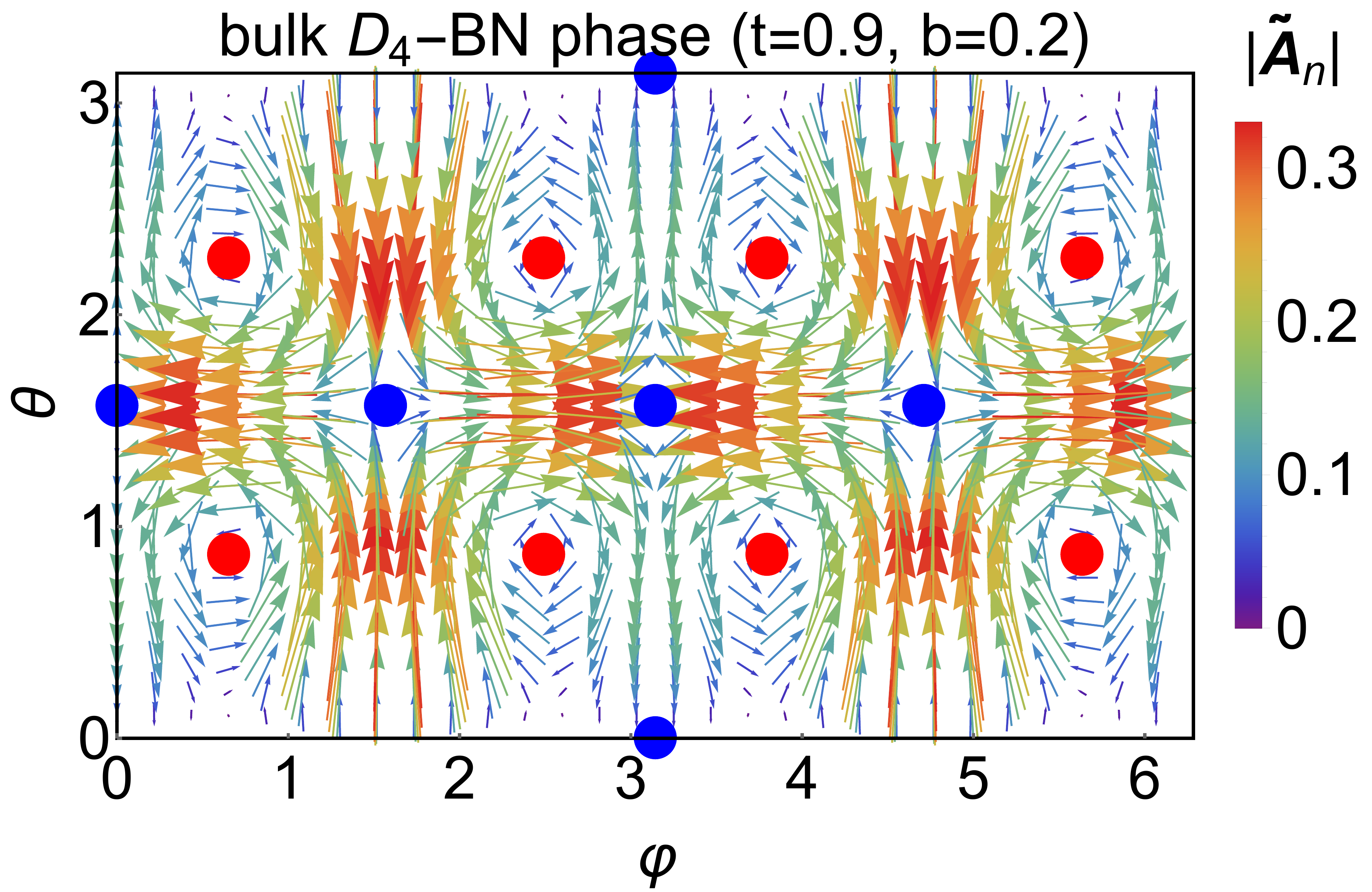}
\caption{Top row: The plots of the dimensionless surface energy density on the surface of the matter in different magnetic fields ($b$) at the temperature $t=0.9$. The region of $\varphi$ is extended to cover $0 \le \varphi < 2\pi$. The first, second, and third columns are for (i) $b=0$ (the UN phase), (ii) $b=0.15$ (the D$_{2}$-BN phase), and (iii) $b=0.2$ (the D$_{4}$-BN phase). Bottom row: We plot $\tilde{\vec{A}}_{n}$ on the $\theta$-$\varphi$ plane. Notice that $\varphi=0$ is identical to $\varphi=2\pi$. The red-filled and blue-filled circles indicate topological defects with changes $+1$ and $-1$, respectively (cf.~Fig.~\ref{fig:Fig_190510}). There are distributed positive and negative charges ($\chi_{\pm}$): +10 and -8 in the UN phase, +12 and -10 in the D$_{2}$-BN phase, and +8 and -6 in the D$_{4}$-BN phase.} 
\label{fig:surface_vector_3P2_modified_v2}
\end{center}
\end{figure}

We consider the surface energy density,
which is defined by the energy density per unit area on the surface.
The surface energy stems from the gradient term and the difference between the potential value near the surface and the potential value in the bulk space.
With the solutions of $f_{\alpha}(x)$ ($\alpha=1,2$) and $g_{\beta}(x)$ ($\beta=1,2,3$) from the EL equations \eqref{eq:EL_f} and \eqref{eq:EL_g}, we obtain the surface energy density expressed as
\begin{eqnarray}
   \sigma(\vec{n})
=
   \int_{0}^{\infty} \Delta f(d;\vec{n}) \mathrm{d}d
=
   \frac{p_{F}^{2}T_{c0}}{2\pi^{2}}
   \tilde{\sigma}(\vec{n}),
\label{eq:surface_energy_density}
\end{eqnarray}
for $\Delta f(d;\vec{n}) = f(d;\vec{n}) - f_{\bulk}$ with the GL free energy density $f_{\bulk}$ in the bulk space ($d \rightarrow \infty$).
In the preceding equation, for convenience, we have defined the dimensionless surface energy density by
\begin{eqnarray}
   \tilde{\sigma}(\vec{n})
=
   \int_{0}^{\infty} \Delta \tilde{f}(x;\vec{n}) \mathrm{d}x,
\end{eqnarray}
for $\Delta \tilde{f}(x;\vec{n}) = \tilde{f}(x;\vec{n}) - \tilde{f}_{\bulk}$ with  the dimensionless GL free energy density $\tilde{f}_{\bulk}$ in the bulk space ($x \rightarrow \infty$).\footnote{Here we recover $\vec{n}$ to emphasize that the surface energy density depends on the direction of the normal vector.}
In Eq.~\eqref{eq:surface_energy_density}, the value of the coefficient is given by
   ${p_{F}^{2}T_{c0}}/({2\pi^{2}}) \simeq 27$ $\mathrm{keV}/\mathrm{fm}^{2}$
for $p_{F}=338$ MeV and $T_{c0}=0.2$ MeV.
We show the numerical results of the distributions of the surface energy density $\tilde{\sigma}(\vec{n})$ on the plane spanned by $\theta$ and $\varphi$ in the top row in Fig.~\ref{fig:surface_vector_3P2_modified_v2}.
For the bulk D$_{4}$-BN phase, for example, the minimum surface energy density [$\sigma(\vec{n})=0$] is realized at the point $\vec{n}=(0,\pm1,0)$ ($\theta=\pi/2$ and $\varphi=\pi/2$, $3\pi/2$).
Notice that zero density is reasonable because there is no change in $f_{\alpha}(x)$ ($\alpha=1,2$) and $g_{\beta}(x)$ ($\beta=1,2,3$), as discussed in the previous section.
In this situation, because the magnetic field is applied along the $x_{2}$ axis,
it is expected that the shape of the surface of the neutron star should be deformed to be an oblate spheroid with the short axis being aligned along the $x_{2}$ axis when the surface shape can be changed to be in balance with the Fermi pressure and the gravity.
In Fig.~\ref{fig:surface_vector_3P2_modified_v2},
we observe that the geometrical distribution of the surface energy density on the sphere obeys the dihedral symmetries as a subgroup of the spherical symmetry: the D$_{4}$ symmetry for the bulk UN phase, the D$_{2}$ symmetry for the bulk D$_{2}$-BN phase, and the D$_{4}$ symmetry for the D$_{4}$-BN phase, as summarized in Table~\ref{table:sigma_average}.

We estimate the averaged values of the surface energy density.
For this purpose, we define the averaged values of $\sigma(\vec{n})$ over $\theta$ and $\varphi$ by
$\sigma_{\mathrm{av}}\equiv({p_{F}^{2}T_{c0}}/({2\pi^{2}}))\tilde{\sigma}_{\mathrm{av}}$ and
\begin{eqnarray}
   \tilde{\sigma}_{\mathrm{av}} \equiv \frac{1}{4\pi} \int_{0}^{\pi} \drm \theta \, \sin\theta \int_{0}^{2\pi} \drm \varphi \, \tilde{\sigma}(\vec{n}).
\end{eqnarray}
From the values shown in Table~\ref{table:sigma_average}, they are almost close to $\sigma_{\mathrm{av}} \approx 0.2$ keV/fm$^{2}$ for each bulk phase.
Among them, the minimum surface energy is provided by the bulk D$_{4}$-BN phase.
With the values of $\sigma_{\mathrm{av}}$, we obtain the surface energy in total, $E_{\mathrm{surf}}=4\pi R^{2} \sigma_{\mathrm{av}}$, for the neutron star with the radius $R$.
Numerically, we obtain $E_{\mathrm{surf}} \approx 6 \times 10^{29}$ erg for $R=10$ km.
The values of $E_{\mathrm{surf}}$ for each bulk phase are summarized in Table~\ref{table:sigma_average}.

\begin{table}[!tb]
\begin{center}
\begin{tabular}{|c|c|c|c|}
\hline
 bulk phase & UN & D$_{2}$-BN & D$_{4}$-BN \\
\hline \hline
 $\tilde{\sigma}_{\mathrm{av}}$ & 0.0095 & 0.0099 & 0.0085 \\  
\hline
 $\sigma_{\mathrm{av}}$ [keV/fm$^{2}$] & 0.26 & 0.27 & 0.23 \\  
\hline
 $E_{\mathrm{surf}}$ [erg] & $6.5\times10^{29}$ & $6.8\times10^{29}$ & $5.8\times10^{29}$ \\  
\hline
 geometrical sym. & D$_{4}$ & D$_{2}$ & D$_{4}$ \\
\hline
\end{tabular}
\end{center}
\caption{The averaged (dimensionless) surface energy density $\sigma_{\mathrm{av}}$ ($\tilde{\sigma}_{\mathrm{av}}$) for the bulk phases (the UN phase, the D$_{2}$-BN phase, and the D$_{4}$-BN phase). $E_{\mathrm{surf}}=4\pi R^{2} \sigma_{\mathrm{av}}$ is the surface energy in total for a neutron star with the radius $R=10$ km. The energy unit conversion $1 \, \mathrm{keV} \approx 2.0 \times 10^{-9} \, \mathrm{erg}$ is used. The last row indicates the geometrical symmetries for the spatial distribution of the energy density on the surface.}
\label{table:sigma_average}
\end{table}%

The directions of the condensate $A$ at the surface have unique topological properties.
In order to see this, 
we define the vector $\vec{A}_{n}(\theta,\varphi) \equiv A(0) \vec{n}$ with the normal vector $\vec{n}$ at the surface [$A(0)$ is a $3\times3$ matrix].

There are two remarks:
First, we remember that the $d$ vector is defined by $A \hat{\vec{p}}$ with $\hat{\vec{p}}=\vec{p}/|\vec{p}|$ for the three-dimensional momentum $\vec{p}$.
Because the axis direction perpendicular to the surface is considered to be a one-dimensional system, we reasonably regard $\hat{\vec{p}}=\vec{n}$, and hence conclude that $\vec{A}_{n}(\theta,\varphi)$ [$\tilde{\vec{A}}_{n}(\theta,\varphi)$] is the same as the $d$ vector at the surface.

\begin{figure}[!tb]
\begin{center}
\includegraphics[scale=0.3]{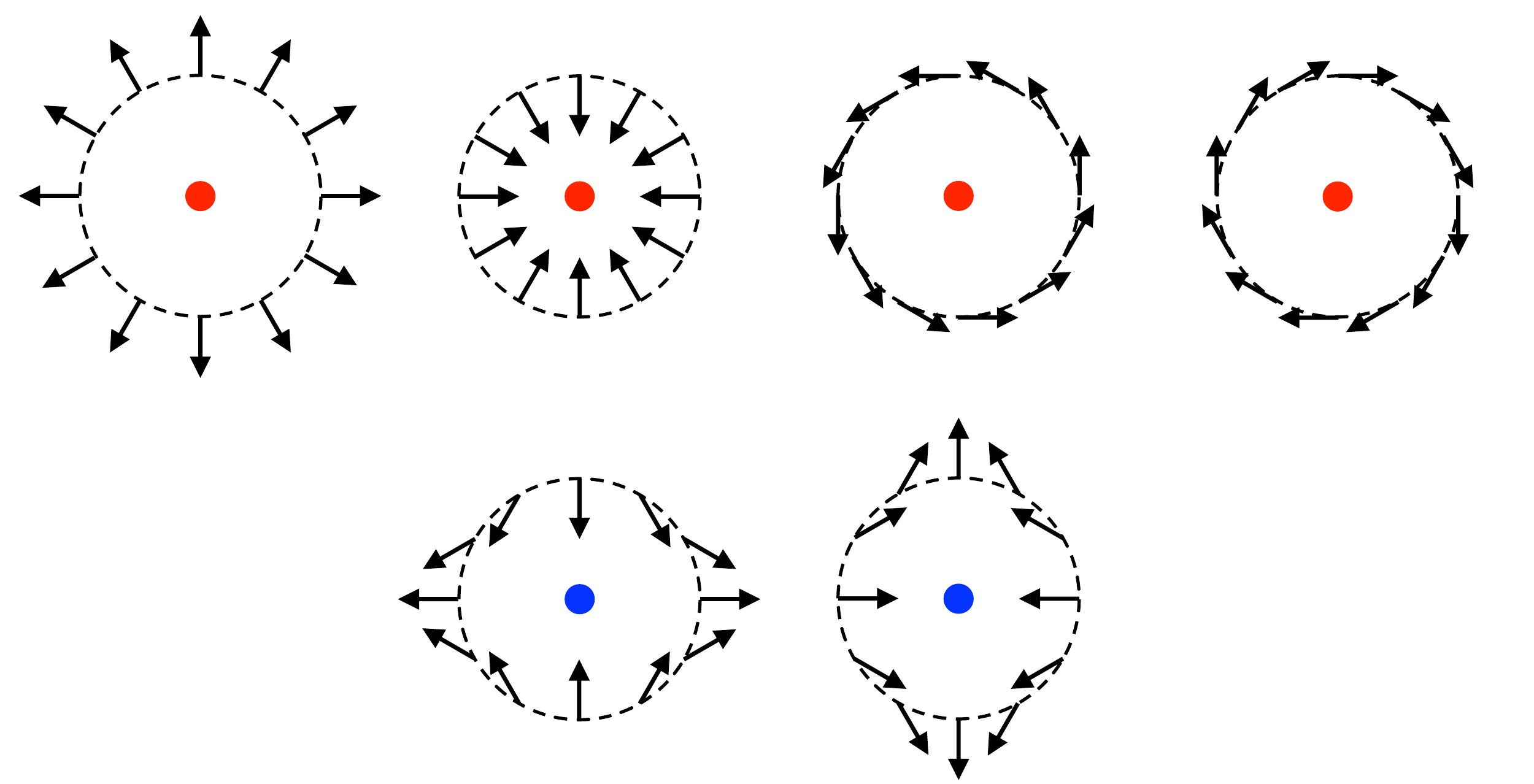}
\caption{The charges of the defects (vortices) on the two-dimensional plane are presented. The red-filled circle (first row) indicates $+1$ charge and the blue-filled circle (second row) indicates $-1$ charge. See also the bottom row in Fig.~\ref{fig:surface_vector_3P2_modified_v2}.} 
\label{fig:Fig_190510}
\end{center}
\end{figure}

Second, we also comment that, more precisely, the vector field $\vec{A}_{n}$ 
parametrizes an order parameter space reduced at the boundary. 
Namely, at the boundary, 
the order parameter space is reduced due to the boundary condition 
to its submanifold,
\begin{eqnarray}
 M_{\rm red} \simeq {S^{1} \times S^{1} \over \mathbb{Z}_{2}},
\end{eqnarray}
where two $S^{1}$'s denote the U(1) phase and the spatial rotation around $\vec{n}$, and $\mathbb{Z}_{2}=\{1,-1\}$ is introduced  to remove identical two points;
a simultaneous transformation of a $\pi$ phase rotation and a $\pi$ spatial rotation around $\vec{n}$ does not change the order parameters and should be removed.
Since we restrict $A$ to be real-valued, $\vec{A}_{n}(\theta,\varphi)$ is a real vector 
parametrizing one $S^1$, and
$\vec{A}_{n}(\theta,\varphi)$ is a two-dimensional vector on the plane orthogonal to the normal vector $\vec{n}$ 
because $\vec{n}\!\cdot\!\vec{A}_{n}(\theta,\varphi)=0$ is induced from the boundary condition \eqref{eq:boundary_condition_1}.
This two-dimensional vector can be expressed by $\vec{A}_{n}(\theta,\varphi)=(T_{c0}/p_{F}) \tilde{\vec{A}}_{n}(\theta,\varphi)$ with
\begin{eqnarray}
   \tilde{\vec{A}}_{n}(\theta,\varphi)
= \tilde{A}_{\theta}(\theta,\varphi) \, \vec{n}_{\theta} + \tilde{A}_{\varphi}(\theta,\varphi) \, \vec{n}_{\varphi},
\end{eqnarray}
for $\vec{n}_{\theta}$ and $\vec{n}_{\varphi}$ defined by 
$\vec{n}_{\theta} \equiv \partial \vec{n} / \partial \theta$ and $\vec{n}_{\varphi} \equiv \partial \vec{n} / \partial \varphi$, respectively.
Notice that $\vec{n}_{\theta}$ and $\vec{n}_{\varphi}$ are the unit vectors perpendicular to each other and also to $\vec{n}$, {\it i.e.}, $\vec{n}\!\cdot\!\vec{n}_{\theta}=\vec{n}\!\cdot\!\vec{n}_{\varphi}=\vec{n}_{\theta}\!\cdot\!\vec{n}_{\varphi}=0$.
We show a plot of
 $\tilde{\vec{A}}_{n}=(\tilde{A}_{\theta},\tilde{A}_{\varphi})$ on the plane spanned by $\theta$ and $\varphi$ ($0 \le \theta < \pi$ and $0 \le \varphi < 2\pi$) in the bottom row in Fig.~\ref{fig:surface_vector_3P2_modified_v2}.
Interestingly, we observe that there exist topological defects (vortices) on the plane.
Each defect has a positive or negative charge $\pm1$.
The charges of the defects are defined according to the circulating directions of the vector fields as shown in Fig.~\ref{fig:Fig_190510}.
We denote the total positive (negative) charge by $\chi_{+}$ ($\chi_{-}$).
In Table~\ref{table:charge_surface}, we summarize the values of $\chi_{+}$ and $\chi_{-}$ for each bulk phase:
$\chi_{+}=10$ and $\chi_{-}=-8$ for the bulk UN phase,
$\chi_{+}=12$ and $\chi_{-}=-10$ for the bulk D$_{2}$-BN phase, and
$\chi_{+}=8$ and $\chi_{-}=-6$ for the bulk D$_{4}$-BN phase.
We define $\chi=\chi_{+}+\chi_{-}$ to denote the sum of $\chi_{+}$ and $\chi_{-}$, which is the difference between the numbers of the defects with a charge $+1$ and charge $-1$.
Importantly, we find that $\chi =2$ is realized uniquely for all the different bulk phases.
This is an inevitable consequence of a topological property of two-dimensional vector fields on a sphere, {\it i.e.}, the Poincar\'{e}-Hopf (hairy ball, no-wind) theorem.
We consider a vector field $\vec{v}$ with isolated zeros on a two-dimensional manifold ${\cal M}$.
Then, the Poincar\'{e}-Hopf theorem requires the relation
\begin{eqnarray}
 \sum_{i} \mathrm{index}_{x_{i}}(\vec{v}) = \chi({\cal M}),
\end{eqnarray}
where $\mathrm{index}_{x_{i}}(\vec{v})$ is the index ($\pm1$ charges) of $\vec{v}$ at the position $x_{i}$ of isolated zeros on ${\cal M}$, $\Sigma_{i}$ indicates a sum over all the isolated zeros ($i$), and $\chi({\cal M})$ is the Euler characteristic of ${\cal M}$.
$\chi({\cal M})=2$ if ${\cal M}$ is a sphere $S^2$ relevant for us.
For example, for the bulk UN phase,
we have $\sum_{i} \mathrm{index}_{x_{i}}(\vec{A}_{n})=8+(-6)=2$, which is indeed identical to $\chi=2$.
This relation is checked for the bulk D$_{2}$-BN phase and for the bulk D$_{4}$-BN phase.
Therefore, we understand that $\chi=2$ should hold for any bulk phases.
On the other hand, $\chi_{+}$ and $\chi_{-}$ can vary according to the symmetries in the bulk phases.
In order for $\chi=2$ to hold,
creations or annihilations should happen for a pair of vortices with charges $\pm1$.
From Table~\ref{table:charge_surface}, we see that two pairs of vortices are created from the bulk UN phase to the bulk D$_{2}$-BN phase,
and that four pairs of vortices are annihilated from the bulk D$_{2}$-BN phase to the bulk D$_{4}$-BN phase.

\begin{table}[!tb]
\begin{center}
\begin{tabular}{|c|c|c|c|}
\hline
 bulk phase & UN & D$_{2}$-BN & D$_{4}$-BN \\
\hline \hline
 positive charge ($\chi_{+}$) & 10 & 12 & 8 \\  
\hline
 negative charge ($\chi_{-}$) & -8 & -10 & -6 \\  
\hline
 total charge ($\chi$) & 2 & 2 & 2 \\  
\hline
\end{tabular}
\end{center}
\caption{The positive and negative charges $\chi_{\pm}$ for the vortices on the plane for the bulk phases (the UN phase, the D$_{2}$-BN phase, and the D$_{4}$-BN phase). We define $\chi=\chi_{+}+\chi_{-}$ to indicate the sum of the positive charges and the negative charges in total. See also the bottom row in Fig.~\ref{fig:surface_vector_3P2_modified_v2}.}
\label{table:charge_surface}
\end{table}%

The final comment is that vortices induced on a boundary are called boojums in the context of liquid crystals
\cite{Urbanski_2017} and the $^{3}$He superfluids~\cite{vollhardt2013superfluid,Volovik:2003fe}. 
It is interesting that these boojums do not extend to the bulk since closed paths winding around $M_{\rm red}$ become trivial loops in the full order parameter space $M$, 
where $M_{\rm red} \subset M$.
We notice that there recently appeared a study on the surface vortices for spinor Bose-Einstein condensates (BECs) with spin-orbital-angular momentum coupling~\cite{cheng2019topological}. The presented results indicate that there are interesting common properties, such as the symmetry on surface, the distribution of topological defects, and so on, as we have discussed in our study. Thus, it would be valuable to investigate further  similarities as well as differences between the neutron $^{3}P_{2}$ superfluids and spinor BECs.

\section{Conclusion and perspectives}

We have studied surface effects of neutron $^{3}P_{2}$ superfluidity in the UN, D$_{2}$-BN, and D$_{4}$-BN phases in neutron stars.
We have supposed the situation in which the neutron $^{3}P_{2}$ superfluid exists in a large ball with a spherical boundary, and introduced a boundary condition suitable for the condensate at the sphere's surface.
Solving the GL equation with the boundary condition,
we have found several interesting properties of the surface effects of the neutron $^{3}P_{2}$ superfluid.
First, we have shown that the symmetry in the bulk space can be restored or broken to other symmetries at the surface.
Second, the distribution of the surface energy density on the sphere has an anisotropy depending on the polar angle.
This will lead to the geometrical deformation of the surface of the neutron $^{3}P_{2}$ superfluidity from the spherical shape to an oblate spheroid.
Third, we have investigated the two-dimensional vector field defined from the condensate at the surface, and have shown that there must exist topological defects (vortices) with the charges $\pm1$ on the sphere.
It should be emphasized that those defects appear in the ground state but not as an excited state.
While the number of the defects (vortices) is dependent on the symmetries in the bulk phases, {\it i.e.}, the UN, D$_{2}$-BN, and D$_{4}$-BN phases,
the difference between the numbers of the defects with a charge $+1$ and charge $-1$ remains topologically invariant ($\chi=2$) thanks to the Poincar\'{e}-Hopf theorem. 
These vortices are called boojums in the context of liquid crystals and $^{3}$He superfluids.

In the present study, we have considered the simplest situation for the boundary condition for neutron $^{3}P_{2}$ superfluidity.
In a more realistic situation, however, the neutron $^{3}P_{2}$ superfluid can interface with a neutron 
$^{1}S_{0}$ superfluid as well as with nuclear crusts composed of a lattice of neutron-rich nuclei.
Deeper inside in neutron stars, a connection of the neutron $^{3}P_{2}$ phase to other exotic phases such as hyperon matter, quark matter, and so on should be interesting. 
One of the important questions to ask is how the topological objects are connected between those phases.
When a neutron star rotates, Abelian quantum vortices in the hadron matter and 
non-Abelian quantum vortices (color magnetic flux tubes)
\cite{Balachandran:2005ev,Nakano:2007dr,Eto:2009kg}
in the quark matter can be connected through colorful boojums (endpoints of vortices)~\cite{Cipriani:2012hr,Alford:2018mqj,Chatterjee:2018nxe,Cherman:2018jir} (see Ref.~\cite{Eto:2013hoa} for a review of non-Abelian quantum vortices).
These boojums are different from those studied in this paper.
Although we have limited ourselves to the BN, D$_{2}$-BN, and D$_{4}$-BN phases for the neutron $^{3}P_{2}$ superfluid,
it will be also important to study the possibilities of cyclic and ferromagnetic phases~\cite{Mermin:1974zz,2010arXiv1001.2072K}.
The former phase can lead to one-third quantized non-Abelian vortices~\cite{Semenoff:2006vv} 
forming a network in collision~\cite{Kobayashi:2008pk}, while  
the latter phase could be relevant to inner structures of magnetars.
In terms of the topological matter, the cyclic and ferromagnetic phases have gapless Weyl fermions in the bulk (Weyl semimetals)~\cite{Mizushima:2016fbn,Mizushima:2017pma}.
The above mentioned subjects should be studied carefully in the future.

\section*{Acknowledgment}
We would like to thank Michikazu Kobayashi and Takeshi Mizushima for discussions.
This work is supported by the Ministry of Education, Culture, Sports, Science (MEXT) Supported Program for the Strategic Research Foundation at Private Universities ``Topological Science" (Grant No. S1511006). 
This work is also supported in part by 
JSPS Grants-in-Aid for Scientific Research [KAKENHI Grant No.~17K05435 (S.Y.), 
No.~19K14713 (C.C.),
No.~16H03984 (M.N.), and No.~18H01217 (M.N.)], 
and also by MEXT KAKENHI Grant-in-Aid for Scientific Research on Innovative Areas ``Topological Materials Science'' No.~15H05855 (M.N.).

\appendix

\section{Euler-Lagrange equations}
\label{sec:EL_appendix}

We present the concrete expressions of the left-hand sides in the EL equations \eqref{eq:EL_f} and \eqref{eq:EL_g}:
\begin{eqnarray}
&&
   - \nabla_{\!d} \frac{\delta {f}}{\delta (\nabla_{\!d}F_{1})}
    + \frac{\delta {f}}{\delta F_{1}}
\nonumber \\ 
&=&
   -\frac{{K}^{(0)}}{4}
   \Bigl(
      2\bigl( 2 - \sin^{2}\theta \, \sin^{2}\varphi \bigr) \nabla_{\!d}^{2}
      F_{1}
   + \bigl( 1 +2 \cos^{2}\theta \bigr)\nabla_{\!d}^{2}
      F_{2}
   + 2
      \cos\theta\sin\theta\sin\varphi
      \nabla_{\!d}^{2}
      G_{1}
    - 2
      \sin^{2}\theta\cos\varphi\sin\varphi
      \nabla_{\!d}^{2}
      G_{3}
   \Bigr)
\nonumber \\ && 
+
   {\alpha}^{(0)} 2 (2 F_{1}+F_{2})
   \nonumber \\ && 
+ {\beta}^{(0)} 4
   (2 F_{1}+F_{2})
   \bigl(
         F_{1}^{2}+F_{1}F_{2}+F_{2}^{2}+G_{1}^{2}+G_{2}^{2}+G_{3}^{2}
   \bigr)
   \nonumber \\ && 
+ {\gamma}^{(0)}24
   \Bigl(
         12 F_{1}^{5}
      + 30 F_{2} F_{1}^{4}
      + 52 F_{2}^{2} F_{1}^{3}
      + 24 G_{1}^{2} F_{1}^{3}
      + 24 G_{2}^{2} F_{1}^{3}
      + 24 G_{3}^{2} F_{1}^{3}
      + 48 F_{2}^{3} F_{1}^{2}
      + 42 F_{2} G_{1}^{2} F_{1}^{2}
      + 36 F_{2} G_{2}^{2} F_{1}^{2}
         \nonumber \\ && \hspace{4em} 
      + 30 F_{2} G_{3}^{2} F_{1}^{2}
      + 26 F_{2}^{4} F_{1}
      + 14 G_{1}^{4} F_{1}
      + 12 G_{2}^{4} F_{1}
      + 14 G_{3}^{4} F_{1}
      + 40 F_{2}^{2} G_{1}^{2} F_{1}
      + 40 F_{2}^{2} G_{2}^{2} F_{1}
      + 24 G_{1}^{2} G_{2}^{2} F_{1}
      + 28 F_{2}^{2} G_{3}^{2} F_{1}
         \nonumber \\ && \hspace{4em} 
      + 20 G_{1}^{2} G_{3}^{2} F_{1}
      + 24 G_{2}^{2} G_{3}^{2} F_{1}
      + 8 F_{2} G_{1} G_{2} G_{3} F_{1}
      + 6 F_{2}^{5}
      + 6 F_{2} G_{1}^{4}
      + 6 F_{2} G_{2}^{4}
      + 8 F_{2} G_{3}^{4}
       - 4 G_{1} G_{2} G_{3}^{3}
      + 12 F_{2}^{3} G_{1}^{2}
         \nonumber \\ && \hspace{4em} 
      + 14 F_{2}^{3} G_{2}^{2}
      + 14 F_{2} G_{1}^{2} G_{2}^{2}
      + 10 F_{2}^{3} G_{3}^{2}
      + 10 F_{2} G_{1}^{2} G_{3}^{2}
      + 10 F_{2} G_{2}^{2} G_{3}^{2}
      + 4 G_{1}^{3} G_{2} G_{3}
      + 4 F_{2}^{2} G_{1} G_{2} G_{3}
      \Bigr)
   \nonumber \\ && 
+ {\delta}^{(0)} 192
   \Bigl(
         16 F_{1}^{7}
      + 56 F_{2} F_{1}^{6}
      + 138 F_{2}^{2} F_{1}^{5}
      + 48 G_{1}^{2} F_{1}^{5}
      + 48 G_{2}^{2} F_{1}^{5}
      + 48 G_{3}^{2} F_{1}^{5}
      + 205 F_{2}^{3} F_{1}^{4}
      + 150 F_{2} G_{1}^{2} F_{1}^{4}
      + 120 F_{2} G_{2}^{2} F_{1}^{4}
         \nonumber \\ && \hspace{4em} 
      + 90 F_{2} G_{3}^{2} F_{1}^{4}
      + 200 F_{2}^{4} F_{1}^{3}
      + 60 G_{1}^{4} F_{1}^{3}
      + 48 G_{2}^{4} F_{1}^{3}
      + 60 G_{3}^{4} F_{1}^{3}
      + 252 F_{2}^{2} G_{1}^{2} F_{1}^{3}
      + 228 F_{2}^{2} G_{2}^{2} F_{1}^{3}
      + 96 G_{1}^{2} G_{2}^{2} F_{1}^{3}
         \nonumber \\ && \hspace{4em} 
      + 132 F_{2}^{2} G_{3}^{2} F_{1}^{3}
      + 72 G_{1}^{2} G_{3}^{2} F_{1}^{3}
      + 96 G_{2}^{2} G_{3}^{2} F_{1}^{3}
      + 48 F_{2} G_{1} G_{2} G_{3} F_{1}^{3}
      + 123 F_{2}^{5} F_{1}^{2}
      + 99 F_{2} G_{1}^{4} F_{1}^{2}
      + 72 F_{2} G_{2}^{4} F_{1}^{2}
         \nonumber \\ && \hspace{4em} 
      + 81 F_{2} G_{3}^{4} F_{1}^{2}
       - 36 G_{1} G_{2} G_{3}^{3} F_{1}^{2}
      + 222 F_{2}^{3} G_{1}^{2} F_{1}^{2}
       + 222 F_{2}^{3} G_{2}^{2} F_{1}^{2}
       + 180 F_{2} G_{1}^{2} G_{2}^{2} F_{1}^{2}
       + 114 F_{2}^{3} G_{3}^{2} F_{1}^{2}
       + 108 F_{2} G_{1}^{2} G_{3}^{2} F_{1}^{2}
         \nonumber \\ && \hspace{4em} 
       + 108 F_{2} G_{2}^{2} G_{3}^{2} F_{1}^{2}
       + 36 G_{1}^{3} G_{2} G_{3} F_{1}^{2}
       + 72 F_{2}^{2} G_{1} G_{2} G_{3} F_{1}^{2}
       + 46 F_{2}^{6} F_{1}
       + 22 G_{1}^{6} F_{1}
       + 16 G_{2}^{6} F_{1}
       + 22 G_{3}^{6} F_{1}
       + 90 F_{2}^{2} G_{1}^{4} F_{1}
         \nonumber \\ && \hspace{4em} 
       + 90 F_{2}^{2} G_{2}^{4} F_{1}
       + 48 G_{1}^{2} G_{2}^{4} F_{1}
       + 72 F_{2}^{2} G_{3}^{4} F_{1}
       + 42 G_{1}^{2} G_{3}^{4} F_{1}
       + 54 G_{2}^{2} G_{3}^{4} F_{1}
        - 24 F_{2} G_{1} G_{2} G_{3}^{3} F_{1}
       + 114 F_{2}^{4} G_{1}^{2} F_{1}
         \nonumber \\ && \hspace{4em} 
       + 126 F_{2}^{4} G_{2}^{2} F_{1}
       + 54 G_{1}^{4} G_{2}^{2} F_{1}
       + 180 F_{2}^{2} G_{1}^{2} G_{2}^{2} F_{1}
       + 66 F_{2}^{4} G_{3}^{2} F_{1}
       + 42 G_{1}^{4} G_{3}^{2} F_{1}
       + 48 G_{2}^{4} G_{3}^{2} F_{1}
       + 108 F_{2}^{2} G_{1}^{2} G_{3}^{2} F_{1}
         \nonumber \\ && \hspace{4em} 
       + 108 F_{2}^{2} G_{2}^{2} G_{3}^{2} F_{1}
       + 108 G_{1}^{2} G_{2}^{2} G_{3}^{2} F_{1}
       + 48 F_{2} G_{1} G_{2}^{3} G_{3} F_{1}
       + 48 F_{2} G_{1}^{3} G_{2} G_{3} F_{1}
       + 48 F_{2}^{3} G_{1} G_{2} G_{3} F_{1}
       + 8 F_{2}^{7}
         \nonumber \\ && \hspace{4em} 
       + 8 F_{2} G_{1}^{6}
       + 8 F_{2} G_{2}^{6}
       + 14 F_{2} G_{3}^{6}
        - 12 G_{1} G_{2} G_{3}^{5}
       + 24 F_{2}^{3} G_{1}^{4}
       + 33 F_{2}^{3} G_{2}^{4}
       + 30 F_{2} G_{1}^{2} G_{2}^{4}
       + 27 F_{2}^{3} G_{3}^{4}
       + 24 F_{2} G_{1}^{2} G_{3}^{4}
         \nonumber \\ && \hspace{4em} 
       + 24 F_{2} G_{2}^{2} G_{3}^{4}
        - 12 G_{1} G_{2}^{3} G_{3}^{3}
        - 12 F_{2}^{2} G_{1} G_{2} G_{3}^{3}
       + 24 F_{2}^{5} G_{1}^{2}
       + 30 F_{2}^{5} G_{2}^{2}
       + 30 F_{2} G_{1}^{4} G_{2}^{2}
       + 60 F_{2}^{3} G_{1}^{2} G_{2}^{2}
       + 18 F_{2}^{5} G_{3}^{2}
         \nonumber \\ && \hspace{4em} 
       + 18 F_{2} G_{1}^{4} G_{3}^{2}
       + 18 F_{2} G_{2}^{4} G_{3}^{2}
       + 36 F_{2}^{3} G_{1}^{2} G_{3}^{2}
       + 36 F_{2}^{3} G_{2}^{2} G_{3}^{2}
       + 54 F_{2} G_{1}^{2} G_{2}^{2} G_{3}^{2}
       + 12 G_{1}^{3} G_{2}^{3} G_{3}
       + 24 F_{2}^{2} G_{1} G_{2}^{3} G_{3}
         \nonumber \\ && \hspace{4em} 
       + 12 G_{1}^{5} G_{2} G_{3}
       + 24 F_{2}^{2} G_{1}^{3} G_{2} G_{3}
       +12 F_{2}^{4} G_{1} G_{2} G_{3}
   \Bigr)
   \nonumber \\ && 
+ {\beta}^{(2)}
   \Bigl(
         2 F_{1} b_{x}^{2}+2 b_{z}^{2} (F_{1}+F_{2})
   \Bigr)
   \nonumber \\ && 
+ {\beta}^{(4)} 
   \Bigl(
         b_{x}^{2}+b_{y}^{2}+b_{z}^{2}
   \Bigr)
   \Bigl(
         2 F_{1} b_{x}^{2}+2 b_{z}^{2} (F_{1}+F_{2})
   \Bigr)
   \nonumber \\ && 
+ {\gamma}^{(2)}
   \Bigl(
       - 12b_{x}^{2}
         \Bigl(
                 8F_{1}^{3}
              + 3F_{1}^{2} F_{2}
              + 2F_{1} F_{2}^{2}
              + 2F_{1} G_{1}^{2}
              + 8F_{1} G_{2}^{2}
              + F_{2} G_{2}^{2}
               - 2G_{1} G_{2} G_{3}
              + 10F_{1} G_{3}^{2}
              + 3F_{2} G_{3}^{2}
         \Bigr)
         \nonumber \\ && \hspace{3em} 
       - 12b_{y}^{2}
         \Bigl(
                 2F_{1} F_{2}^{2}
              + F_{2}^{3}
              + 4F_{1} G_{1}^{2}
              + F_{2} G_{1}^{2}
              + 4F_{1} G_{2}^{2}
              + 3F_{2} G_{3}^{2}
         \Bigr)
         \nonumber \\ && \hspace{3em} 
       - 12b_{z}^{2}
         \Bigl(
                 8F_{1}^{3}
              + 21F_{1}^{2} F_{2}
              + 20F_{1}F_{2}^{2}
              + 7F_{2}^{3}
              + 10F_{1}G_{1}^{2}
              + 7F_{2} G_{1}^{2}
              + 8F_{1} G_{2}^{2}
              + 7F_{2} G_{2}^{2}
              + 2G_{1} G_{2} G_{3}
              + 2F_{1} G_{3}^{2}
              + 2F_{2} G_{3}^{2}
         \Bigr)
   \Bigr),
\nonumber \\
\end{eqnarray}
for $F_{1}$,
\begin{eqnarray}
&&
   - \nabla_{\!d} \frac{\delta {f}}{\delta (\nabla_{\!d}F_{2})}
   + \frac{\delta {f}}{\delta F_{2}}
\nonumber \\ 
&=&
   -\frac{{K}^{(0)}}{4}
   \Bigl(
        \bigl( 1 +2 \cos^{2}\theta \bigr)\nabla_{\!d}^{2}
        F_{1}
     + 2\bigl( 2 - \sin^{2}\theta \, \cos^{2}\varphi \bigr) \nabla_{\!d}^{2}
        F_{2}
      - 2
        \sin^{2}\theta\cos\varphi\sin\varphi
        \nabla_{\!d}^{2}
        G_{3}
     + 2
        \cos\theta\sin\theta\cos\varphi
        \nabla_{\!d}^{2}
        G_{2}
   \Bigr)
\nonumber \\ && 
+
{\alpha}^{(0)}
   2 (F_{1}+2 F_{2}) 
   \nonumber \\ && 
+{\beta}^{(0)}
   4 (F_{1}+2 F_{2})
   \bigl(F_{1}^{2}+F_{2}
   F_{1}+F_{2}^{2}+G_{1}^{2}+G_{2}^{2}+G_{3}^{2}\bigr)
   \nonumber \\ && 
+{\gamma}^{(0)}
   24
   \Bigl(
           6 F_{1}^{5}
         +26 F_{2} F_{1}^{4}
         +48 F_{2}^{2} F_{1}^{3}
         +14 G_{1}^{2} F_{1}^{3}
         +12 G_{2}^{2} F_{1}^{3}
         +10 G_{3}^{2} F_{1}^{3}
         +52 F_{2}^{3} F_{1}^{2}
         +40 F_{2} G_{1}^{2} F_{1}^{2}
         +40 F_{2} G_{2}^{2} F_{1}^{2}
         +28 F_{2} G_{3}^{2} F_{1}^{2}
            \nonumber \\ && \hspace{4em} 
         +4 G_{1} G_{2} G_{3} F_{1}^{2}
         +30 F_{2}^{4} F_{1}
         +6 G_{1}^{4} F_{1}
         +6 G_{2}^{4} F_{1}
         +8 G_{3}^{4} F_{1}
         +36 F_{2}^{2} G_{1}^{2} F_{1}
         +42 F_{2}^{2} G_{2}^{2} F_{1}
         +14 G_{1}^{2} G_{2}^{2} F_{1}
         +30 F_{2}^{2} G_{3}^{2} F_{1}
            \nonumber \\ && \hspace{4em} 
         +10 G_{1}^{2} G_{3}^{2} F_{1}
         +10 G_{2}^{2} G_{3}^{2} F_{1}
         +8 F_{2} G_{1} G_{2} G_{3} F_{1}
         +12 F_{2}^{5}
         +12 F_{2} G_{1}^{4}
         +14 F_{2} G_{2}^{4}
         +14 F_{2} G_{3}^{4}
         -4 G_{1} G_{2} G_{3}^{3}
            \nonumber \\ && \hspace{4em} 
         +24 F_{2}^{3} G_{1}^{2}
         +24 F_{2}^{3} G_{2}^{2}
         +24 F_{2} G_{1}^{2} G_{2}^{2}
         +24 F_{2}^{3} G_{3}^{2}
         +24 F_{2} G_{1}^{2} G_{3}^{2}
         +20 F_{2} G_{2}^{2} G_{3}^{2}
         +4 G_{1} G_{2}^{3} G_{3}
   \Bigr)
   \nonumber \\ && 
+{\delta}^{(0)}
   192
   \Bigl(
           8 F_{1}^{7}
         +46 F_{2} F_{1}^{6}
         +123 F_{2}^{2} F_{1}^{5}
         +30 G_{1}^{2} F_{1}^{5}
         +24 G_{2}^{2} F_{1}^{5}
         +18 G_{3}^{2} F_{1}^{5}
         +200 F_{2}^{3} F_{1}^{4}
         +126 F_{2} G_{1}^{2} F_{1}^{4}
         +114 F_{2} G_{2}^{2} F_{1}^{4}
           \nonumber \\ && \hspace{4em} 
         +66 F_{2} G_{3}^{2} F_{1}^{4}
         +12 G_{1} G_{2} G_{3} F_{1}^{4}
         +205 F_{2}^{4} F_{1}^{3}
         +33 G_{1}^{4} F_{1}^{3}
         +24 G_{2}^{4} F_{1}^{3}
         +27 G_{3}^{4} F_{1}^{3}
         +222 F_{2}^{2} G_{1}^{2} F_{1}^{3}
         +222 F_{2}^{2} G_{2}^{2} F_{1}^{3}
           \nonumber \\ && \hspace{4em} 
         +60 G_{1}^{2} G_{2}^{2} F_{1}^{3}
         +114 F_{2}^{2} G_{3}^{2} F_{1}^{3}
         +36 G_{1}^{2} G_{3}^{2} F_{1}^{3}
         +36 G_{2}^{2} G_{3}^{2} F_{1}^{3}
         +48 F_{2} G_{1} G_{2} G_{3} F_{1}^{3}
         +138 F_{2}^{5} F_{1}^{2}
         +90 F_{2} G_{1}^{4} F_{1}^{2}
           \nonumber \\ && \hspace{4em} 
         +90 F_{2} G_{2}^{4} F_{1}^{2}
         +72 F_{2} G_{3}^{4} F_{1}^{2}
         -12 G_{1} G_{2} G_{3}^{3} F_{1}^{2}
         +228 F_{2}^{3} G_{1}^{2} F_{1}^{2}
         +252 F_{2}^{3} G_{2}^{2} F_{1}^{2}
         +180 F_{2} G_{1}^{2} G_{2}^{2} F_{1}^{2}
         +132 F_{2}^{3} G_{3}^{2} F_{1}^{2}
           \nonumber \\ && \hspace{4em} 
         +108 F_{2} G_{1}^{2} G_{3}^{2} F_{1}^{2}
         +108 F_{2} G_{2}^{2} G_{3}^{2} F_{1}^{2}
         +24 G_{1} G_{2}^{3} G_{3} F_{1}^{2}
         +24 G_{1}^{3} G_{2} G_{3} F_{1}^{2}
         +72 F_{2}^{2} G_{1} G_{2} G_{3} F_{1}^{2}
         +56 F_{2}^{6} F_{1}
         +8 G_{1}^{6} F_{1}
           \nonumber \\ && \hspace{4em} 
         +8 G_{2}^{6} F_{1}
         +14 G_{3}^{6} F_{1}
         +72 F_{2}^{2} G_{1}^{4} F_{1}
         +99 F_{2}^{2} G_{2}^{4} F_{1}
         +30 G_{1}^{2} G_{2}^{4} F_{1}
         +81 F_{2}^{2} G_{3}^{4} F_{1}
         +24 G_{1}^{2} G_{3}^{4} F_{1}
         +24 G_{2}^{2} G_{3}^{4} F_{1}
           \nonumber \\ && \hspace{4em} 
         -24 F_{2} G_{1} G_{2} G_{3}^{3} F_{1}
         +120 F_{2}^{4} G_{1}^{2} F_{1}
         +150 F_{2}^{4} G_{2}^{2} F_{1}
         +30 G_{1}^{4} G_{2}^{2} F_{1}
         +180 F_{2}^{2} G_{1}^{2} G_{2}^{2} F_{1}
         +90 F_{2}^{4} G_{3}^{2} F_{1}
         +18 G_{1}^{4} G_{3}^{2} F_{1}
           \nonumber \\ && \hspace{4em} 
         +18 G_{2}^{4} G_{3}^{2} F_{1}
         +108 F_{2}^{2} G_{1}^{2} G_{3}^{2} F_{1}
         +108 F_{2}^{2} G_{2}^{2} G_{3}^{2} F_{1}
         +54 G_{1}^{2} G_{2}^{2} G_{3}^{2} F_{1}
         +48 F_{2} G_{1} G_{2}^{3} G_{3} F_{1}
         +48 F_{2} G_{1}^{3} G_{2} G_{3} F_{1}
           \nonumber \\ && \hspace{4em} 
         +48 F_{2}^{3} G_{1} G_{2} G_{3} F_{1}
         +16 F_{2}^{7}+16 F_{2} G_{1}^{6}
         +22 F_{2} G_{2}^{6}
         +22 F_{2} G_{3}^{6}
         -12 G_{1} G_{2} G_{3}^{5}
         +48 F_{2}^{3} G_{1}^{4}
         +60 F_{2}^{3} G_{2}^{4}
         +54 F_{2} G_{1}^{2} G_{2}^{4}
           \nonumber \\ && \hspace{4em} 
         +60 F_{2}^{3} G_{3}^{4}
         +54 F_{2} G_{1}^{2} G_{3}^{4}
         +42 F_{2} G_{2}^{2} G_{3}^{4}
         -12 G_{1}^{3} G_{2} G_{3}^{3}
         -36 F_{2}^{2} G_{1} G_{2} G_{3}^{3}
         +48 F_{2}^{5} G_{1}^{2}
         +48 F_{2}^{5} G_{2}^{2}
         +48 F_{2} G_{1}^{4} G_{2}^{2}
           \nonumber \\ && \hspace{4em} 
         +96 F_{2}^{3} G_{1}^{2} G_{2}^{2}
         +48 F_{2}^{5} G_{3}^{2}
         +48 F_{2} G_{1}^{4} G_{3}^{2}
         +42 F_{2} G_{2}^{4} G_{3}^{2}
         +96 F_{2}^{3} G_{1}^{2} G_{3}^{2}
         +72 F_{2}^{3} G_{2}^{2} G_{3}^{2}
         +108 F_{2} G_{1}^{2} G_{2}^{2} G_{3}^{2}
           \nonumber \\ && \hspace{4em} 
         +12 G_{1} G_{2}^{5} G_{3}
         +12 G_{1}^{3} G_{2}^{3} G_{3}
         +36 F_{2}^{2} G_{1} G_{2}^{3} G_{3}
   \Bigr)
   \nonumber \\ && 
+{\beta}^{(2)}
   \bigl(2 F_{2} b_{y}^{2}+2 b_{z}^{2} (F_{1}+F_{2})\bigr)   
   \nonumber \\ && 
+{\beta}^{(4)}
   \bigl(b_{x}^{2}+b_{y}^{2}+b_{z}^{2}\bigr)
   \bigl(2 F_{2} b_{y}^{2}+2 b_{z}^{2} (F_{1}+F_{2})\bigr) 
   \nonumber \\ && 
+{\gamma}^{(2)}
   \Bigl(
         - 12b_{x}^{2}
           \bigl(
                    F_{1}^{3}
                 + 2F_{1}^{2} F_{2}
                 + F_{1} G_{2}^{2}
                 + 4F_{2} G_{2}^{2}
                 + 3F_{1} G_{3}^{2}
                 + 4F_{2} G_{3}^{2}
           \bigr) 
           \nonumber \\ && \hspace{3em} 
         - 12b_{y}^{2}
           \bigl(
                   2F_{1}^{2} F_{2}
                + 3F_{1} F_{2}^{2}
                + 8F_{2}^{3}
                + F_{1} G_{1}^{2}
                + 8F_{2} G_{1}^{2}
                + 2F_{2} G_{2}^{2}
                 - 2G_{1} G_{2} G_{3}
                + 3F_{1} G_{3}^{2}
                + 10F_{2} G_{3}^{2}
           \bigr)
           \nonumber \\ && \hspace{3em} 
         - 12b_{z}^{2}
           \bigl(
                    7F_{1}^{3}
                 + 20F_{1}^{2} F_{2}
                 + 21F_{1} F_{2}^{2}
                 + 8F_{2}^{3}
                 + 7F_{1} G_{1}^{2}
                 + 8F_{2} G_{1}^{2}
                 + 7F_{1} G_{2}^{2}
                 + 10 F_{2} G_{2}^{2}
                 + 2G_{1} G_{2} G_{3}
                 + 2F_{1} G_{3}^{2}
                 + 2F_{2} G_{3}^{2}
           \bigr)
   \Bigr),
\nonumber \\
\end{eqnarray}
for $F_{2}$,
\begin{eqnarray}
&&
   - \nabla_{\!d} \frac{\delta {f}}{\delta (\nabla_{\!d}G_{1})}
   + \frac{\delta {f}}{\delta G_{1}}
\nonumber \\ 
&=&
   -\frac{{K}^{(0)}}{4}
   \Bigl(
          2
          \cos\theta\sin\theta\sin\varphi
          \nabla_{\!d}^{2}
          F_{1}
       + 2
          \bigl( 2 - \sin^{2}\theta \cos^{2}\varphi \bigr) \nabla_{\!d}^{2}
          G_{1}
       + 2
          \sin^{2}\theta\cos\varphi\sin\varphi
          \nabla_{\!d}^{2}
          G_{2}
       + 2
          \cos\theta\sin\theta\cos\varphi
          \nabla_{\!d}^{2}
          G_{3}
   \Bigr)
\nonumber \\ && 
+
  {\alpha}^{(0)} 4 G_{1}
   \nonumber \\ && 
+{\beta}^{(0)} 8 G_{1}
   \bigl(
            F_{1}^{2}+F_{2} F_{1}
            +F_{2}^{2}+G_{1}^{2}
            +G_{2}^{2}+G_{3}^{2}
    \bigr) 
   \nonumber \\ && 
+{\gamma}^{(0)}
   24
   \Bigl(
           12 G_{1}^{5}
         +28 F_{1}^{2} G_{1}^{3}
         +24 F_{2}^{2} G_{1}^{3}
         +24 G_{2}^{2} G_{1}^{3}
         +24 G_{3}^{2} G_{1}^{3}
         +24 F_{1} F_{2} G_{1}^{3}
         +12 F_{1} G_{2} G_{3} G_{1}^{2}
         +12 F_{1}^{4} G_{1}
         +12 F_{2}^{4} G_{1}
            \nonumber \\ && \hspace{4em} 
         +12 G_{2}^{4} G_{1}
         +12 G_{3}^{4} G_{1}
         +24 F_{1} F_{2}^{3} G_{1}
         +40 F_{1}^{2} F_{2}^{2} G_{1}
         +24 F_{1}^{2} G_{2}^{2} G_{1}
         +24 F_{2}^{2} G_{2}^{2} G_{1}
         +28 F_{1} F_{2} G_{2}^{2} G_{1}
            \nonumber \\ && \hspace{4em} 
         +20 F_{1}^{2} G_{3}^{2} G_{1}
         +24 F_{2}^{2} G_{3}^{2} G_{1}
         +32 G_{2}^{2} G_{3}^{2} G_{1}
         +20 F_{1} F_{2} G_{3}^{2} G_{1}
         +28 F_{1}^{3} F_{2} G_{1}
         -4 F_{1} G_{2} G_{3}^{3}
         -4 F_{2} G_{2} G_{3}^{3}
         +4 F_{2} G_{2}^{3} G_{3}
            \nonumber \\ && \hspace{4em} 
         +4 F_{1} F_{2}^{2} G_{2} G_{3}
         +4 F_{1}^{2} F_{2} G_{2} G_{3}
   \Bigr)
   \nonumber \\ && 
+{\delta}^{(0)}
   384
   \Bigl(
           8 G_{1}^{7}
         +33 F_{1}^{2} G_{1}^{5}
         +24 F_{2}^{2} G_{1}^{5}
         +24 G_{2}^{2} G_{1}^{5}
         +24 G_{3}^{2} G_{1}^{5}
         +24 F_{1} F_{2} G_{1}^{5}
         +30 F_{1} G_{2} G_{3} G_{1}^{4}
         +30 F_{1}^{4} G_{1}^{3}
         +24 F_{2}^{4} G_{1}^{3}
            \nonumber \\ && \hspace{4em} 
         +24 G_{2}^{4} G_{1}^{3}
         +24 G_{3}^{4} G_{1}^{3}
         +48 F_{1} F_{2}^{3} G_{1}^{3}
         +90 F_{1}^{2} F_{2}^{2} G_{1}^{3}
         +54 F_{1}^{2} G_{2}^{2} G_{1}^{3}
         +48 F_{2}^{2} G_{2}^{2} G_{1}^{3}
         +60 F_{1} F_{2} G_{2}^{2} G_{1}^{3}
         +42 F_{1}^{2} G_{3}^{2} G_{1}^{3}
            \nonumber \\ && \hspace{4em} 
         +48 F_{2}^{2} G_{3}^{2} G_{1}^{3}
         +72 G_{2}^{2} G_{3}^{2} G_{1}^{3}
         +36 F_{1} F_{2} G_{3}^{2} G_{1}^{3}
         +66 F_{1}^{3} F_{2} G_{1}^{3}
         -18 F_{2} G_{2} G_{3}^{3} G_{1}^{2}
         +18 F_{1} G_{2}^{3} G_{3} G_{1}^{2}
         +18 F_{2} G_{2}^{3} G_{3} G_{1}^{2}
            \nonumber \\ && \hspace{4em} 
         +18 F_{1}^{3} G_{2} G_{3} G_{1}^{2}
         +36 F_{1} F_{2}^{2} G_{2} G_{3} G_{1}^{2}
         +36 F_{1}^{2} F_{2} G_{2} G_{3} G_{1}^{2}
         +8 F_{1}^{6} G_{1}
         +8 F_{2}^{6} G_{1}
         +8 G_{2}^{6} G_{1}
         +8 G_{3}^{6} G_{1}
         +24 F_{1} F_{2}^{5} G_{1}
            \nonumber \\ && \hspace{4em} 
         +57 F_{1}^{2} F_{2}^{4} G_{1}
         +24 F_{1}^{2} G_{2}^{4} G_{1}
         +27 F_{2}^{2} G_{2}^{4} G_{1}
         +30 F_{1} F_{2} G_{2}^{4} G_{1}
         +21 F_{1}^{2} G_{3}^{4} G_{1}
         +27 F_{2}^{2} G_{3}^{4} G_{1}
         +36 G_{2}^{2} G_{3}^{4} G_{1}
            \nonumber \\ && \hspace{4em} 
         +24 F_{1} F_{2} G_{3}^{4} G_{1}
         +74 F_{1}^{3} F_{2}^{3} G_{1}
         +63 F_{1}^{4} F_{2}^{2} G_{1}
         +24 F_{1}^{4} G_{2}^{2} G_{1}
         +24 F_{2}^{4} G_{2}^{2} G_{1}
         +60 F_{1} F_{2}^{3} G_{2}^{2} G_{1}
         +90 F_{1}^{2} F_{2}^{2} G_{2}^{2} G_{1}
            \nonumber \\ && \hspace{4em} 
         +60 F_{1}^{3} F_{2} G_{2}^{2} G_{1}
         +18 F_{1}^{4} G_{3}^{2} G_{1}
         +24 F_{2}^{4} G_{3}^{2} G_{1}
         +36 G_{2}^{4} G_{3}^{2} G_{1}
         +36 F_{1} F_{2}^{3} G_{3}^{2} G_{1}
         +54 F_{1}^{2} F_{2}^{2} G_{3}^{2} G_{1}
         +54 F_{1}^{2} G_{2}^{2} G_{3}^{2} G_{1}
            \nonumber \\ && \hspace{4em} 
         +54 F_{2}^{2} G_{2}^{2} G_{3}^{2} G_{1}
         +54 F_{1} F_{2} G_{2}^{2} G_{3}^{2} G_{1}
         +36 F_{1}^{3} F_{2} G_{3}^{2} G_{1}
         +30 F_{1}^{5} F_{2} G_{1}
         -6 F_{1} G_{2} G_{3}^{5}
         -6 F_{2} G_{2} G_{3}^{5}
         -6 F_{1} G_{2}^{3} G_{3}^{3}
            \nonumber \\ && \hspace{4em} 
         -6 F_{1}^{3} G_{2} G_{3}^{3}
         -6 F_{2}^{3} G_{2} G_{3}^{3}
         -6 F_{1} F_{2}^{2} G_{2} G_{3}^{3}
         -6 F_{1}^{2} F_{2} G_{2} G_{3}^{3}
         +6 F_{2} G_{2}^{5} G_{3}
         +6 F_{2}^{3} G_{2}^{3} G_{3}
         +12 F_{1} F_{2}^{2} G_{2}^{3} G_{3}
            \nonumber \\ && \hspace{4em} 
         +12 F_{1}^{2} F_{2} G_{2}^{3} G_{3}
         +6 F_{1} F_{2}^{4} G_{2} G_{3}
         +12 F_{1}^{2} F_{2}^{3} G_{2} G_{3}
         +12 F_{1}^{3} F_{2}^{2} G_{2} G_{3}
         +6 F_{1}^{4} F_{2} G_{2} G_{3}
   \Bigr)
   \nonumber \\ && 
+{\beta}^{(2)} 2
   \bigl( b_{y}^{2}+b_{z}^{2} \bigr) G_{1}
   \nonumber \\ && 
+{\beta}^{(4)} 2
   \bigl(b_{x}^{2}+b_{y}^{2}+b_{z}^{2}\bigr)
   \bigl( b_{y}^{2}+b_{z}^{2} \bigr) G_{1}
   \nonumber \\ && 
+{\gamma}^{(2)}
   \Bigl(
       - 12b_{x}^{2}
         \bigl(
                2F_{1}^{2} G_{1}
             + 4G_{1} G_{2}^{2}
              - 2F_{1} G_{2} G_{3}
             + 4G_{1} G_{3}^{2}
         \bigr)
         \nonumber \\ && \hspace{3em} 
      - 12b_{y}^{2}
         \bigl(
                4F_{1}^{2} G_{1}
             + 2F_{1} F_{2} G_{1}
             + 8F_{2}^{2} G_{1}
             + 8G_{1}^{3}
             + 4G_{1} G_{2}^{2}
              - 2F_{2} G_{2} G_{3}
             + 8G_{1} G_{3}^{2}
         \bigr)
         \nonumber \\ && \hspace{3em} 
      - 12b_{z}^{2}
         \bigl(
                 10F_{1}^{2} G_{1}
              + 14F_{1} F_{2} G_{1}
              + 8F_{2}^{2} G_{1}
              + 8G_{1}^{3}
              + 8G_{1} G_{2}^{2}
              + 2F_{1} G_{2} G_{3}
              + 2F_{2} G_{2} G_{3}
              + 4G_{1} G_{3}^{2}
         \bigr)
   \Bigr),
\end{eqnarray}
for $G_{1}$,
\begin{eqnarray}
&&
   - \nabla_{\!d} \frac{\delta {f}}{\delta (\nabla_{\!d}G_{2})}
   + \frac{\delta {f}}{\delta G_{2}}
\nonumber \\ 
&=&
   -\frac{{K}^{(0)}}{4}
   \Bigl(
          2
          \cos\theta\sin\theta\cos\varphi
          \nabla_{\!d}^{2}
          F_{2}
       + 2
          \sin^{2}\theta\cos\varphi\sin\varphi
          \nabla_{\!d}^{2}
          G_{1}
       + 2
          \bigl( 2 - \sin^{2}\theta \sin^{2}\varphi \bigr) \nabla_{\!d}^{2}
          G_{2}
       + 2
          \cos\theta\sin\theta\sin\varphi
          \nabla_{\!d}^{2}
          G_{3}
   \Bigr)
\nonumber \\ && 
+{\alpha}^{(0)}
  4 G_{2}
   \nonumber \\ && 
+{\beta}^{(0)}
  8 G_{2}
  \bigl(F_{1}^{2}+F_{2} F_{1}+F_{2}^{2}+G_{1}^{2}+G_{2}^{2}+G_{3}^{2}\bigr)
   \nonumber \\ && 
+{\gamma}^{(0)}
   24
   \Bigl(
         12 G_{2}^{5}
        +24 F_{1}^{2} G_{2}^{3}
        +28 F_{2}^{2} G_{2}^{3}
        +24 G_{1}^{2} G_{2}^{3}
        +24 G_{3}^{2} G_{2}^{3}
        +24 F_{1} F_{2} G_{2}^{3}
        +12 F_{2} G_{1} G_{3} G_{2}^{2}
        +12 F_{1}^{4} G_{2}
        +12 F_{2}^{4} G_{2}
          \nonumber \\ && \hspace{4em} 
        +12 G_{1}^{4} G_{2}
        +12 G_{3}^{4} G_{2}
        +28 F_{1} F_{2}^{3} G_{2}
        +40 F_{1}^{2} F_{2}^{2} G_{2}
        +24 F_{1}^{2} G_{1}^{2} G_{2}
        +24 F_{2}^{2} G_{1}^{2} G_{2}
        +28 F_{1} F_{2} G_{1}^{2} G_{2}
        +24 F_{1}^{2} G_{3}^{2} G_{2}
          \nonumber \\ && \hspace{4em} 
        +20 F_{2}^{2} G_{3}^{2} G_{2}
        +32 G_{1}^{2} G_{3}^{2} G_{2}
        +20 F_{1} F_{2} G_{3}^{2} G_{2}
        +24 F_{1}^{3} F_{2} G_{2}
        -4 F_{1} G_{1} G_{3}^{3}
        -4 F_{2} G_{1} G_{3}^{3}
        +4 F_{1} G_{1}^{3} G_{3}
        +4 F_{1} F_{2}^{2} G_{1} G_{3}
          \nonumber \\ && \hspace{4em} 
        +4 F_{1}^{2} F_{2} G_{1} G_{3}
   \Bigr)
   \nonumber \\ && 
+{\delta}^{(0)}
   384
   \bigl(
         8 G_{2}^{7}
       +24 F_{1}^{2} G_{2}^{5}
       +33 F_{2}^{2} G_{2}^{5}
       +24 G_{1}^{2} G_{2}^{5}
       +24 G_{3}^{2} G_{2}^{5}
       +24 F_{1} F_{2} G_{2}^{5}
       +30 F_{2} G_{1} G_{3} G_{2}^{4}
       +24 F_{1}^{4} G_{2}^{3}
       +30 F_{2}^{4} G_{2}^{3}
         \nonumber \\ && \hspace{4em} 
       +24 G_{1}^{4} G_{2}^{3}
       +24 G_{3}^{4} G_{2}^{3}
       +66 F_{1} F_{2}^{3} G_{2}^{3}
       +90 F_{1}^{2} F_{2}^{2} G_{2}^{3}
       +48 F_{1}^{2} G_{1}^{2} G_{2}^{3}
       +54 F_{2}^{2} G_{1}^{2} G_{2}^{3}
       +60 F_{1} F_{2} G_{1}^{2} G_{2}^{3}
       +48 F_{1}^{2} G_{3}^{2} G_{2}^{3}
         \nonumber \\ && \hspace{4em} 
       +42 F_{2}^{2} G_{3}^{2} G_{2}^{3}
       +72 G_{1}^{2} G_{3}^{2} G_{2}^{3}
       +36 F_{1} F_{2} G_{3}^{2} G_{2}^{3}
       +48 F_{1}^{3} F_{2} G_{2}^{3}
       -18 F_{1} G_{1} G_{3}^{3} G_{2}^{2}
       +18 F_{1} G_{1}^{3} G_{3} G_{2}^{2}
       +18 F_{2} G_{1}^{3} G_{3} G_{2}^{2}
         \nonumber \\ && \hspace{4em} 
       +18 F_{2}^{3} G_{1} G_{3} G_{2}^{2}
       +36 F_{1} F_{2}^{2} G_{1} G_{3} G_{2}^{2}
       +36 F_{1}^{2} F_{2} G_{1} G_{3} G_{2}^{2}
       +8 F_{1}^{6} G_{2}
       +8 F_{2}^{6} G_{2}
       +8 G_{1}^{6} G_{2}
       +8 G_{3}^{6} G_{2}
       +30 F_{1} F_{2}^{5} G_{2}
         \nonumber \\ && \hspace{4em} 
       +63 F_{1}^{2} F_{2}^{4} G_{2}
       +27 F_{1}^{2} G_{1}^{4} G_{2}
       +24 F_{2}^{2} G_{1}^{4} G_{2}
       +30 F_{1} F_{2} G_{1}^{4} G_{2}
       +27 F_{1}^{2} G_{3}^{4} G_{2}
       +21 F_{2}^{2} G_{3}^{4} G_{2}
       +36 G_{1}^{2} G_{3}^{4} G_{2}
         \nonumber \\ && \hspace{4em} 
       +24 F_{1} F_{2} G_{3}^{4} G_{2}
       +74 F_{1}^{3} F_{2}^{3} G_{2}
       +57 F_{1}^{4} F_{2}^{2} G_{2}
       +24 F_{1}^{4} G_{1}^{2} G_{2}
       +24 F_{2}^{4} G_{1}^{2} G_{2}
       +60 F_{1} F_{2}^{3} G_{1}^{2} G_{2}
       +90 F_{1}^{2} F_{2}^{2} G_{1}^{2} G_{2}
         \nonumber \\ && \hspace{4em} 
       +60 F_{1}^{3} F_{2} G_{1}^{2} G_{2}
       +24 F_{1}^{4} G_{3}^{2} G_{2}
       +18 F_{2}^{4} G_{3}^{2} G_{2}
       +36 G_{1}^{4} G_{3}^{2} G_{2}
       +36 F_{1} F_{2}^{3} G_{3}^{2} G_{2}
       +54 F_{1}^{2} F_{2}^{2} G_{3}^{2} G_{2}
       +54 F_{1}^{2} G_{1}^{2} G_{3}^{2} G_{2}
         \nonumber \\ && \hspace{4em} 
       +54 F_{2}^{2} G_{1}^{2} G_{3}^{2} G_{2}
       +54 F_{1} F_{2} G_{1}^{2} G_{3}^{2} G_{2}
       +36 F_{1}^{3} F_{2} G_{3}^{2} G_{2}
       +24 F_{1}^{5} F_{2} G_{2}
       -6 F_{1} G_{1} G_{3}^{5}
       -6 F_{2} G_{1} G_{3}^{5}
       -6 F_{2} G_{1}^{3} G_{3}^{3}
         \nonumber \\ && \hspace{4em} 
       -6 F_{1}^{3} G_{1} G_{3}^{3}
       -6 F_{2}^{3} G_{1} G_{3}^{3}
       -6 F_{1} F_{2}^{2} G_{1} G_{3}^{3}
       -6 F_{1}^{2} F_{2} G_{1} G_{3}^{3}
       +6 F_{1} G_{1}^{5} G_{3}
       +6 F_{1}^{3} G_{1}^{3} G_{3}
         \nonumber \\ && \hspace{4em} 
       +12 F_{1} F_{2}^{2} G_{1}^{3} G_{3}
       +12 F_{1}^{2} F_{2} G_{1}^{3} G_{3}
       +6 F_{1} F_{2}^{4} G_{1} G_{3}
       +12 F_{1}^{2} F_{2}^{3} G_{1} G_{3}
       +12 F_{1}^{3} F_{2}^{2} G_{1} G_{3}
       +6 F_{1}^{4} F_{2} G_{1} G_{3}
   \Bigr)
   \nonumber \\ && 
+{\beta}^{(2)} 2
  \bigl( b_{x}^{2}+b_{z}^{2} \bigr) G_{2}
   \nonumber \\ && 
+{\beta}^{(4)} 2
  \bigl(b_{x}^{2}+b_{y}^{2}+b_{z}^{2}\bigr)
  \bigl( b_{x}^{2}+b_{z}^{2} \bigr) G_{2}
   \nonumber \\ && 
+{\gamma}^{(2)}
   \Bigl(
        - 12b_{x}^{2}
           \bigl(
                    8F_{1}^{2} G_{2}
                 + 2F_{1} F_{2} G_{2}
                 + 4F_{2}^{2} G_{2}
                 + 4G_{1}^{2} G_{2}
                 + 8G_{2}^{3}
                  - 2F_{1}G_{1}G_{3}
                 + 8G_{2} G_{3}^{2} 
            \bigr)
            \nonumber \\ && \hspace{3em} 
         - 12b_{y}^{2}
            \bigl(
                    2F_{2}^{2} G_{2}
                 + 4G_{1}^{2} G_{2}
                  - 2F_{2} G_{1} G_{3}
                + 4G_{2} G_{3}^{2}
            \bigr)
            \nonumber \\ && \hspace{3em} 
         - 12b_{z}^{2}
            \bigl(
                    8F_{1}^{2} G_{2}
                 + 14F_{1} F_{2} G_{2}
                 + 10F_{2}^{2} G_{2}
                 + 8G_{1}^{2} G_{2}
                 + 8G_{2}^{3}
                 + 2F_{1} G_{1} G_{3}
                 + 2F_{2} G_{1} G_{3}
                 + 4G_{2} G_{3}^{2}
            \bigr)
   \Bigr),
\end{eqnarray}
for $G_{2}$, and
\begin{eqnarray}
&&
   - \nabla_{\!d} \frac{\delta {f}}{\delta (\nabla_{\!d}G_{3})}
   + \frac{\delta {f}}{\delta G_{3}}
\nonumber \\ 
&=&
   -\frac{{K}^{(0)}}{4}
   \Bigl(
       - 2
         \sin^{2}\theta\cos\varphi\sin\varphi
         \nabla_{\!d}^{2}
         F_{1}
       - 2
         \sin^{2}\theta\cos\varphi\sin\varphi
         \nabla_{\!d}^{2}
         F_{2}
         \nonumber \\ && \hspace{2em} 
      + 2
         \cos\theta\sin\theta\cos\varphi
         \nabla_{\!d}^{2}
         G_{1}
      + 2
         \cos\theta\sin\theta\sin\varphi
         \nabla_{\!d}^{2}
         G_{2}
      + 2\bigl(1+\sin^{2}\theta\bigr)
         \nabla_{\!d}
         G_{3}
   \Bigr)
\nonumber \\ && 
+{\alpha}^{(0)} 4 G_{3}
   \nonumber \\ && 
+{\beta}^{(0)}
   8 G_{3}
   \bigl(F_{1}^{2}+F_{2} F_{1}+F_{2}^{2}+G_{1}^{2}+G_{2}^{2}+G_{3}^{2}\bigr)
   \nonumber \\ && 
+{\gamma}^{(0)}
   24
   \Bigl(
           12 G_{3}^{5}
          +28 F_{1}^{2} G_{3}^{3}
          +28 F_{2}^{2} G_{3}^{3}
          +24 G_{1}^{2} G_{3}^{3}
          +24 G_{2}^{2} G_{3}^{3}
          +32 F_{1} F_{2} G_{3}^{3}
          -12 F_{1} G_{1} G_{2} G_{3}^{2}
          -12 F_{2} G_{1} G_{2} G_{3}^{2}
          +12 F_{1}^{4} G_{3}
            \nonumber \\ && \hspace{4em} 
          +12 F_{2}^{4} G_{3}
          +12 G_{1}^{4} G_{3}
          +12 G_{2}^{4} G_{3}
          +20 F_{1} F_{2}^{3} G_{3}
          +28 F_{1}^{2} F_{2}^{2} G_{3}
          +20 F_{1}^{2} G_{1}^{2} G_{3}
          +24 F_{2}^{2} G_{1}^{2} G_{3}
          +20 F_{1} F_{2} G_{1}^{2} G_{3}
            \nonumber \\ && \hspace{4em} 
          +24 F_{1}^{2} G_{2}^{2} G_{3}
          +20 F_{2}^{2} G_{2}^{2} G_{3}
          +32 G_{1}^{2} G_{2}^{2} G_{3}
          +20 F_{1} F_{2} G_{2}^{2} G_{3}
          +20 F_{1}^{3} F_{2} G_{3}
          +4 F_{2} G_{1} G_{2}^{3}
          +4 F_{1} G_{1}^{3} G_{2}
            \nonumber \\ && \hspace{4em} 
          +4 F_{1} F_{2}^{2} G_{1} G_{2}
          +4 F_{1}^{2} F_{2} G_{1} G_{2}
   \Bigr)
   \nonumber \\ && 
+{\delta}^{(0)}
   384
   \Bigl(
           8 G_{3}^{7}
         +33 F_{1}^{2} G_{3}^{5}
         +33 F_{2}^{2} G_{3}^{5}
         +24 G_{1}^{2} G_{3}^{5}
         +24 G_{2}^{2} G_{3}^{5}
         +42 F_{1} F_{2} G_{3}^{5}
         -30 F_{1} G_{1} G_{2} G_{3}^{4}
         -30 F_{2} G_{1} G_{2} G_{3}^{4}
         +30 F_{1}^{4} G_{3}^{3}
           \nonumber \\ && \hspace{4em} 
         +30 F_{2}^{4} G_{3}^{3}
         +24 G_{1}^{4} G_{3}^{3}
         +24 G_{2}^{4} G_{3}^{3}
         +54 F_{1} F_{2}^{3} G_{3}^{3}
         +72 F_{1}^{2} F_{2}^{2} G_{3}^{3}
         +42 F_{1}^{2} G_{1}^{2} G_{3}^{3}
         +54 F_{2}^{2} G_{1}^{2} G_{3}^{3}
         +48 F_{1} F_{2} G_{1}^{2} G_{3}^{3}
           \nonumber \\ && \hspace{4em} 
         +54 F_{1}^{2} G_{2}^{2} G_{3}^{3}
         +42 F_{2}^{2} G_{2}^{2} G_{3}^{3}
         +72 G_{1}^{2} G_{2}^{2} G_{3}^{3}
         +48 F_{1} F_{2} G_{2}^{2} G_{3}^{3}
         +54 F_{1}^{3} F_{2} G_{3}^{3}
         -18 F_{1} G_{1} G_{2}^{3} G_{3}^{2}
         -18 F_{2} G_{1}^{3} G_{2} G_{3}^{2}
           \nonumber \\ && \hspace{4em} 
         -18 F_{1}^{3} G_{1} G_{2} G_{3}^{2}
         -18 F_{2}^{3} G_{1} G_{2} G_{3}^{2}
         -18 F_{1} F_{2}^{2} G_{1} G_{2} G_{3}^{2}
         -18 F_{1}^{2} F_{2} G_{1} G_{2} G_{3}^{2}
         +8 F_{1}^{6} G_{3}
         +8 F_{2}^{6} G_{3}
         +8 G_{1}^{6} G_{3}
           \nonumber \\ && \hspace{4em} 
         +8 G_{2}^{6} G_{3}
         +18 F_{1} F_{2}^{5} G_{3}
         +33 F_{1}^{2} F_{2}^{4} G_{3}
         +21 F_{1}^{2} G_{1}^{4} G_{3}
         +24 F_{2}^{2} G_{1}^{4} G_{3}
         +18 F_{1} F_{2} G_{1}^{4} G_{3}
         +24 F_{1}^{2} G_{2}^{4} G_{3}
           \nonumber \\ && \hspace{4em} 
         +21 F_{2}^{2} G_{2}^{4} G_{3}
         +36 G_{1}^{2} G_{2}^{4} G_{3}
         +18 F_{1} F_{2} G_{2}^{4} G_{3}
         +38 F_{1}^{3} F_{2}^{3} G_{3}
         +33 F_{1}^{4} F_{2}^{2} G_{3}
         +18 F_{1}^{4} G_{1}^{2} G_{3}
         +24 F_{2}^{4} G_{1}^{2} G_{3}
           \nonumber \\ && \hspace{4em} 
         +36 F_{1} F_{2}^{3} G_{1}^{2} G_{3}
         +54 F_{1}^{2} F_{2}^{2} G_{1}^{2} G_{3}
         +36 F_{1}^{3} F_{2} G_{1}^{2} G_{3}
         +24 F_{1}^{4} G_{2}^{2} G_{3}
         +18 F_{2}^{4} G_{2}^{2} G_{3}
         +36 G_{1}^{4} G_{2}^{2} G_{3}
         +36 F_{1} F_{2}^{3} G_{2}^{2} G_{3}
           \nonumber \\ && \hspace{4em} 
         +54 F_{1}^{2} F_{2}^{2} G_{2}^{2} G_{3}
         +54 F_{1}^{2} G_{1}^{2} G_{2}^{2} G_{3}
         +54 F_{2}^{2} G_{1}^{2} G_{2}^{2} G_{3}
         +54 F_{1} F_{2} G_{1}^{2} G_{2}^{2} G_{3}
         +36 F_{1}^{3} F_{2} G_{2}^{2} G_{3}
         +18 F_{1}^{5} F_{2} G_{3}
           \nonumber \\ && \hspace{4em} 
         +6 F_{2} G_{1} G_{2}^{5}
         +6 F_{1} G_{1}^{3} G_{2}^{3}
         +6 F_{2} G_{1}^{3} G_{2}^{3}
         +6 F_{2}^{3} G_{1} G_{2}^{3}
         +12 F_{1} F_{2}^{2} G_{1} G_{2}^{3}
         +12 F_{1}^{2} F_{2} G_{1} G_{2}^{3}
         +6 F_{1} G_{1}^{5} G_{2}
         +6 F_{1}^{3} G_{1}^{3} G_{2}
           \nonumber \\ && \hspace{4em} 
         +12 F_{1} F_{2}^{2} G_{1}^{3} G_{2}
         +12 F_{1}^{2} F_{2} G_{1}^{3} G_{2}
         +6 F_{1} F_{2}^{4} G_{1} G_{2}
         +12 F_{1}^{2} F_{2}^{3} G_{1} G_{2}
         +12 F_{1}^{3} F_{2}^{2} G_{1} G_{2}
         +6 F_{1}^{4} F_{2} G_{1} G_{2}
   \Bigr)
   \nonumber \\ && 
+{\beta}^{(2)} 2\bigl( b_{x}^{2}+b_{y}^{2} \bigr) G_{3}
   \nonumber \\ && 
+{\beta}^{(4)} 2
   \bigl(b_{x}^{2}+b_{y}^{2}+b_{z}^{2}\bigr)
   \bigl( b_{x}^{2}+b_{y}^{2} \bigr) G_{3}
   \nonumber \\ && 
+{\gamma}^{(2)}
   \Bigl(
          - 12b_{x}^{2}
            \bigl(
                    - 2F_{1} G_{1} G_{2}
                   + 10F_{1}^{2} G_{3}
                   + 6F_{1} F_{2} G_{3}
                   + 4F_{2}^{2} G_{3}
                   + 4G_{1}^{2} G_{3}
                   + 8G_{2}^{2} G_{3}
                   + 8G_{3}^{3}
             \bigr)
             \nonumber \\ && \hspace{3em} 
          - 12b_{y}^{2}
             \bigl(
                    - 2F_{2} G_{1} G_{2}
                   + 4F_{1}^{2} G_{3}
                   + 6F_{1} F_{2} G_{3}
                   + 10F_{2}^{2} G_{3}
                   + 8G_{1}^{2} G_{3}
                   + 4G_{2}^{2} G_{3}
                   + 8G_{3}^{3}
             \bigr)
             \nonumber \\ && \hspace{3em} 
          - 12b_{z}^{2}
             \bigl(
                      2F_{1} G_{1} G_{2}
                   + 2F_{2} G_{1} G_{2}
                   + 2F_{1}^{2} G_{3}
                   + 4F_{1} F_{2} G_{3}
                   + 2F_{2}^{2} G_{3}
                   + 4G_{1}^{2} G_{3}
                   + 4G_{2}^{2} G_{3}
             \bigr)
   \Bigr),
\end{eqnarray}
for $G_{3}$.

\section{Symmetries}
\label{sec:symmetries}

We summarize the properties of the symmetries of the order parameter $A$ in Eq.~\eqref{eq:A_gs}.
First of all, we remember that $A$ possesses the symmetry
\begin{eqnarray}
 A(\vec{x}) \rightarrow e^{i\alpha} O(\theta,\vec{n}) A(\tilde{\vec{x}}) O^{t}(\theta',\vec{n}'),
\label{eq:A_symmetry_boundary}
\end{eqnarray}
in the Lagrangian,
where $e^{i\alpha} \in \mathrm{U}(1)$ and $O(\theta,\vec{n})$, $O(\theta',\vec{n}') \in \mathrm{SO}(3)$ with $\vec{n}$ ($\vec{n}'$) the rotation axis and $\theta$ ($\theta'$) the rotation angle around $\vec{n}$ ($\vec{n}'$).
Here, $O(\theta,\vec{n})$ is the rotation in the spin space, and $O(\theta',\vec{n}')$ is the rotation in the real space.
$\tilde{\vec{x}}$ is the vector rotated by $O(\theta',\vec{n}')$ from $\vec{x}$.
The symmetries \eqref{eq:A_symmetry_boundary} are spontaneously broken to subgroups when the state is in the nematic phase as presented in Eq.~\eqref{eq:A_gs}.
Instead, there exist the O(2), D$_{2}$, and D$_{4}$ symmetries,
and the corresponding phases are called the UN ($r=-1/2$), D$_{2}$-BN ($-1<r<-1/2$), and D$_{4}$-BN ($r=-1$) phases, respectively.
We will give the concrete forms of those symmetries in the following.

\subsection{UN phase ($r=-1/2$)}

For $r=-1/2$, the order parameter $A(\vec{x})$ is written as
\begin{eqnarray}
  A_{\mathrm{UN}}(\vec{x})
=
A_{0}
\left(
\begin{array}{ccc}
 -1/2 & 0  & 0  \\
 0 & -1/2  & 0  \\
 0 & 0 & 1  
\end{array}
\right).
\label{eq:tau_gs_UN}
\end{eqnarray}
This is invariant under the rotation around the $x_{3}$-axis
$  A_{\mathrm{UN}}(\vec{x})
\rightarrow
  O(\theta) A_{\mathrm{UN}}(\vec{x}') O^{t}(\theta)
$
with $e^{i\alpha}=1$ ($\alpha=0$),
where $O(\theta) \in \mathrm{O}(2)$ is the rotation operator
\begin{eqnarray}
O(\theta)
=
\left(
\begin{array}{ccc}
 \cos \theta & -\sin \theta  & 0  \\
 \sin \theta & \cos \theta  & 0  \\
 0 & 0  & 1   
\end{array}
\right),
\end{eqnarray}
with the rotation angle $\theta$ ($0\le\theta<2\pi$).
This is the uniaxial nematic (UN) phase.
The rotation in the spin space and the rotation in the real space are locked to each other, $O(\theta,\vec{n})=O(\theta',\vec{n}')$ with $\vec{n}=\vec{n}'$ and $\theta=\theta'$. The locking also occurs for the BN phase.

\subsection{D$_{2}$-BN phase ($-1< r < -1/2$)}

For $-1< r < -1/2$, the order parameter $A(\vec{x})$ is written as
\begin{eqnarray}
  A_{\mathrm{D}_{2}\mathrm{BN}}(\vec{x})
=
A_{0}
\left(
\begin{array}{ccc}
 r & 0  & 0  \\
 0 & -1-r  & 0  \\
 0 & 0 & 1  
\end{array}
\right).
\label{eq:A_gs_D2BN}
\end{eqnarray}
This is invariant under the D$_{2}$ symmetry.
The generators of the D$_{2}$ group are given by
\begin{eqnarray}
   \bigl\{O\bigr\}
= \bigl\{\vec{1}_{3},I_{1},I_{2},I_{3}\bigr\},
\end{eqnarray}
with $e^{i\alpha}=1$ ($\alpha=0$),
$\vec{1}_{3}$ is a unit matrix,
and $I_{1}$, $I_{2}$, and $I_{3}$ are defined by
\begin{eqnarray}
 \vec{1}_{3}
=
\left(
\begin{array}{ccc}
 1 & 0 & 0  \\
 0 & 1 & 0  \\
 0 & 0 & 1  
\end{array}
\right),
\hspace{0.5em}
 I_{1}
=
\left(
\begin{array}{ccc}
 1 & 0 & 0  \\
 0 & -1 & 0  \\
 0 & 0 & -1  
\end{array}
\right),
\hspace{0.5em}
 I_{2}
=
\left(
\begin{array}{ccc}
 -1 & 0 & 0  \\
 0 & 1 & 0  \\
 0 & 0 & -1  
\end{array}
\right),
\hspace{0.5em}
 I_{3}
=
\left(
\begin{array}{ccc}
 -1 & 0 & 0  \\
 0 & -1 & 0  \\
 0 & 0 & 1  
\end{array}
\right).
\end{eqnarray}
$I_{i}$ ($i=1,2,3$) indicates a $\pi$ rotation around the $i$th axis.
We confirm easily that $A_{\mathrm{D}_{2}\mathrm{BN}}(\vec{x})$ is invariant under the transformation $A_{\mathrm{D}_{2}\mathrm{BN}}(\vec{x}) \rightarrow O A_{\mathrm{D}_{2}\mathrm{BN}}(\vec{x}') O^{t}$ ($O \in \mathrm{D}_{2}$).
This phase is called the D$_{2}$-biaxial nematic (D$_{2}$-BN) phase.

\subsection{D$_{4}$-BN phase ($r=-1$)}

For $r=-1/2$, the order parameter $A$ is written as
\begin{eqnarray}
  A_{\mathrm{D}_{4}\mathrm{BN}}
=
A_{0}
\left(
\begin{array}{ccc}
 -1 & 0  & 0  \\
 0 & 0  & 0  \\
 0 & 0 & 1  
\end{array}
\right),
\label{eq:tau_gs_D4BN}
\end{eqnarray}
This is invariant under the D$_{4}$ symmetry.
The generators of the D$_{4}$ group are given by
\begin{eqnarray}
   \bigl\{e^{i\alpha},O\bigr\}
=\bigl\{(1,\vec{1}_{3}),\hspace{0.1em}(-1,R_{2}),\hspace{0.1em}(1,I_{2}),\hspace{0.1em}(-1,I_{2}R_{2}),\hspace{0.1em}(1,I_{1}),\hspace{0.1em}(1,I_{3}),\hspace{0.1em}(-1,I_{1}R_{2}),\hspace{0.1em}(-1,I_{3}R_{2})\bigr\},
\end{eqnarray}
where $\vec{1}_{3}$ and $I_{i}$ ($i=1,2,3$) have been defined in the D$_{2}$ group, and $R_{2}$ is newly defined by
\begin{eqnarray}
 R_{2}
=
\left(
\begin{array}{ccc}
 0 & 0 & -1  \\
 0 & 0 & 0  \\
 1 & 0 & 0  
\end{array}
\right).
\end{eqnarray}
$R_{2}$ indicates the $\pi/2$ rotation around the second axis ($x_{2}$ axis).
It should be noted that the phase $\bigl\{e^{i\alpha}\bigr\}=\bigl\{1,-1\bigr\}\in \mathbb{Z}_{2}$ ($\alpha=0,\pi$) is locked with the spin rotation and the spatial rotation.
We confirm easily that $A_{\mathrm{D}_{4}\mathrm{BN}}$ is invariant under the transformation $A_{\mathrm{D}_{4}\mathrm{BN}}(\vec{x})\rightarrow e^{i\alpha} O A_{\mathrm{D}_{4}\mathrm{BN}}(\vec{x}') O^{t}$ [$(e^{i\alpha},O) \in \mathrm{D}_{4}$].
This is the D$_{4}$-biaxial nematic (D$_{4}$-BN) phase.

\section{Numerical results for the profile functions}
\label{sec:symmetry_surface_results}

\begin{figure}[p] 
\begin{center}
\includegraphics[scale=0.18]{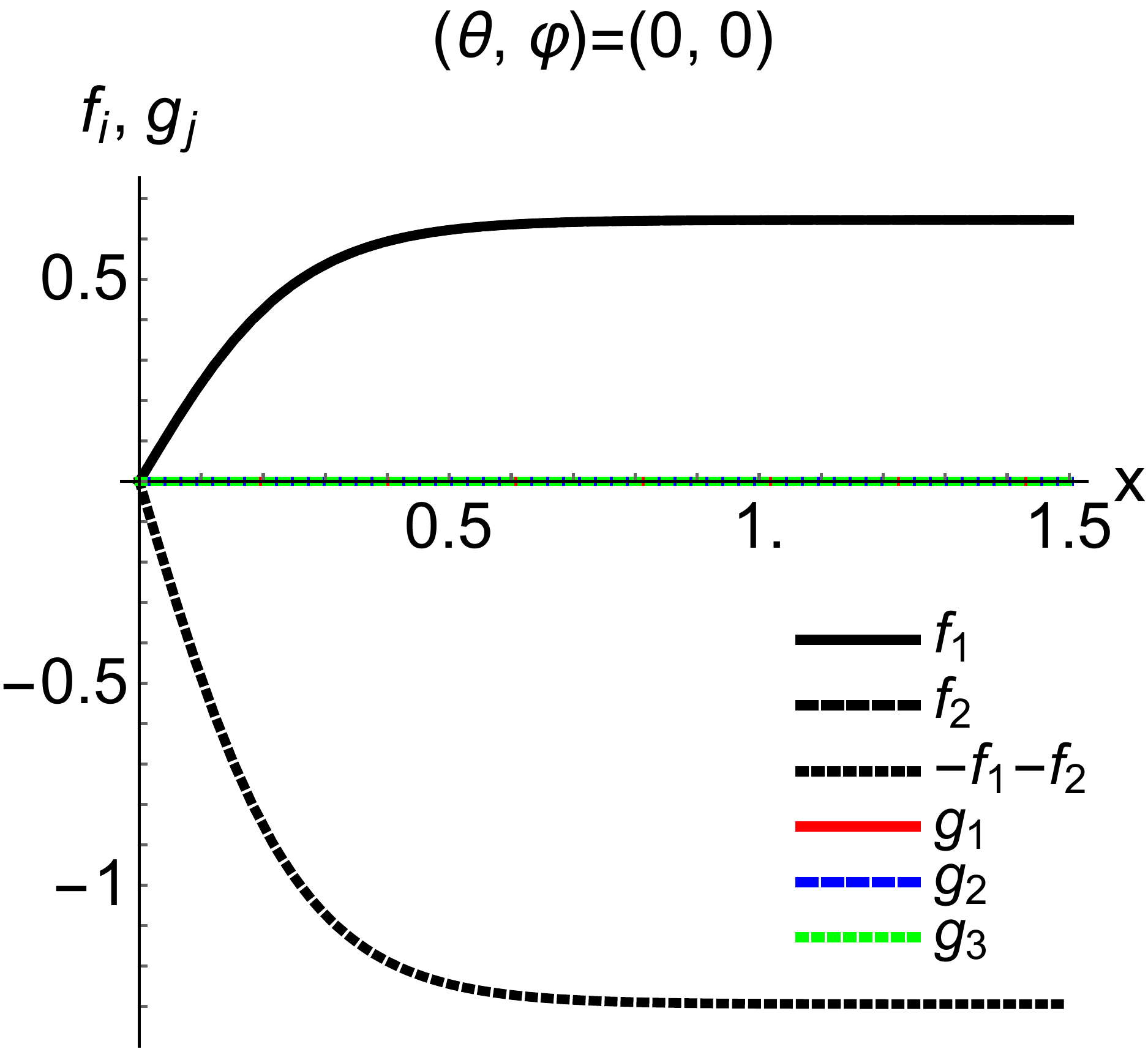}
\\ \vspace{2em} 
\includegraphics[scale=0.18]{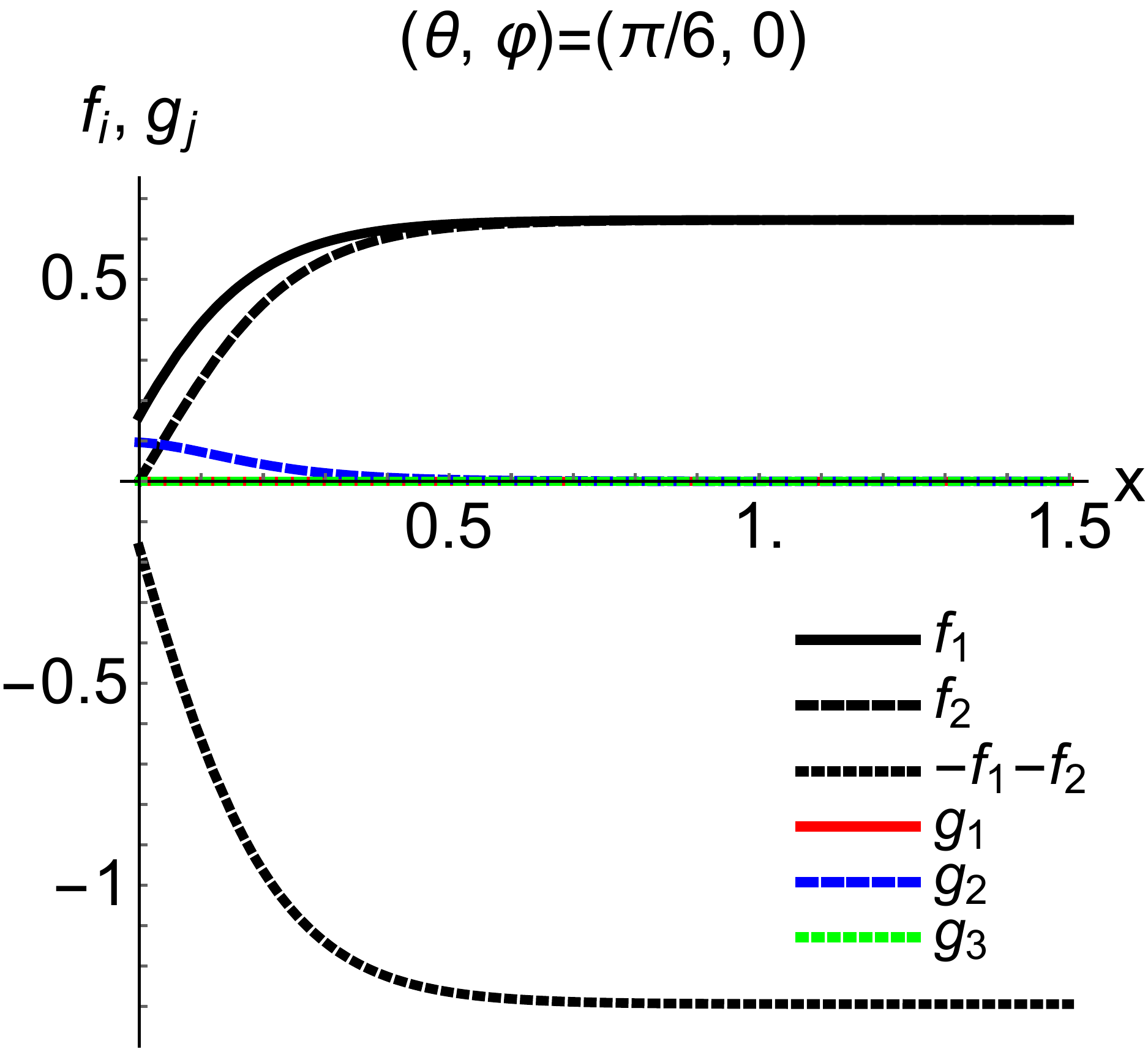}
\hspace{1em} %
\includegraphics[scale=0.18]{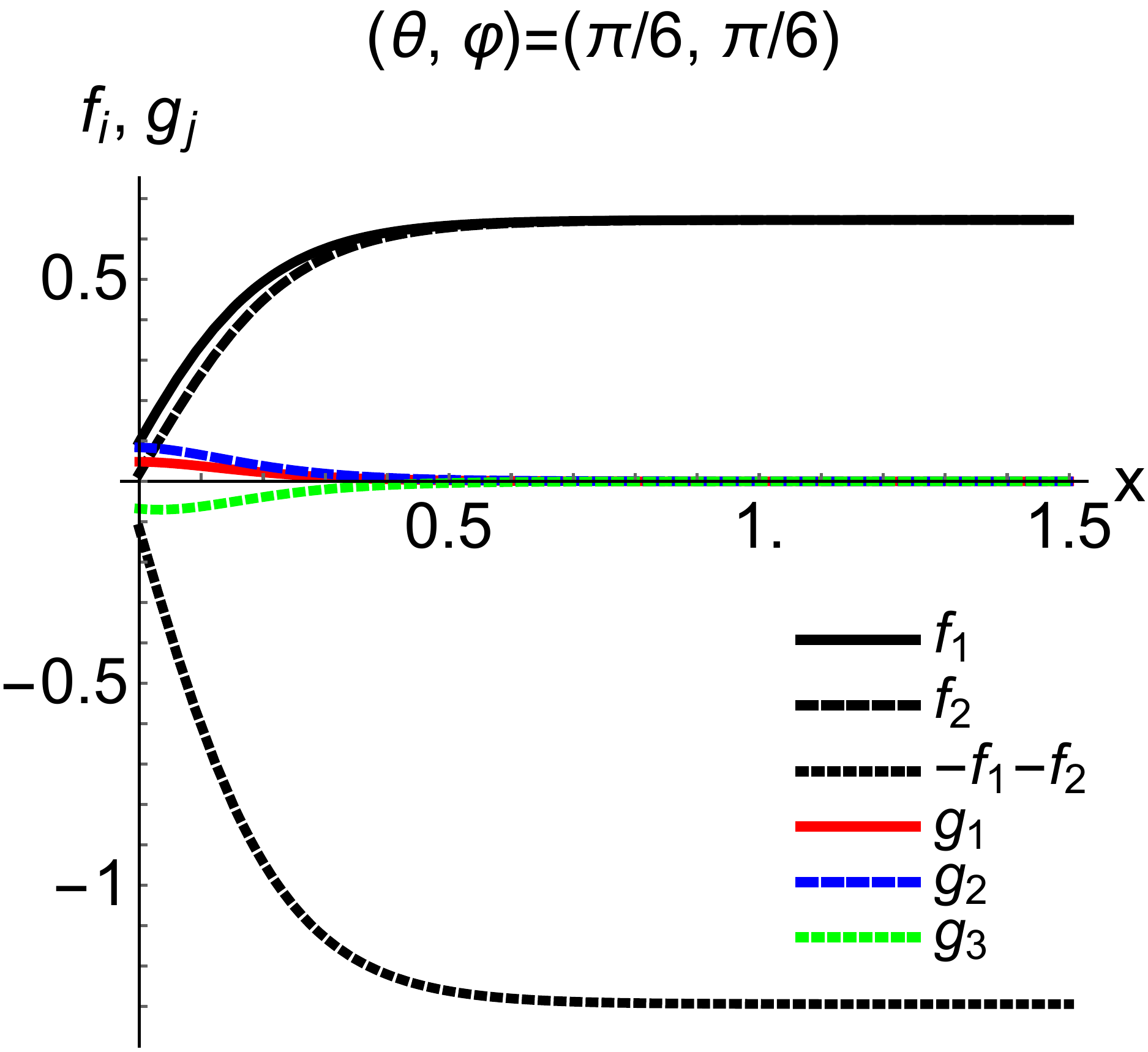}
\hspace{1em} %
\includegraphics[scale=0.18]{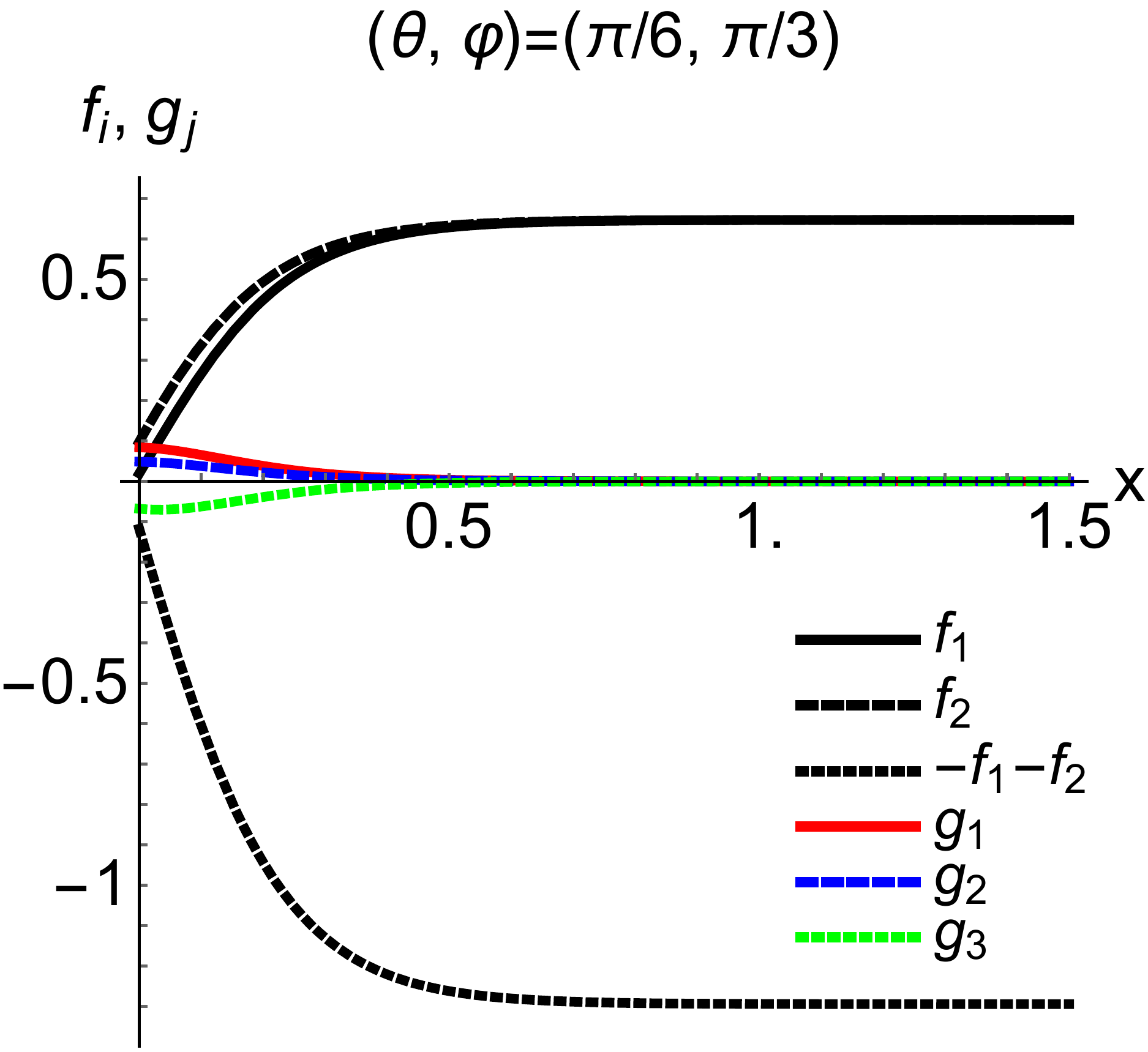}
\hspace{1em} %
\includegraphics[scale=0.18]{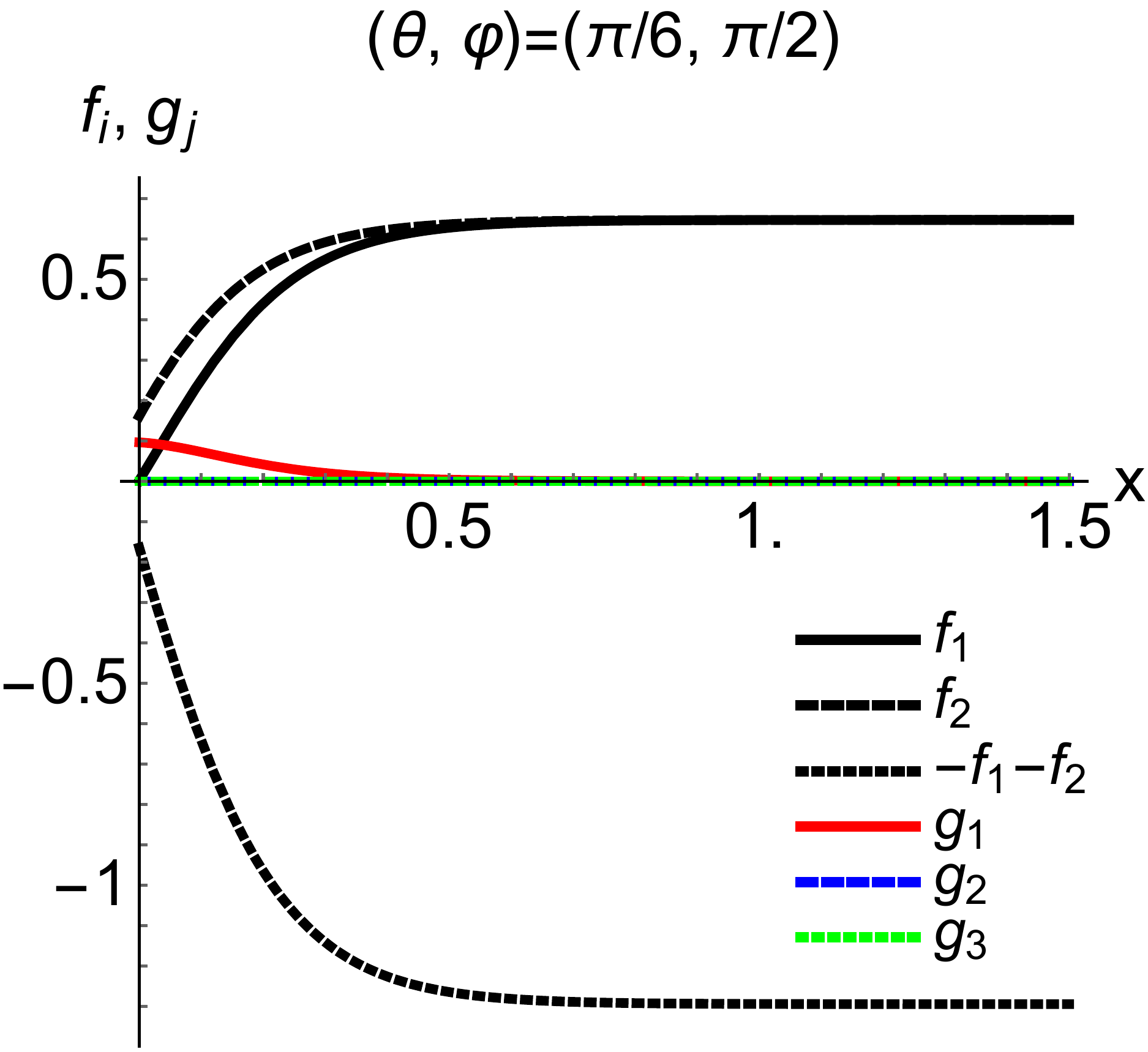}
\\ \vspace{2em} 
\includegraphics[scale=0.18]{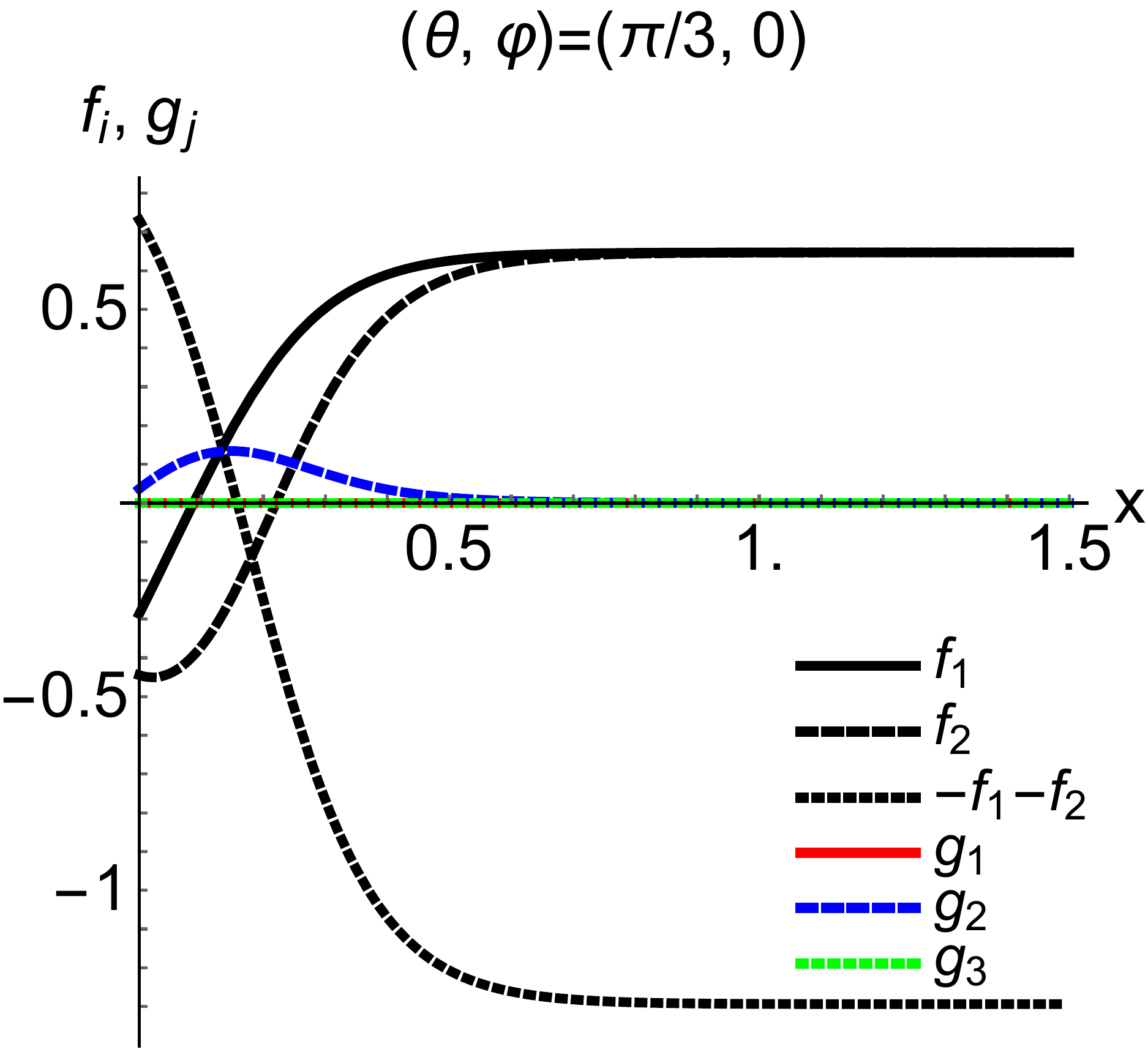}
\hspace{1em} %
\includegraphics[scale=0.18]{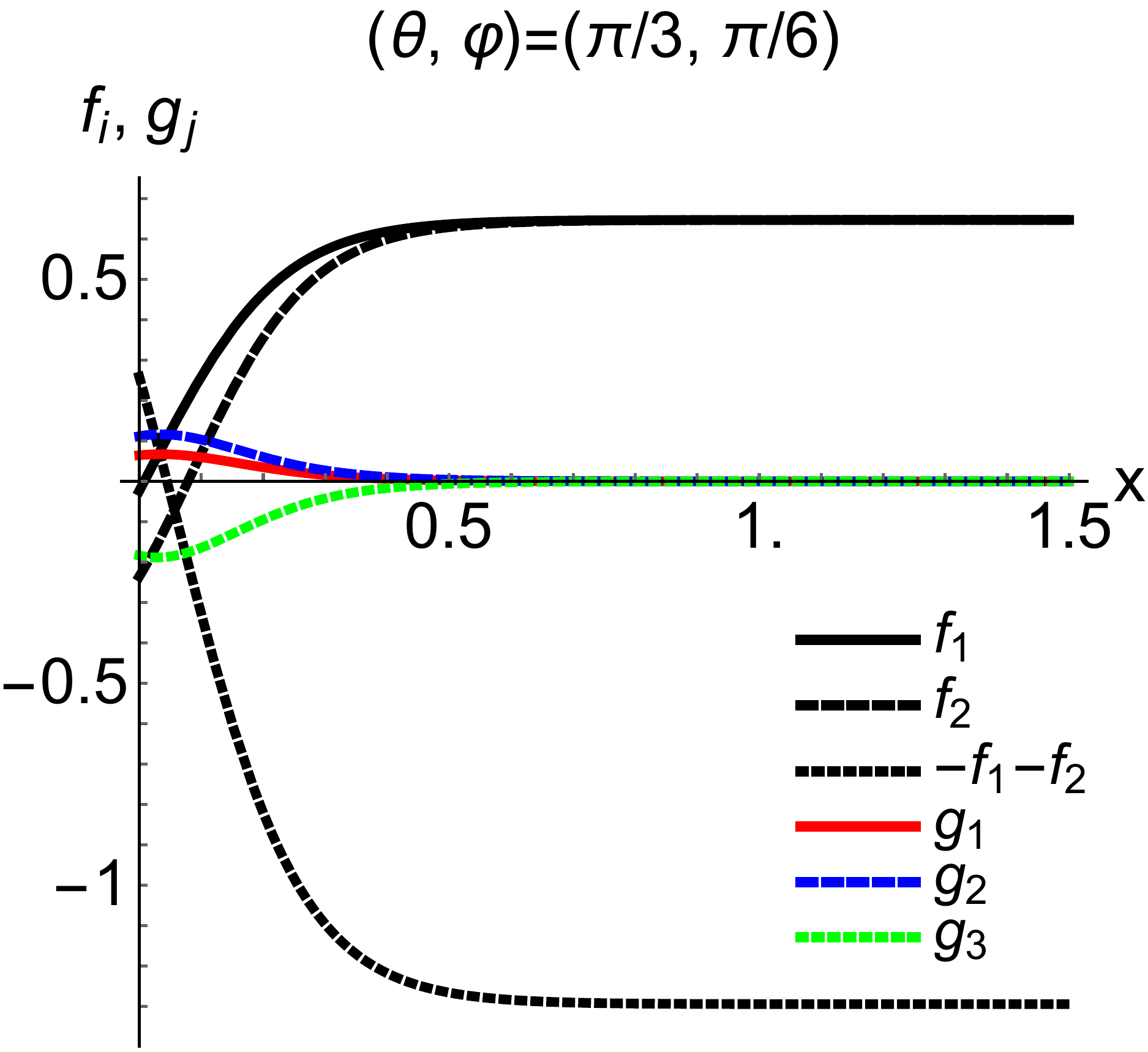}
\hspace{1em} %
\includegraphics[scale=0.18]{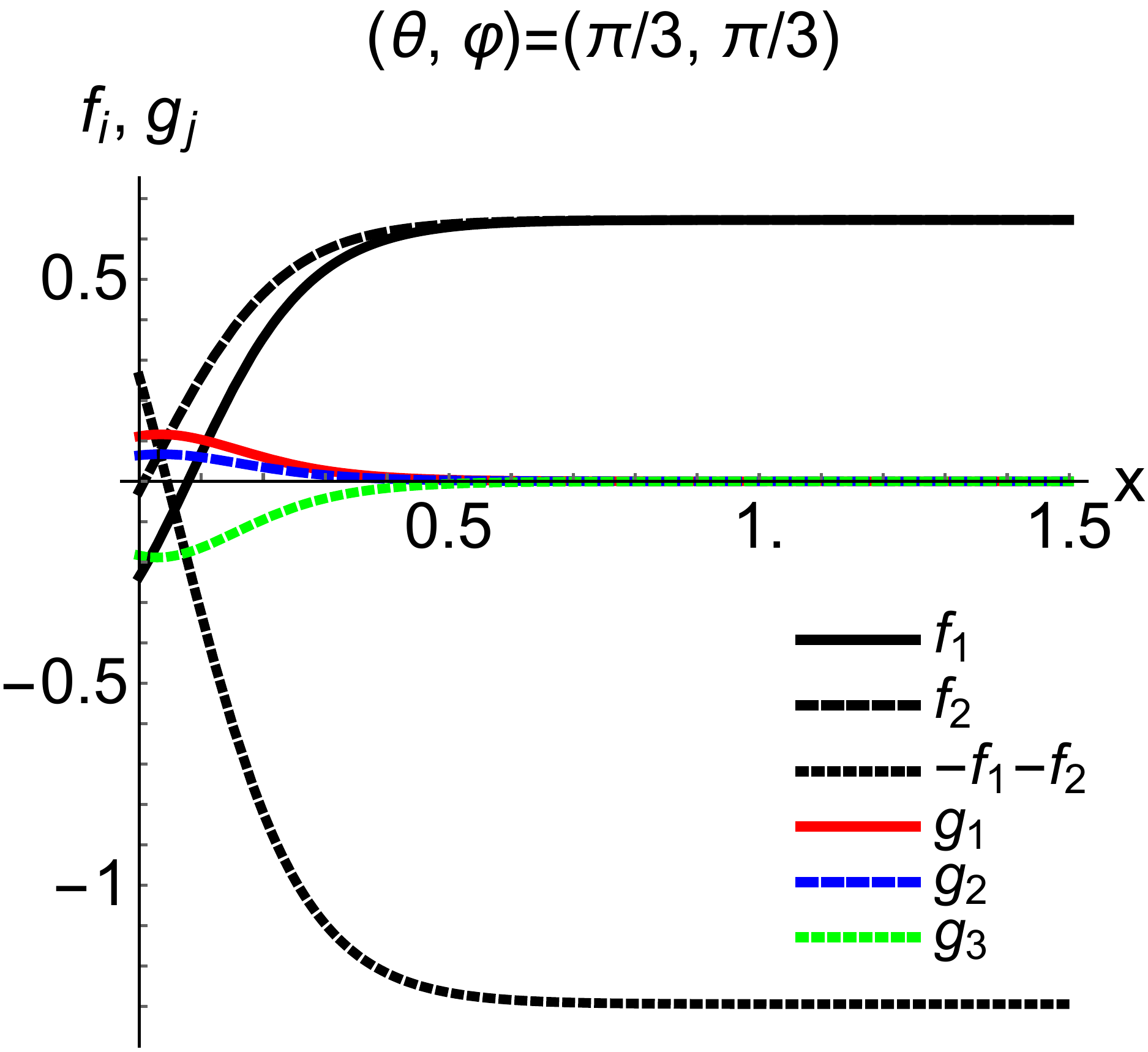}
\hspace{1em} %
\includegraphics[scale=0.18]{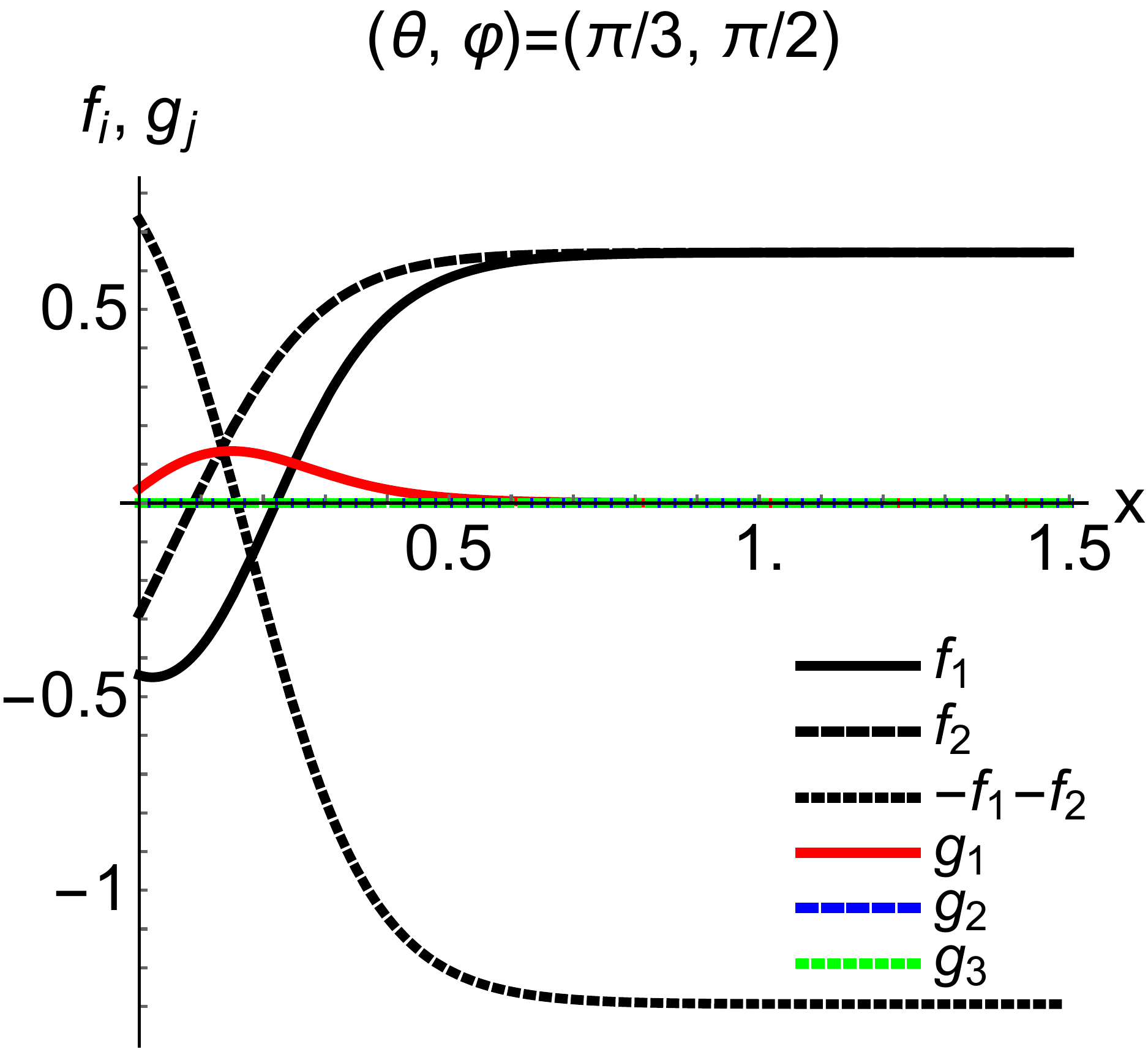}
\\ \vspace{2em} 
\includegraphics[scale=0.18]{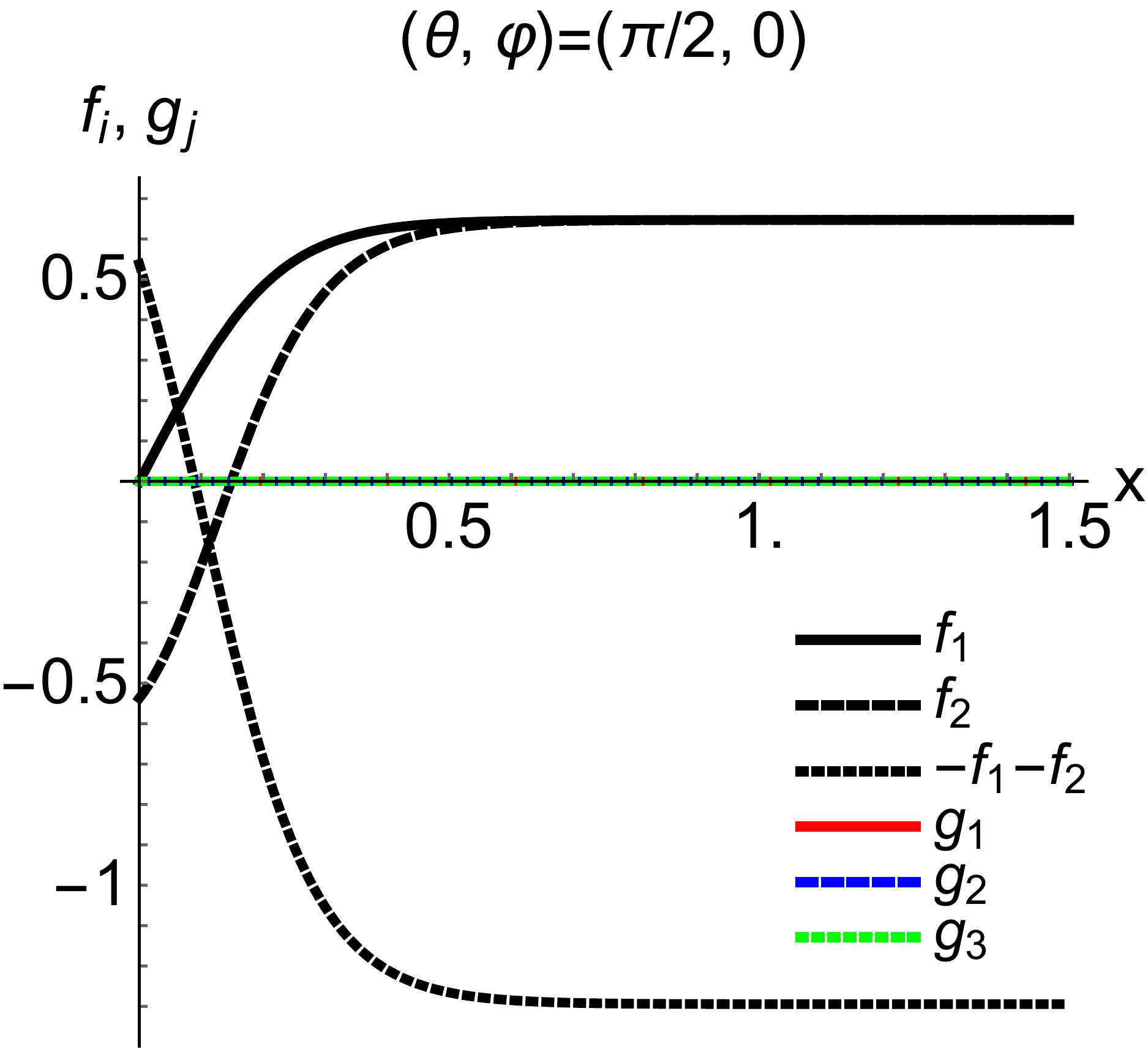}
\hspace{1em} %
\includegraphics[scale=0.18]{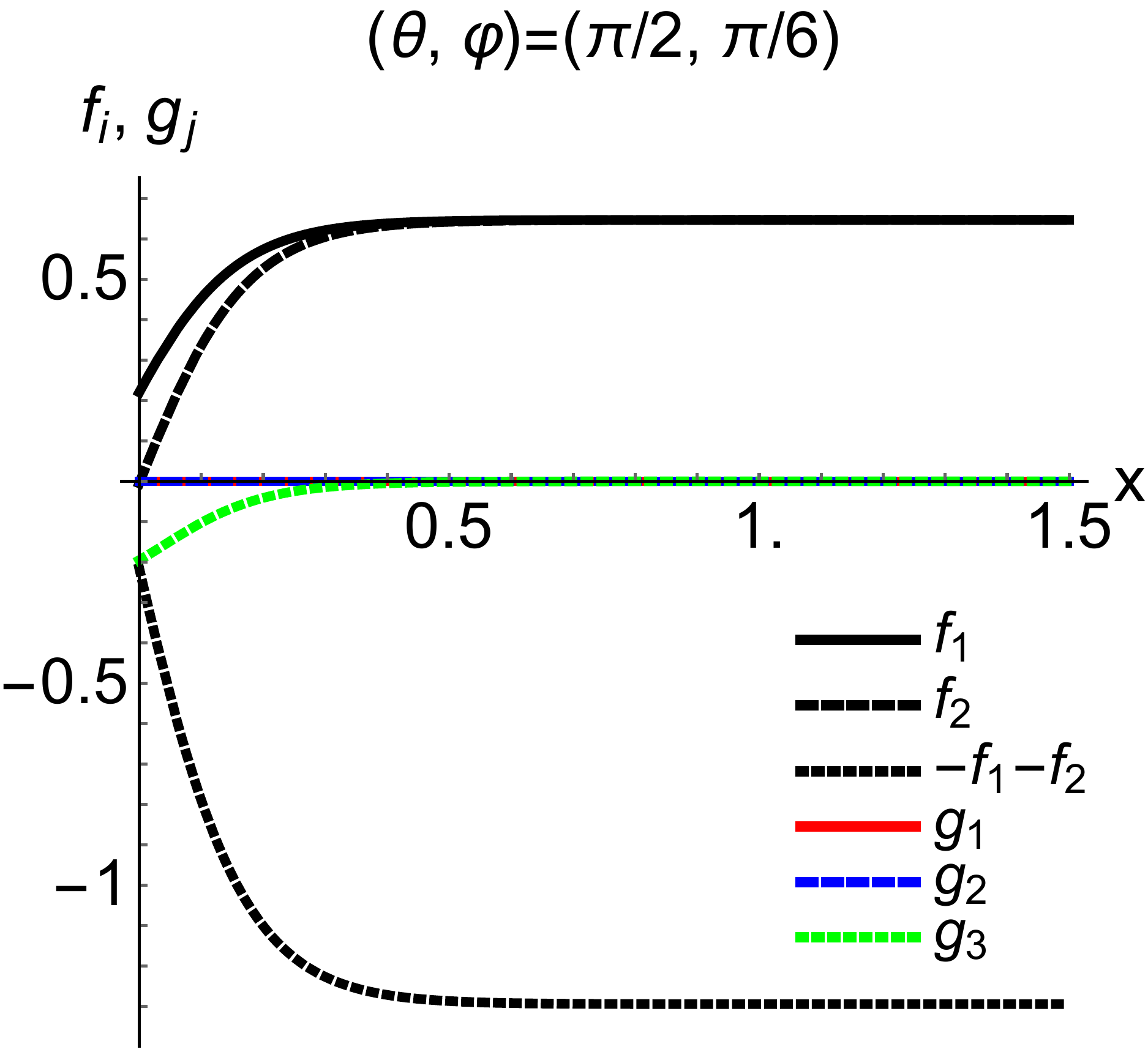}
\hspace{1em} %
\includegraphics[scale=0.18]{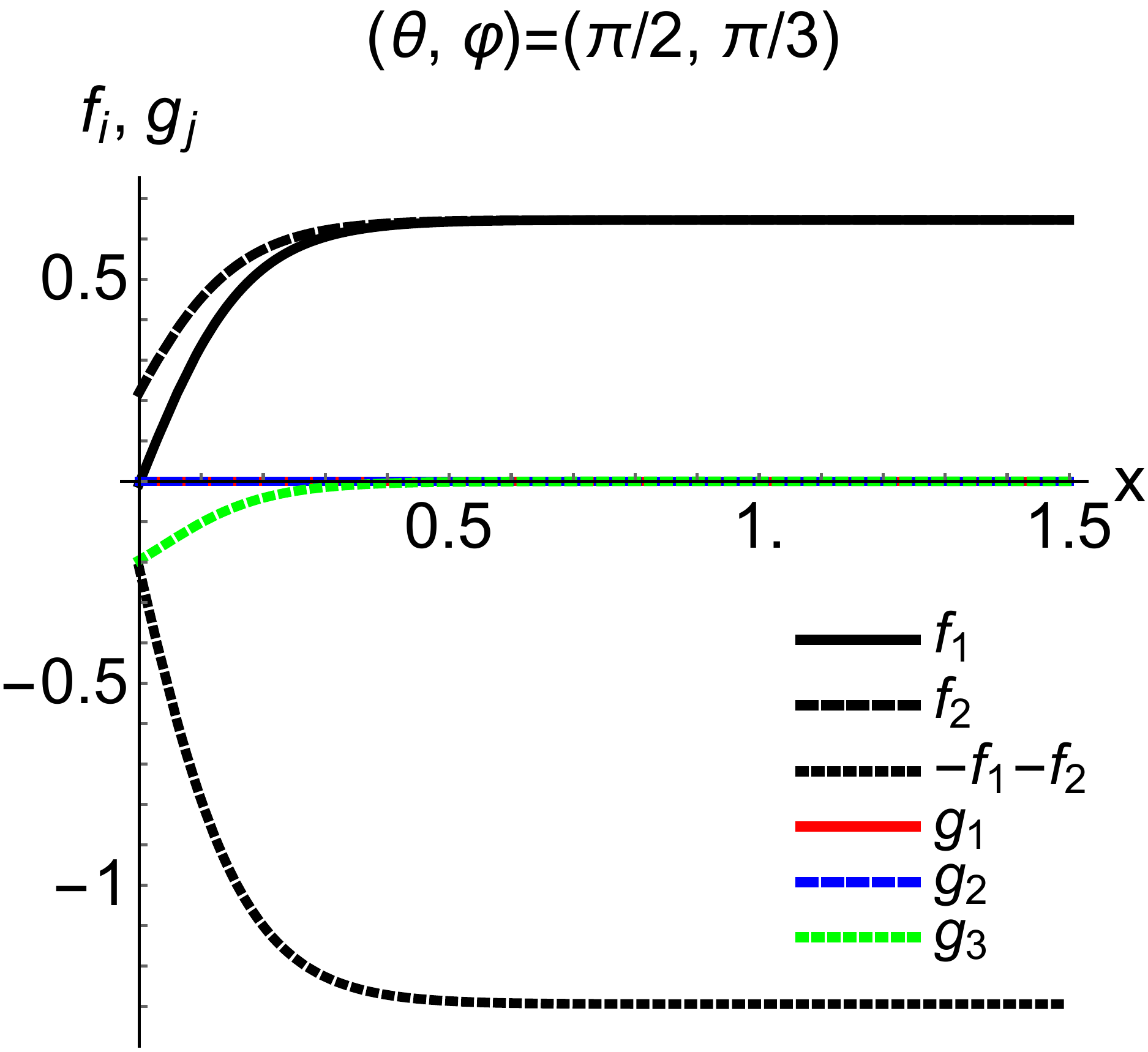}
\hspace{1em} %
\includegraphics[scale=0.18]{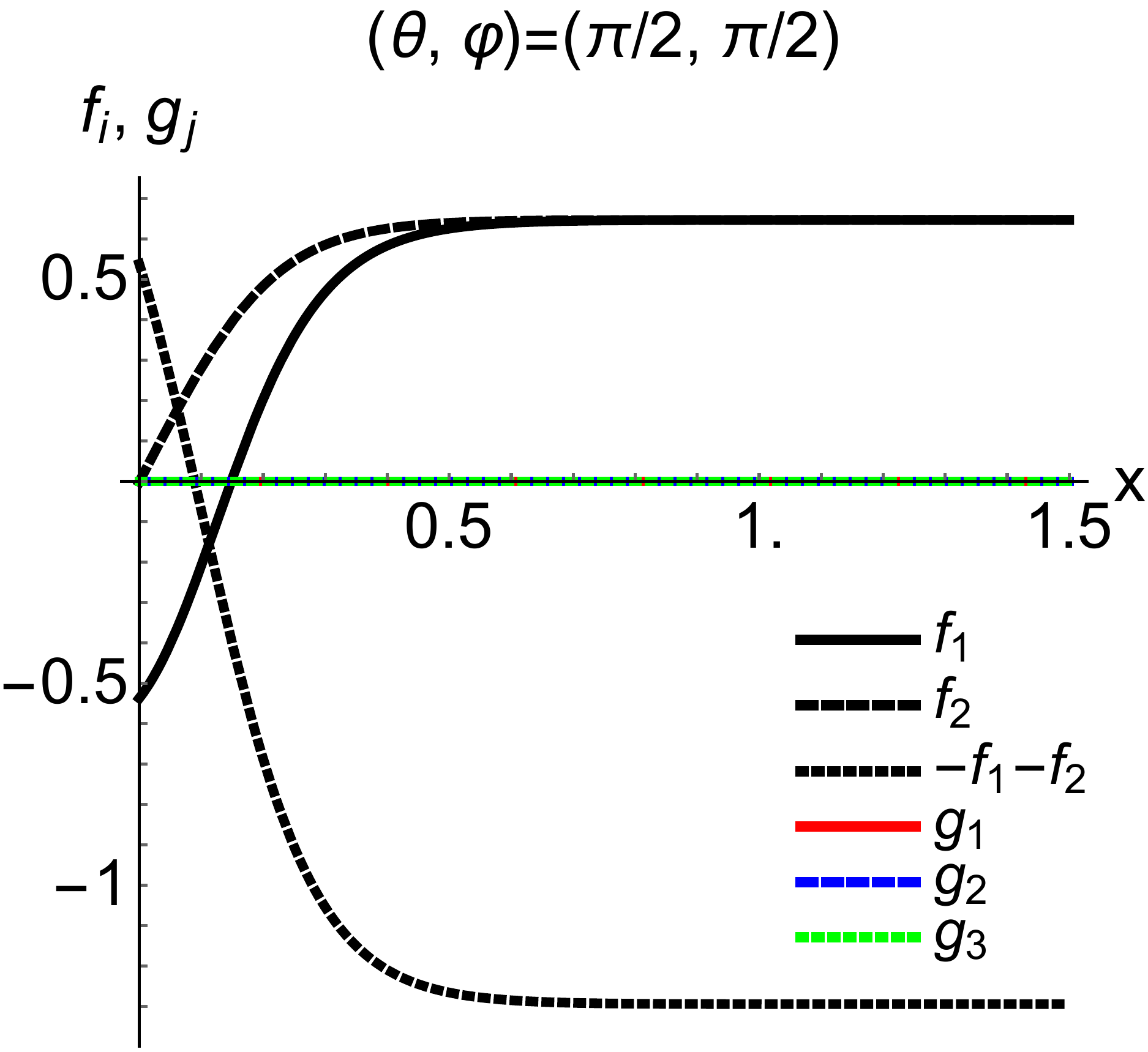}
\caption{The plots of the profile functions $f_{1}(x)$, $f_{2}(x)$, $g_{1}(x)$, $g_{2}(x)$, $g_{3}(x)$ for the bulk condition (i) $(f_{1}^{\bulk},f_{2}^{\bulk})=(0.64,0.64)$ at $t=0.9$ and $b=0$ (the bulk UN phase).}
\label{fig:profile_functions_3P2_bulk1}
\end{center}
\end{figure}

\begin{figure}[p] 
\begin{center}
\includegraphics[scale=0.18]{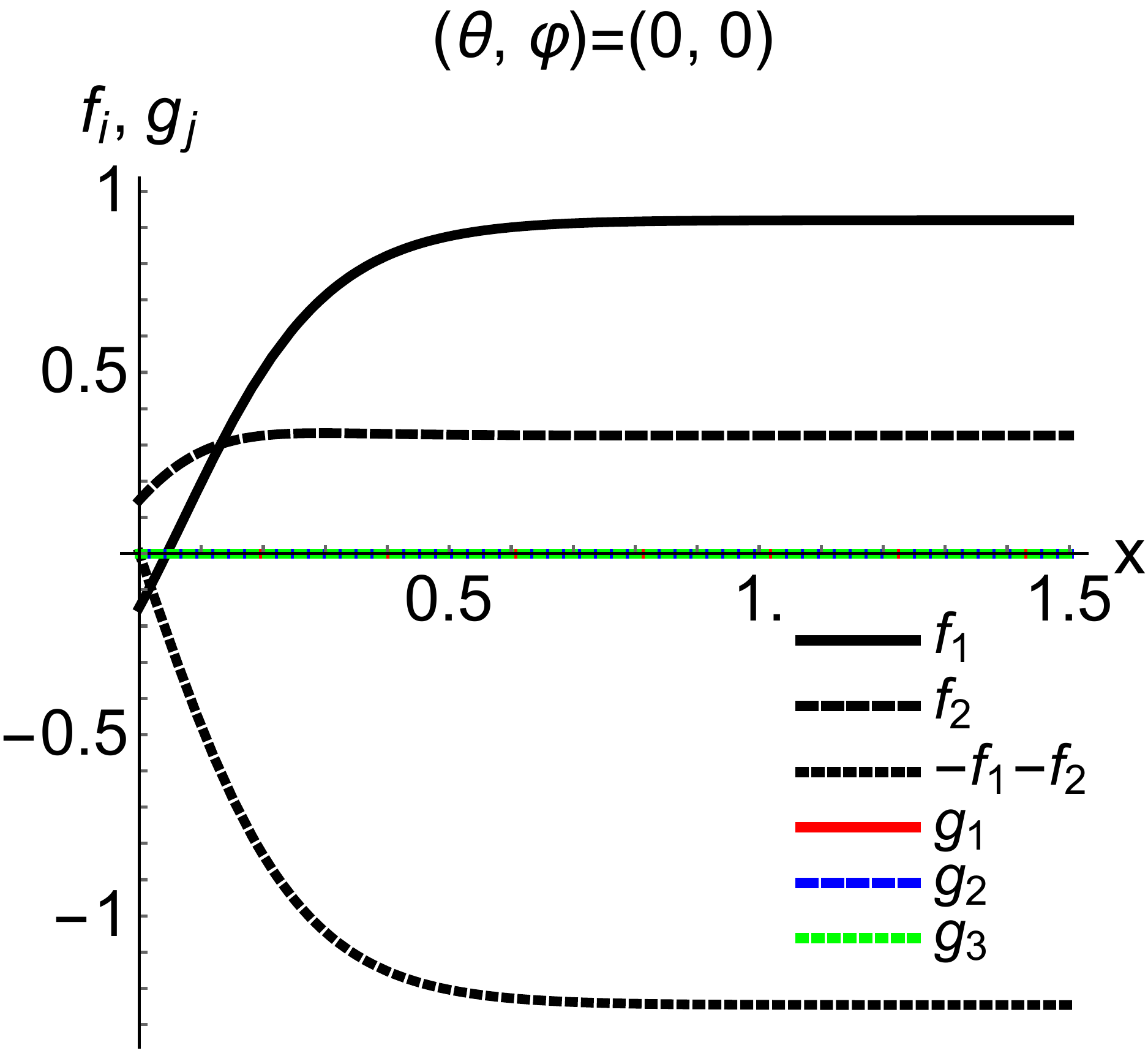}
\\ \vspace{2em} 
\includegraphics[scale=0.18]{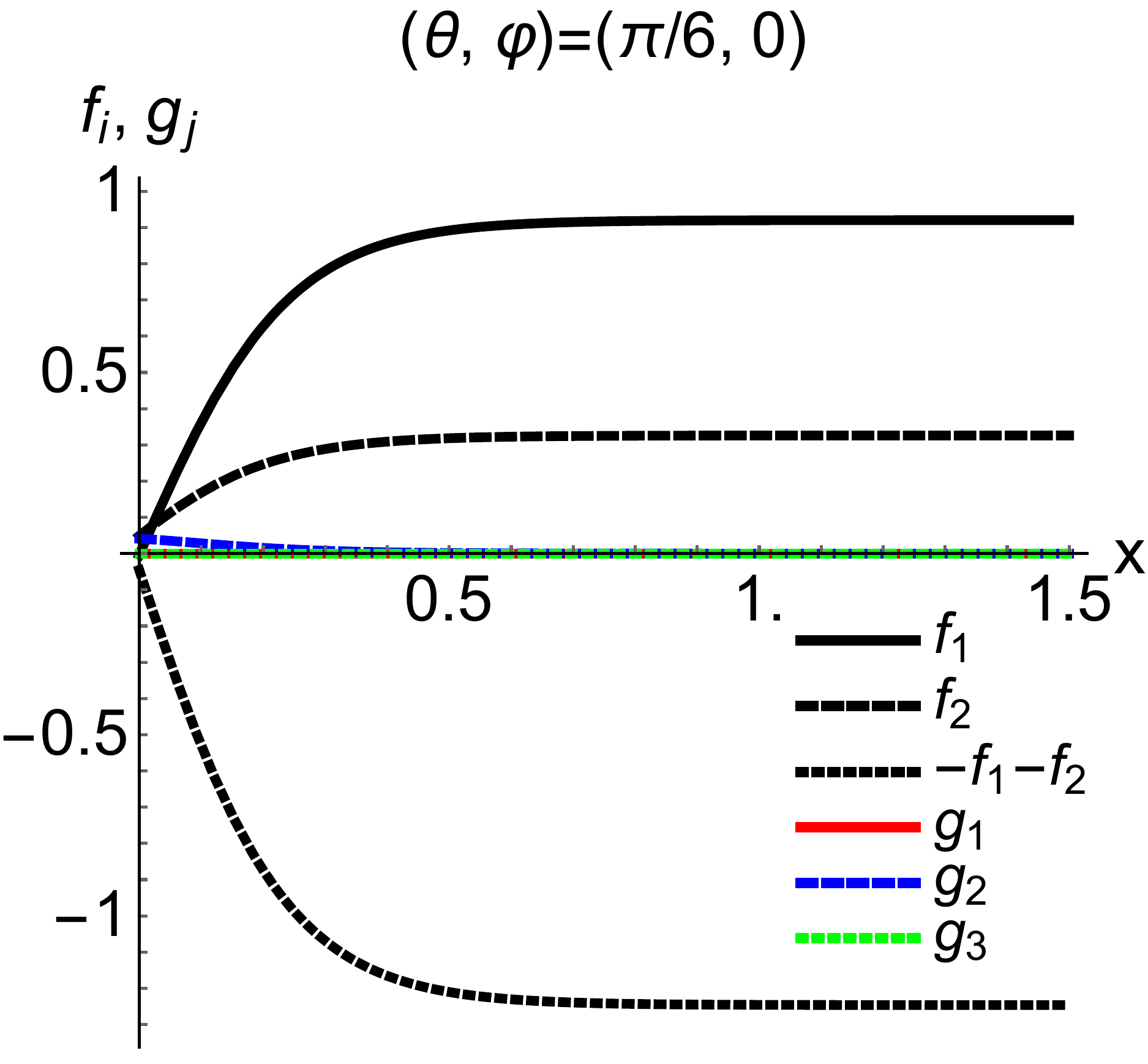}
\hspace{1em} %
\includegraphics[scale=0.18]{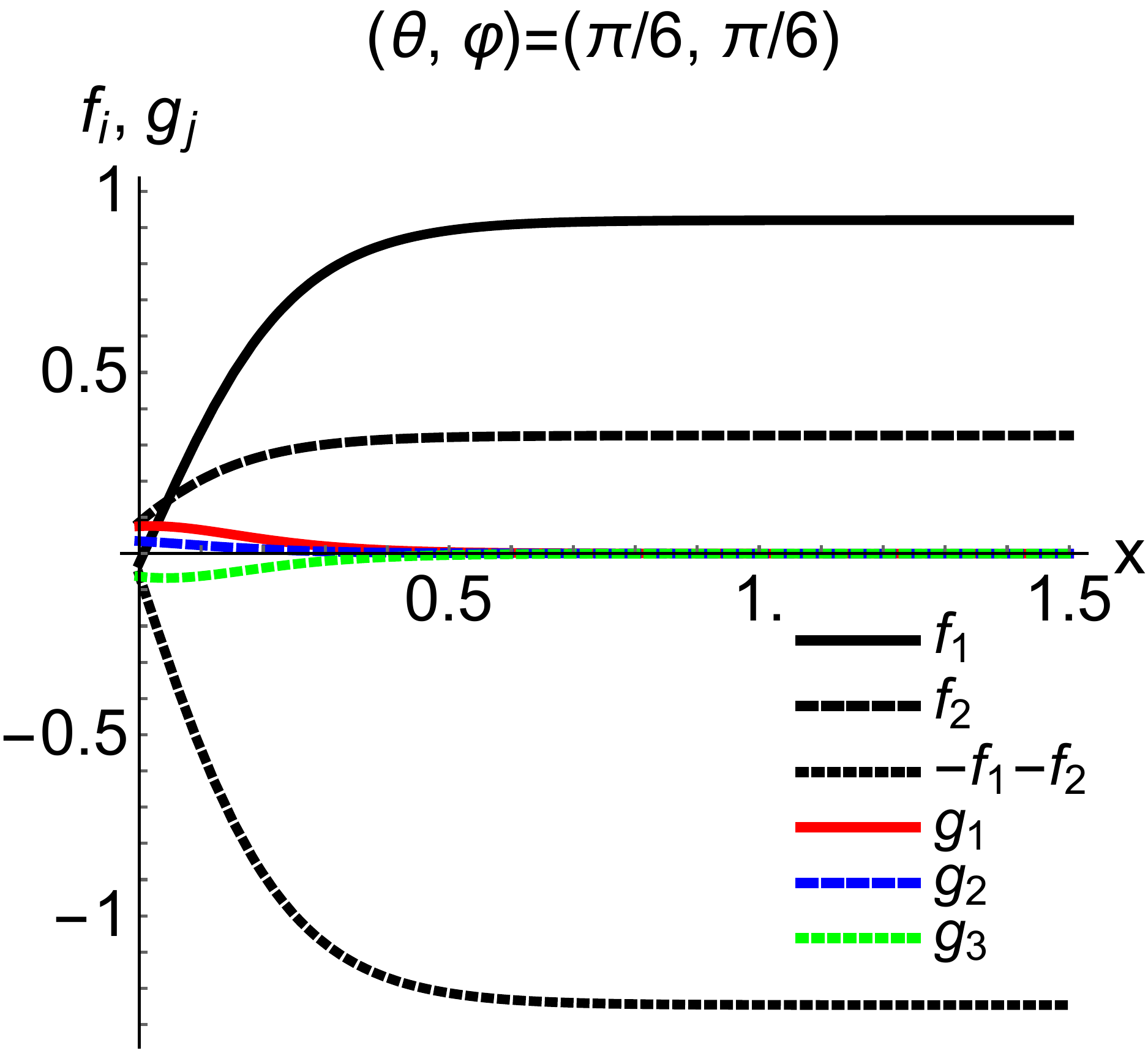}
\hspace{1em} %
\includegraphics[scale=0.18]{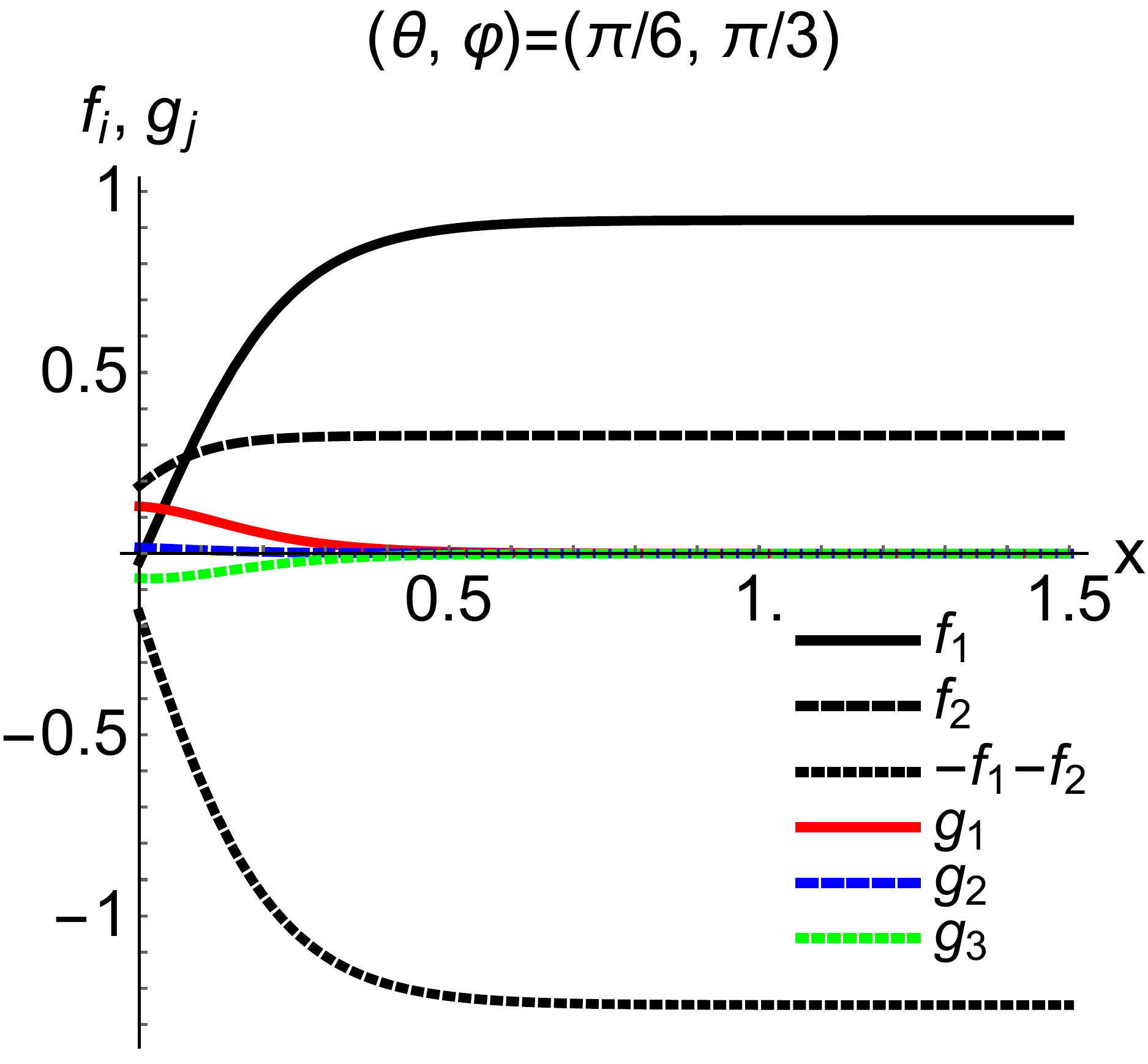}
\hspace{1em} %
\includegraphics[scale=0.18]{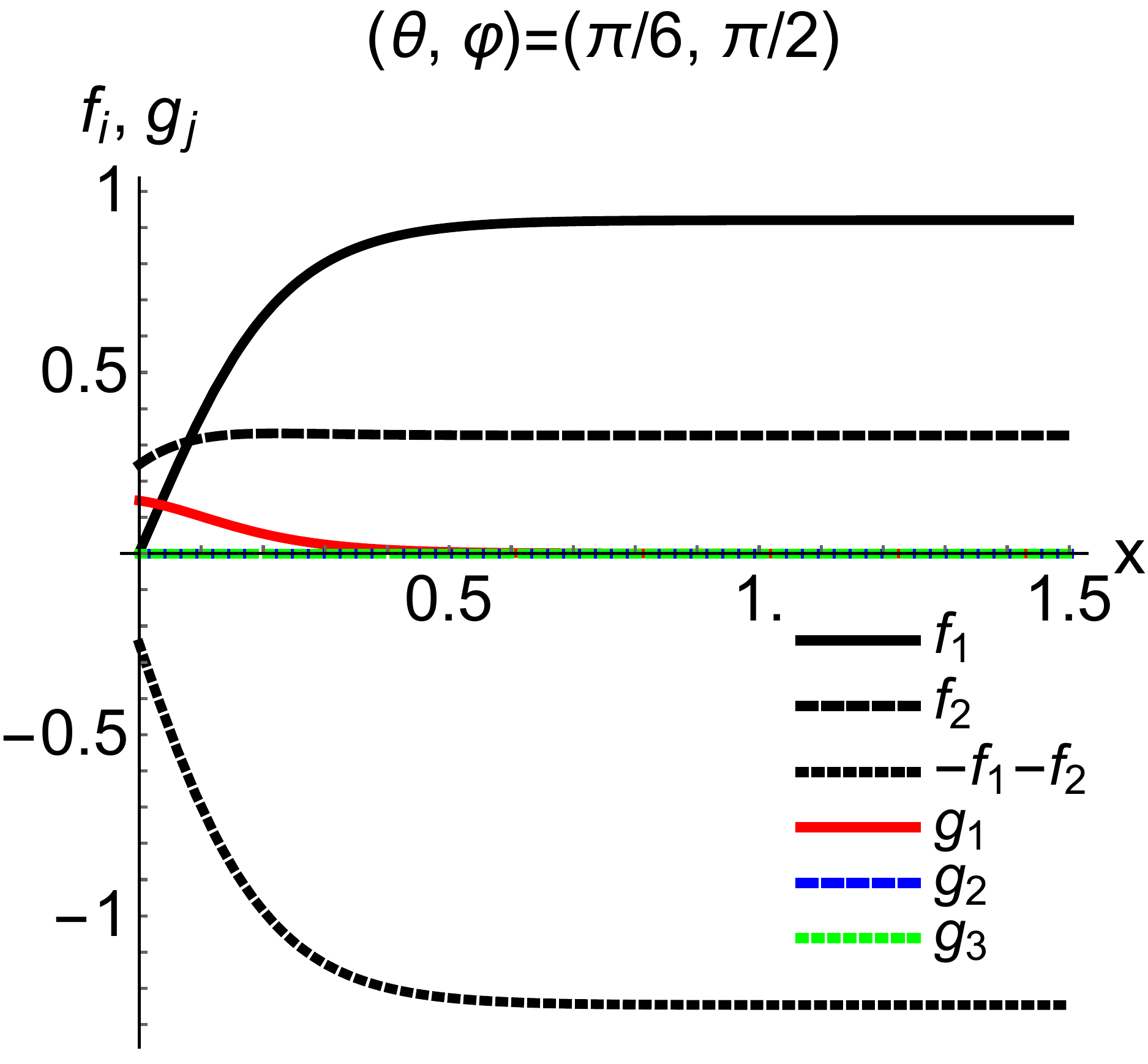}
\\ \vspace{2em} 
\includegraphics[scale=0.18]{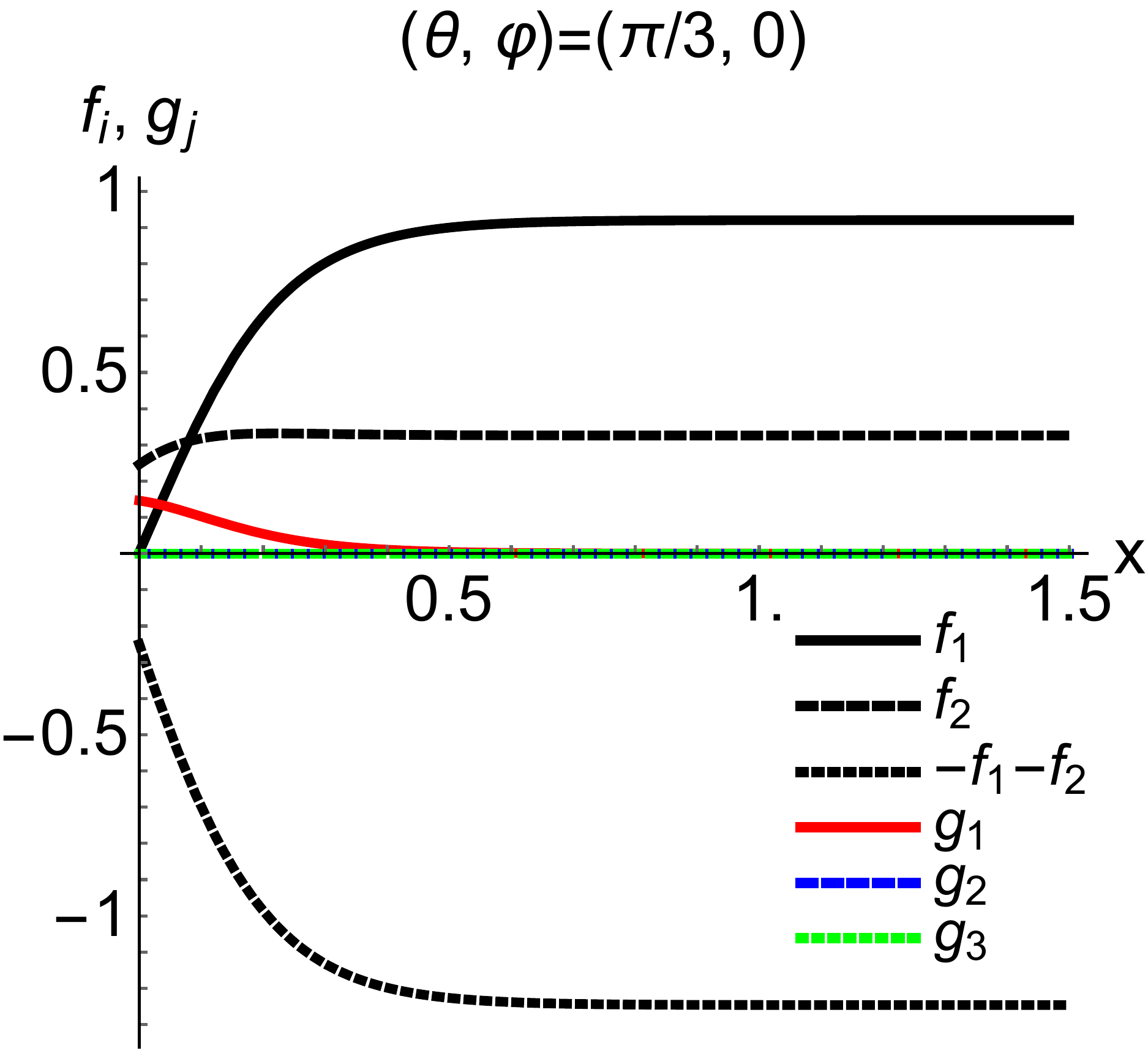}
\hspace{1em} %
\includegraphics[scale=0.18]{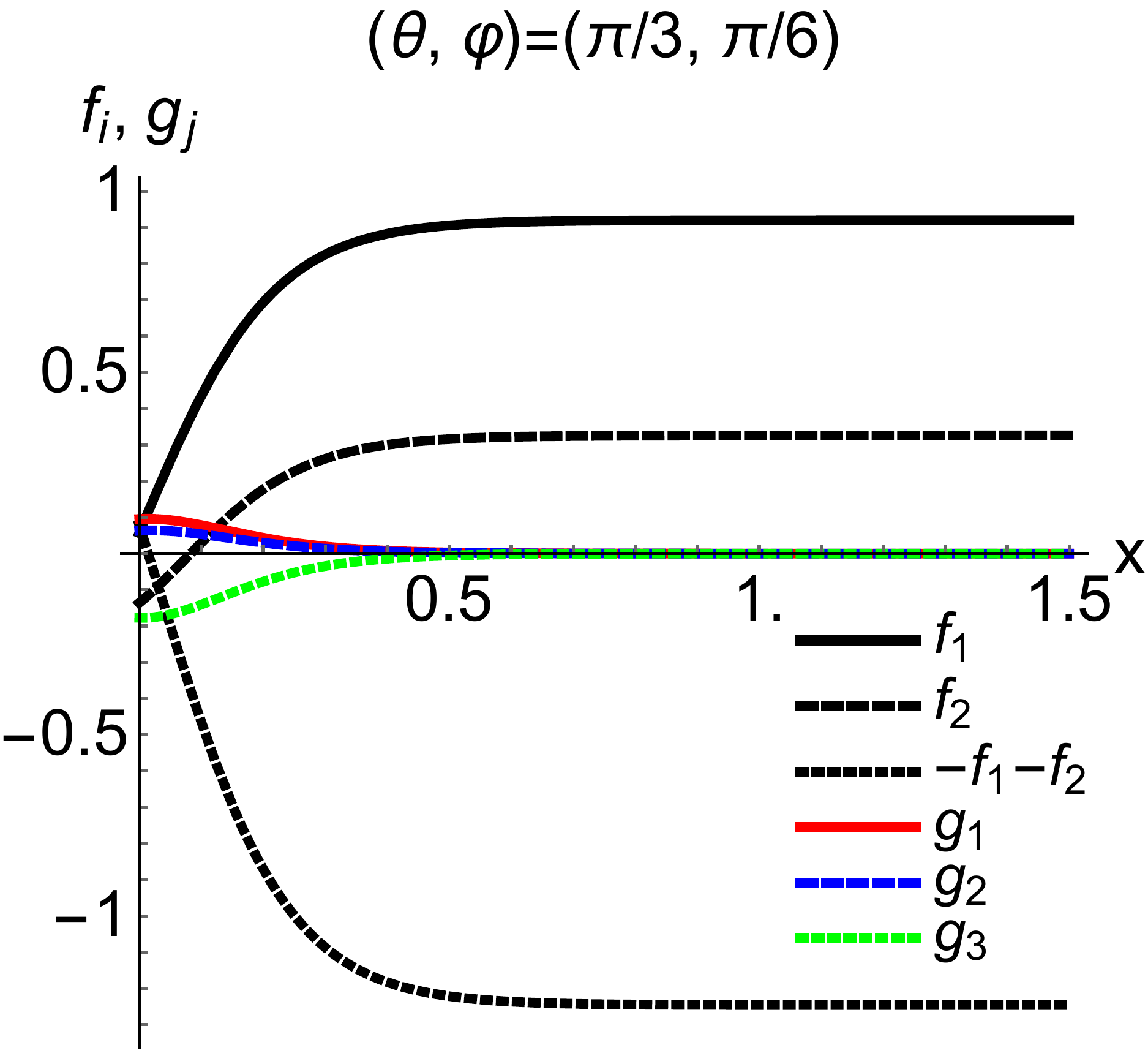}
\hspace{1em} %
\includegraphics[scale=0.18]{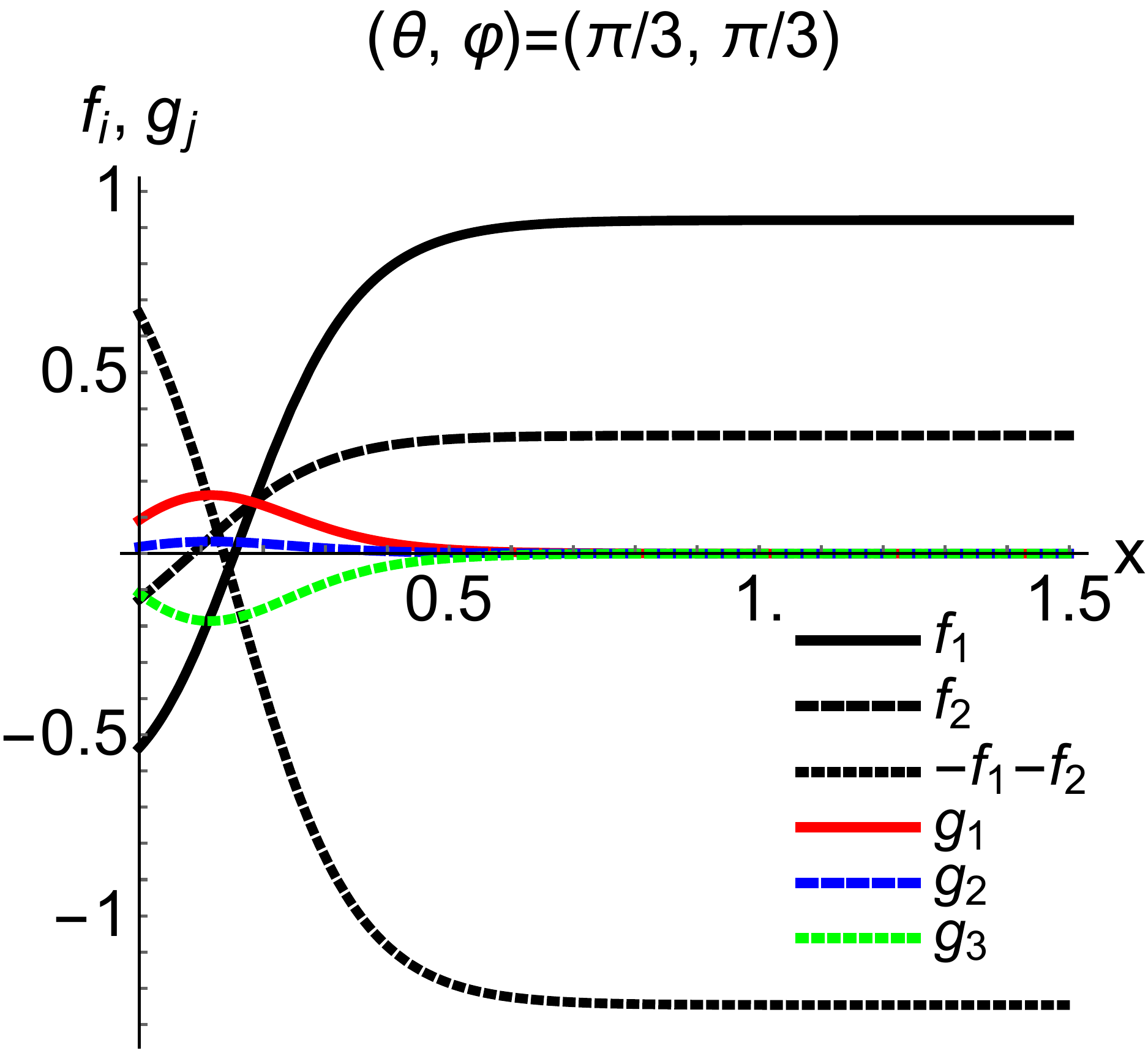}
\hspace{1em} %
\includegraphics[scale=0.18]{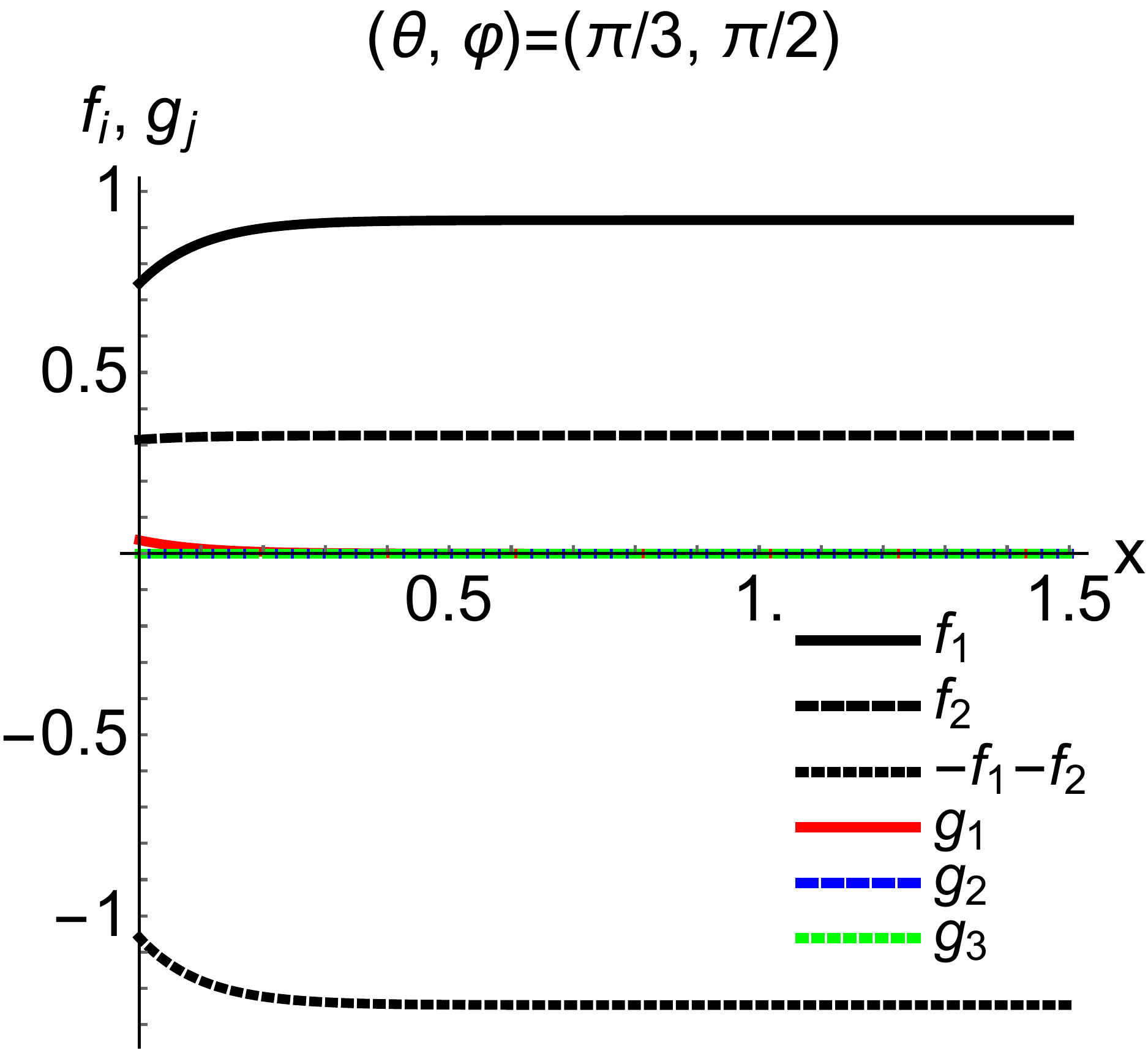}
\\ \vspace{2em} 
\includegraphics[scale=0.18]{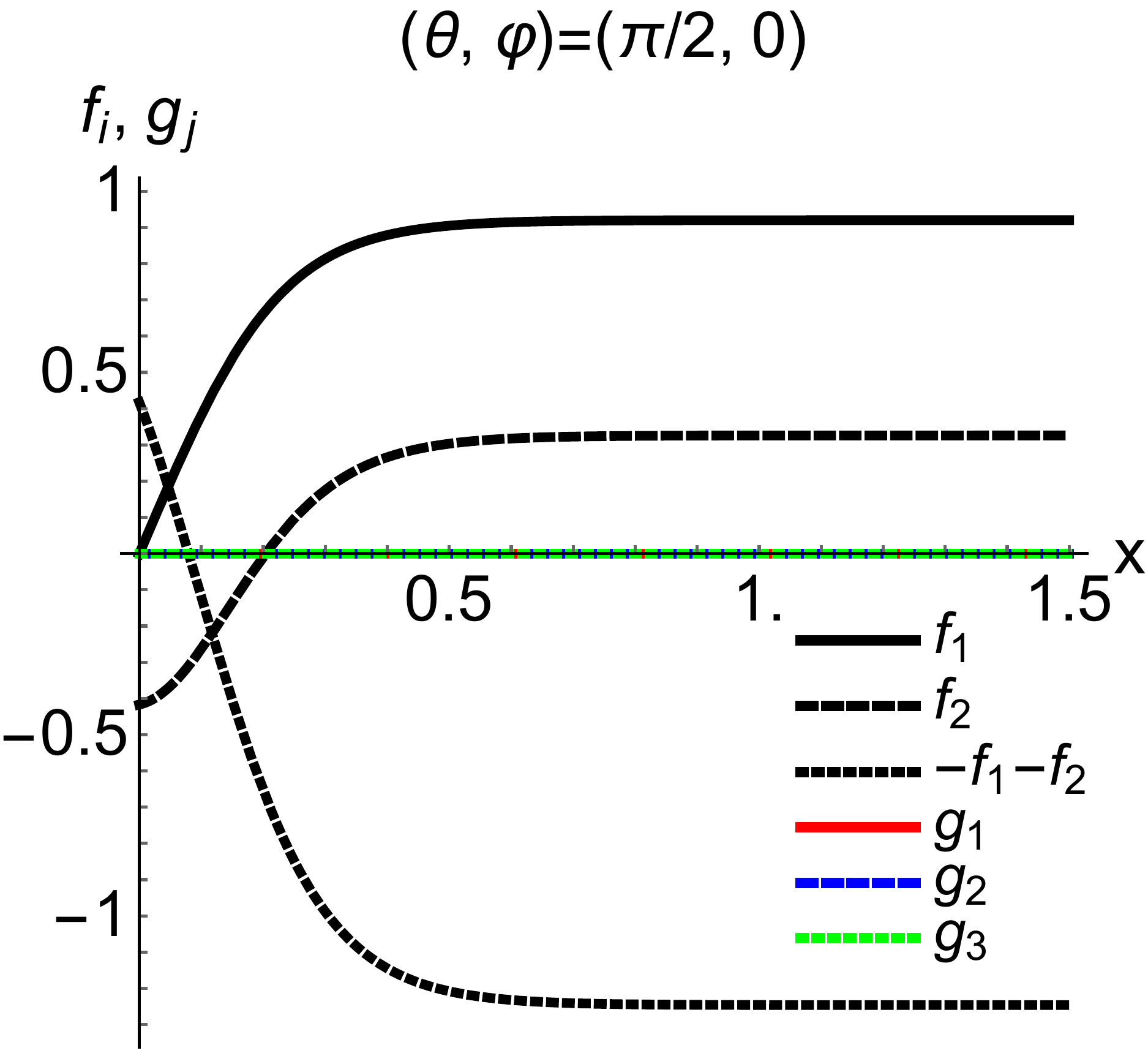}
\hspace{1em} %
\includegraphics[scale=0.18]{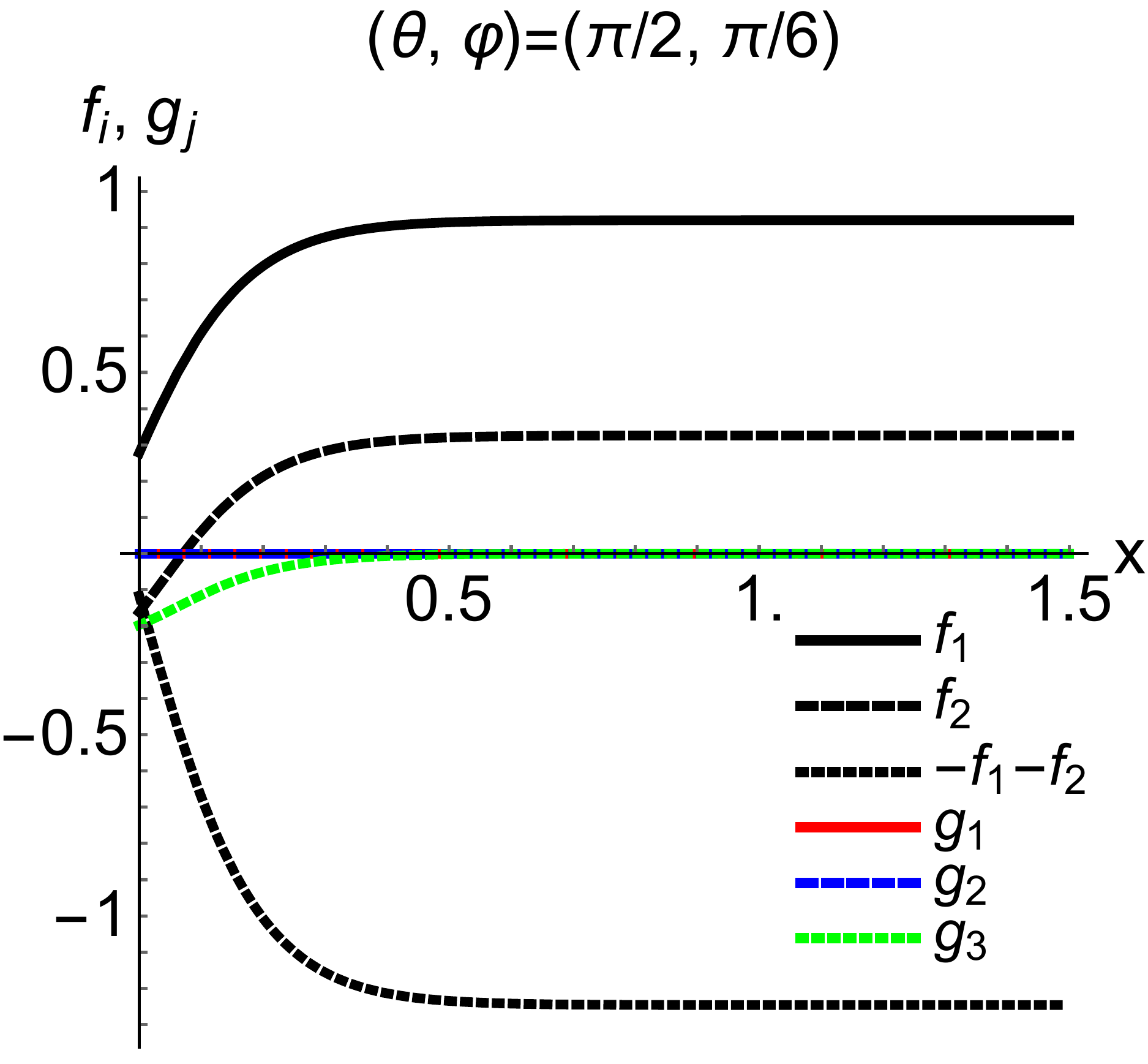}
\hspace{1em} %
\includegraphics[scale=0.18]{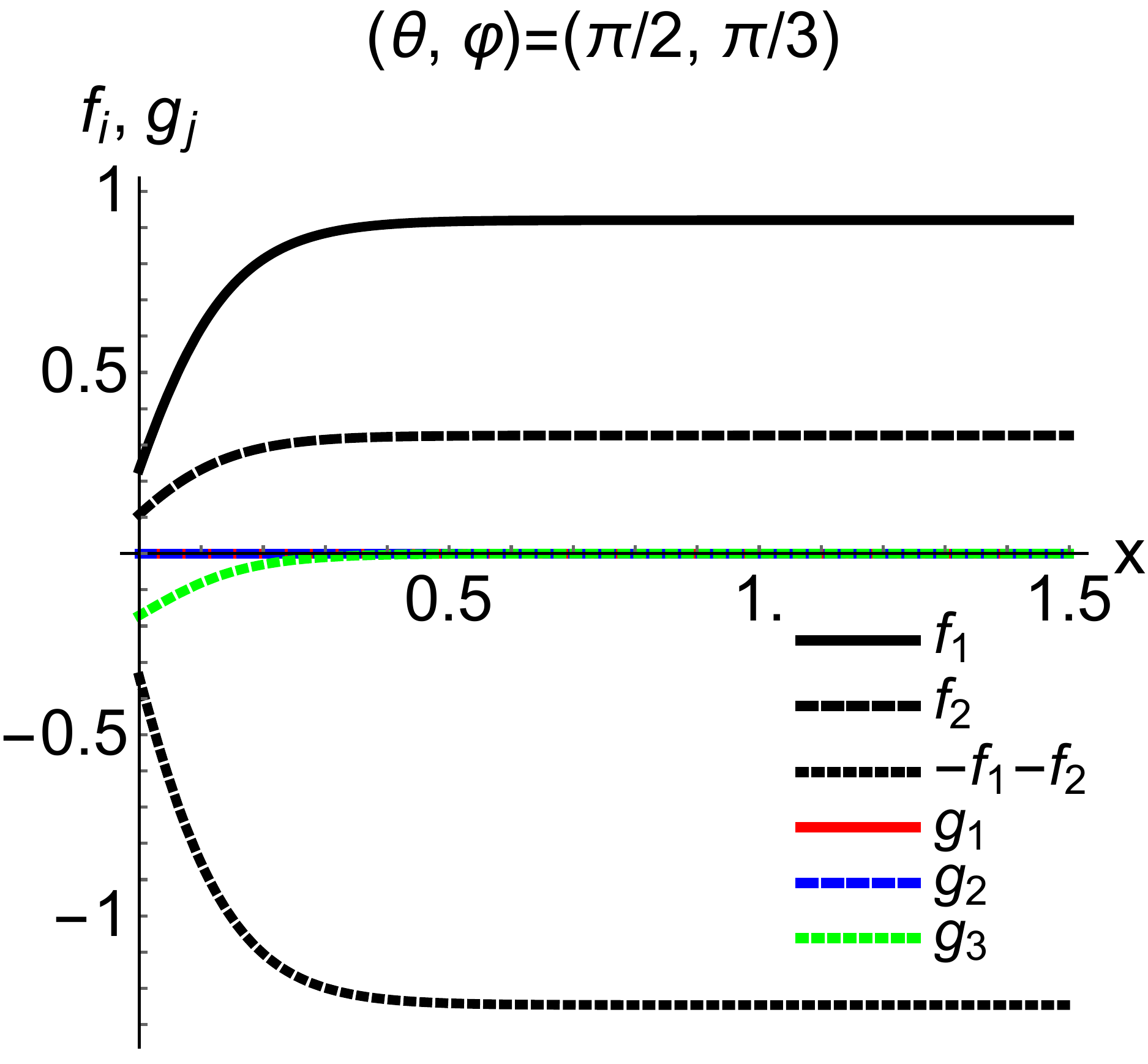}
\hspace{1em} %
\includegraphics[scale=0.18]{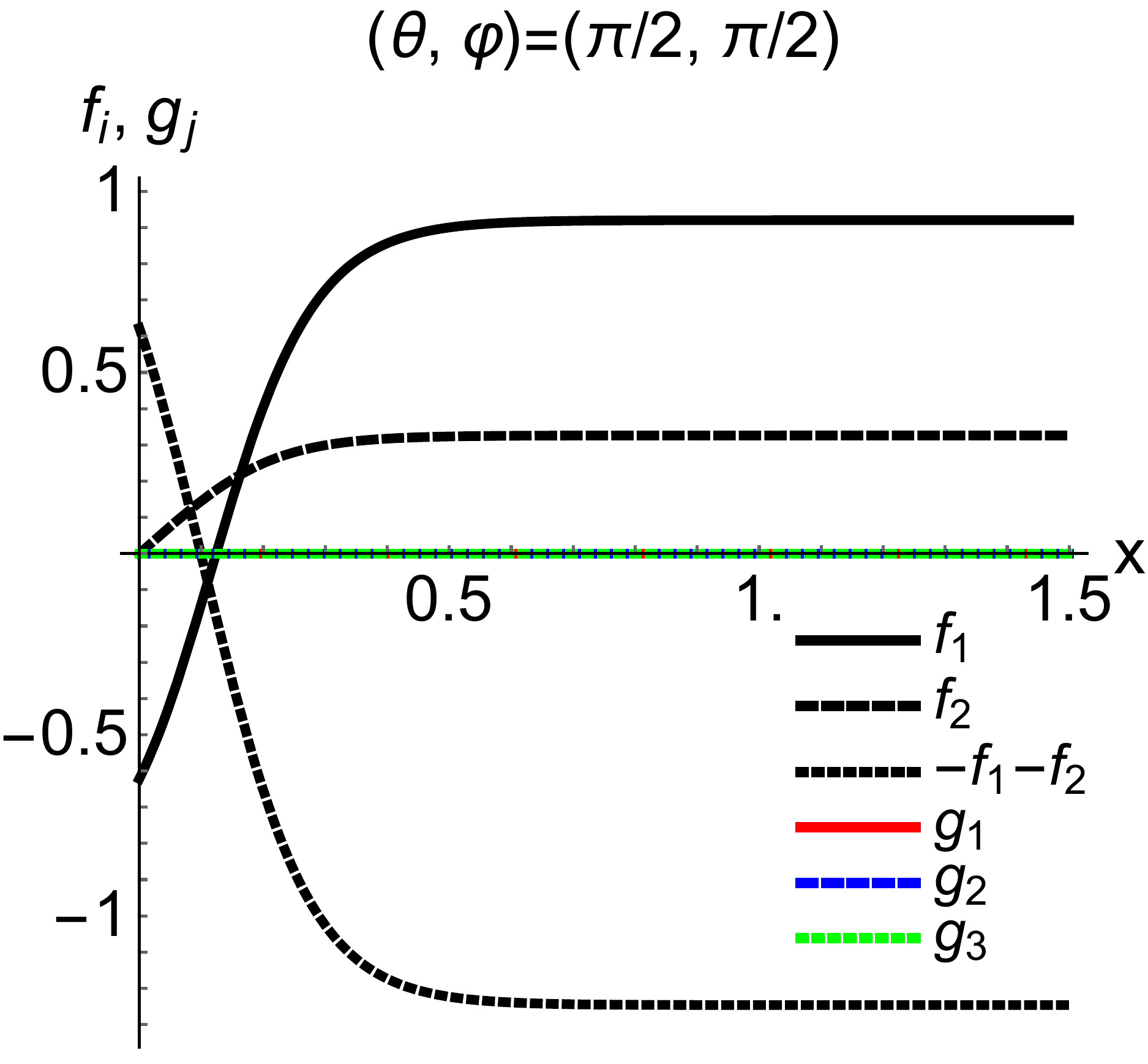}
\caption{The plots of the profile functions $f_{1}(x)$, $f_{2}(x)$, $g_{1}(x)$, $g_{2}(x)$, $g_{3}(x)$ for the bulk condition (ii) $(f_{1}^{\bulk},f_{2}^{\bulk})=(0.92,0.32)$ at $t=0.9$ and $b=0.15$ (the bulk D$_{2}$-BN phase).}
\label{fig:profile_functions_3P2_bulk2}
\end{center}
\end{figure}

\begin{figure}[p] 
\begin{center}
\includegraphics[scale=0.18]{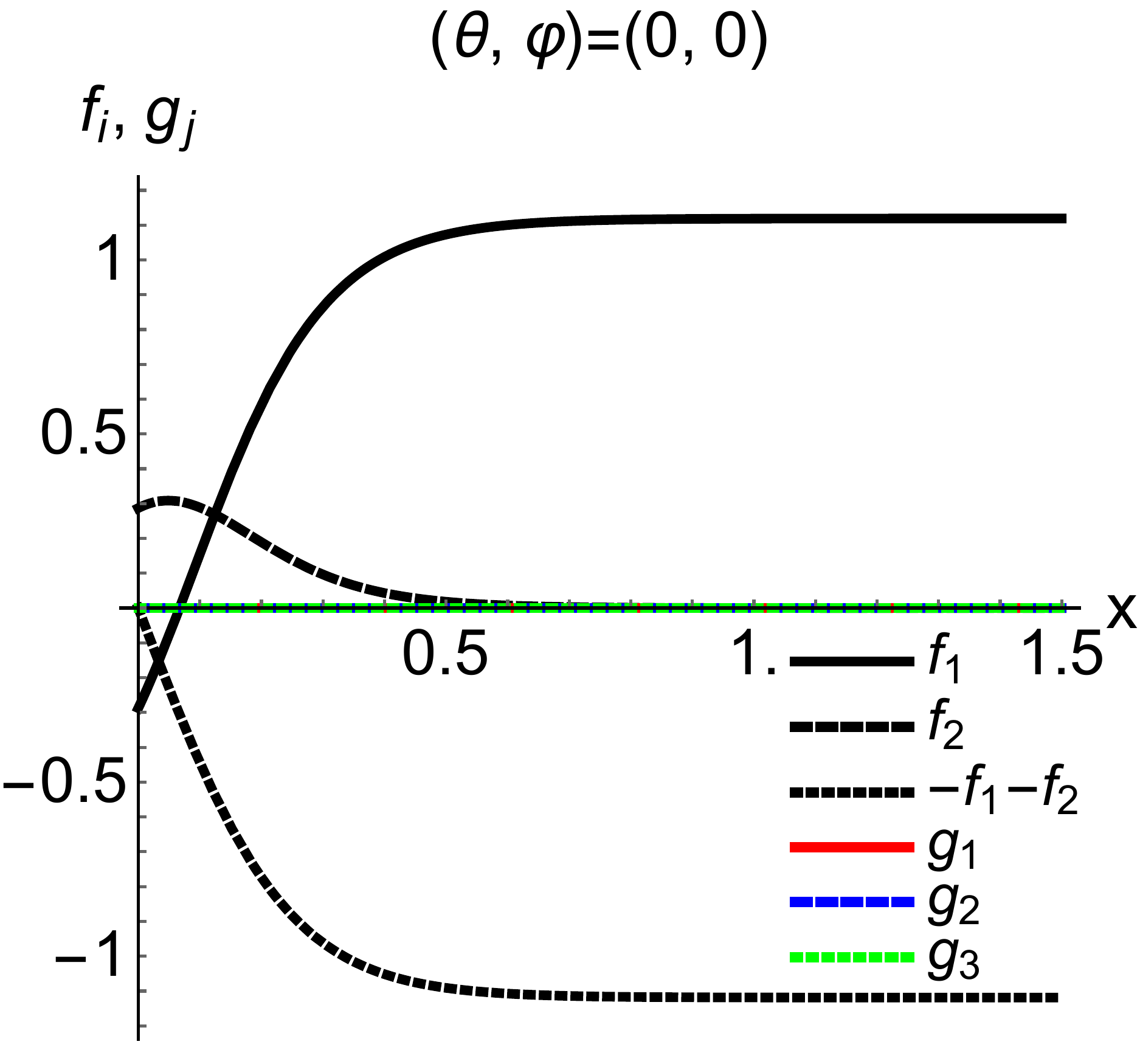}
\\ \vspace{2em} 
\includegraphics[scale=0.18]{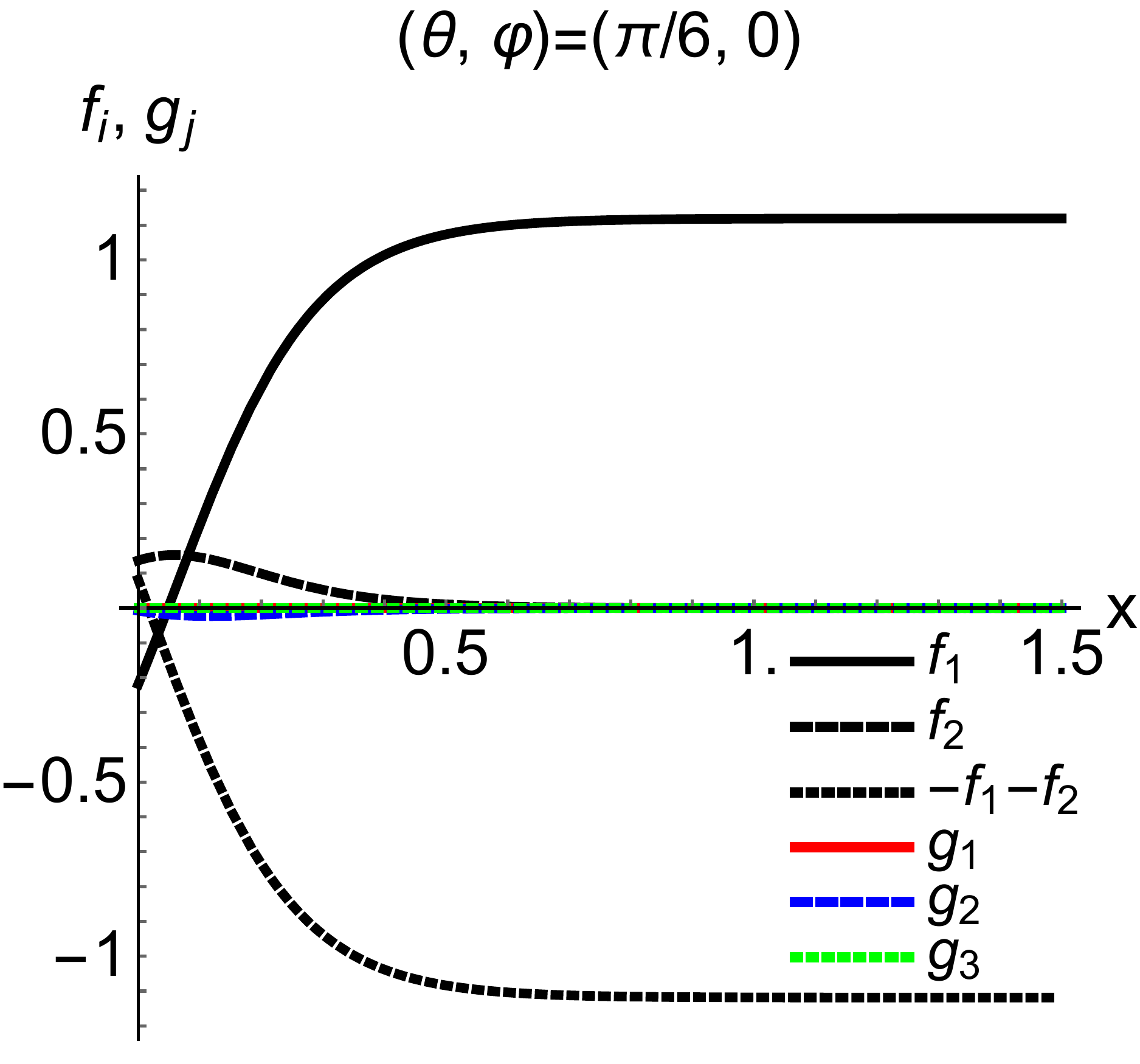}
\hspace{1em} %
\includegraphics[scale=0.18]{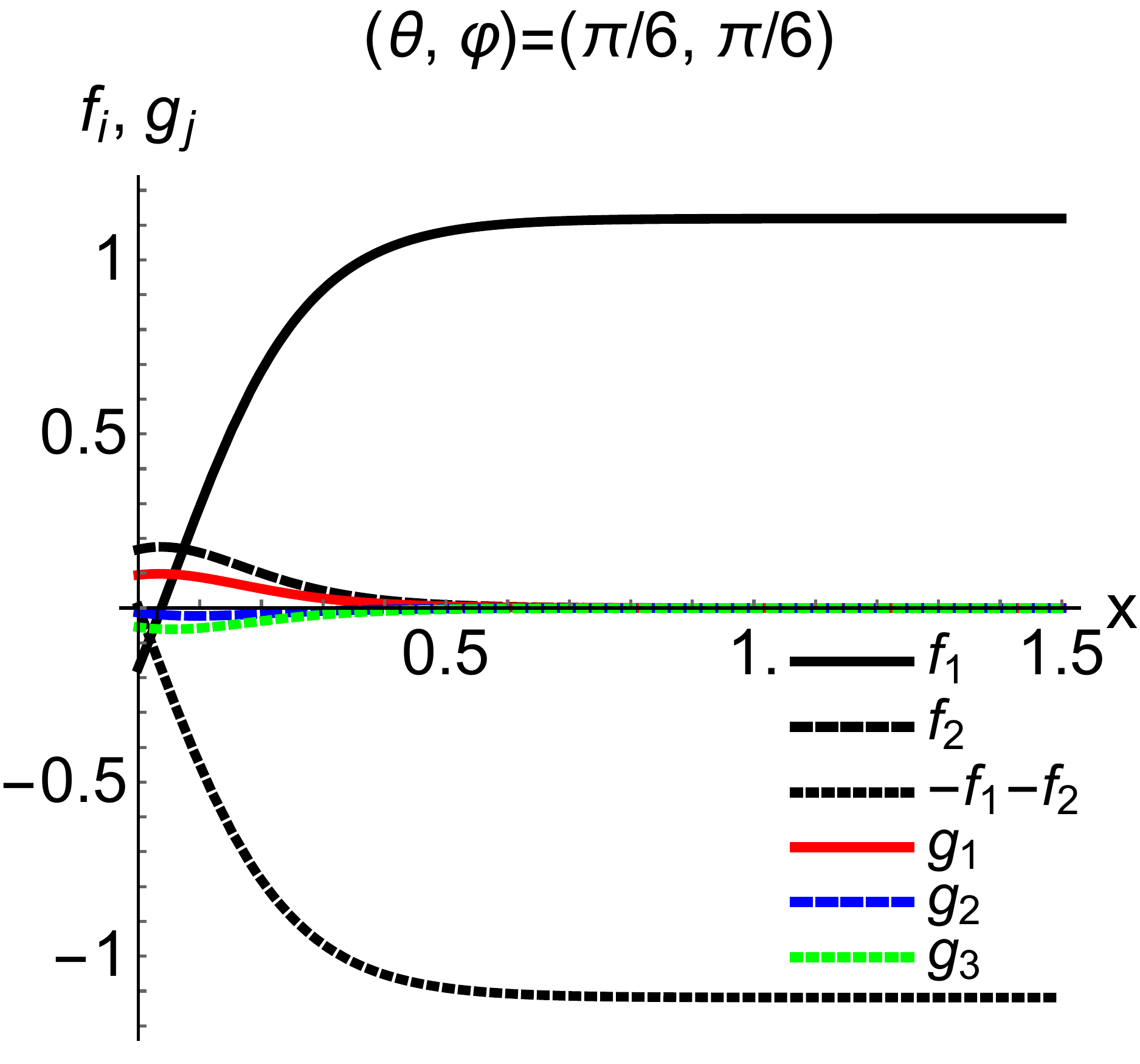}
\hspace{1em} %
\includegraphics[scale=0.18]{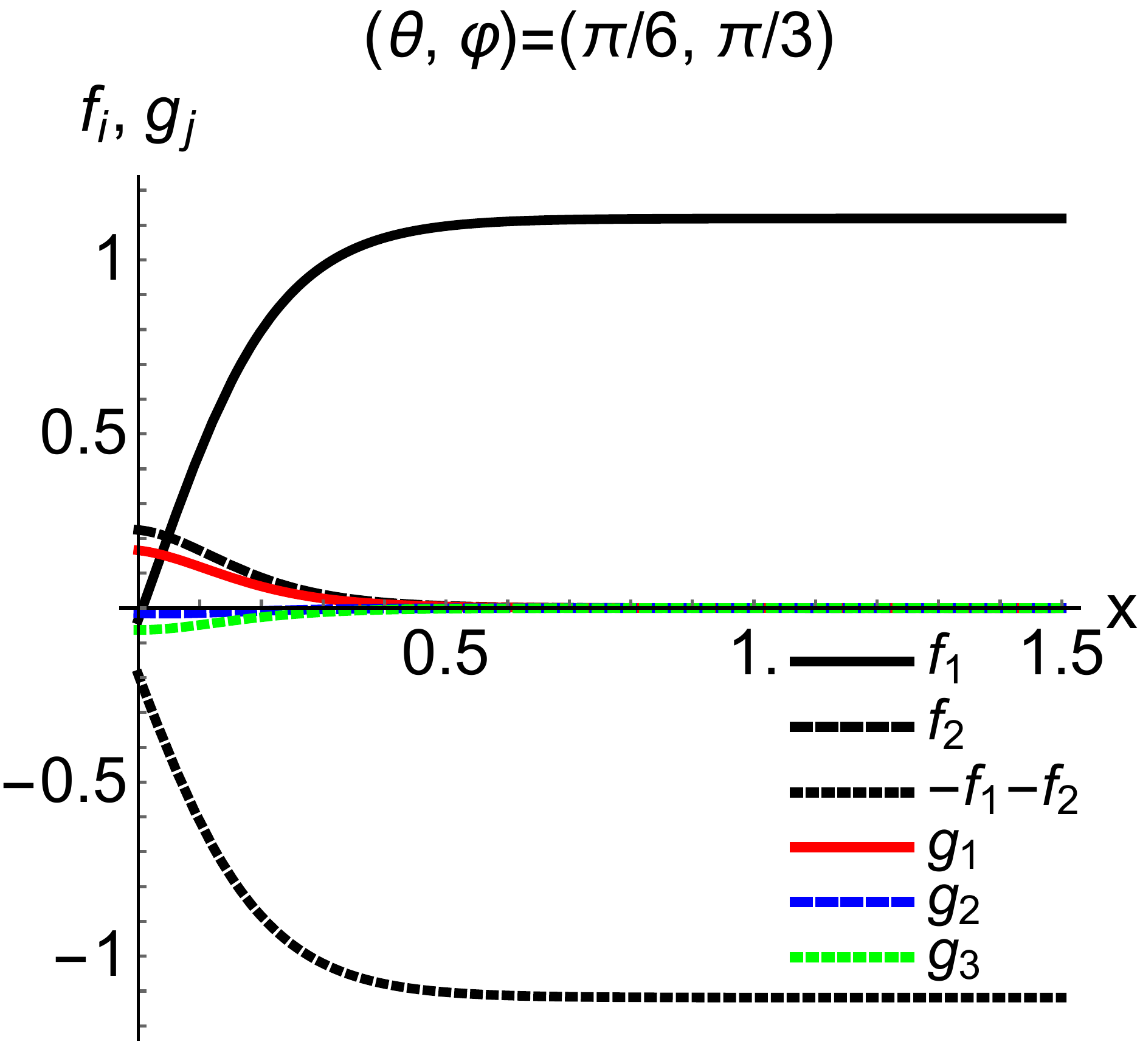}
\hspace{1em} %
\includegraphics[scale=0.18]{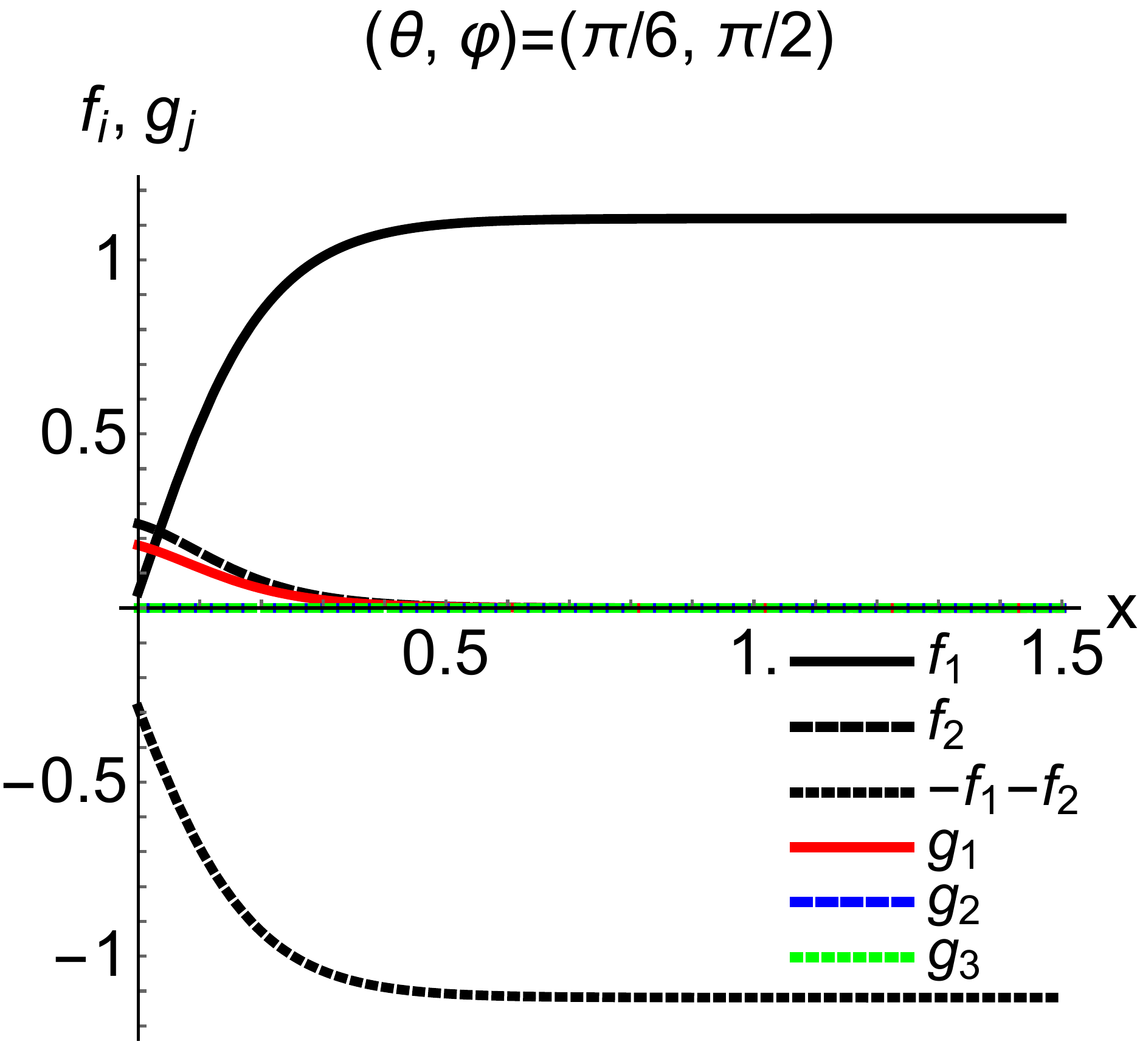}
\\ \vspace{2em} 
\includegraphics[scale=0.18]{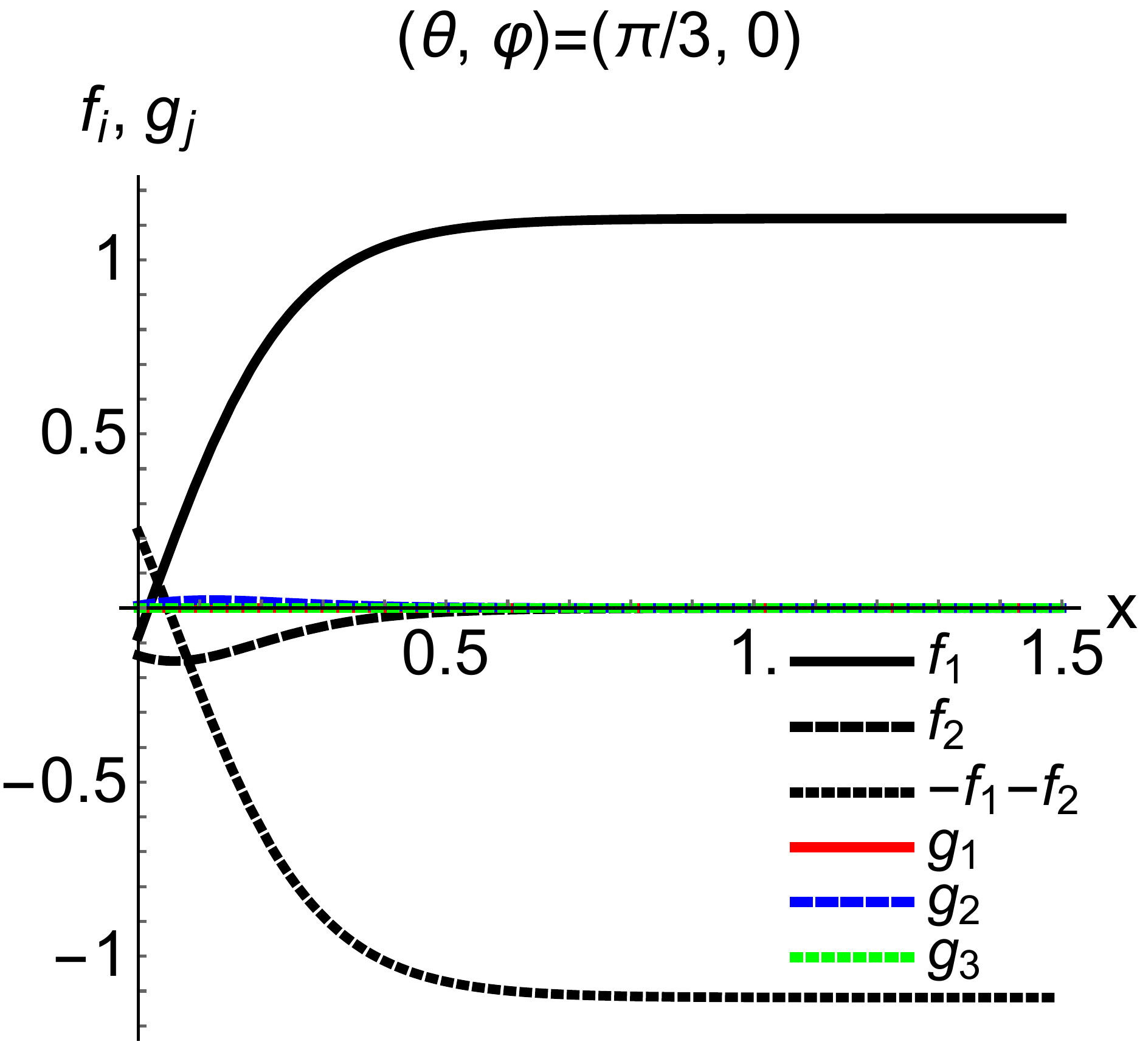}
\hspace{1em} %
\includegraphics[scale=0.18]{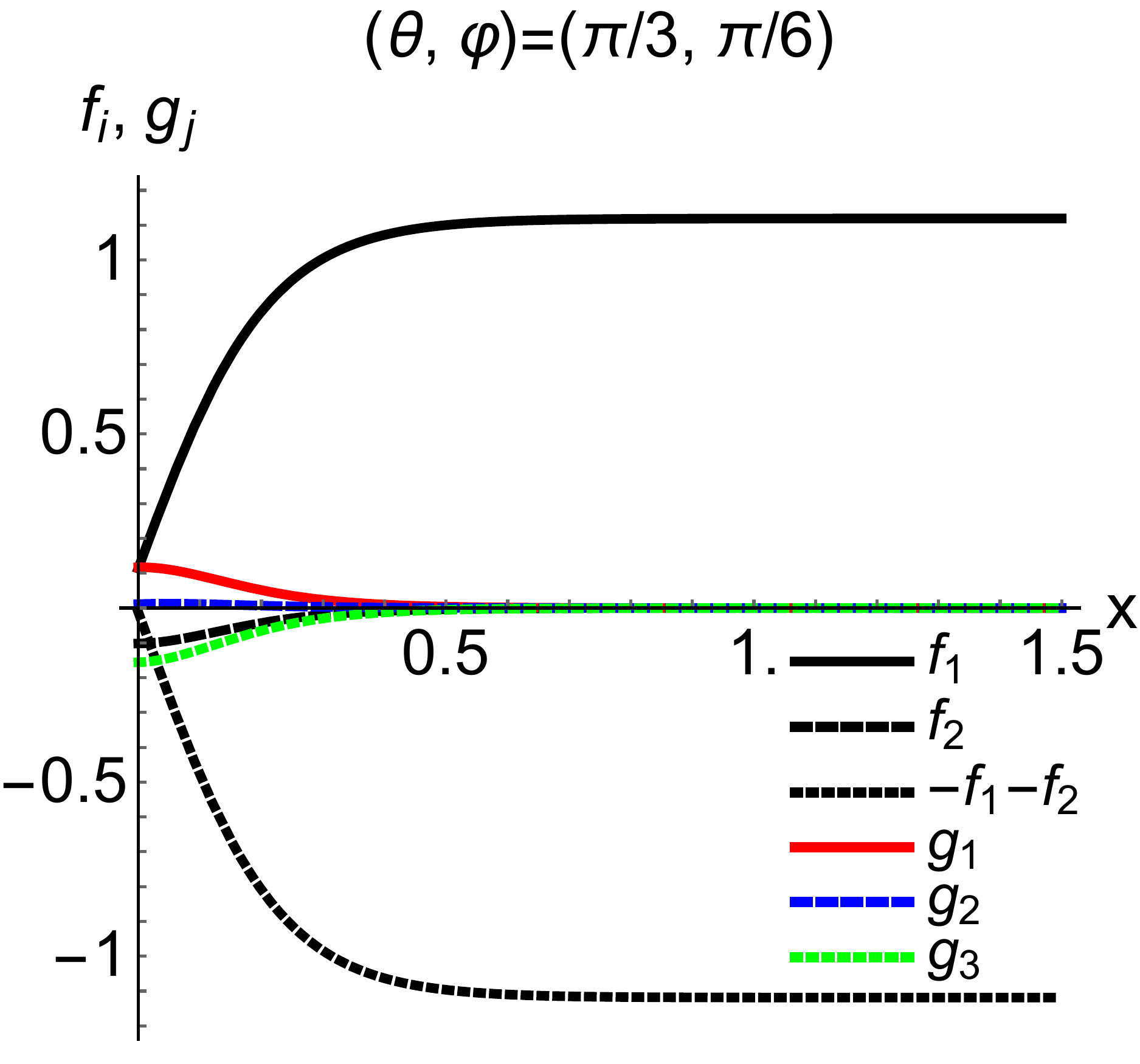}
\hspace{1em} %
\includegraphics[scale=0.18]{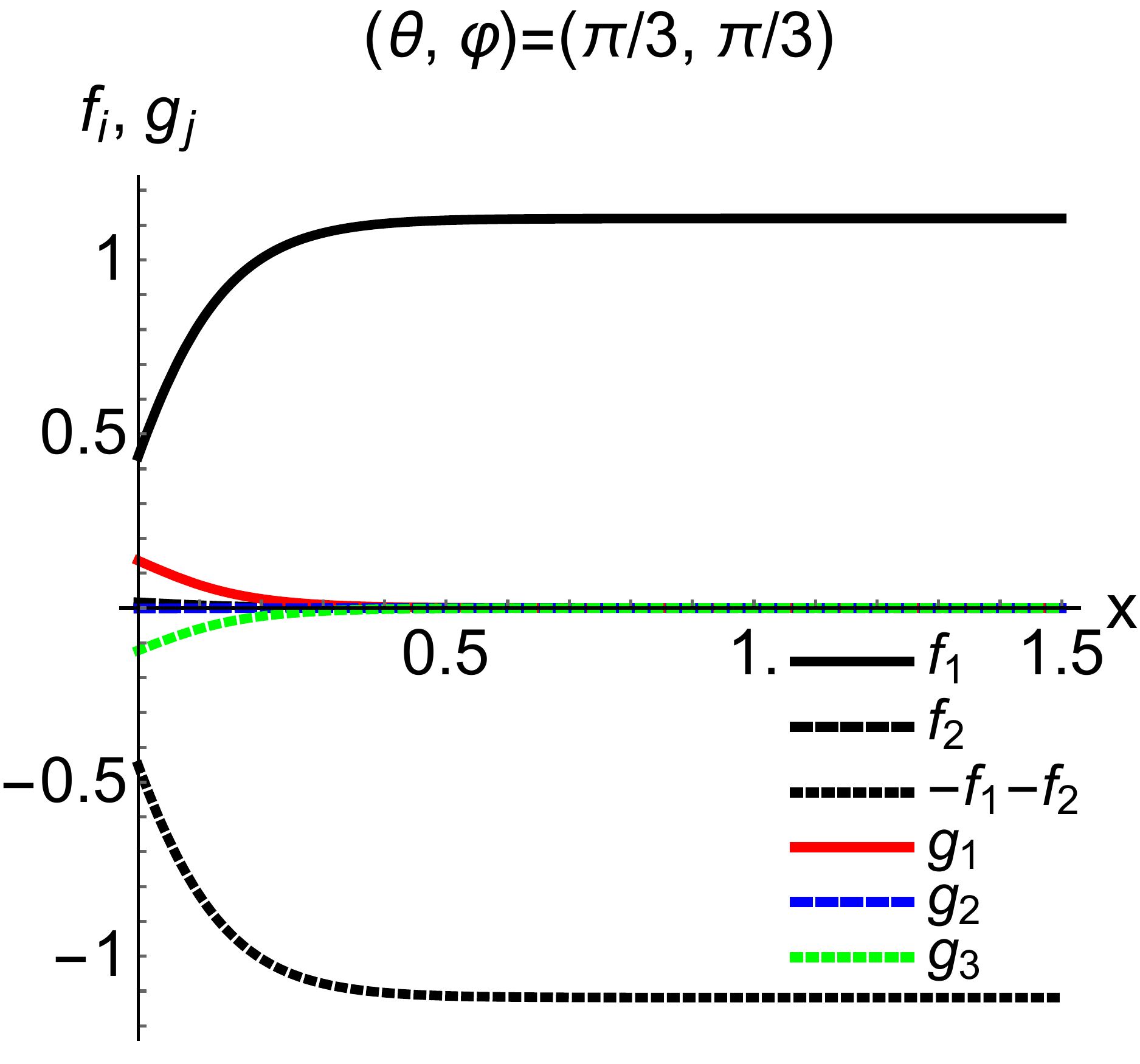}
\hspace{1em} %
\includegraphics[scale=0.18]{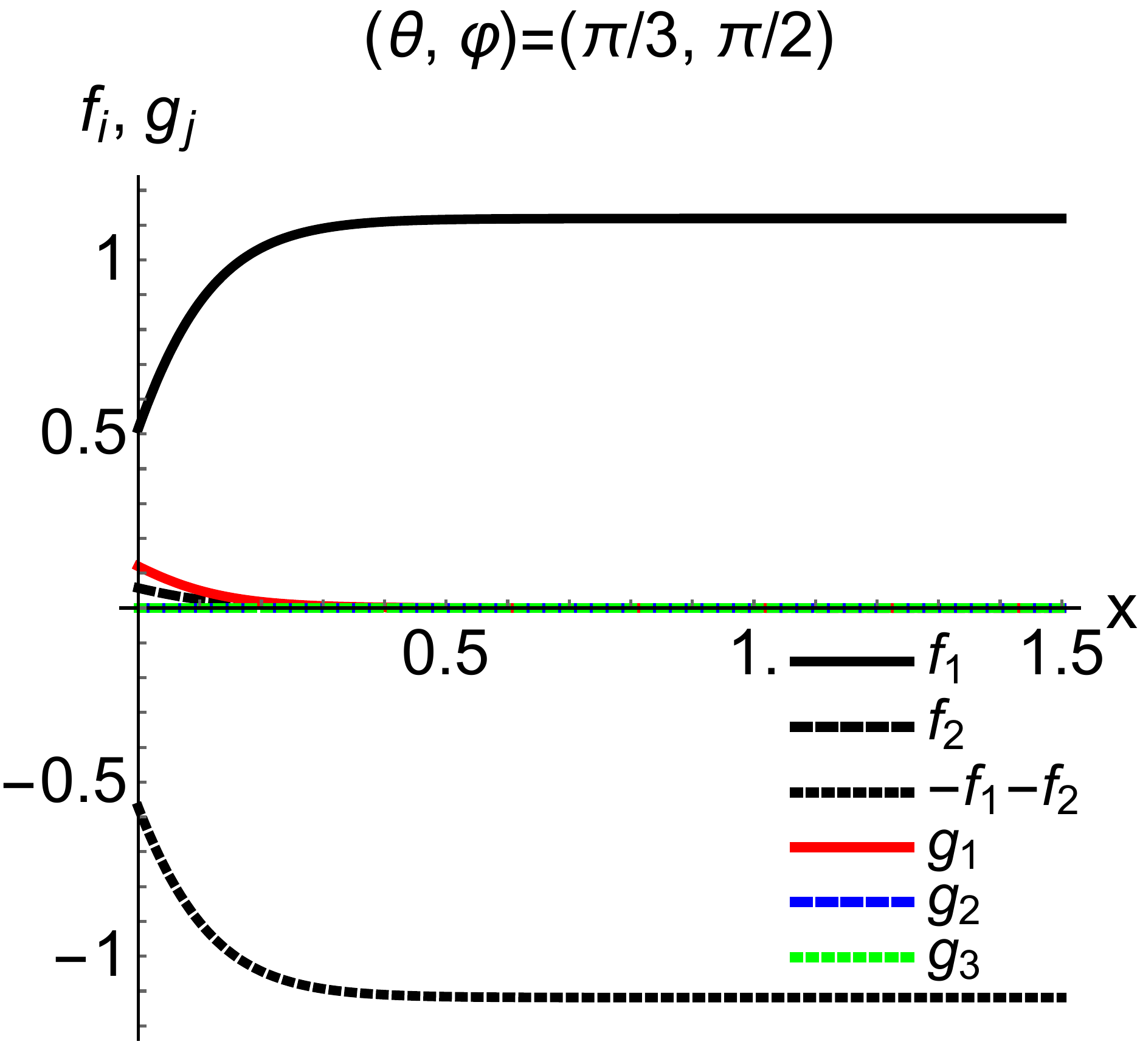}
\\ \vspace{2em} 
\includegraphics[scale=0.18]{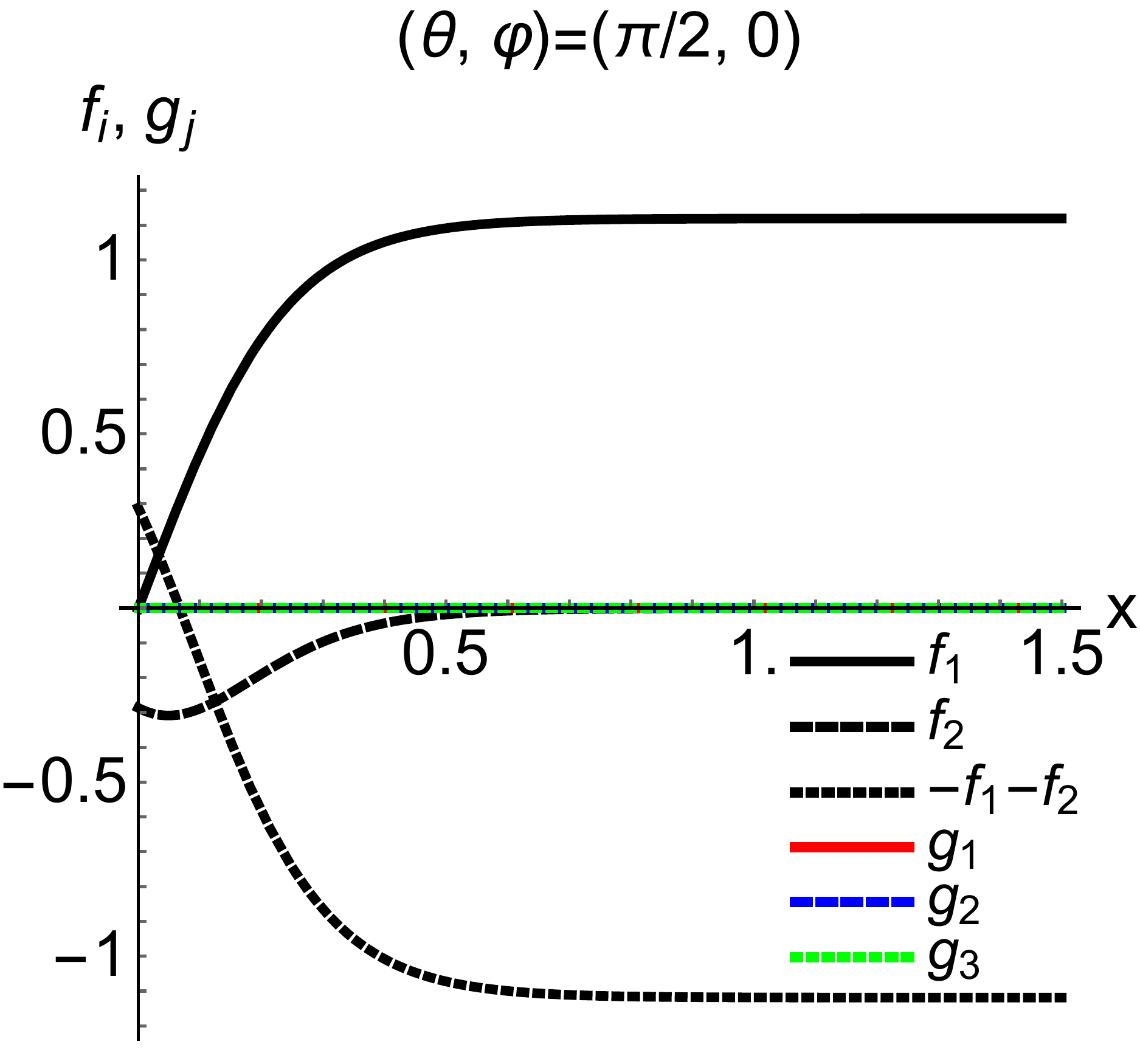}
\hspace{1em} %
\includegraphics[scale=0.18]{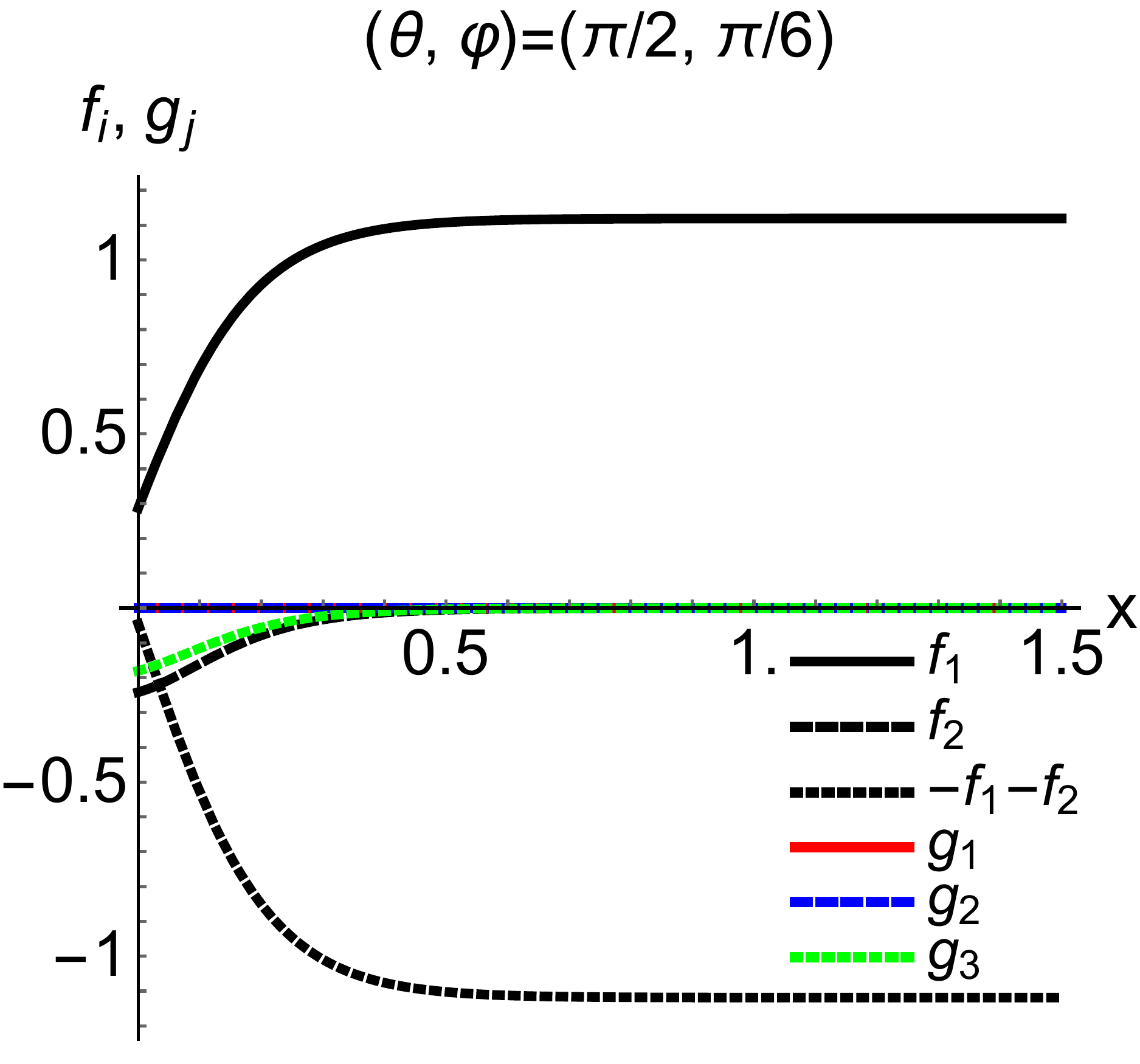}
\hspace{1em} %
\includegraphics[scale=0.18]{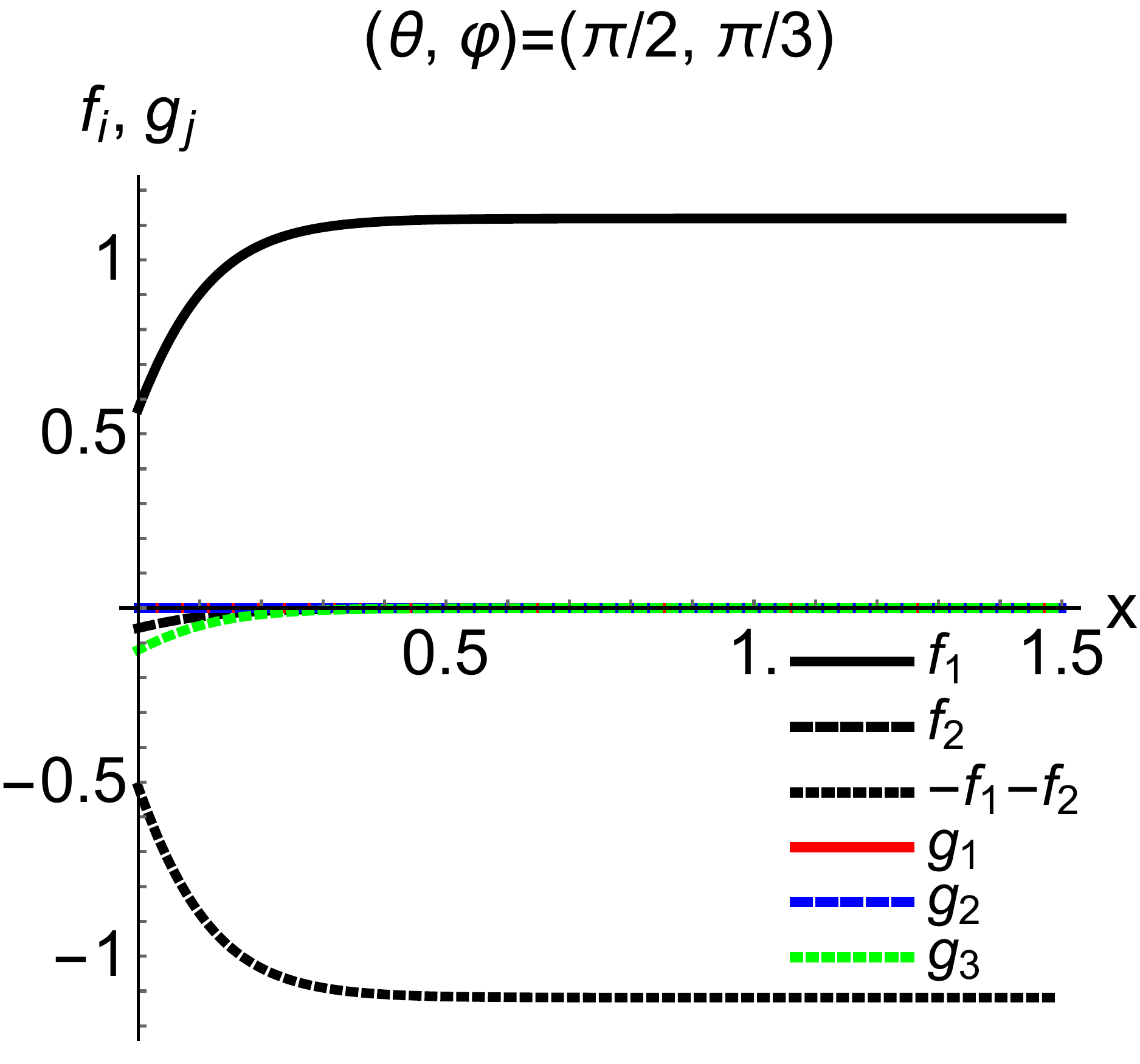}
\hspace{1em} %
\includegraphics[scale=0.18]{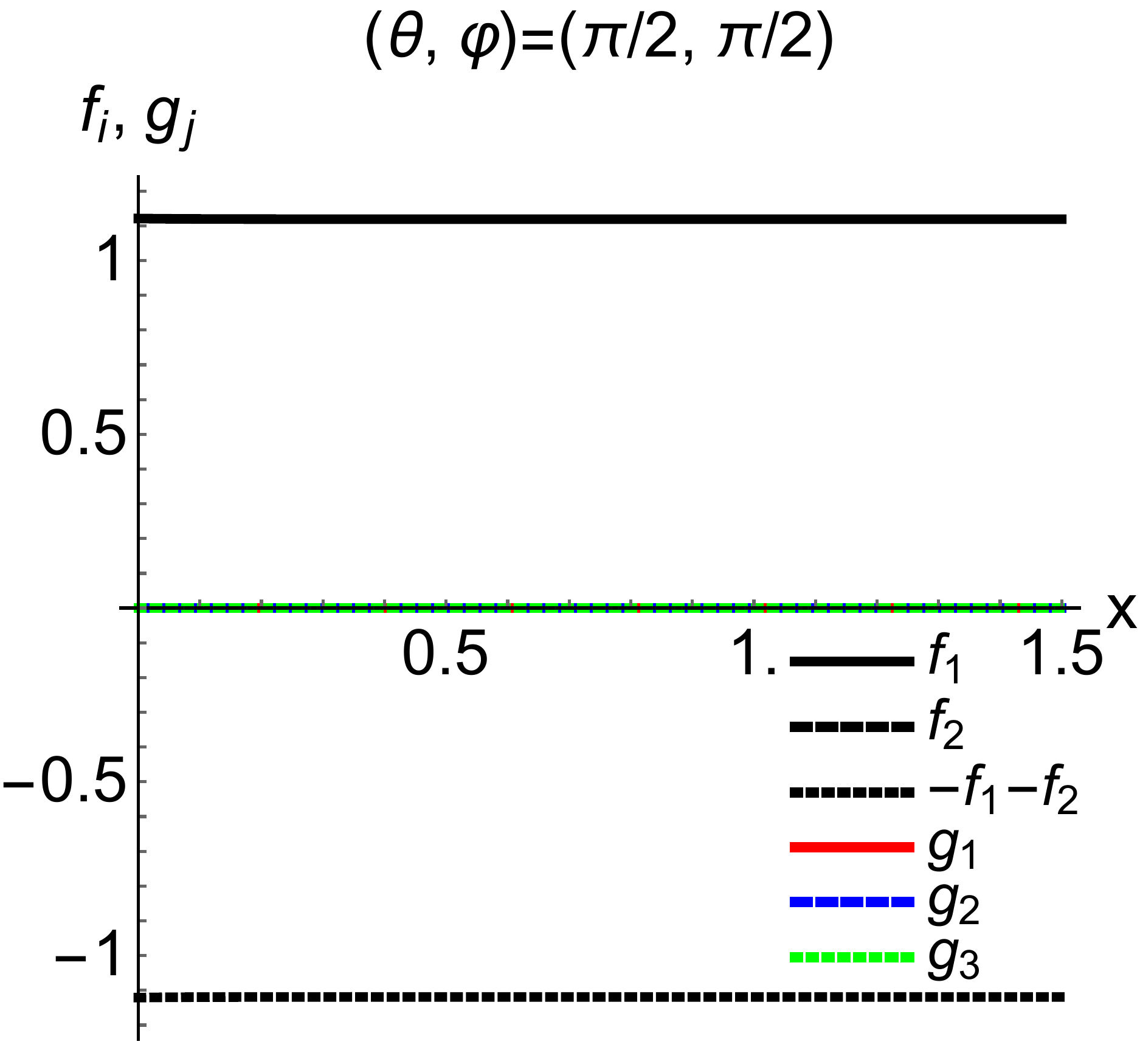}
\caption{The plots of the profile functions $f_{1}(x)$, $f_{2}(x)$, $g_{1}(x)$, $g_{2}(x)$, $g_{3}(x)$ for the bulk condition (iii) $(f_{1}^{\bulk},f_{2}^{\bulk})=(1.12,0)$ at $t=0.9$ and $b=0.2$ (the bulk D$_{4}$-BN phase).}
\label{fig:profile_functions_3P2_bulk3}
\end{center}
\end{figure}

\begin{figure}[p] 
\begin{center}
\includegraphics[scale=0.18]{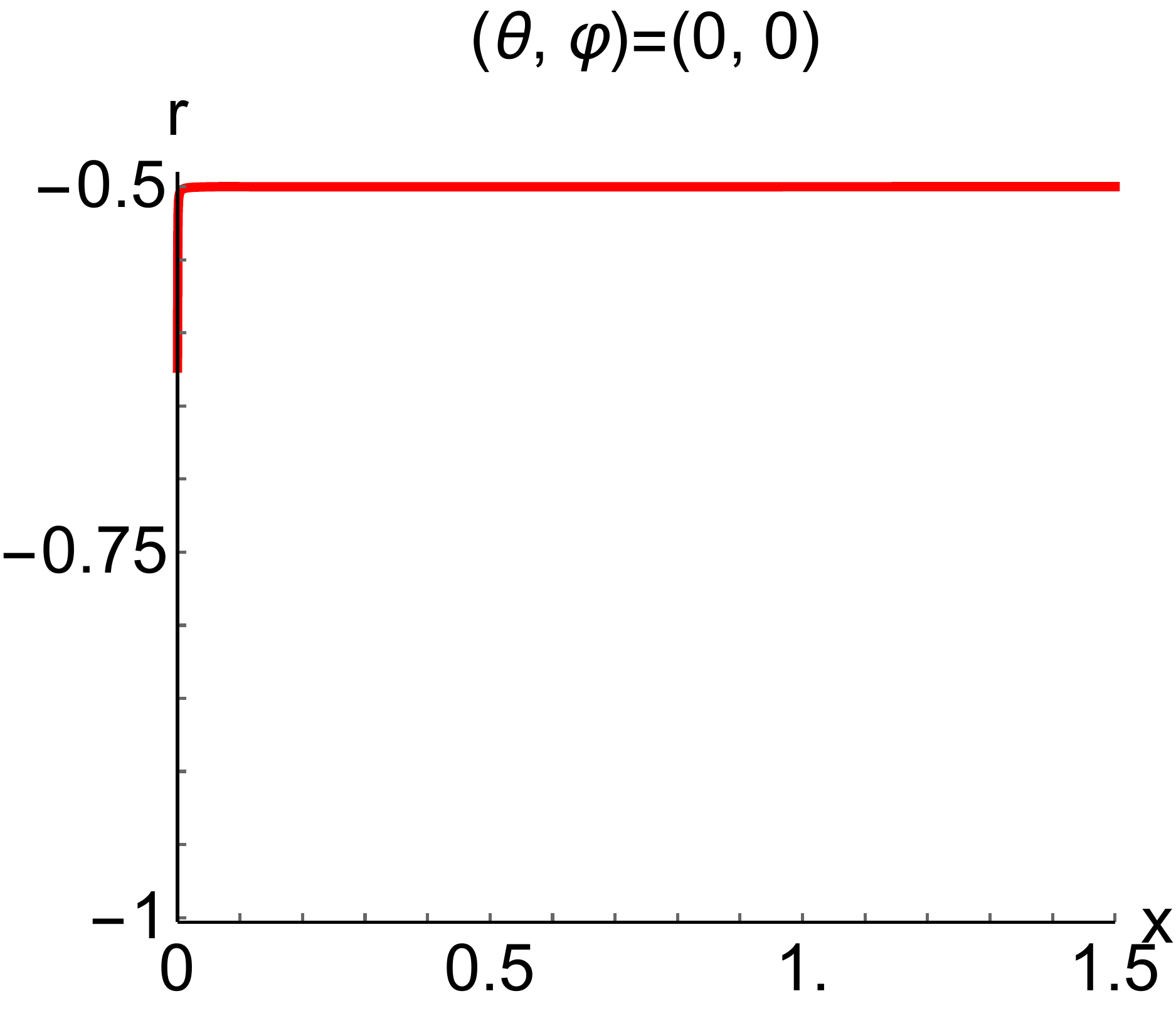}
\\ \vspace{2em} 
\includegraphics[scale=0.18]{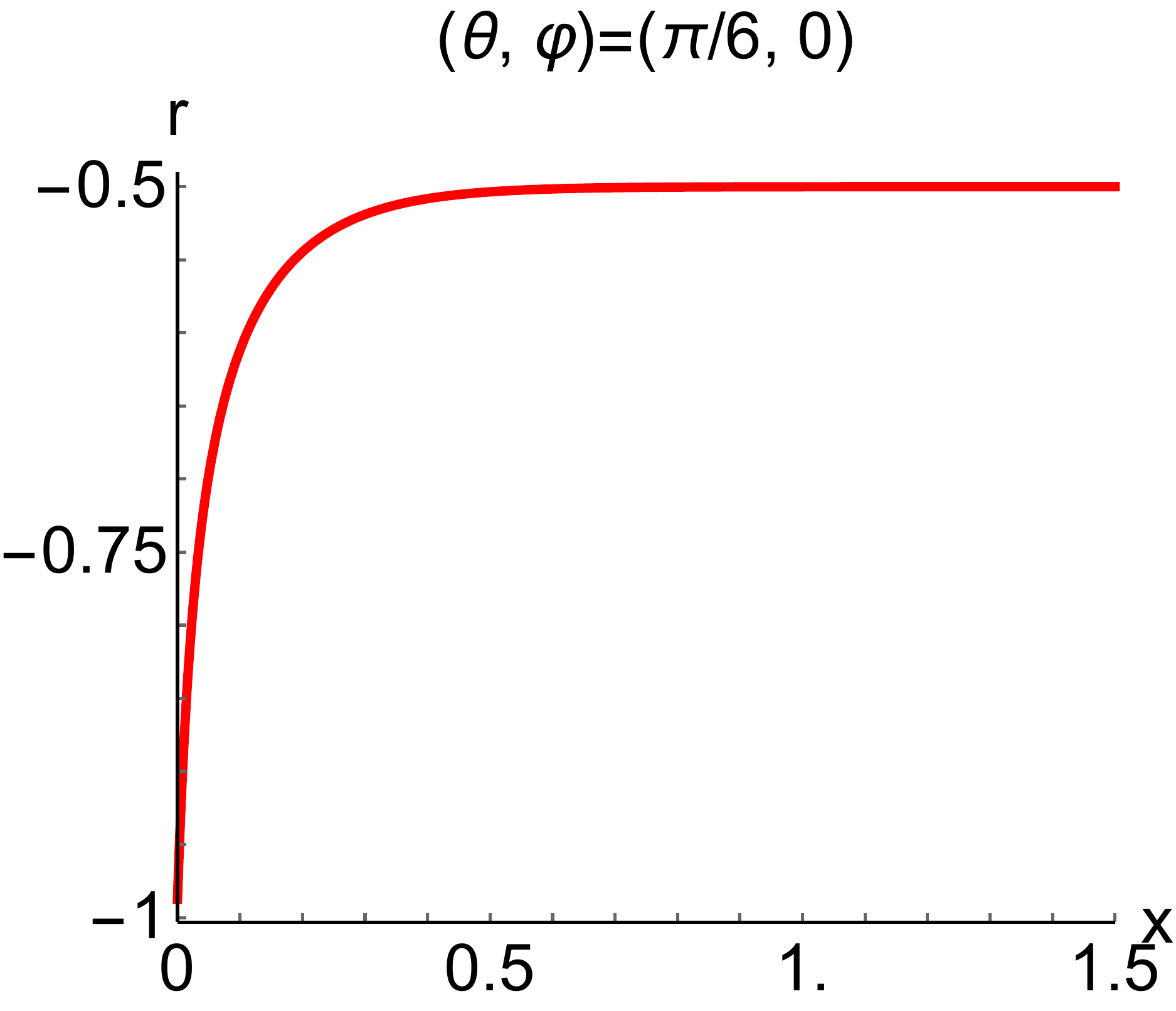}
\hspace{1em} %
\includegraphics[scale=0.18]{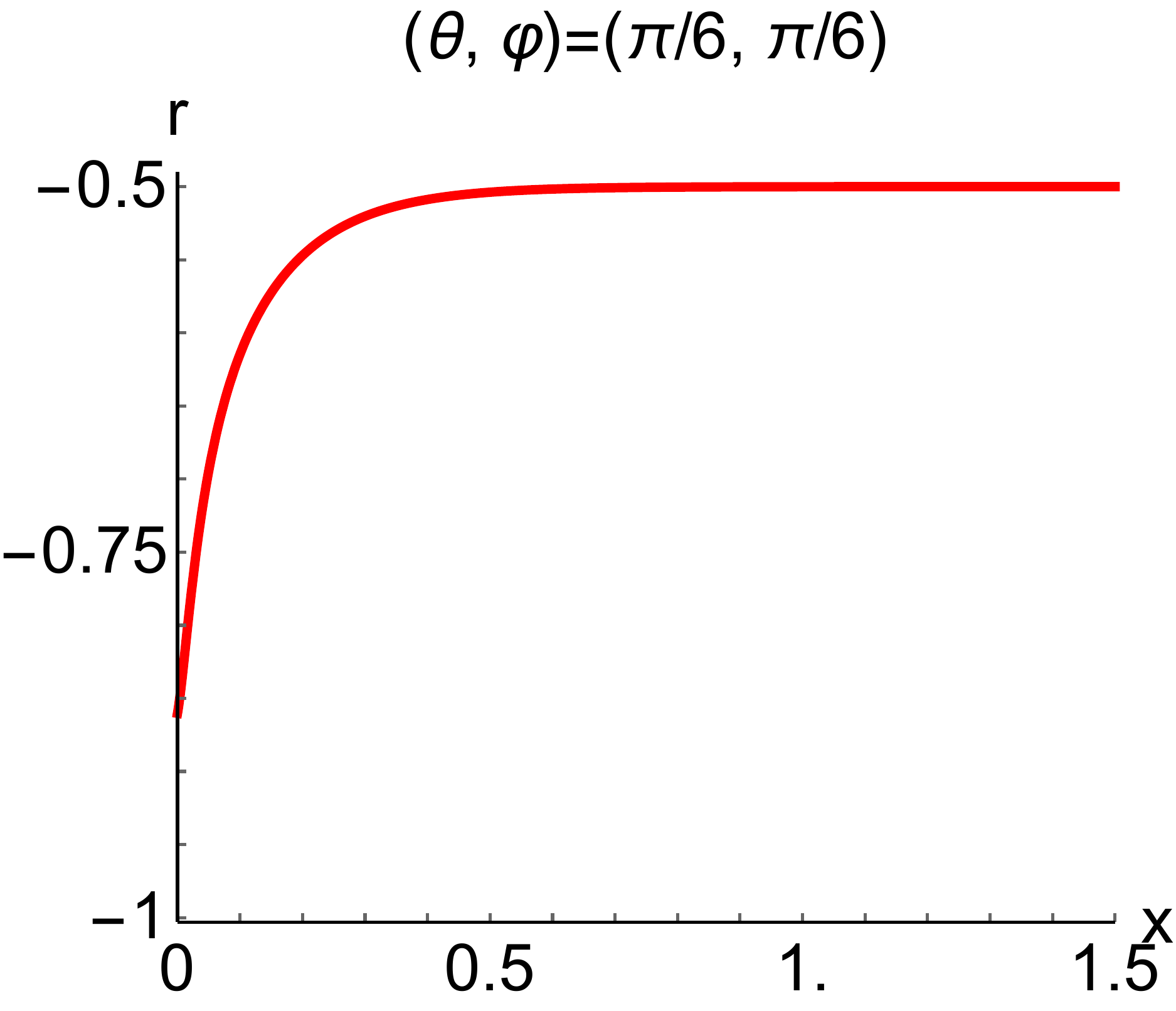}
\hspace{1em} %
\includegraphics[scale=0.18]{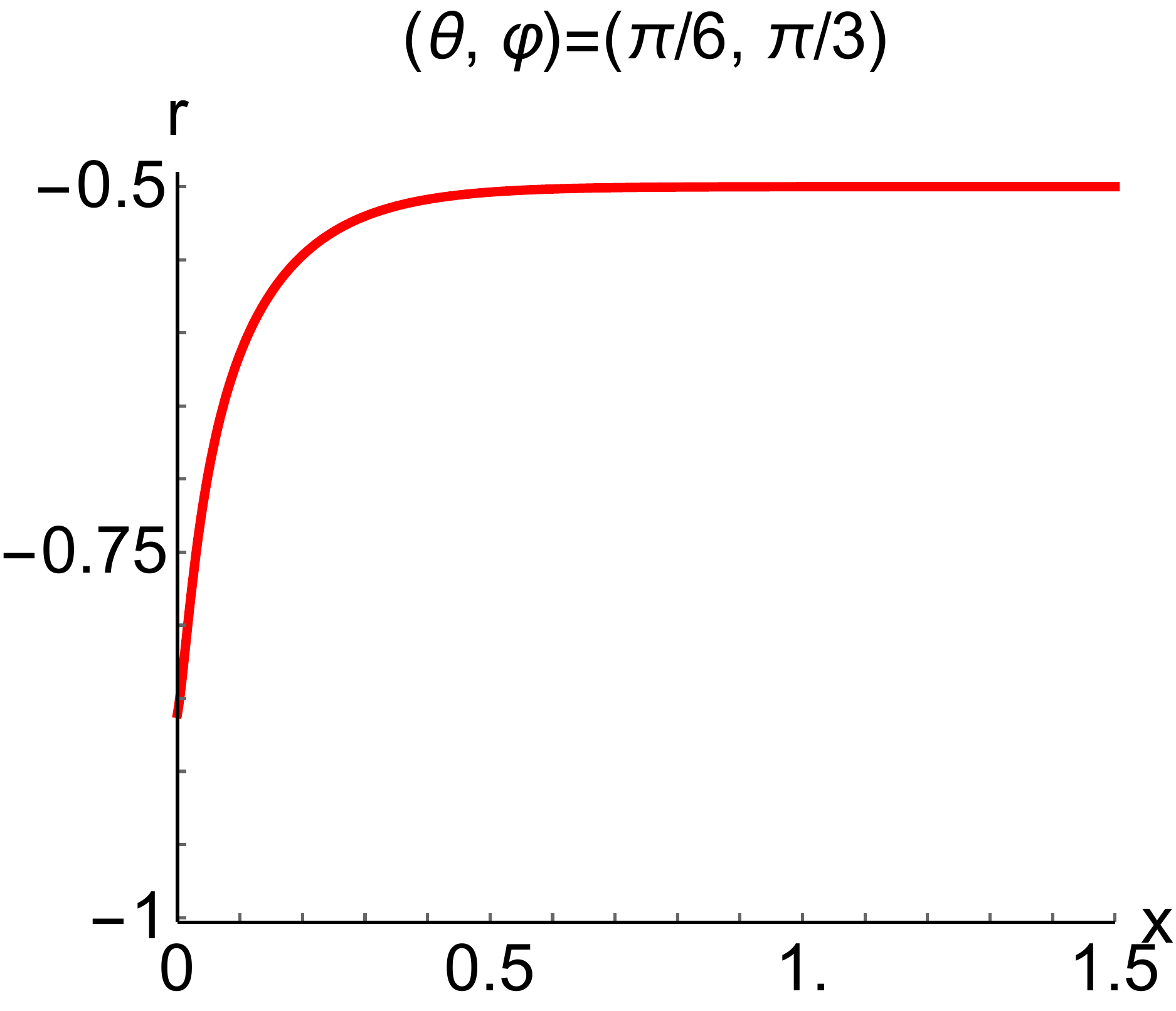}
\hspace{1em} %
\includegraphics[scale=0.18]{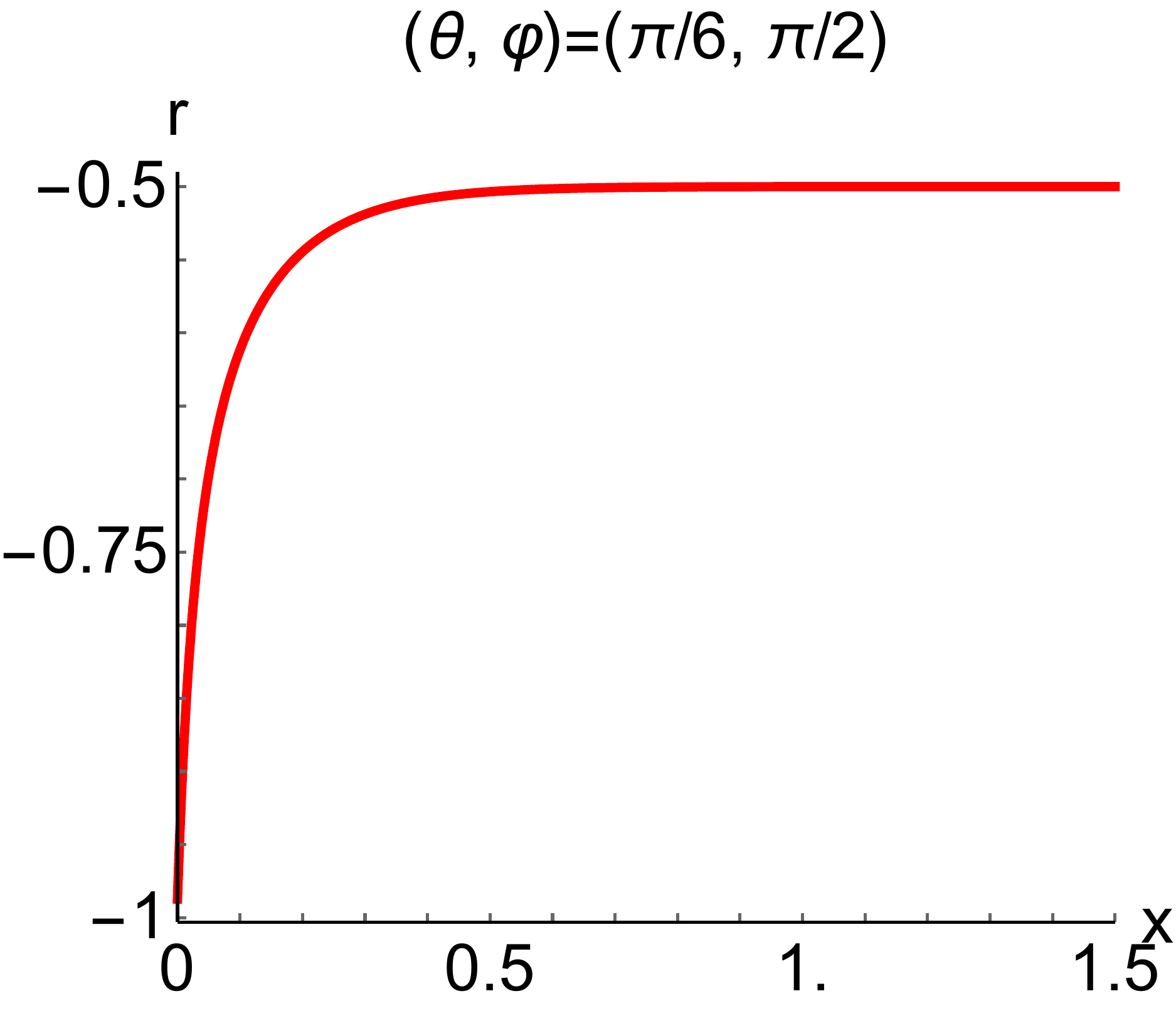}
\\ \vspace{2em} 
\includegraphics[scale=0.18]{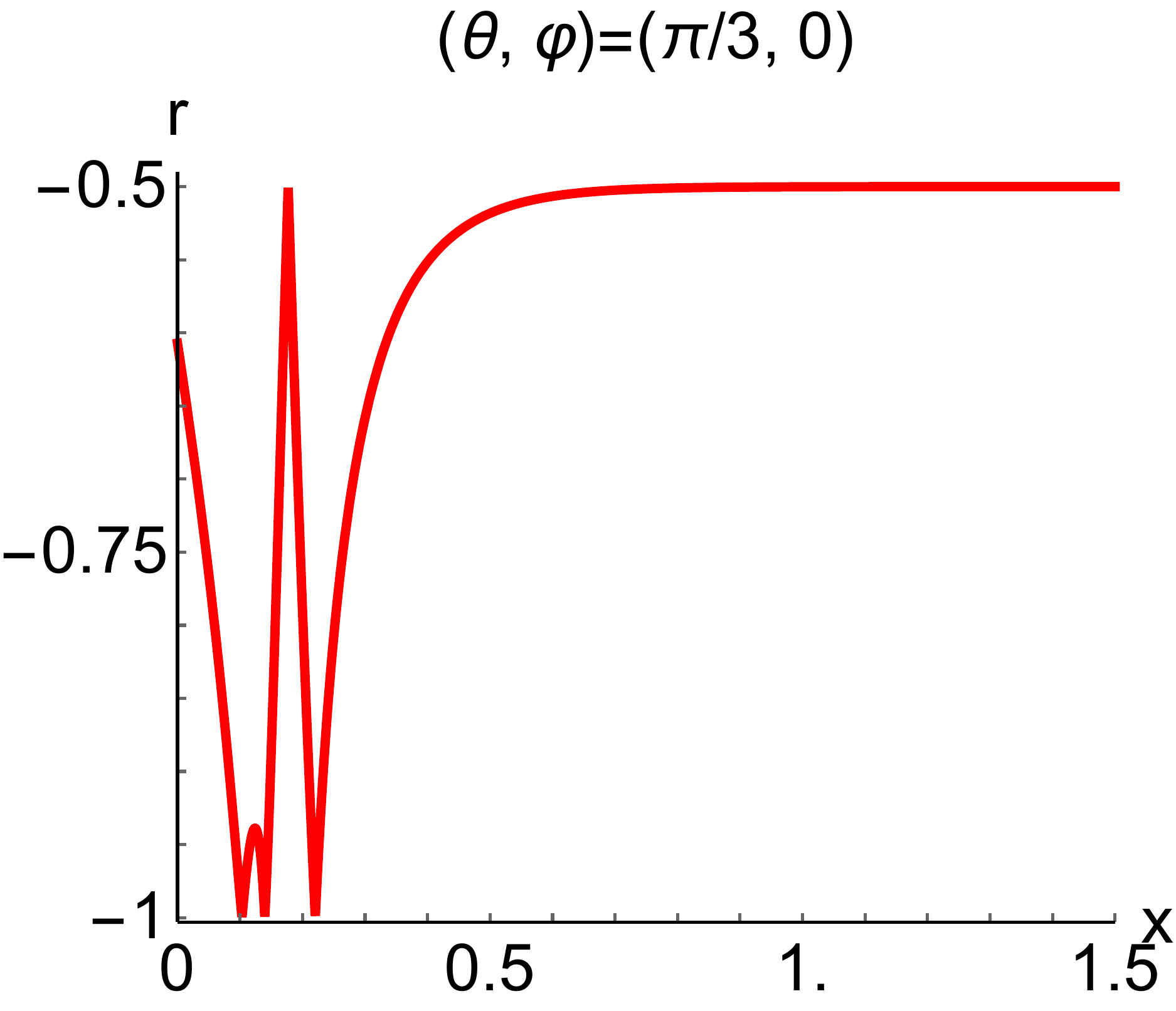}
\hspace{1em} %
\includegraphics[scale=0.18]{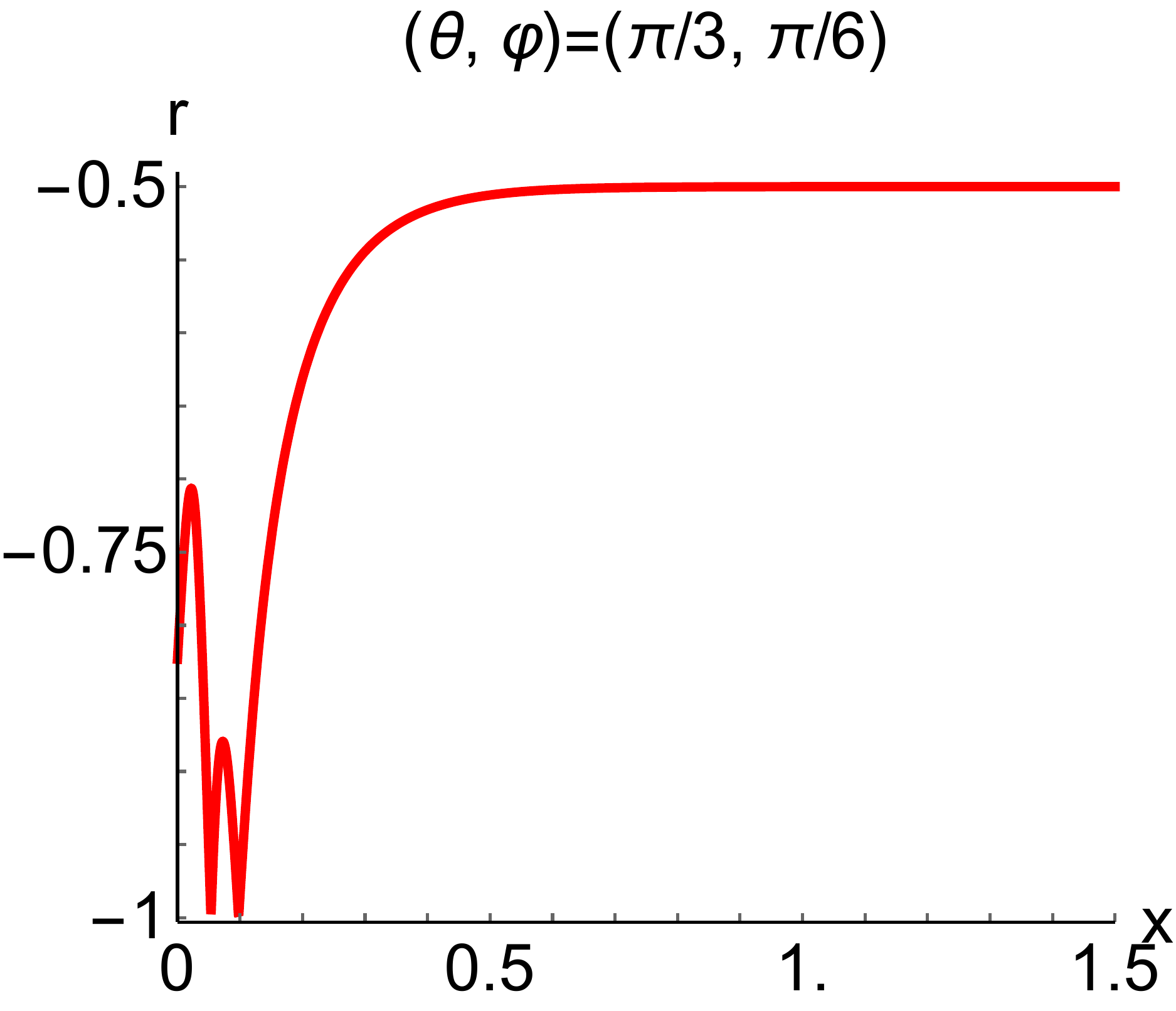}
\hspace{1em} %
\includegraphics[scale=0.18]{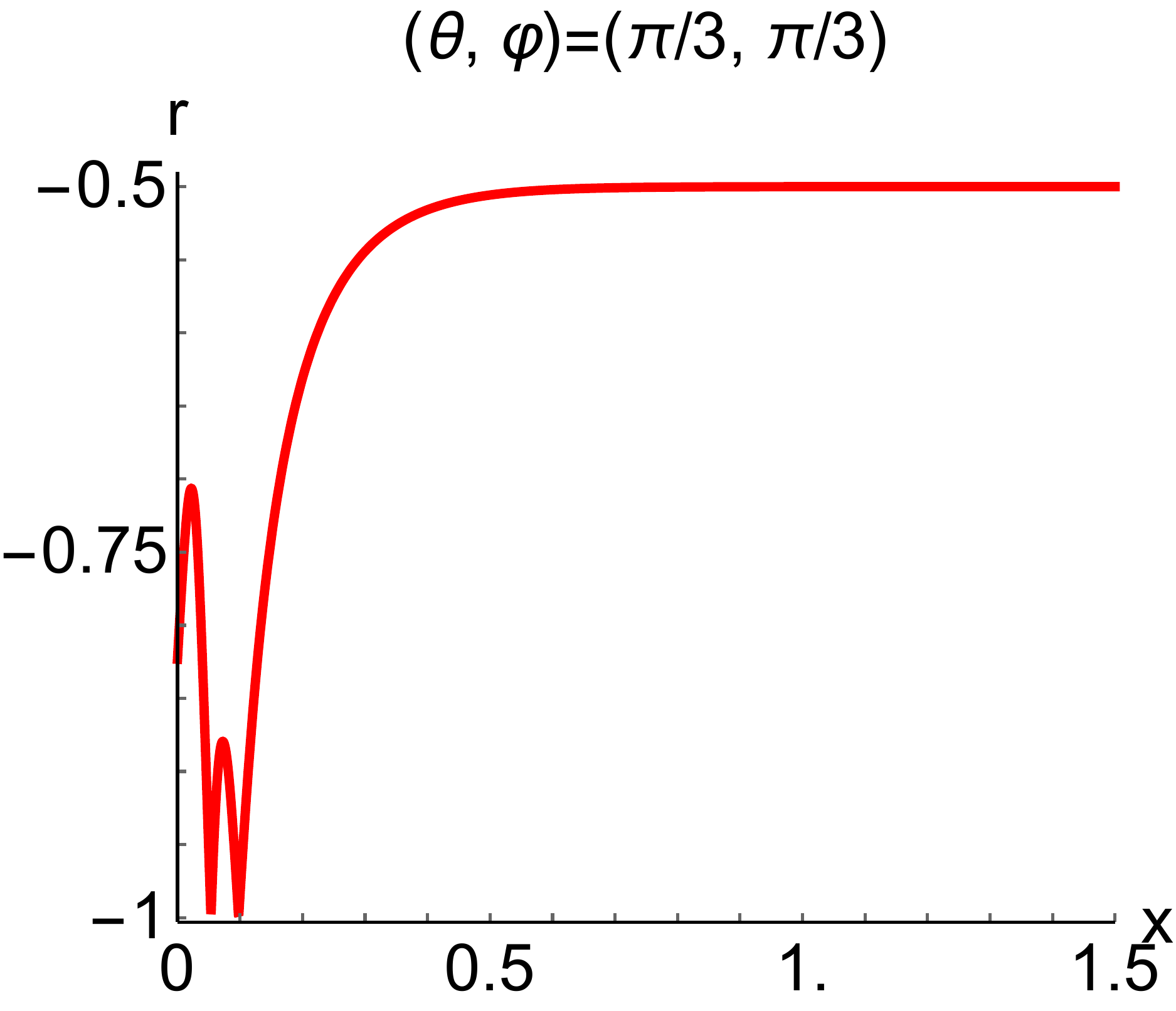}
\hspace{1em} %
\includegraphics[scale=0.18]{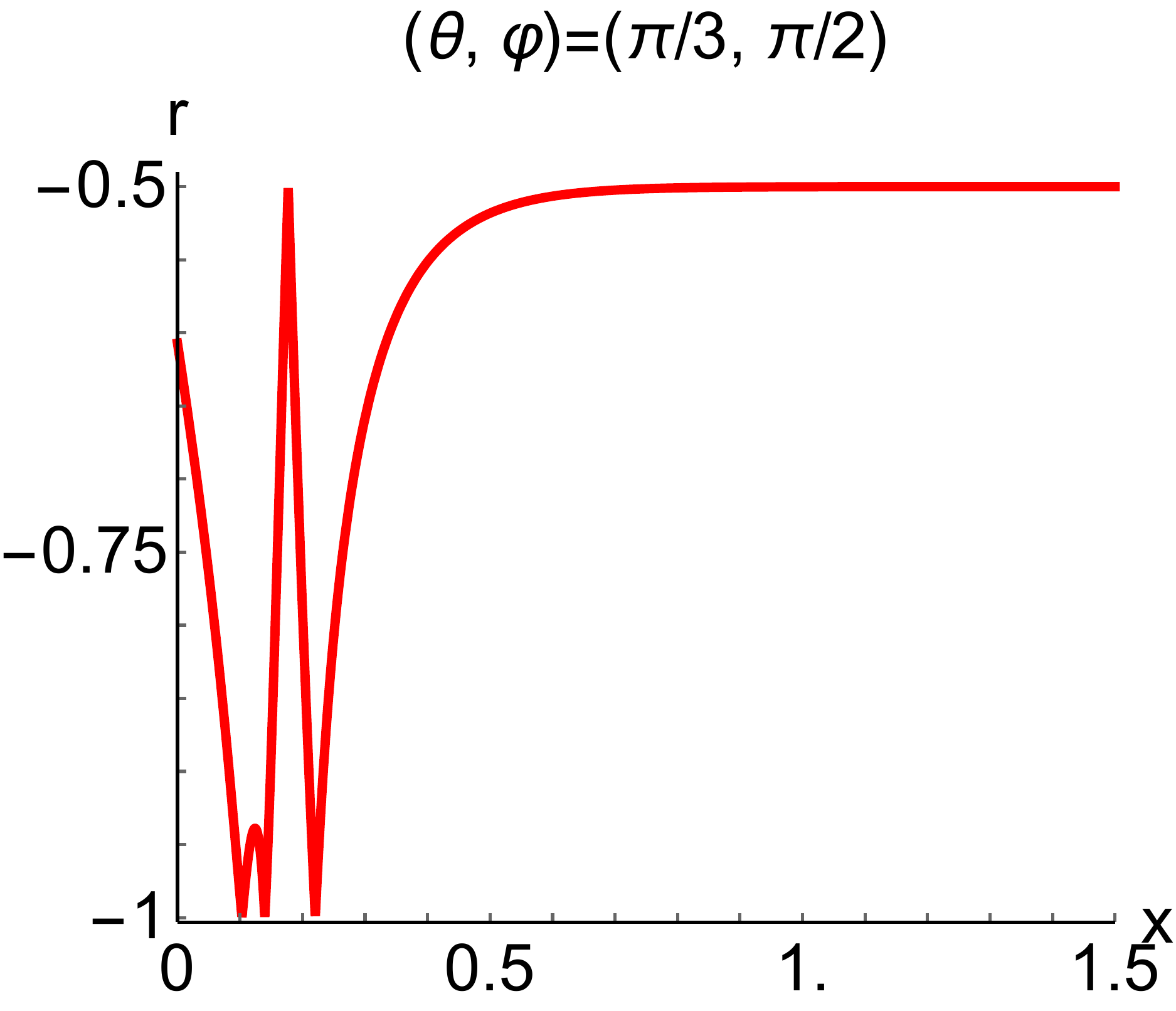}
\\ \vspace{2em} 
\includegraphics[scale=0.18]{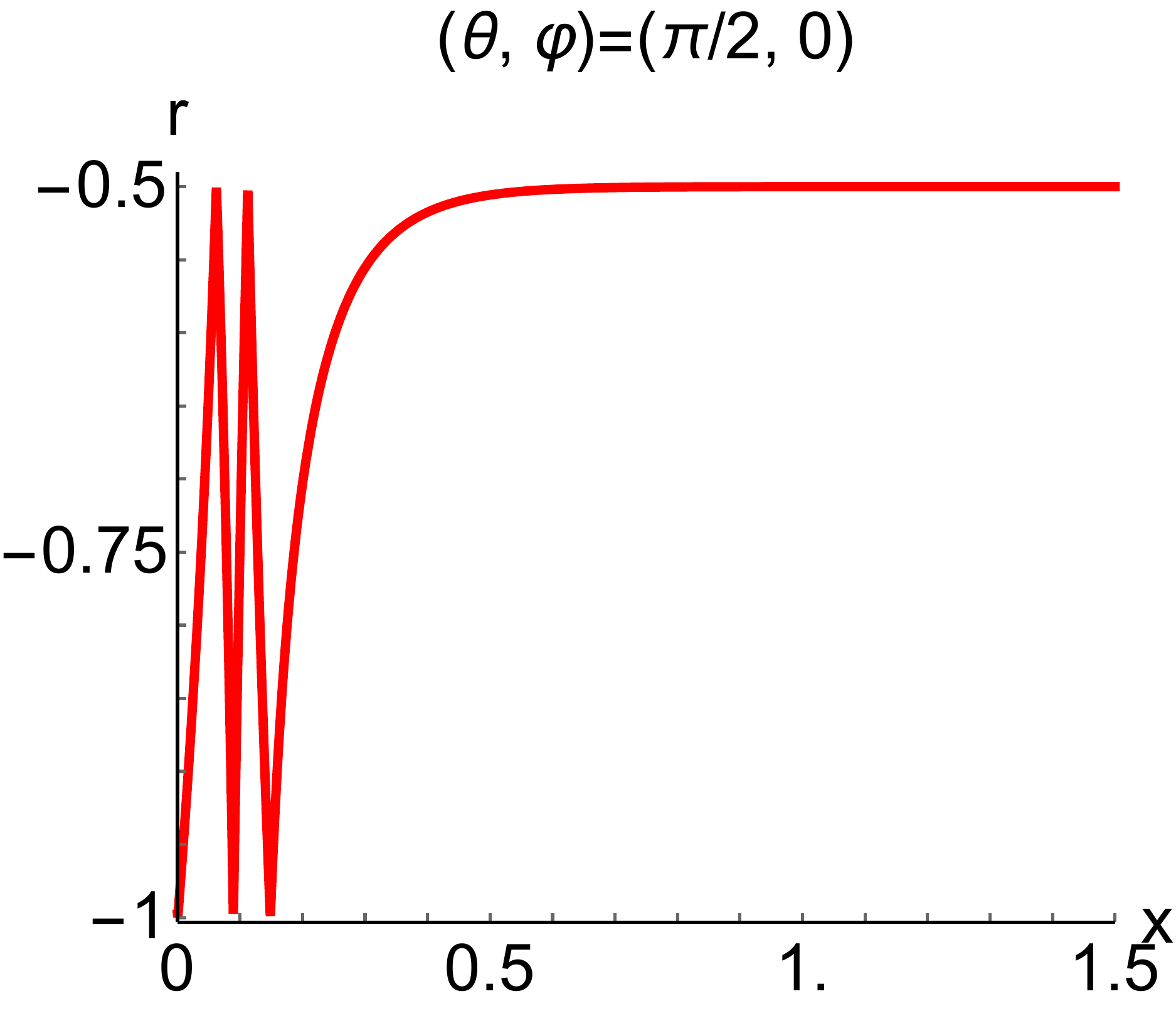}
\hspace{1em} %
\includegraphics[scale=0.18]{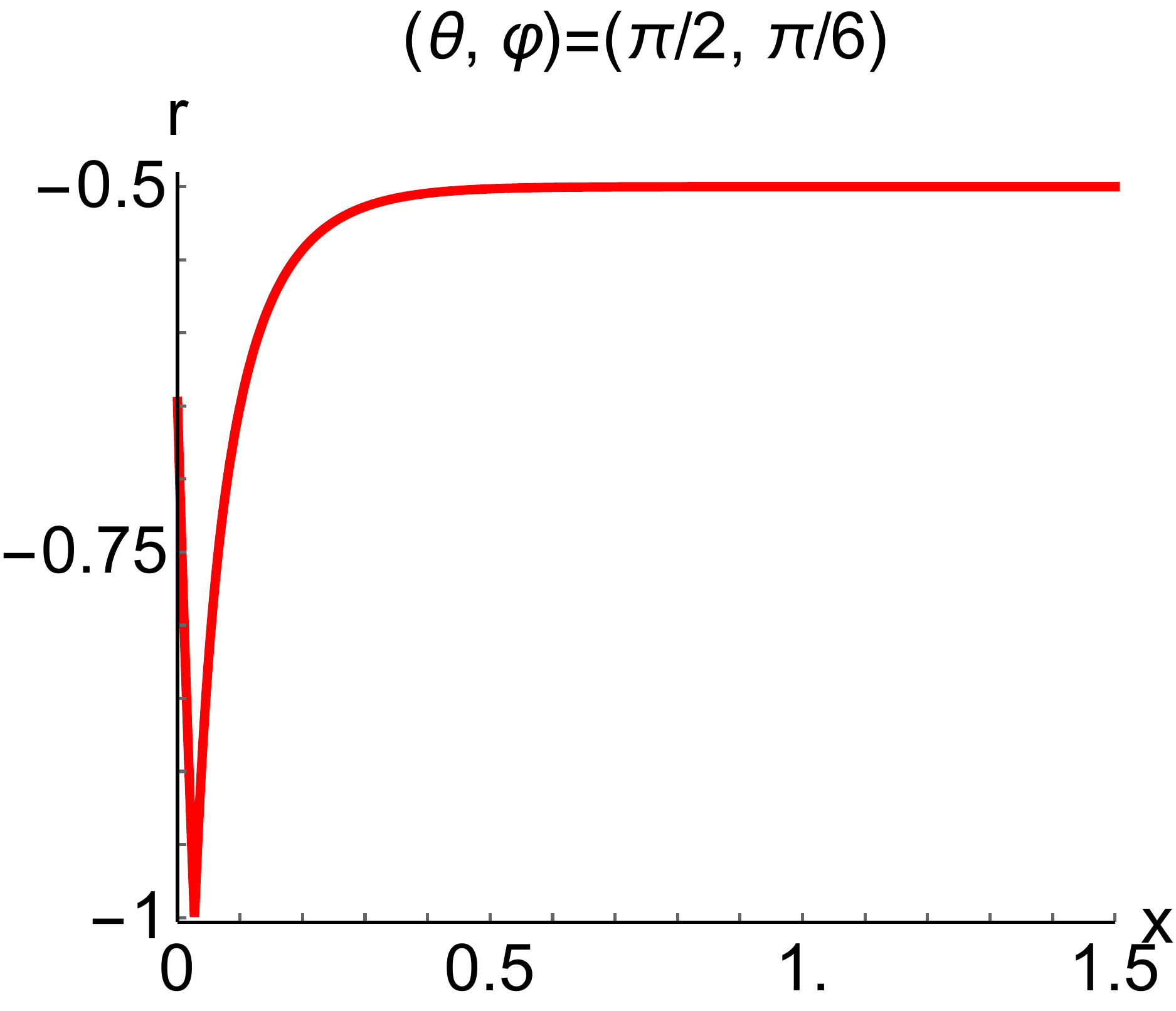}
\hspace{1em} %
\includegraphics[scale=0.18]{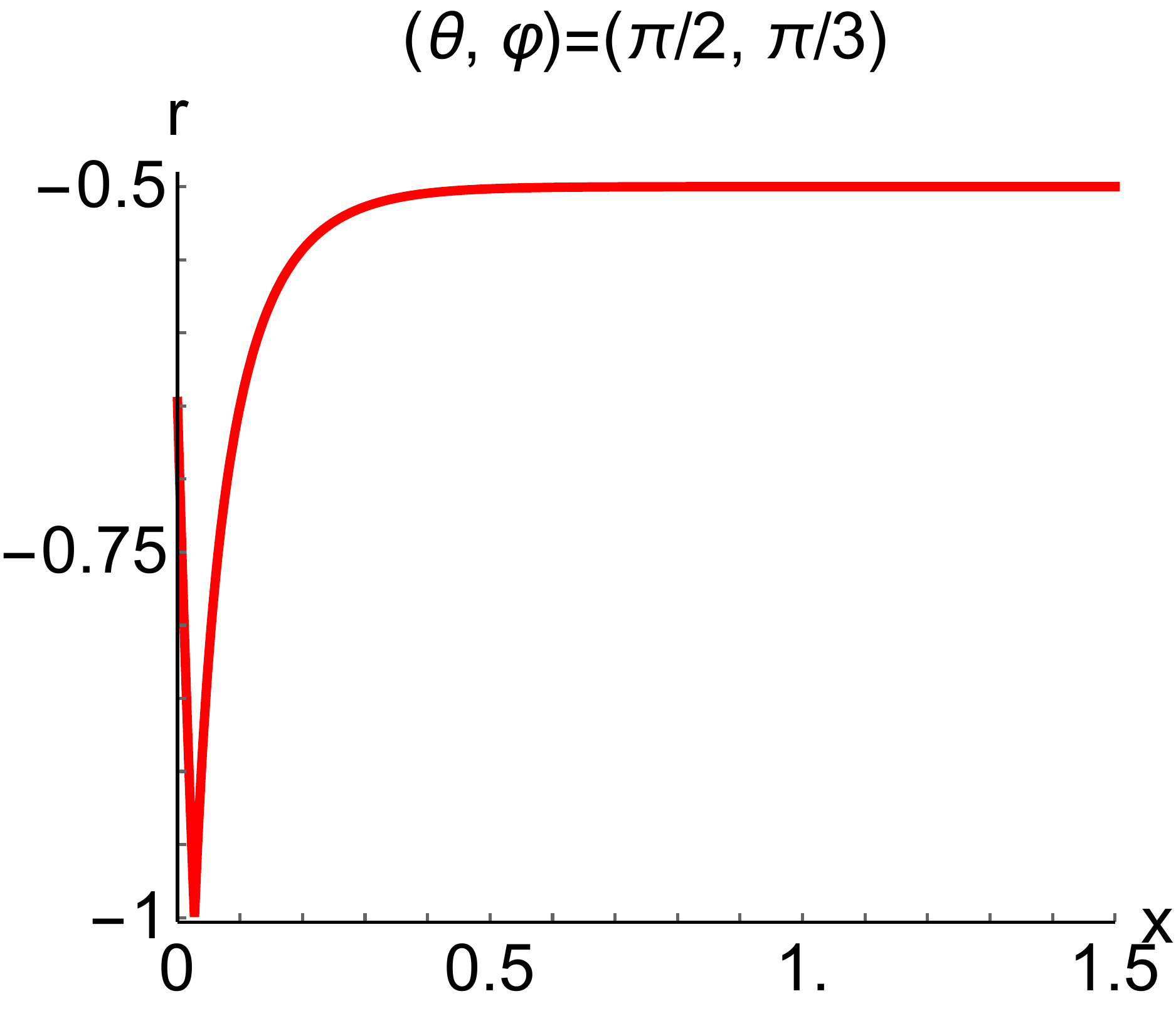}
\hspace{1em} %
\includegraphics[scale=0.18]{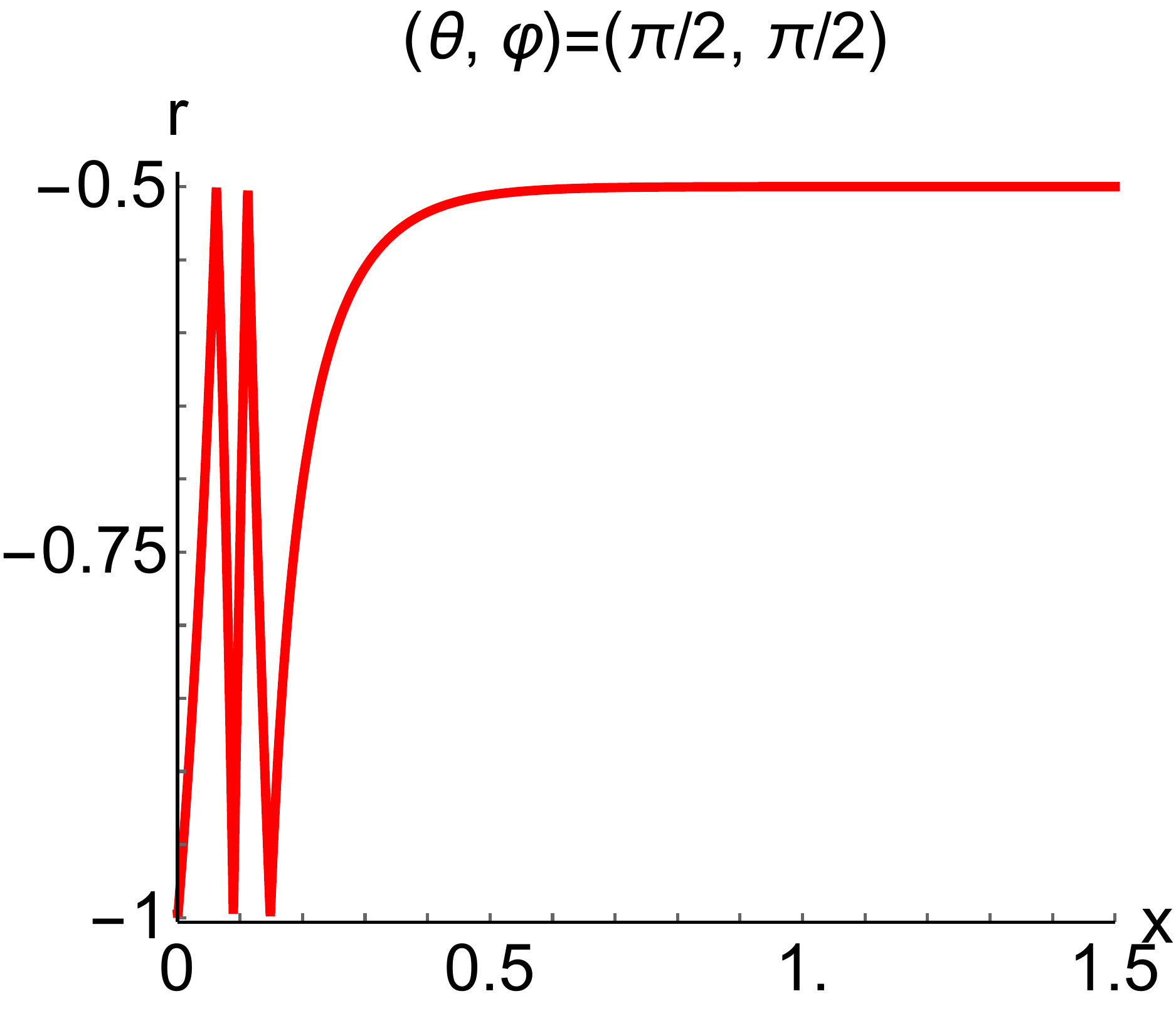}
\caption{The plots of $r(x)$ for the bulk condition (i) $(f_{1}^{\bulk},f_{2}^{\bulk})=(0.64,0.64)$ at $t=0.9$ and $b=0$ (the bulk UN phase).}
\label{fig:r_3P2_bulk1}
\end{center}
\end{figure}

\begin{figure}[p] 
\begin{center}
\includegraphics[scale=0.18]{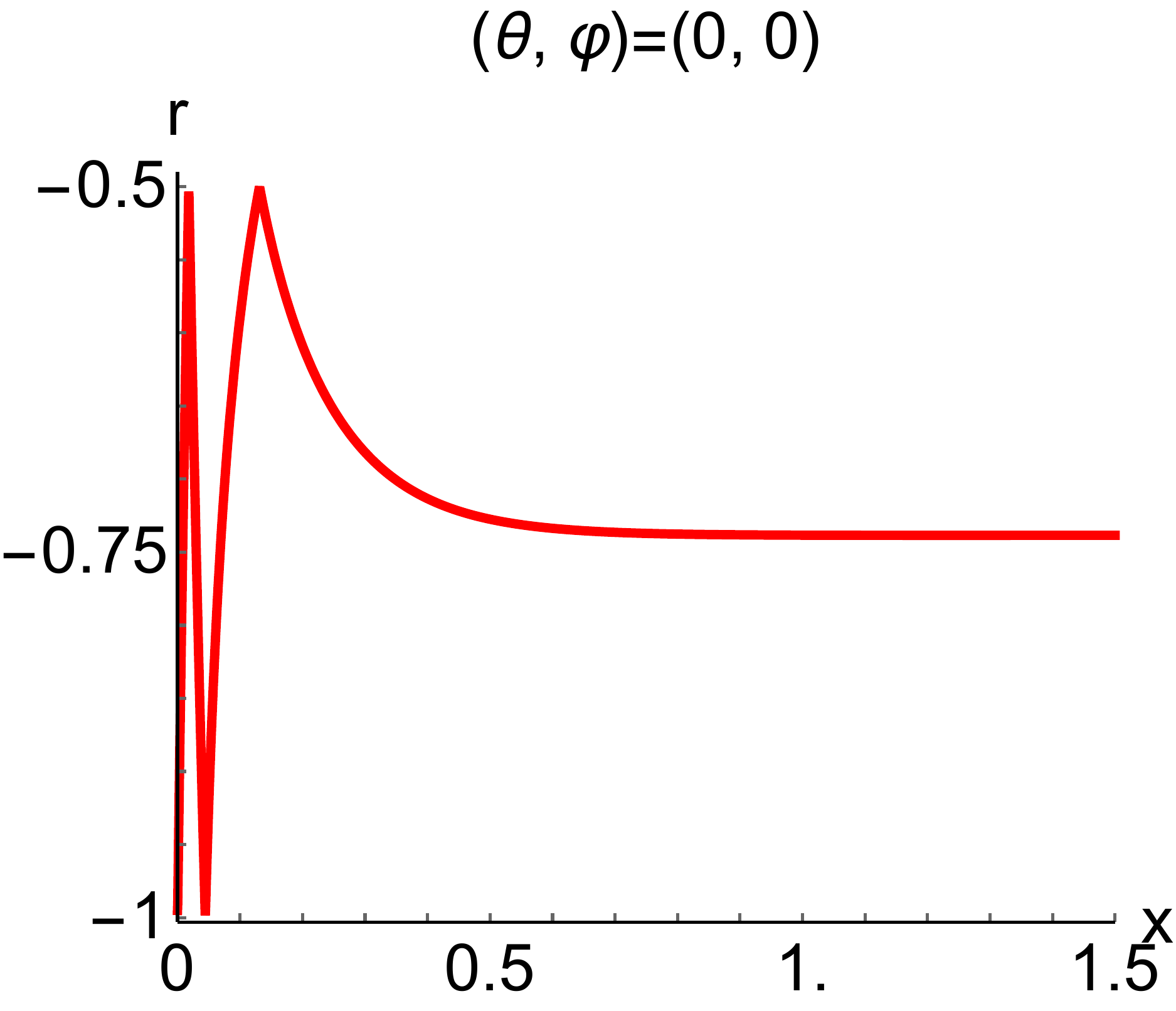}
\\ \vspace{2em} 
\includegraphics[scale=0.18]{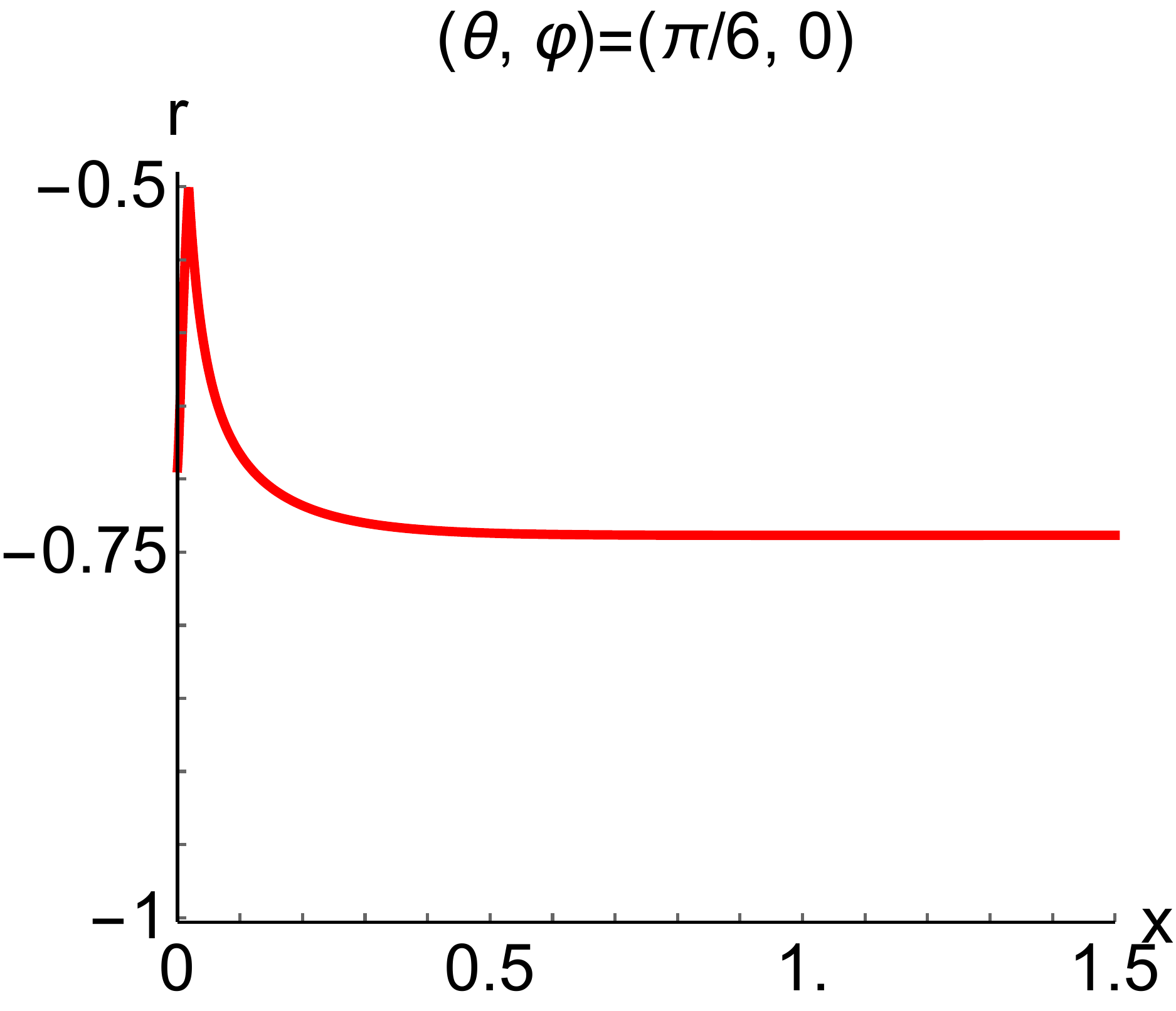}
\hspace{1em} %
\includegraphics[scale=0.18]{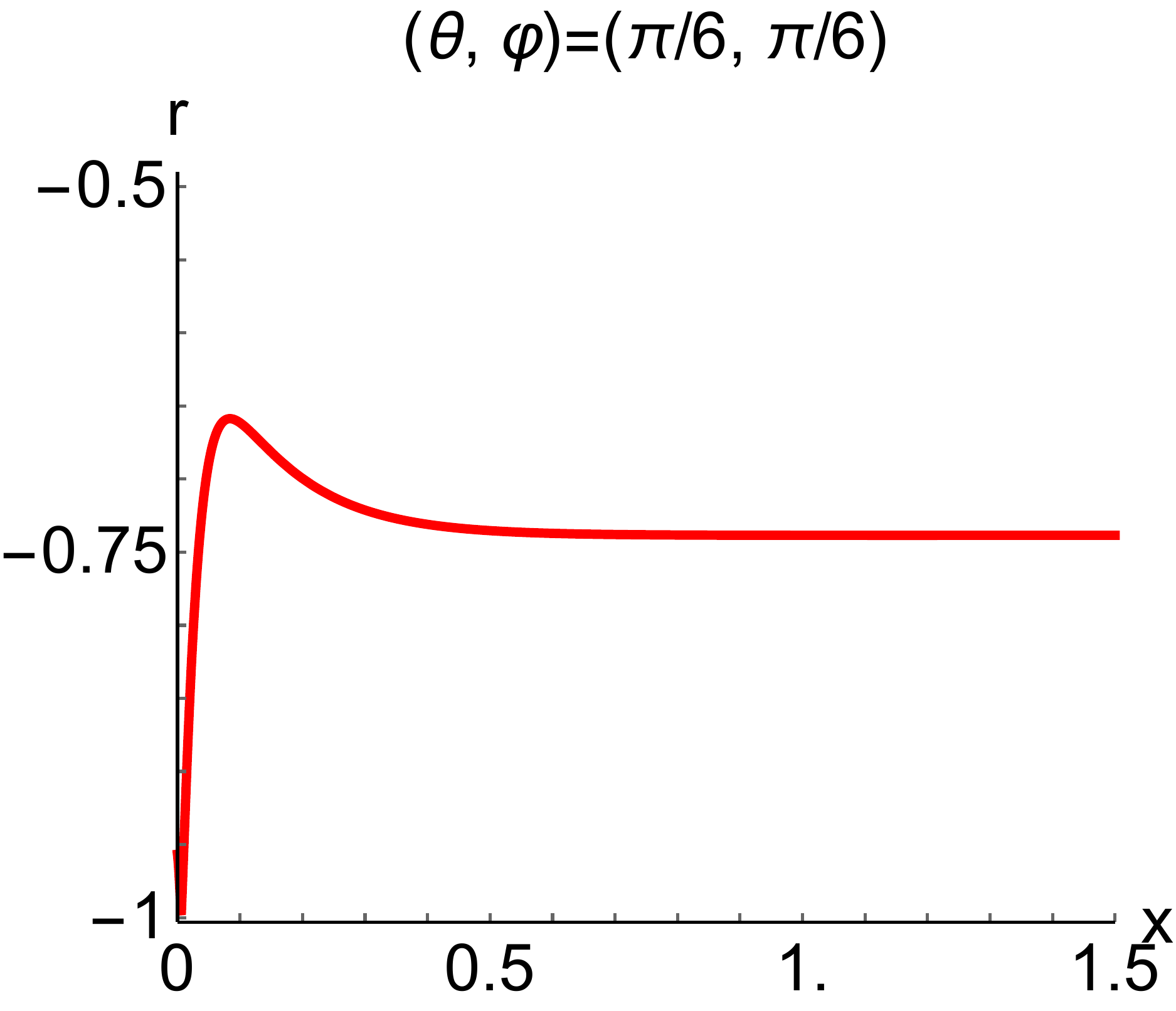}
\hspace{1em} %
\includegraphics[scale=0.18]{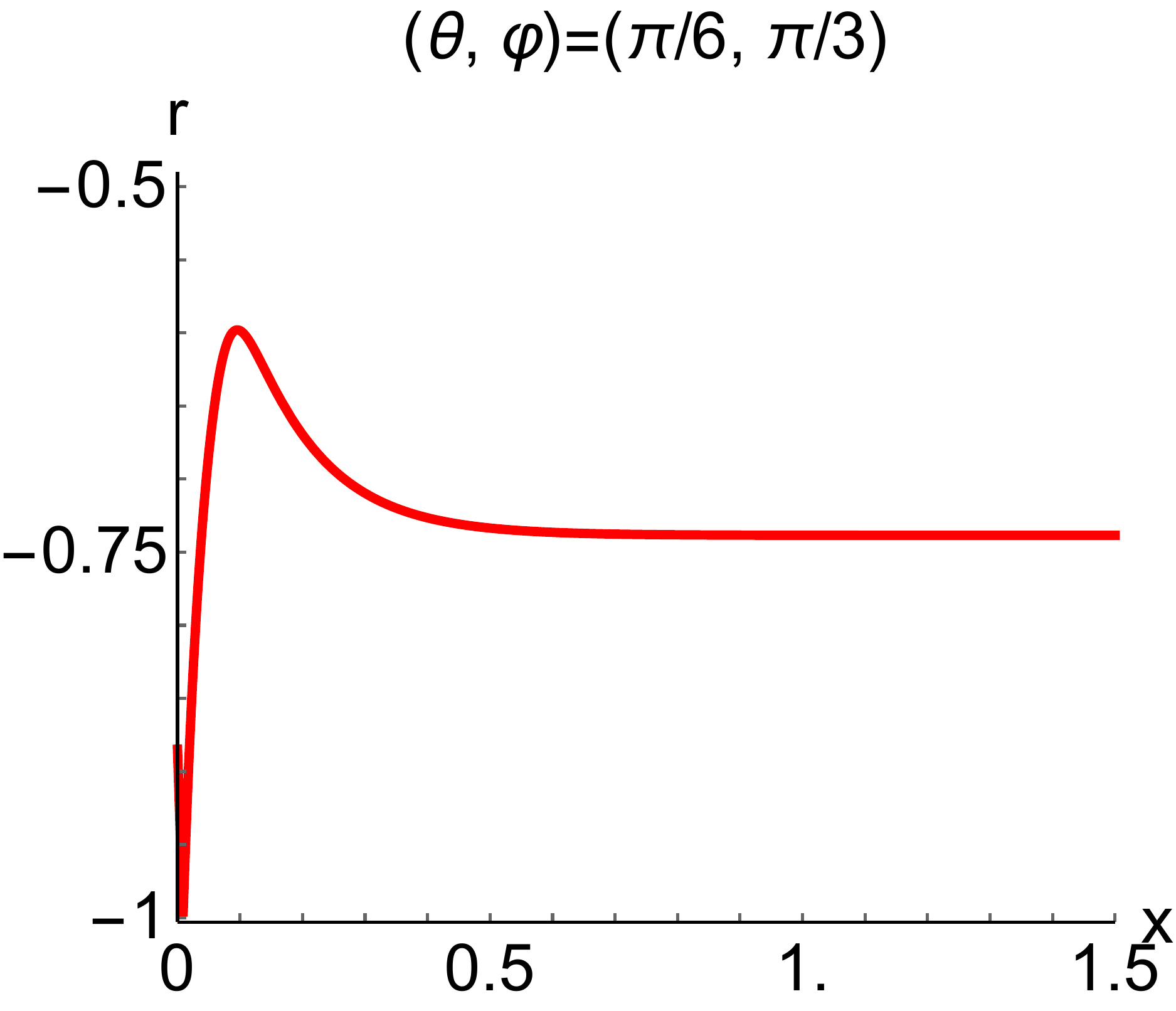}
\hspace{1em} %
\includegraphics[scale=0.18]{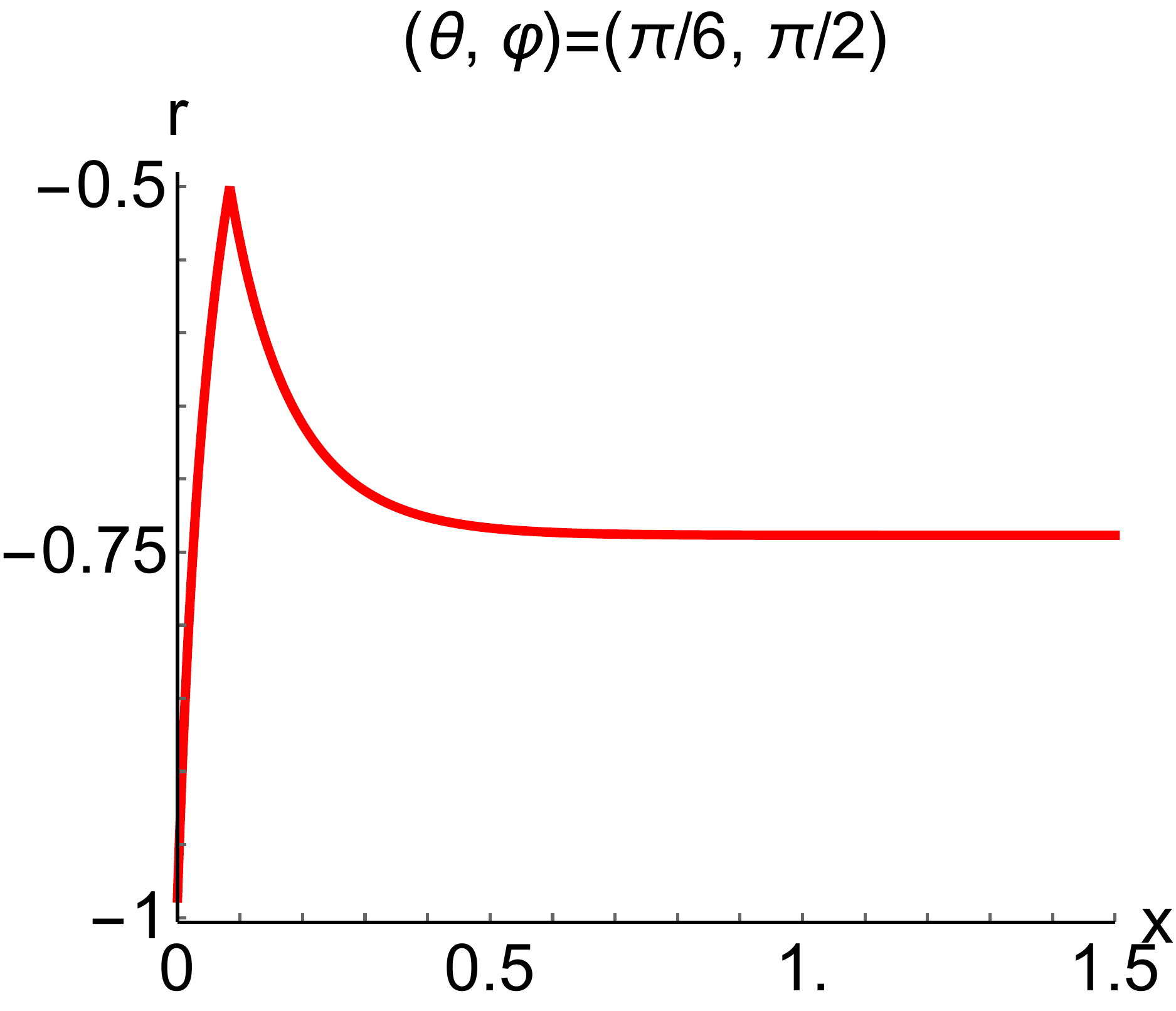}
\\ \vspace{2em} 
\includegraphics[scale=0.18]{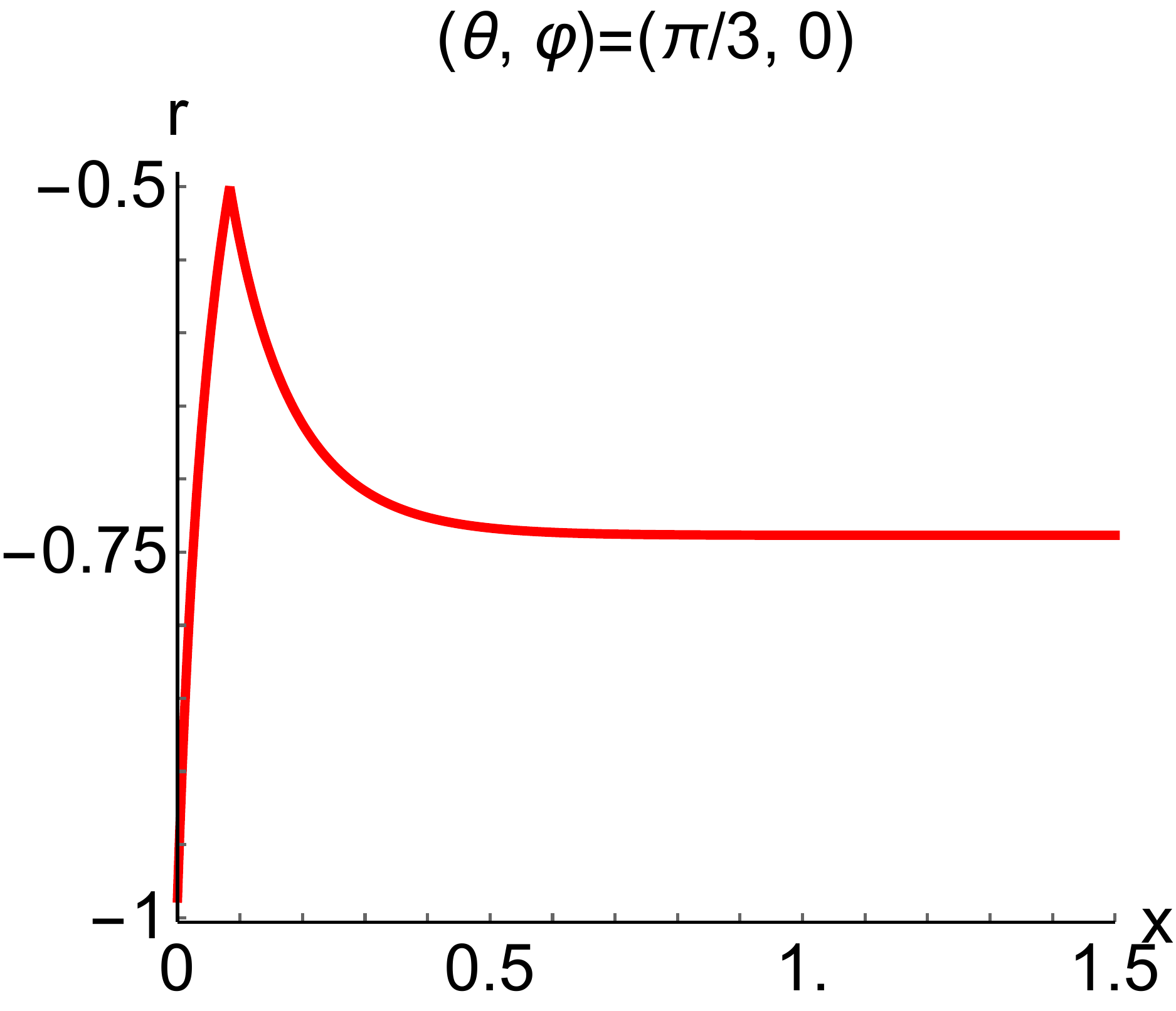}
\hspace{1em} %
\includegraphics[scale=0.18]{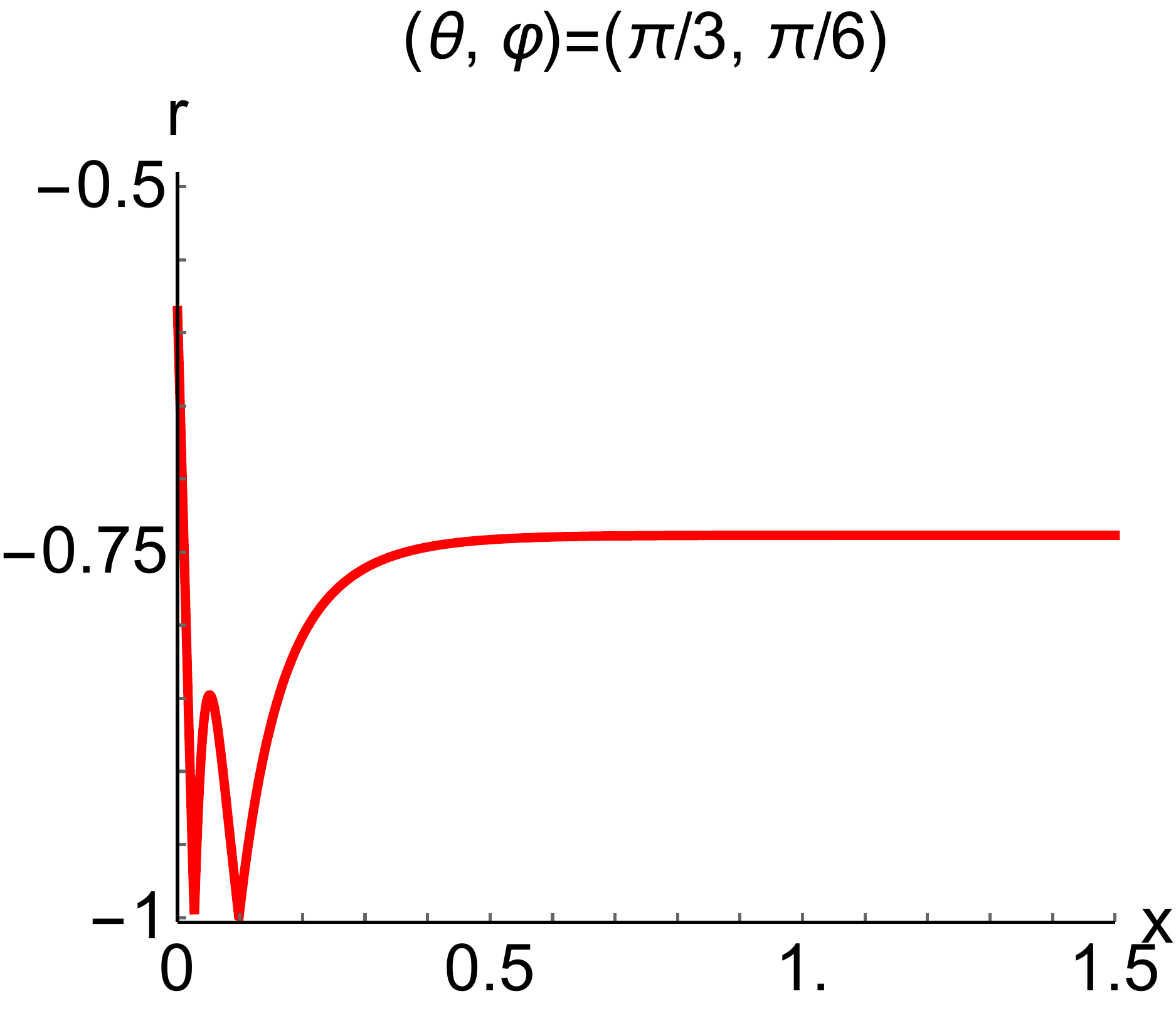}
\hspace{1em} %
\includegraphics[scale=0.18]{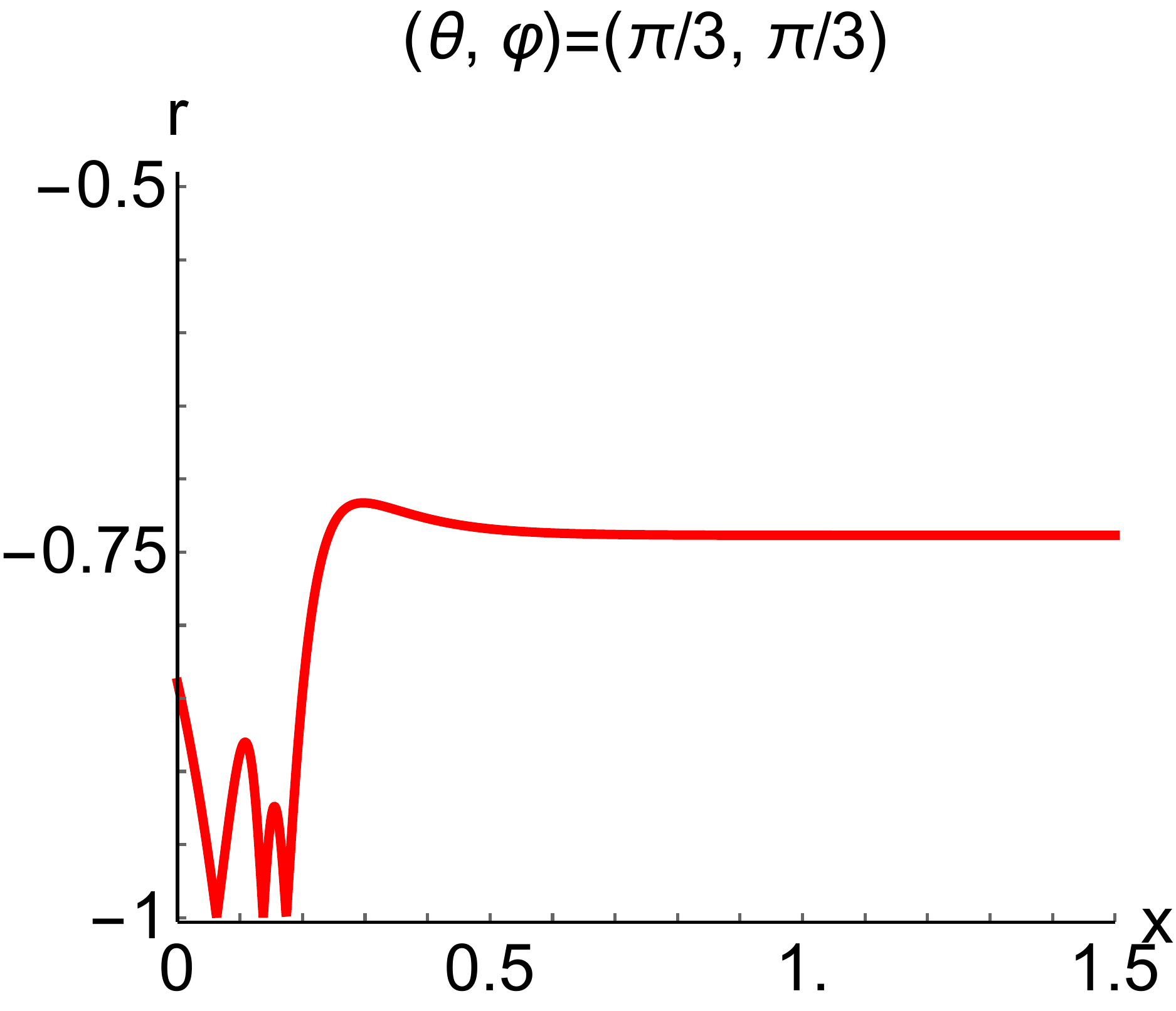}
\hspace{1em} %
\includegraphics[scale=0.18]{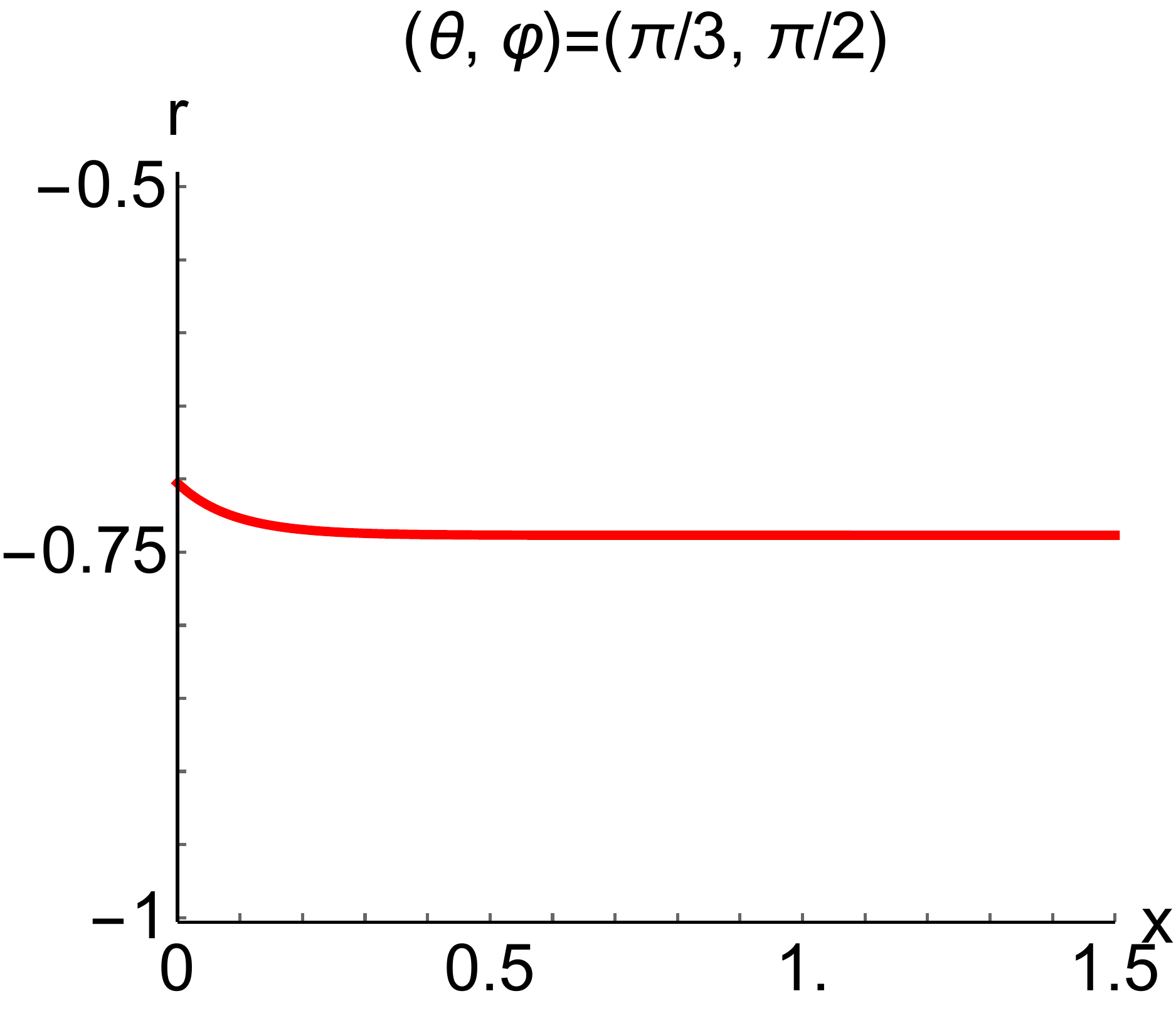}
\\ \vspace{2em} 
\includegraphics[scale=0.18]{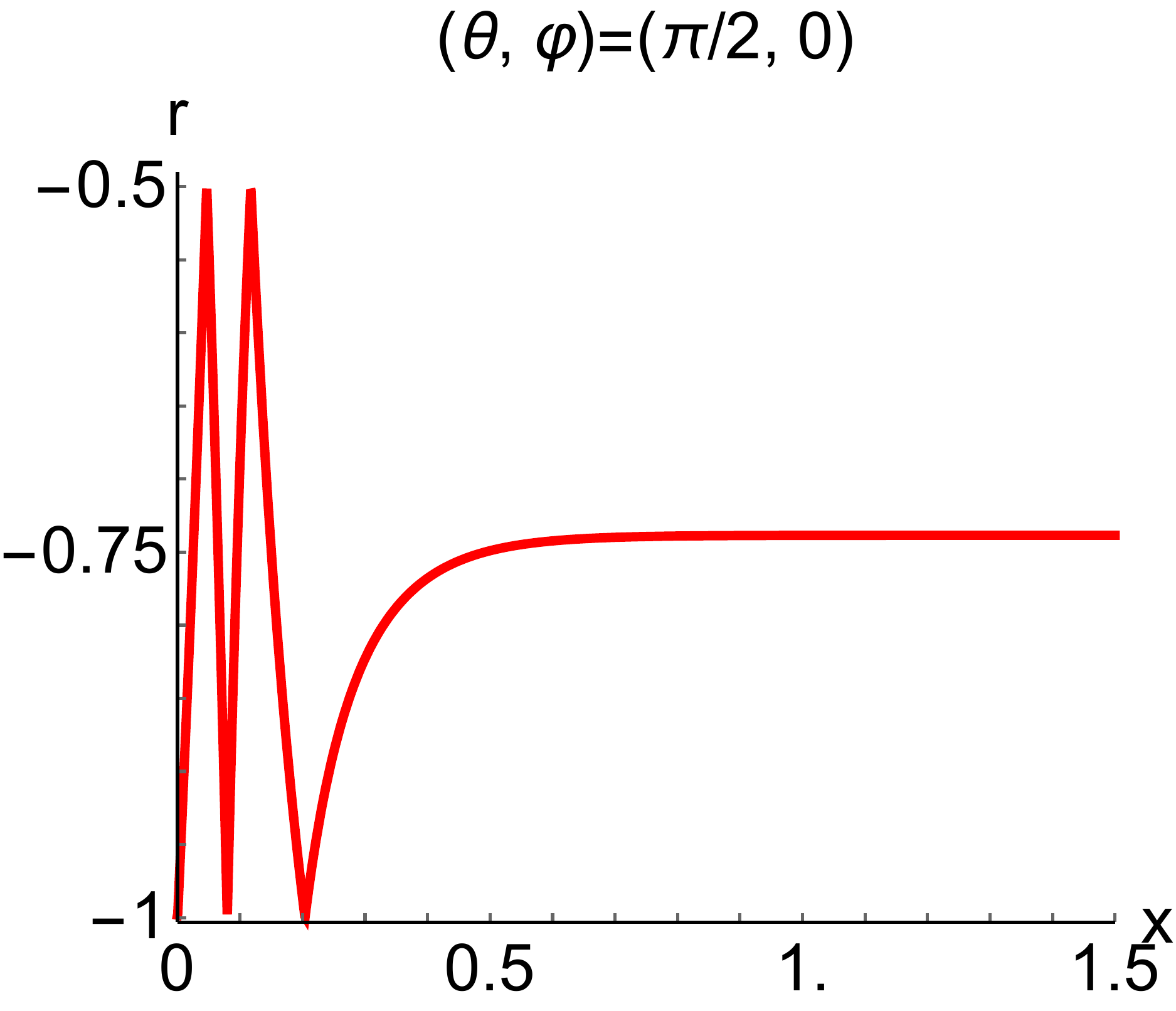}
\hspace{1em} %
\includegraphics[scale=0.18]{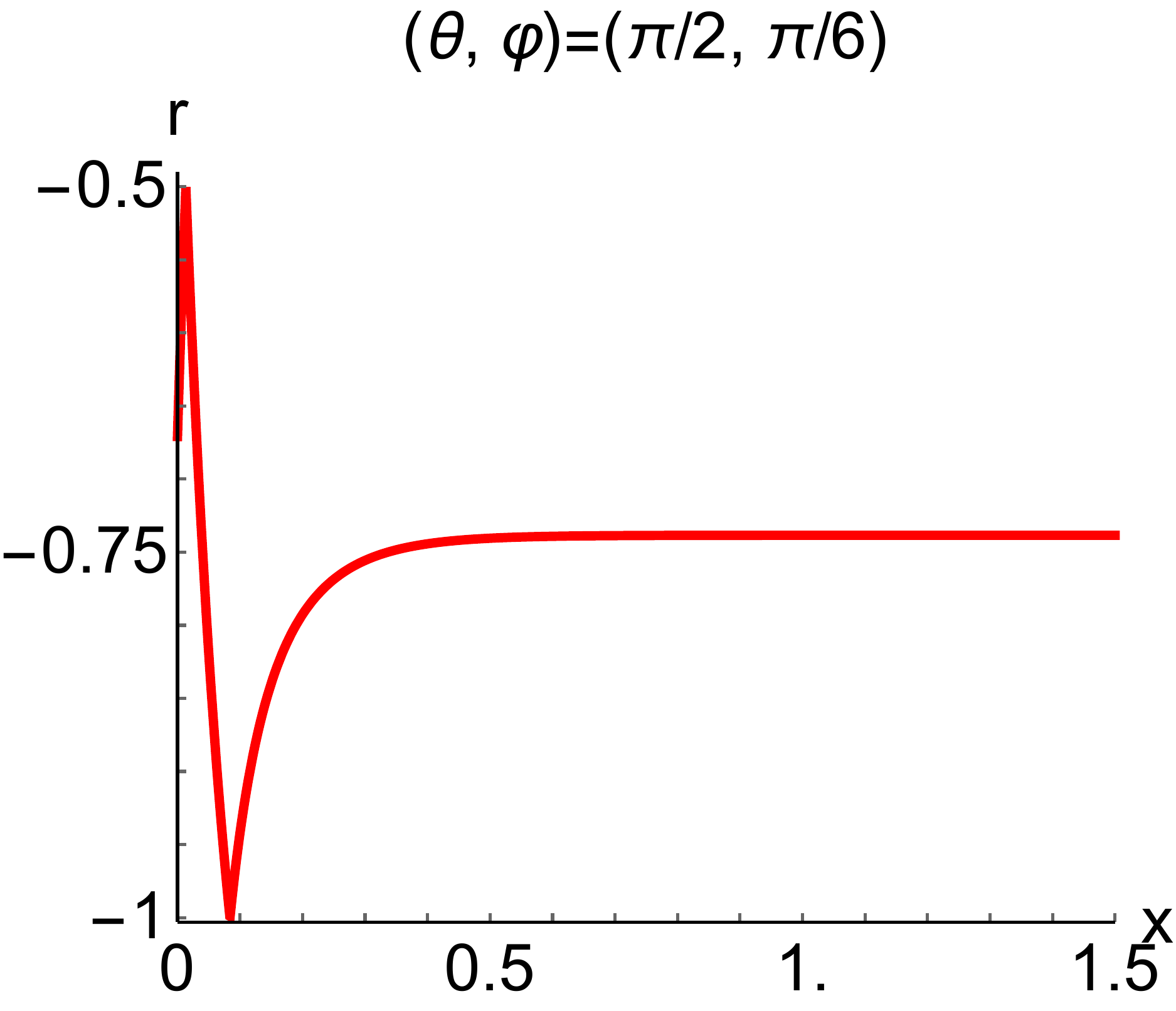}
\hspace{1em} %
\includegraphics[scale=0.18]{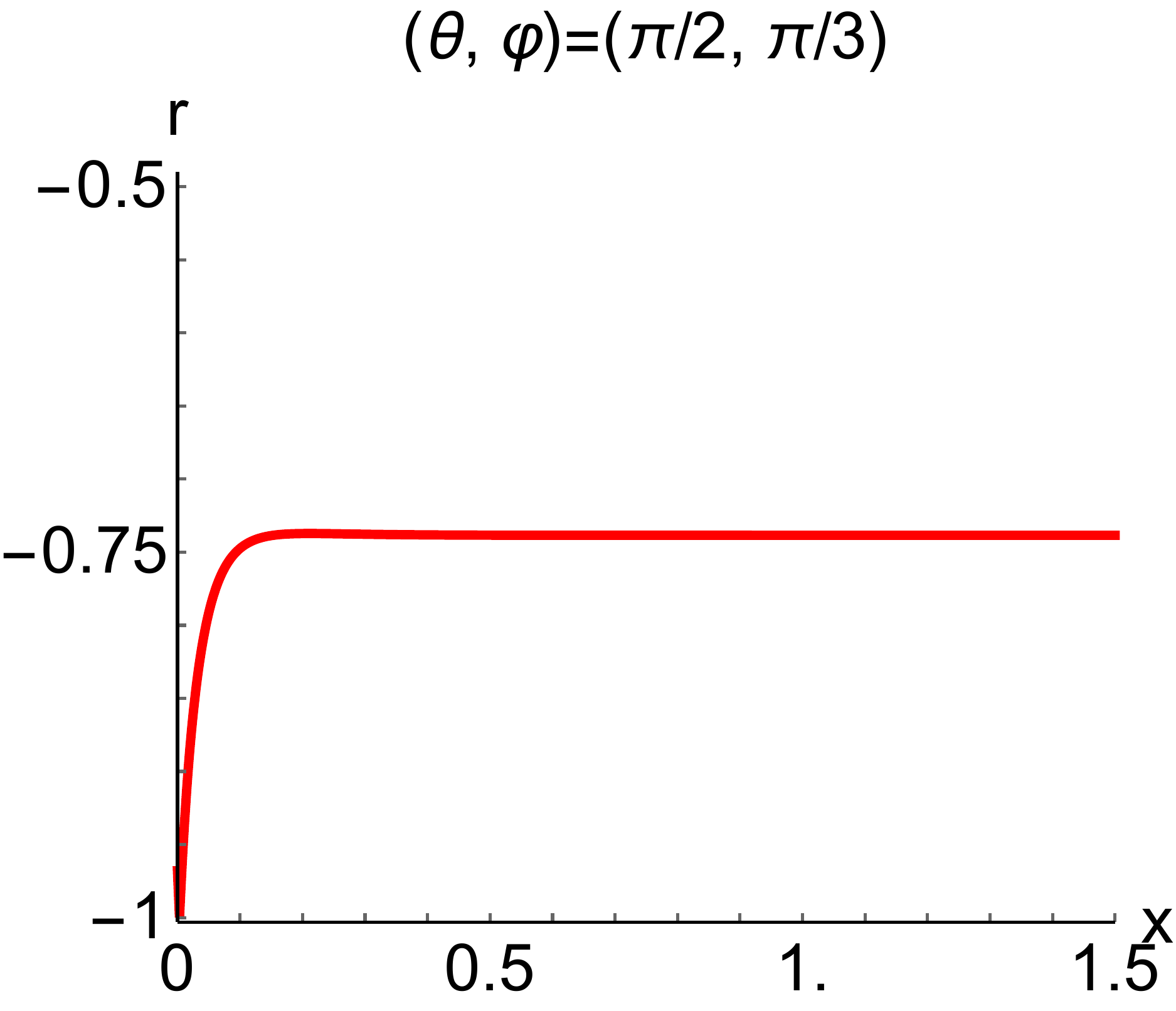}
\hspace{1em} %
\includegraphics[scale=0.18]{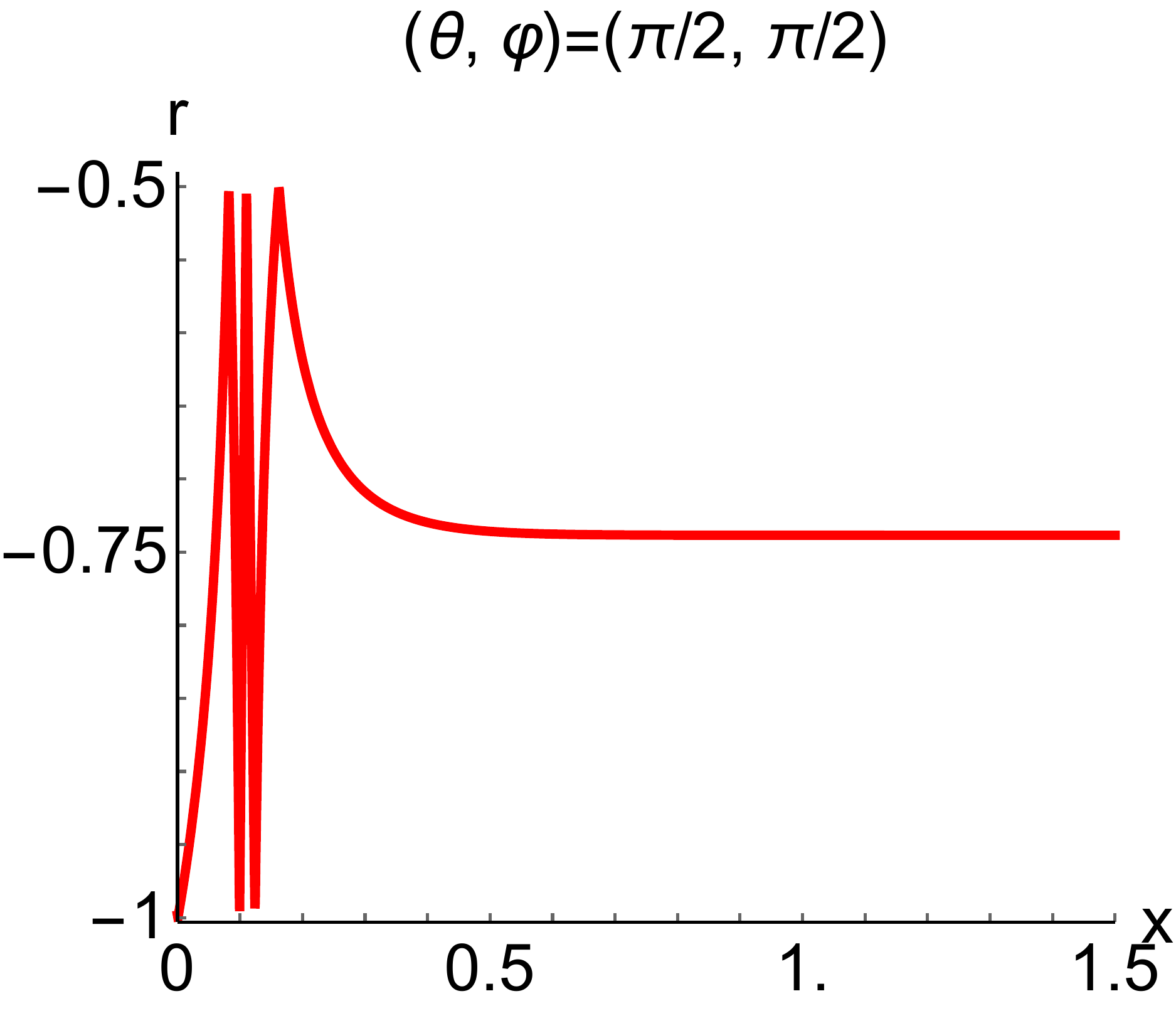}
\caption{The plots of $r(x)$ for the bulk condition (ii) $(f_{1}^{\bulk},f_{2}^{\bulk})=(0.92,0.32)$ at $t=0.9$ and $b=0.15$ (the bulk D$_{2}$-BN phase).}
\label{fig:r_3P2_bulk2}
\end{center}
\end{figure}

\begin{figure}[p] 
\begin{center}
\includegraphics[scale=0.18]{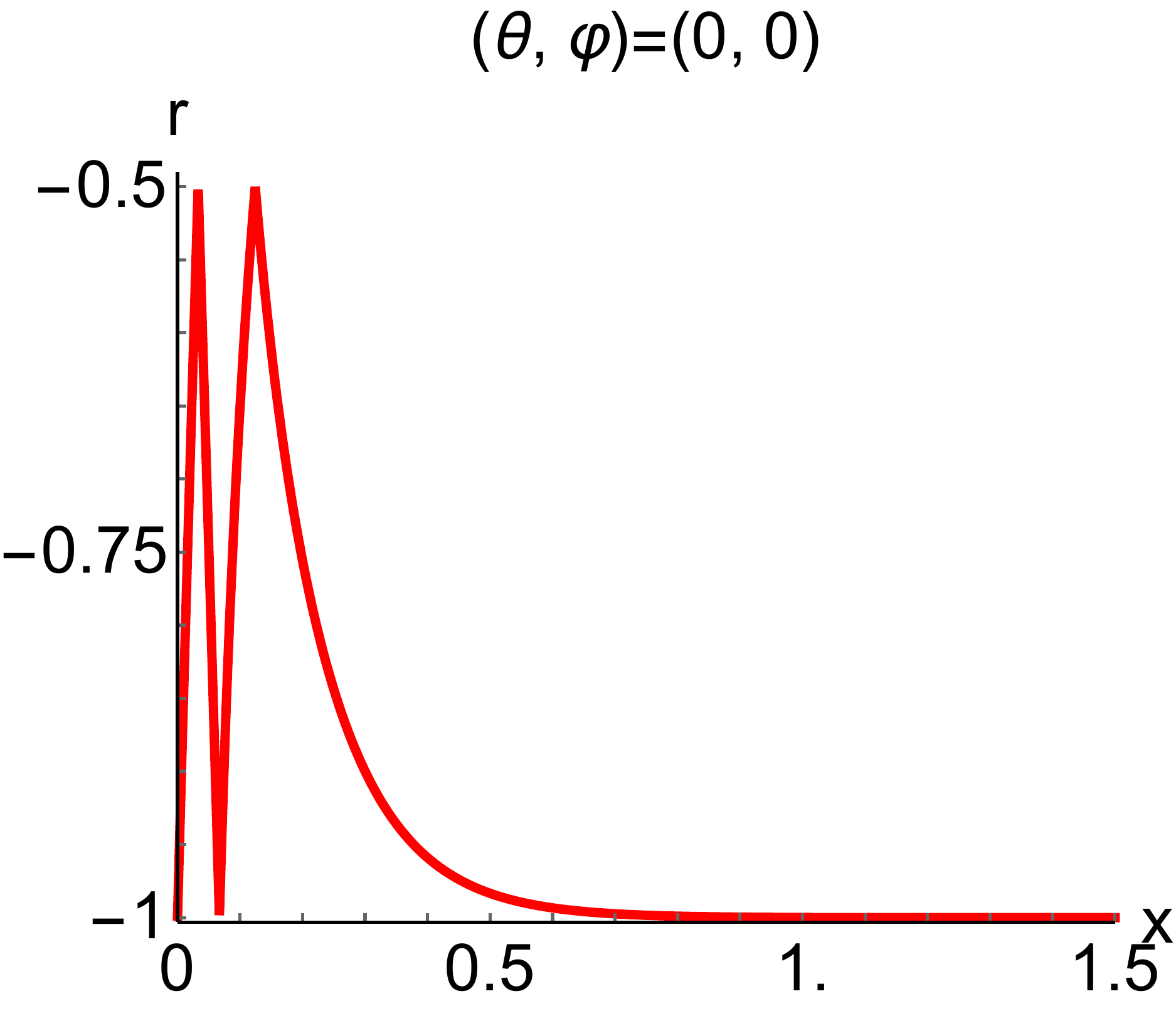}
\\ \vspace{2em} 
\includegraphics[scale=0.18]{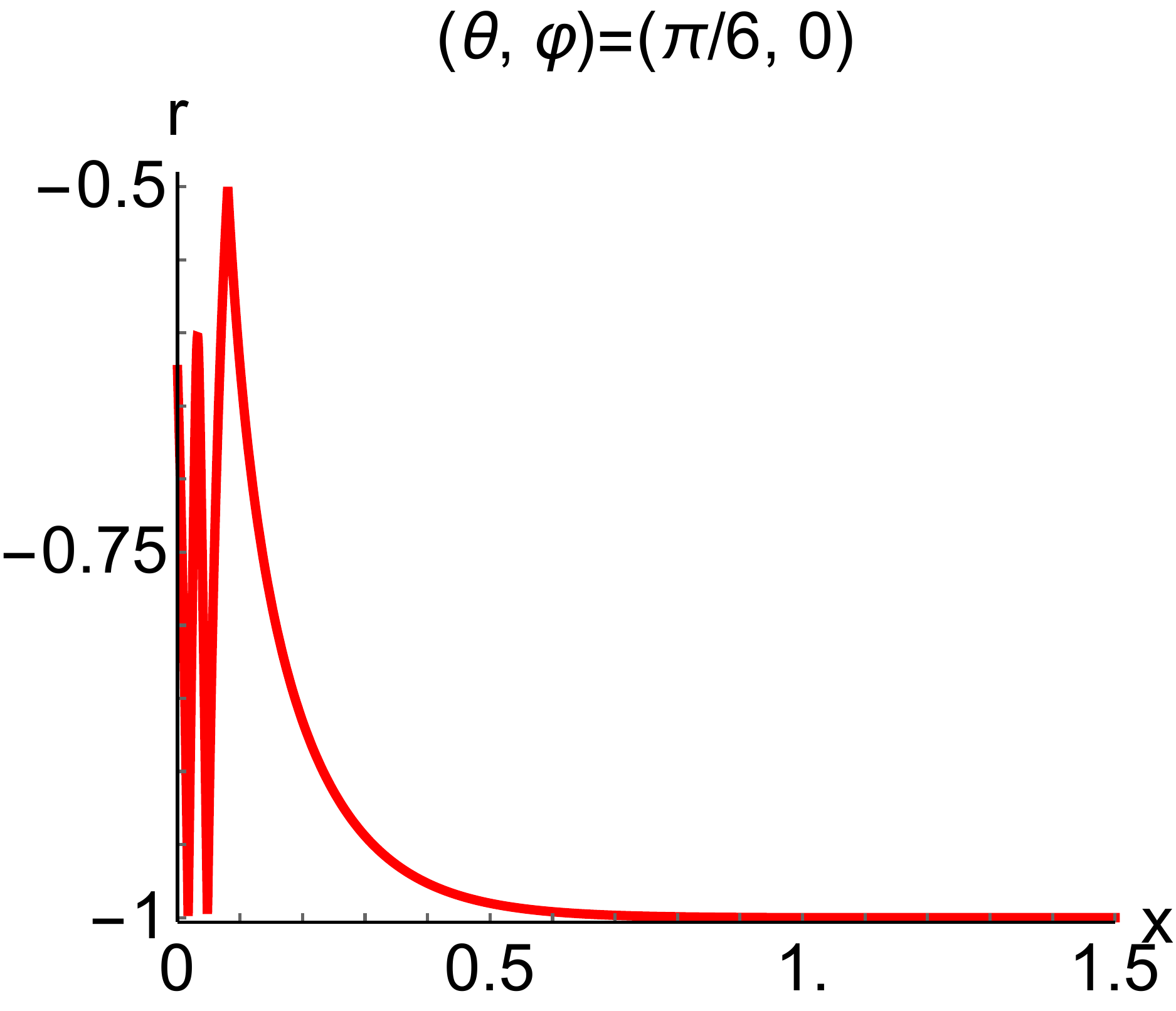}
\hspace{1em} %
\includegraphics[scale=0.18]{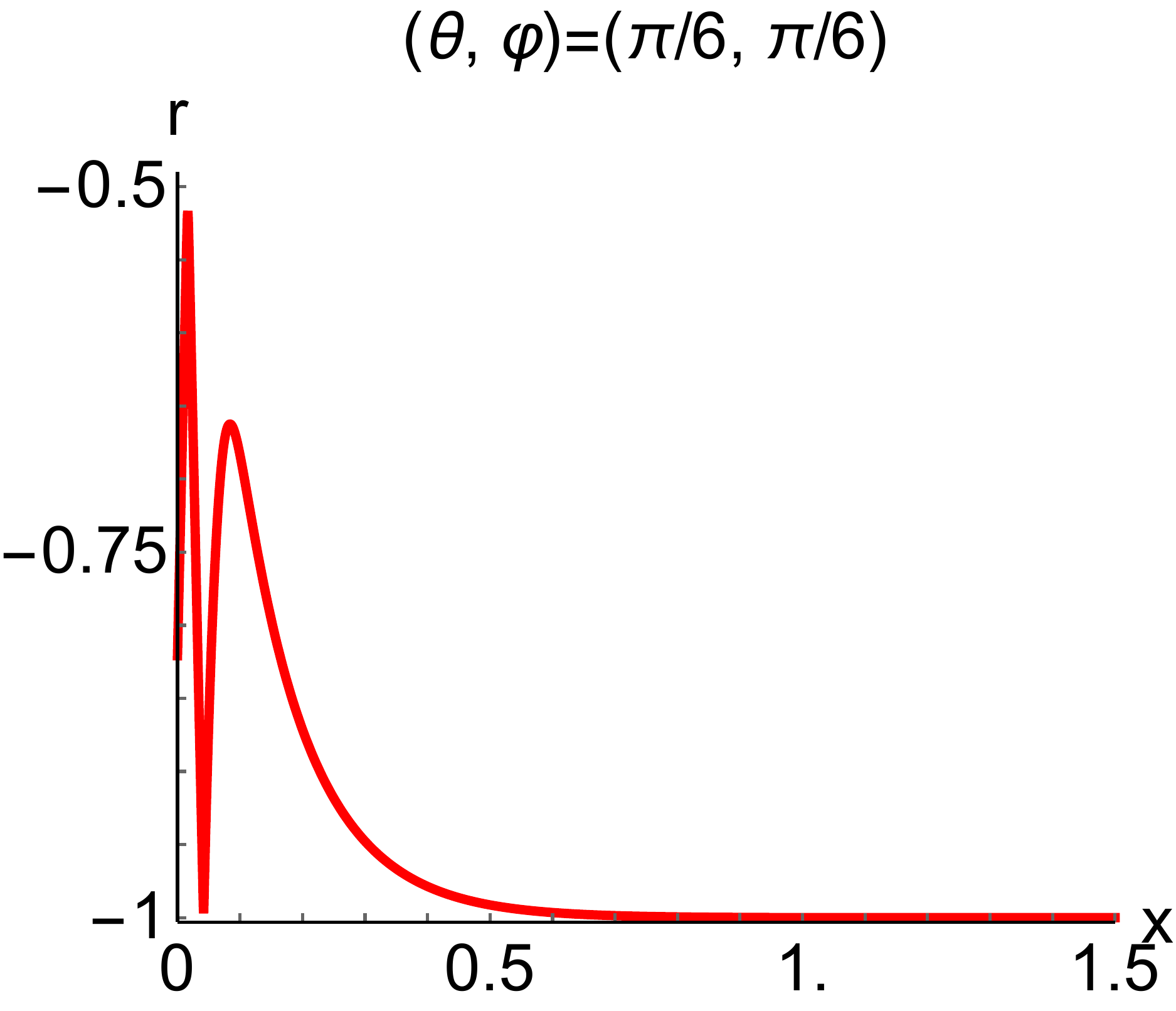}
\hspace{1em} %
\includegraphics[scale=0.18]{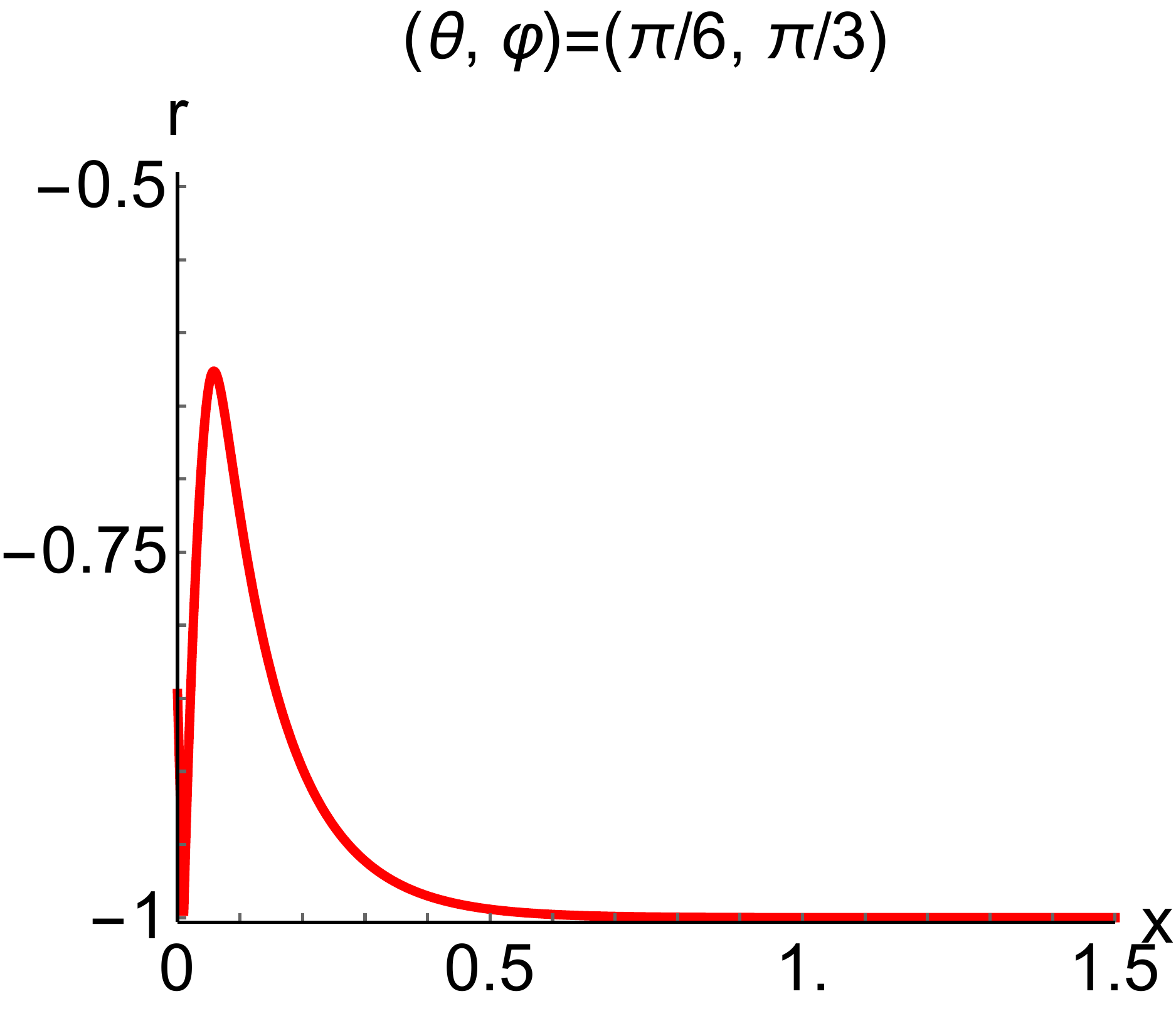}
\hspace{1em} %
\includegraphics[scale=0.18]{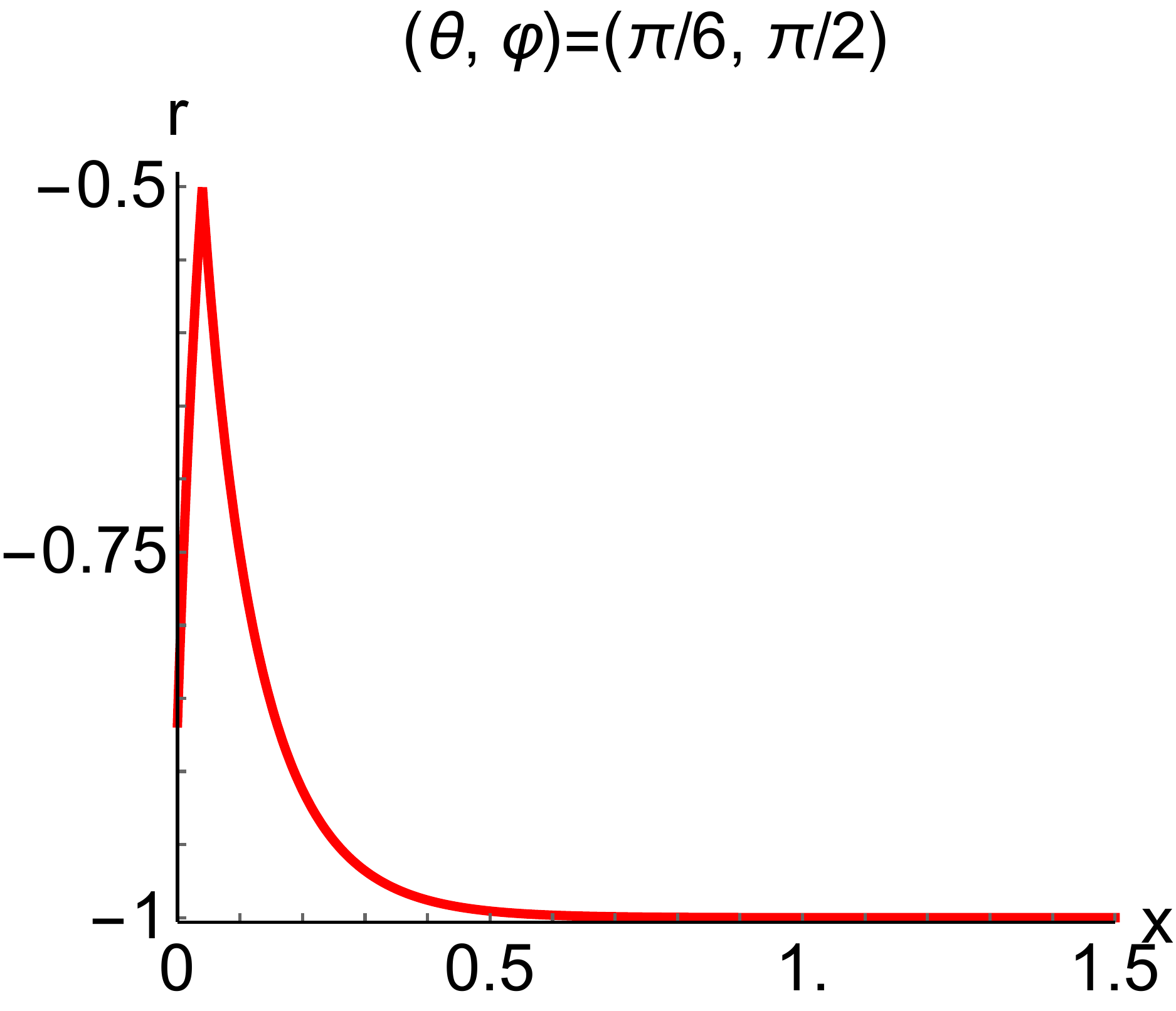}
\\ \vspace{2em} 
\includegraphics[scale=0.18]{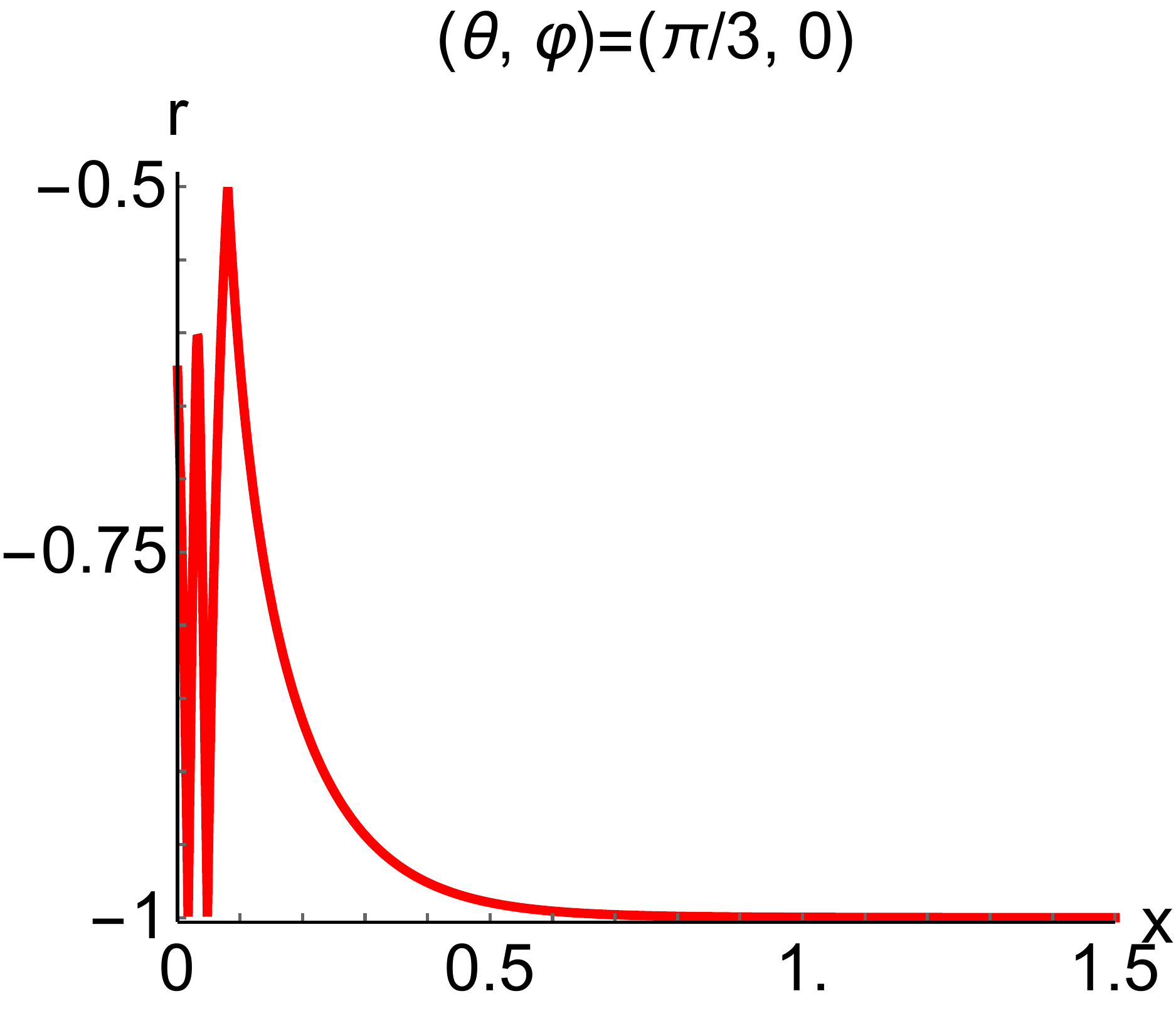}
\hspace{1em} %
\includegraphics[scale=0.18]{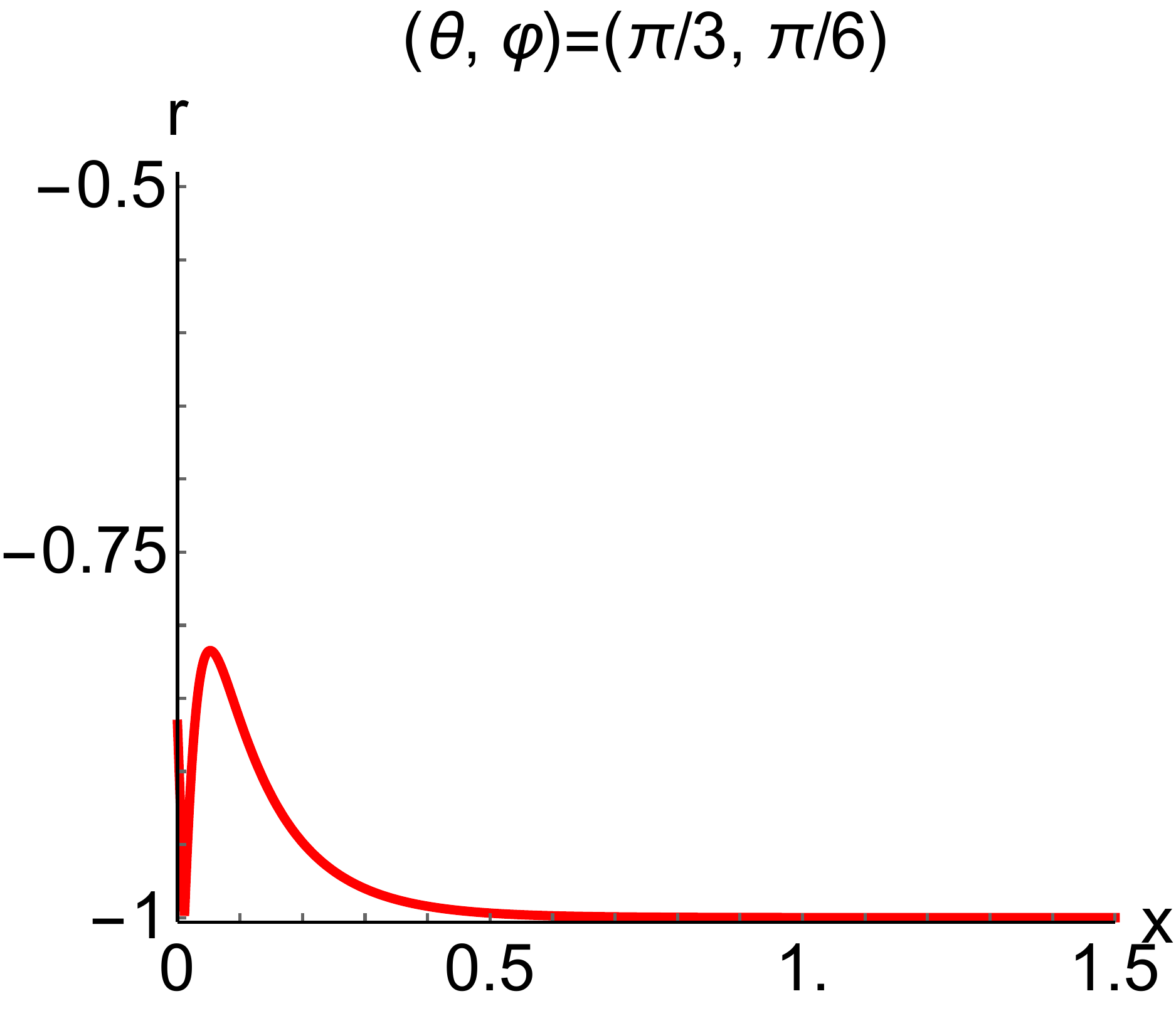}
\hspace{1em} %
\includegraphics[scale=0.18]{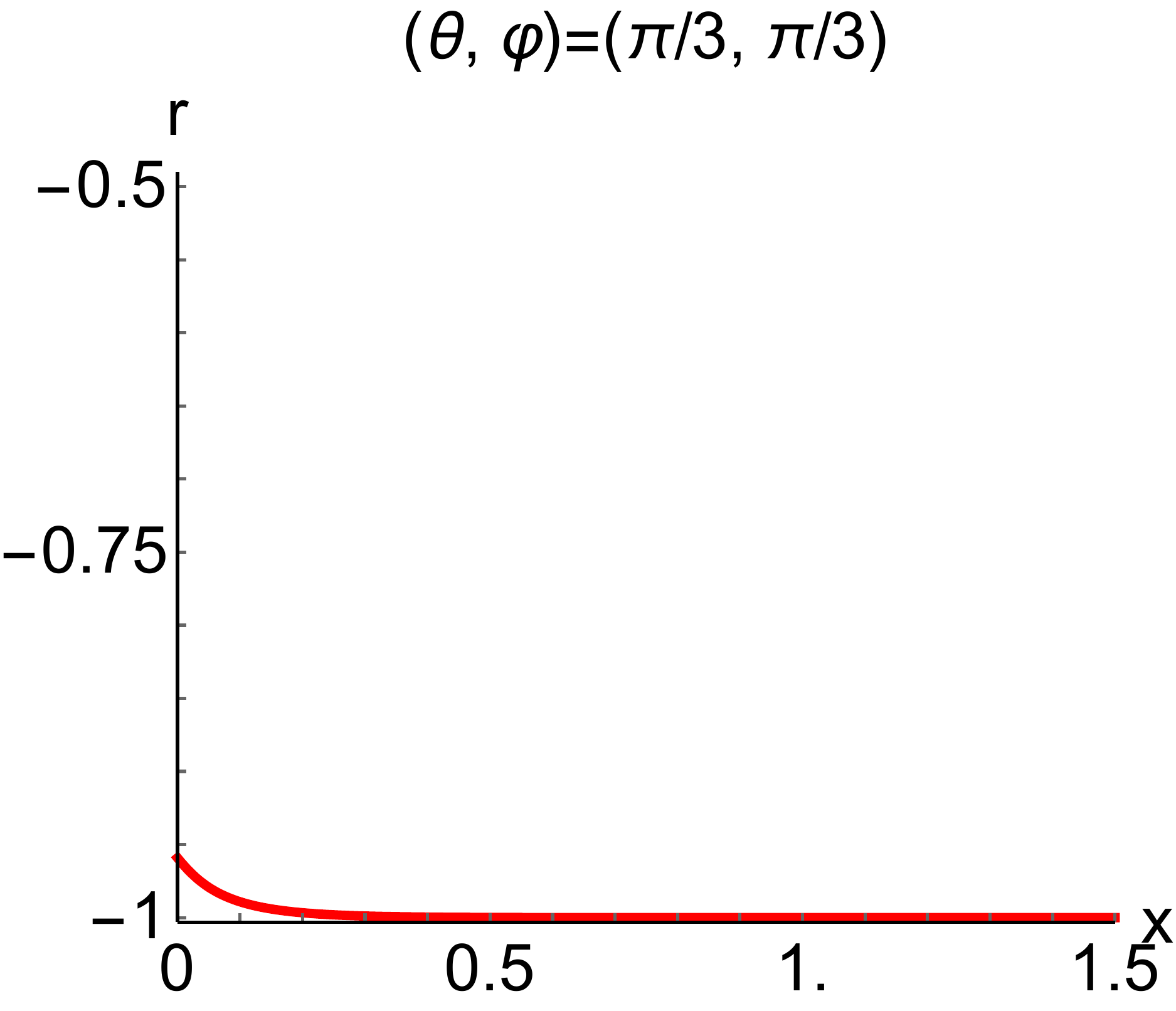}
\hspace{1em} %
\includegraphics[scale=0.18]{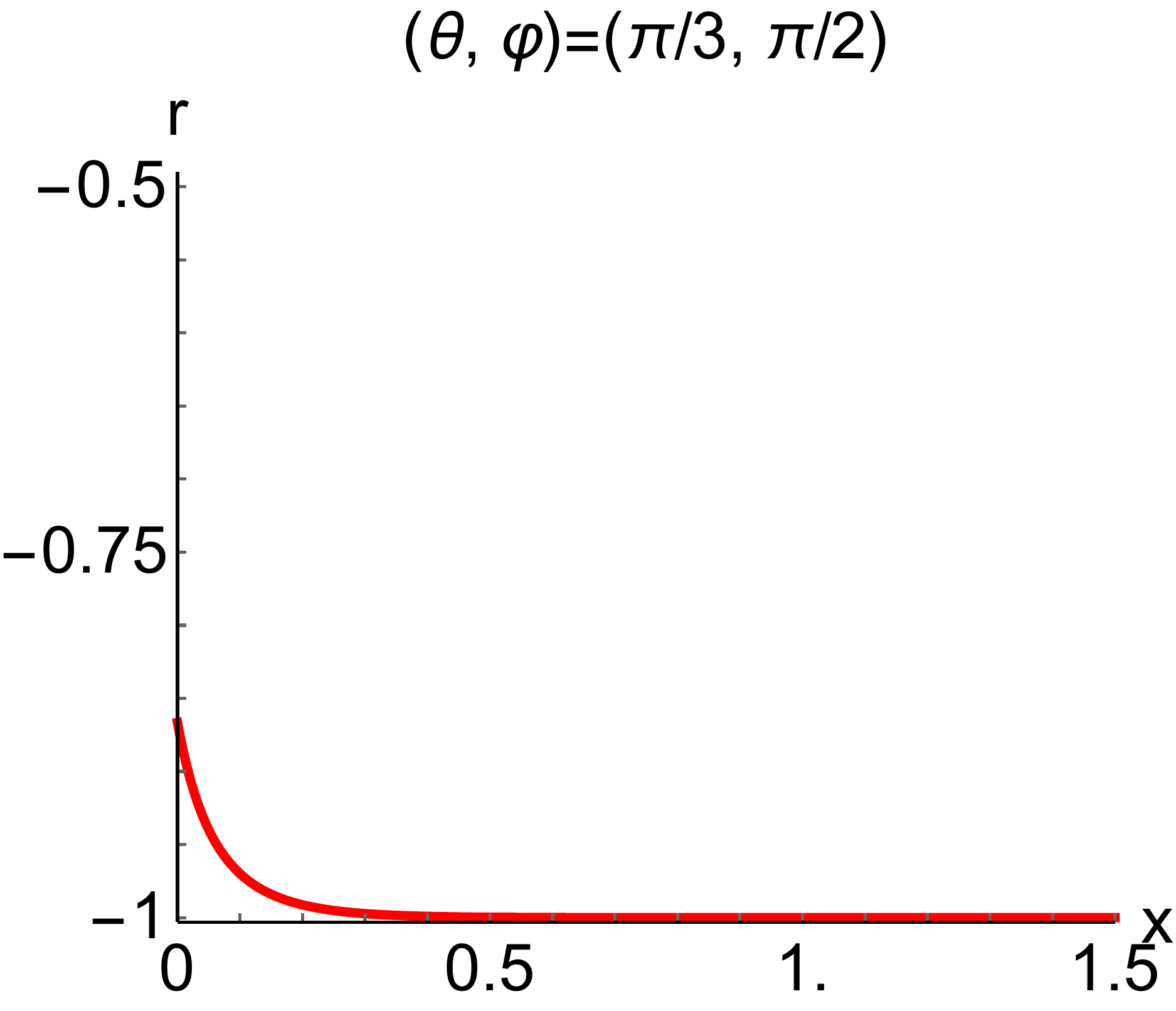}
\\ \vspace{2em} 
\includegraphics[scale=0.18]{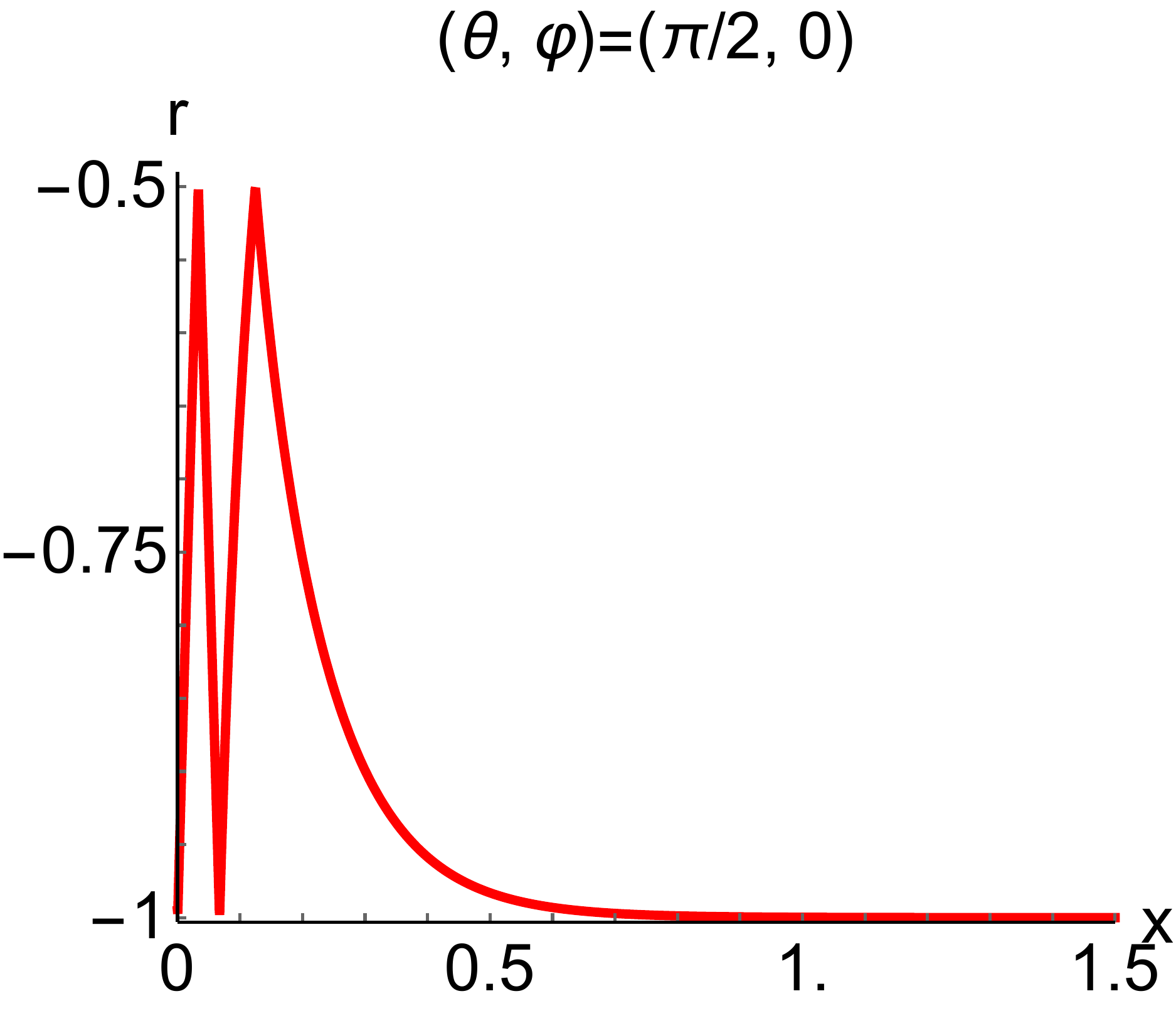}
\hspace{1em} %
\includegraphics[scale=0.18]{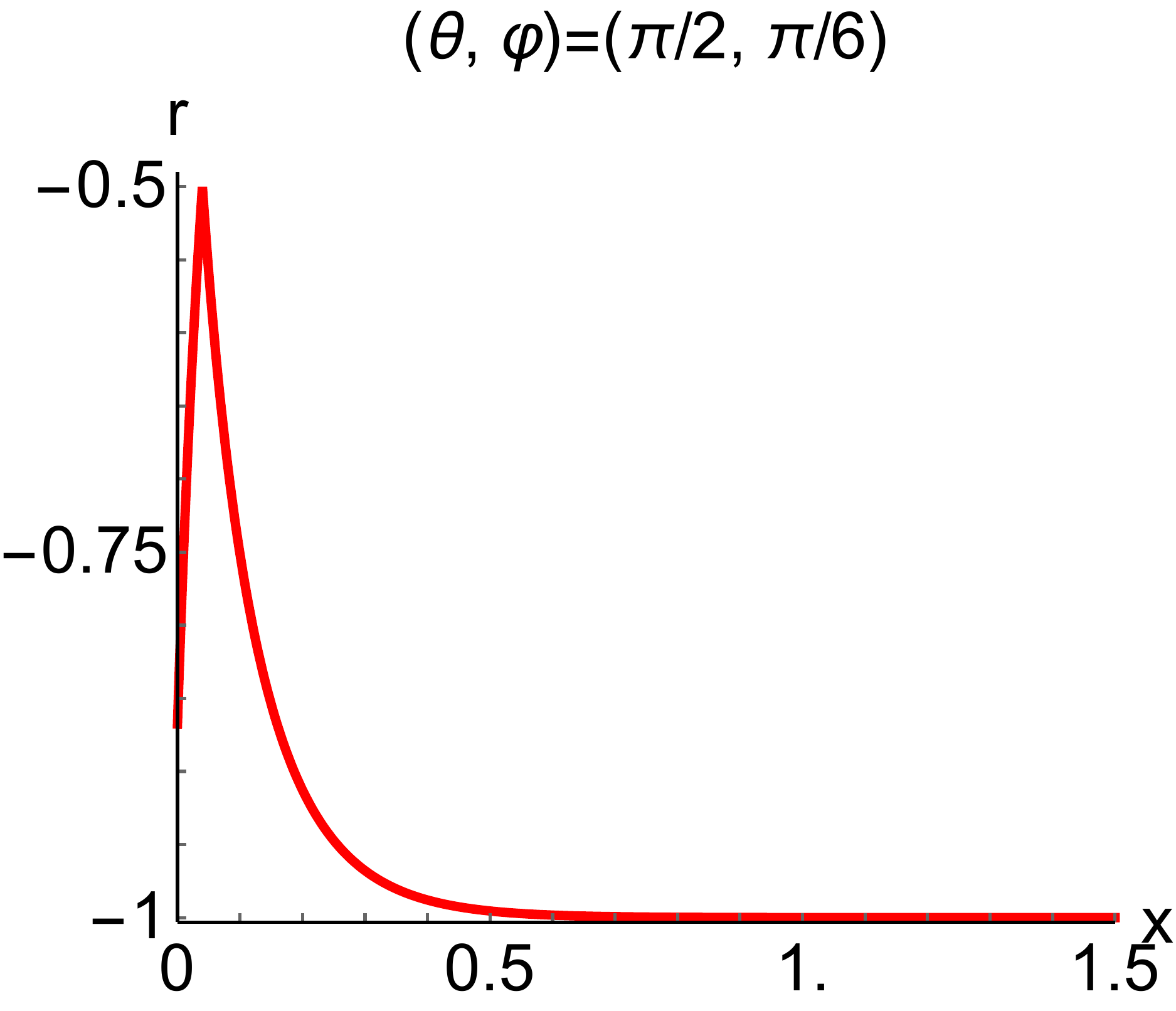}
\hspace{1em} %
\includegraphics[scale=0.18]{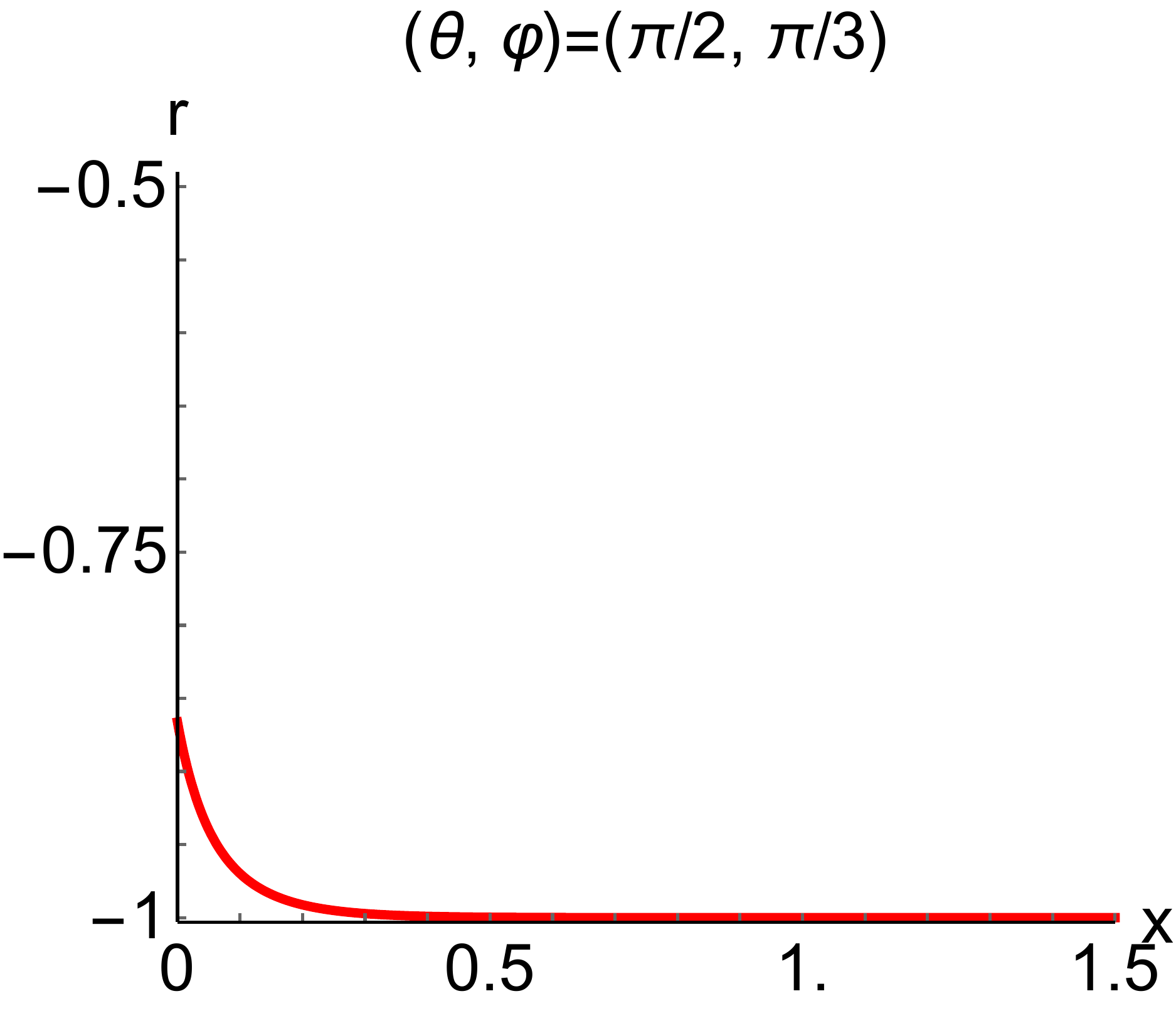}
\hspace{1em} %
\includegraphics[scale=0.18]{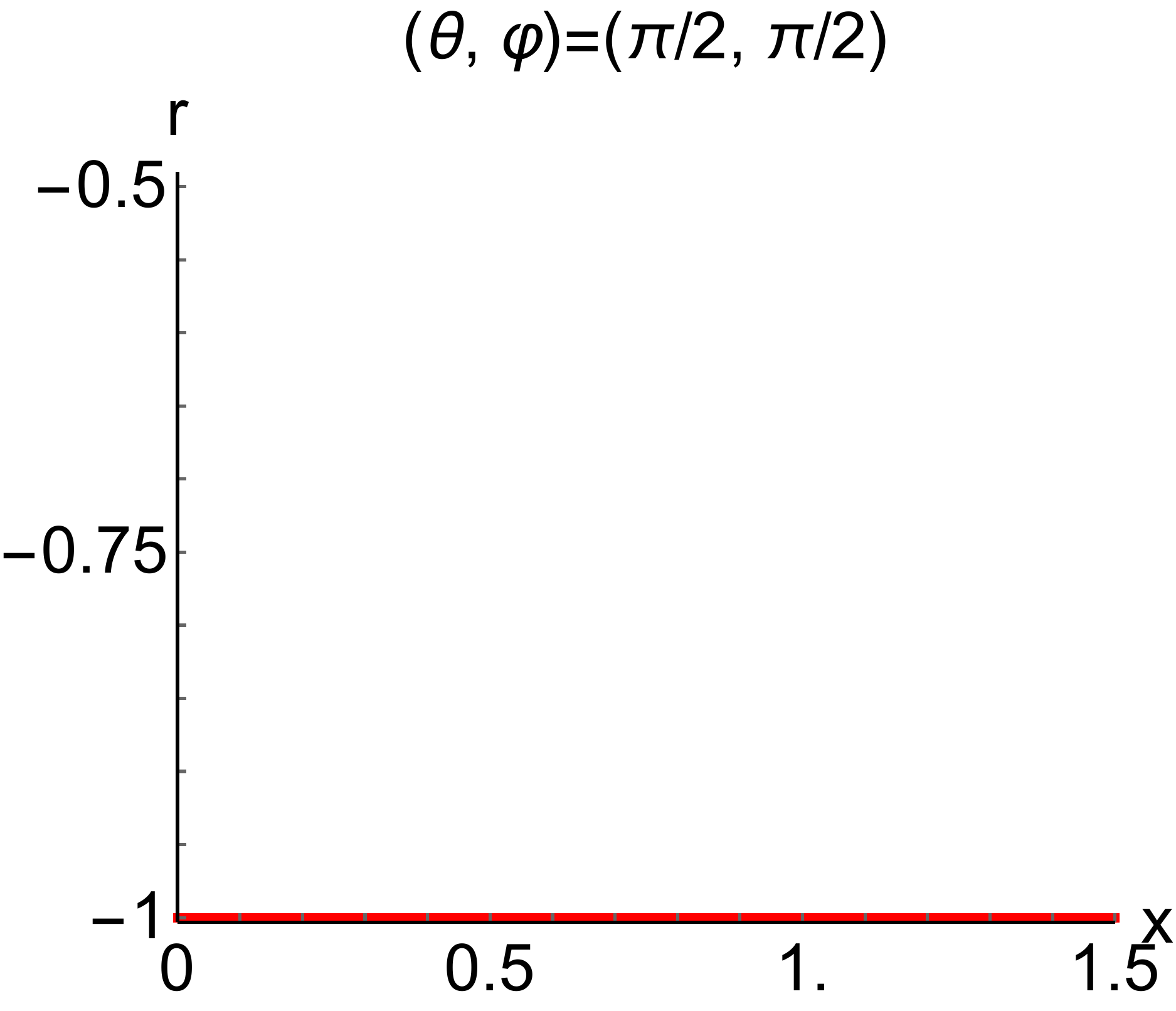}
\caption{The plots of $r(x)$ for the bulk condition (iii) $(f_{1}^{\bulk},f_{2}^{\bulk})=(1.12,0)$ at $t=0.9$ and $b=0.2$ (the bulk D$_{4}$-BN phase).}
\label{fig:r_3P2_bulk3}
\end{center}
\end{figure}

From Sec.~\ref{sec:symmetry_surface}, we summarize the numerical results for $f_{\alpha}(x)$ ($\alpha=1,2,3$), $g_{\beta}(x)$ ($\beta=1,2,3$) in Figs.~\ref{fig:profile_functions_3P2_bulk1}, \ref{fig:profile_functions_3P2_bulk2}, and \ref{fig:profile_functions_3P2_bulk3} and $r(x)$ in Figs.~\ref{fig:r_3P2_bulk1}, \ref{fig:r_3P2_bulk2}, and \ref{fig:r_3P2_bulk3} for the UN, D$_{2}$-BN, and D$_{4}$-BN phases in the bulk.

\bibliography{reference}

\end{document}